To the blessed memory of my parents
who taught me to think.

# World of Movable Objects

## *Preface*

Suppose that you are sitting at the writing table. In front of you there are books, text-books, pens, pencils, pieces of paper with some short notes, and a lot of other things. You need to do something, you have to organize your working place, and for this you will start moving the things around the table. There are no restrictions on moving all these things and there are no regulations on how they must be placed. You can put some items side by side or atop each other, you can put several books in an accurate pile, or you can move the bunch of small items to one side in a single movement and forget about them, because you do not need them at the moment. You do not need any instructions for the process of organizing your working place; you simply start putting the things in such an order, which is the best for you at this particular moment. Later, if something does not satisfy you or if you do not need some items in their places, you will move some of the items around and rearrange everything in whatever order you need without even paying attention to this process. You make your working place comfortable for your work at each moment and according to your wish.

The majority of those who are going to read this text nearly forgot everything about paper, books, hand writing… A personal computer became the only instrument and the working place for millions of people. Screens of our computers are occupied with different programs; the area of each program is filled with many different objects. Whenever you need to do something, you start the needed program and in each of them you know exactly, what you want to do and what you can do. Did it ever come to your attention that in any program you know exactly the whole set of possible movements and actions? Have you ever understood that the set of allowed actions is extremely small and you try to do your work within a strictly limited area of allowed steps? Those limits are the same in all the programs; that is why there are no questions about the fairness of such situation. You can press a button; you can select a line or several lines in a list; you can do several other typical things, but you never go out of the standard actions which you do in any other program. You and everyone else know exactly what they are allowed to do. Other things are neither tried nor discussed. They simply do not exist.

Now try to forget for a minute that for many years you were taught what you could do. Let us say that you know, how to use the mouse (press – move – release), but there are no restrictions on what you are allowed to do with a mouse. Switch off the rules of your behaviour “in programs” according to which all your work was going for years.

Try the new set of rules:

- You can press ANY object and move it to any new location.
- You can press the border of any object and change its size in the same easy way; there are no limitations on such resizing (except some obvious and natural).
- A lot of objects you can reconfigure in the same simple way by moving one side or another.

It must be obvious to you that the reaction of a program on your clicking a button does not depend on the screen position of this button, so if you change the size or location of a button or a list it is not going to change any code that is linked with clicking a button, selecting a line in the list or making a choice between several positions. All the programs are still going to work according to their purposes. Buttons and lists represent a tiny part of objects that occupy the screens. Now make one more step and imagine the programs, in which ALL the objects are under your control in exactly the same way. You can do whatever you want with the objects, while the programs continue to work according to their purposes.

What would you say about such a world? Do not be in a hurry with your answer. You never worked with such programs; you would better try before giving an answer. This book is about the screen world of movable and resizable objects. Those objects can be of very different shapes, behaviour, and origin; that is why there are more than 80 examples in the accompanying program. Some of those examples are simple; others are in reality complex and big enough applications by themselves. And there is <u>not a single unmovable object in all of them</u>.

The world of movable objects – the world of *user-driven applications*.

# Contents

# User-driven applications

This is the first chapter in the second part of the book. All the previous chapters were about HOW to turn different objects into movable / resizable. Further on I will write more about turning into movable / resizable some of the most complex objects used for scientific, engineering, and financial plotting, but beginning from this place and especially in this chapter I am going to write much more about WHAT FOR all the objects have to be turned into movable.

Let us quickly remember, what was discussed in the first part of the book.

- The idea and an algorithm of turning any screen object into movable / resizable was introduced.
- The method of turning objects into movable was demonstrated on a variety of objects starting from the simplest and then dealing with more and more complex: rectangles, polygons, circles, rings, different figures with holes…
- Movement restrictions were considered.
- After graphical objects, the same request for movability was applied to controls.
- From solitary elements to the groups. The wide variety of groups with different behaviour was developed.

I think that everyone would agree that this required a lot of work to be done. Some of the examples look interesting and can be implemented in applications just in the way they are demonstrated. Others can be used in one or another application after some minor adjustments. But do you really think that turning one or another object of the current applications into movable was the real goal of all this work?

I never tried to hide the fact that I started my work on the movability of objects not as an attempt to invent the general algorithm for all kinds of objects. I will write about it in more details further on, because these details are very important for understanding of the new applications, but the triggering thing of this work was the demand for movable / resizable plots for scientific programs. I work on design of scientific and engineering applications throughout all my professional life. It became obvious to me several years ago that:

- There is a long period of stagnation in design of scientific / engineering applications.
- The only chance to end this stagnation and move on is to use absolutely different main idea of design.
- The control over applications has to be passed to the users.
- The only way to do this is to turn all the plots into movable / resizable.

I had to invent an algorithm and I had to check it on some simpler models, like rectangles, but the main purpose was the movability of the scientific plots. When the movability of the plots became good enough to be used in real programs and some time after it, while I was constantly improving the algorithm, I was thinking more and more that that idea of movability could not be bound only by the original area. Applying of the designed technique to other objects and the spread of the idea of movability on all the objects turned out to be much more important than the algorithm itself.[*]

From my point of view, the consequences of designing the applications on the basis of movable / resizable objects are much more important than the real algorithm, which is used. You can use my algorithm or you can use any other. If you use any algorithm invented by yourself or by somebody else, if you start to design applications on the basis of movable / resizable objects, you will eventually come to the same ideas and principles, which constitute what I call the **user-driven applications**.

Applications, based on movable / resizable elements, differ so much from what we got used to for many years that for many people it is difficult to accept them at first. This is the nature of all the human beings: if we see the new detail, which differs not more than for 10-15 percent from the usual things, we can estimate it without any bias. But if we are introduced to something that is nearly 100 percent different from the familiar things, then we have big problems not only in understanding it. We reject such things even without trying. It is not the prejudice about one or another idea; such reaction against the absolutely new thing (technique, idea, device) is deep inside the human psychology.

---

[*] The historical parallel of the consequences, which are much more important than the algorithm itself. Newton and Leibniz worked on the invention of calculus independently. It looks like Leibniz had a slightly better approach, but Newton's claims for priority sounds much louder, because he demonstrated excellent samples of using the new instrument. I am not saying a word about the priority of one or another. The most important thing is the widespread use of calculus; it is much more important than the identification of stones in the base of this mathematical instrument.

I have a clear understanding of this psychological problem in introducing the world of new programs. A lot of programmers have looked at the new applications, trying to understand how this new programs work and how these ideas can be used. I have also met other specialist, who simply refused to spend a minute of their extremely valuable time to look at something that they never saw. It was a bit funny to watch such a reaction, but nobody can be forced into looking at the new type of programs.

I think that you are aware, how difficult it was for people to accept the idea that not the Sun was going around the Earth, but vise worth. Experience and each day observation show to everyone that the Sun is going around the Earth; it is an AXIOM, and axioms are usually not discussed or doubted; they are simply accepted without any second thought.

In programming we have such type of axiom, which was never doubted since the beginning of the programming era: <u>any application is used according to the developer's understanding of the task and the scenario that was coded at the stage of the design</u>. After reading this, you will definitely announce a statement: "Certainly. How else can it be?" Well, the Sun was going around the Earth for centuries. There were no doubts about it. Yet, it turned out to be not correct.

All programs of standard design work strictly according to their developers' scenario. (My programs, including the one, which accompanies this book, do not obey that axiom, but the amount of such programs is still negligible.) When you develop any program for your personal use, then the design under the underlined axiom is not a problem: if you need something different from the program, you change it. The problem starts, when the number of users starts growing. People have different ideas and preferences; they have different demands, especially to interface.

Computers were invented as pure calculators. Powerful, then more powerful and even more, but for many years they were only pure calculators and nothing else. When the main and the only goal of a program is to give out the results of some calculations, then the only important thing is the correctness of results. The results depend on the initial equations and their coding, so there is nothing wrong if such program works strictly according to the developers' ideas. As this was the only type of programs for decades, then the main rule of their design became an axiom.

The invention of graphical display significantly improved the process of data and results analysis (people are much better in understanding the graph than the column of numbers), but did not change the main idea of the programs. Then the era of new programs started; the problems began to grow like an avalanche. Even in the specialized scientific / engineering programs the part of calculations was shrinking very quickly. The visualization became the dominant part even in such applications; there are also a lot of programs without any calculations at all, but with the requirements for a wide variety of data and results presentation and for different types of man – machine interfaces. The questions of interface became the most important in design of programs. When an application is used by hundreds, thousands or millions of users, then it is impossible to imagine that all of them would agree with the proposed view and decisions imposed by designers. Yet the new programs are still developed on the basis of the same axiom, which is absolutely inappropriate for them.

The interface of our programs has hardly changed throughout the last 15 – 18 years. The most popular programming languages change, the environment for the programmers' work is absolutely different, but the results look the same. Only by the view of some controls and several other tiny details the specialist can distinguish the 1994 application from the 2009 program. But these changes are simply seasonal adjustments and nothing else. Why are there no real changes for so many years? I think that to uncover the roots of the problem you have to look 20 years back, because the programming philosophy, under which the interface design continued to change throughout all these years, was incorporated at that time.. Somewhere 20 years ago in one of the hot discussions on the interface design, its future changes and the vector of development, I heard such a declaration of the most popular view: "The clients will have to like whatever we will give them". Rude expression? Maybe, but only from the point of political correctness. If you skip from all the manuals the standard phrases about the priority of the clients' interests (it is only a mantra, which has to be declared; nothing else) and look at what those clients really get, then you will immediately see that it is an exact declaration of how a lot (a majority) of designers continue to think.

The history of different branches of science and engineering starts at different moments and go on with different speeds, but looks like all of them make the same steps in their evolution. Compare that statement about the programs with the famous declaration from Henry Ford: "Any customer can have a car painted any color that he wants so long as it is black". The automakers turned away from that famous motto years ago. I think that it is time for programmers to change their view on the clients. Not in declarations but in reality. Users must decide about the view of the programs they are working with. Not to ask the developers to change one or another feature when the next version is distributed, but to make the decisions about the view of applications and their performance just sitting in front of the computer. Not the changes in view, which developers included into the predetermined list of allowed adjustments, but any changes they, USERS, would like to make.

I hear too often that the good <u>specialists</u> in design (they call themselves experts) <u>know much better</u> than any user, how the program must look and work, what kind of interface must be organized. Each time I hear such a statement, I ask a simple question: "Better for whom?" Does a user have the right to decide, what is better personally for him? Or is he looked at as a total idiot, who cannot estimate his own feelings and say a word about his own preferences? My work is aimed at not only giving each user this chance, but giving him an easy way of changing any program according to his personal wish.

There is an obvious and very funny flaw in that underlined experts' statement. The experts in design declare that because of their huge experience they know better than any user, what he really needs. They look at the users of their programs as the crowd that must be directed and instructed up to the tiny details; from such point of view the users are too dull to decide for themselves the questions of organizing the screen view of the applications. But the same experts in design spend a lot (a bigger part) of their work time as the users of other applications; by the developers of these applications, those users-"experts" are considered as dull as everyone else. Do not you see the funny side of it?

How many times were you really mad even with the most popular applications, because in one or another situation the program was not going on according to your expectations but did something different? I am sure that the huge efforts were made to design those applications in the best way so that you would be always satisfied with their work. And still you are mad with their response on your commands from time to time. The developers are bad? No, they are excellent. You wanted to do something stupid? I doubt. Both sides are correct, but they have different views, preferences, experience, and habits. It is an absolutely wrong idea to declare that only one view on design is correct and to force the opponent to change his view and to agree with your vision of the best design (in such way the developers always have the upper hand).

To solve the problem of the single interface enforced on users with different views, the adaptive interface was introduced. Did it solve the problems? No, only soften them a bit. The popular dynamic layout* did not change anything at all except for some extremely simple situations. If you have a resizable form with a single control inside, then with the resizing of the form this control changes its size so as to fill the same percent of the form's area or keep the constant spaces between the control and the borders of the form (classical anchoring). This is very easy to implement and the results will satisfy nearly everyone, though I mentioned two possibilities even for this extremely primitive case. But what about two controls inside? They can resize both, or only the first one, or only the second. And the programmer has two choices: either to put into code the variant he personally prefers (or the marketing department of his company declares to be the best) or to add an instrument which allows users to select one of three cases. For the case of only two controls you get three possible variants. What if you have not two, but three, four or more controls inside? Do you think that you would be capable to organize a selection between all the possible variants and would like to introduce the users to such a system? I doubt. The only solution that is left for you is to put some variant into the code and explain to the users that this is the best variant, which they would like. That is how all the popular (and less popular) programs are designed. Does it differ from that declaration, which was announced 20 years ago? "The clients will have to like whatever we will give them".

In the previous chapter *Groups of elements* I wrote about the **Form_PersonalData.cs** (**figure 14.16**, 20 pages back from here). That is a real application to deal with a lot of personal data. The form contains 23 controls accompanied by 16 pieces of comments and organized into several groups. If you have to write such an application, what else except the dynamic layout can you use to organize a good view for changing requests? Do you think that dynamic layout would allow you to design such a program with the interface, which thousands or millions of users would like and admire? You can push thousands of people to work with such a program, if it is sold in a package with computers, but there is no way that they would like its interface. They can adjust to it, if there are no other choices, but are you going to call this interface friendly only because the size of several controls is going to change according to the size of the form?

What is common for any form of adaptive interface (it can be called dynamic layout or by any other name) that somewhere inside the program there is a model of the users' behaviour. This model is either fixed at the moment of design or can be refined throughout the work of an application; the second case is slightly better. The most important thing is that this model was put inside by the designer; after it an application tries to work according to this model as best, as it can. The adaptive interface always looked at as a friendly interface. It can be so, if the user absolutely shares the designer's view on the interface, but it is a rare situation. More often than not the program becomes obtrusively friendly, or even overwhelmingly obtrusive. Millions of people around the world are getting mad with the applications that are changing in the way they do not want them to change. But the users have no choices: the model of their behaviour and what was thought to be the best for them was already fixed.

By the main idea of their design all applications can be divided into two groups: instruments and toys.

- If the applications are the instruments to solve more and more sophisticated tasks, then, being their designer, you work on the improvement of the current instruments and the design of the new instruments, without which the new tasks cannot be solved.

* Several years ago Charles Petzold wrote in [4] that "dynamic layout is … an important part of future Windows user-interface design philosophy". Petzold is not only a good author, but he knows very well, what is going to be cooked at the Microsoft's kitchen. That statement is definitely not the Petzold's speculation on the item, but the Microsoft's decision. An extremely wrong one.

- If the programs are simply some kind of toys to amuse the public and push her into buying new toys every year, then the best policy is to add a couple of colorful strips and buttons from time to time and by means of very aggressive marketing declare that this is a huge step of evolution. (While in reality there is absolutely nothing new.) Unfortunately, that is the way the market of programs began to move years ago.

What is the difference between the toys and instruments in the world of applications? If you can do only whatever was predetermined by a designer and fixed in code, then it is a toy; if a designer thought about possibilities and provided them without restrictions, then it is an instrument. The toys can be very interesting and sophisticated; they can allow you to do a lot of things; I like the good toys. But there are a lot of things that cannot be done with any kind of toys, but only with the instruments. So what is the main difference between the programming toys and instruments? The control over the behaviour of the inner elements and the scenario of the whole work. To turn the applications from being toys into instruments, you have to give users the full control of all the screen elements, thus giving them the full control of an application. The goal of an application is not going to change at all: the *Calculator* continues to be a *Calculator* regardless of colors, fonts, sizes, and positions of all the buttons. (I hope that all the readers understand it and agree with it.) So giving the full control over the inner world of an application to users does not change the goal of an application and does not crash it. It is still the same *Calculator*, but any user can rearrange it in such a way, which is the best personally for him. And user can do it at any moment again and again, whenever and in any way he wants to change it. (*Calculator* is one of the user-driven applications that are demonstrated further on.)

Was there any moment throughout the history of programming, when users get much more control over the screen objects then before? Yes, it happened once, when the multi-window operating systems introduced the idea of movable windows at the upper level. It was a huge improvement in simultaneous dealing with several programs, but that was the first and the last step in this direction. Are you aware that it has happened 20 and plus years ago and from that time there was nothing else? The proposed switch from currently used applications to user-driven applications means bigger step for users than that old step from DOS to Windows.

We move the real objects in our real life all the time, because we need them to be positioned slightly different at different moments. Our decisions about the relocation and placement of the real objects depend on many different things, but whenever we think that another place is better, we move the thing (if it is not too heavy) without thinking an extra second. Just move it and that is all; if we do not like it, we will move the thing again. Absolutely the same thing happens to the screen objects: <u>if they were easily movable</u>, we would move them around the screen very often and place them at the most suitable position as we estimate it at each moment. I see this happening all the time with the user-driven applications.

Certainly, some people understood the need of movable (by users!) screen objects years ago. Because of the importance of this task, the movable objects appeared from time to time in different programs. The main problem was that it happened occasionally, when some clever programmers designed such an algorithm for a particular class of objects. I think that because of their complexity those algorithms were never applied to other objects. (I was also doing such things years ago and my old algorithm was too complicated to be used in other situations.) There were no algorithms that could be easily used for different types of objects and there were no attempts to design applications totally on the basis of movable objects. Those movable objects in the older applications were only good solutions for particular cases. I want to emphasize and underline again two things:

- There were no algorithms that allowed to turn into movable / resizable the objects of an arbitrary shape and origin.
- There were no systems or applications totally designed on movable / resizable elements.

Only an algorithm that can be easily applied to any object allows to design the applications entirely on movable / resizable elements. I emphasize and underline again and again, especially for those, who continue to grumble “We have seen the movable objects before; there is nothing new in it.”

1. The new thing is not in moving one or another object, or even the objects of some special class. ALL the objects in an application must be movable without exceptions.
2. All kinds of objects must be turned into movable / resizable by the same algorithm. This algorithm must be easy to use with all the classes, with all shapes of objects, and under all the circumstances. Otherwise the algorithm is unsuitable. That is why I have designed all those examples (be sure, I have more), which were demonstrated and discussed in the first part of the book. To show that a wide variety of graphical objects, controls, and different kind of groups can be turned into movable and resizable.
3. When such algorithm is applied to all the objects in an application, only then it turns into a different type of program – a **user-driven application**.

User-driven applications have some general rules. To become acquainted with these rules, let us have a look at a simple program of data selection.

## Selection of years

File: **Form_YearsSelection.cs**
Menu position: *Applications – Selection of years*

Let us consider a well known case, when you have a limited set of items of which you need to select some subset. Such selection can be implemented for different types of objects, but for this example I decided to stop on the selection of years. The design of such program is simple (**figure 15.1**). The years from the predetermined range are represented by the lines of one `ListView` control. Any number of lines in any combination can be selected in this list. When any number of lines is selected, press the *Select* button to copy them into another `ListView` control. If any mistake was done during this selection of years, then any combination of the unneeded lines can be highlighted in the second list and deleted from there by clicking the *Exclude* button. I do not think that anyone would have any problems on the implementation of such a task.

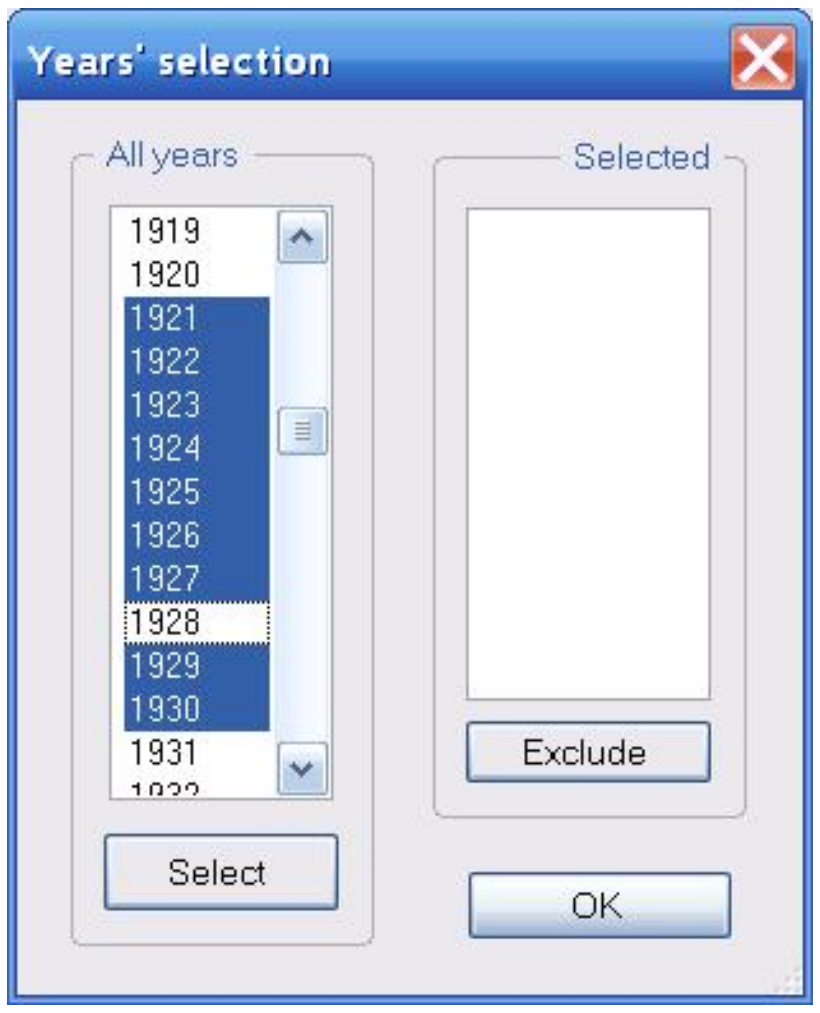

**Fig.15.1** Selection of years

The developers would have no problems with writing the code for such application, but the questions can be (and they certainly will be) brought up by the users on proposed design: half of the users would like to see the full list of items on the left side and the selected items on the right; another half of users would prefer those lists to be positioned in the mirror way. There is also a small percentage of those, who would like to see both lists in one column; they have the right to demand such positioning. In addition, there is an old problem of the best position for OK button. Three big groups of designers prefer to place it in the right-top corner, or the right-bottom corner, or in the middle at the bottom of a form. Members of each group would never agree with anything else except their view.

The task is so primitive that every developer will simply design this form in the way he personally prefers and will ignore any other suggestions. (“The clients will have to like whatever we will give them.”) As a huge respect to the users, the dynamic layout can be applied to this form; the sizes of the lists would be changed according to the used font or resizing of the form. I am sure that you can remember either designing similar thing or using it in one or another application and there was nothing else except what I have mentioned.

But are you sure that this is really the right way to design even such simple programs?

Let us consider the task itself. It is only about the selection of years and nothing else. This selection works correctly regardless of the positions of all the controls and their sizes. As a designer you are responsible for correct work of all the operations (selection, deselection, and saving). If you would insist that application must look exactly like you would like it to be, then stop talking about the user-friendly interface of your programs. If you would agree that it would be nice to allow each user to change the view of the program in such a way as he personally wants, then ask yourself a very simple question: “What has to be changed to give users the full control of the form’s view?”

The answer is obvious: the screen elements must become movable and resizable by the users.

Why was not it done yet? Sluggishness of minds; I do not see any other reason. No ruling boss from the big corporation has declared the possibility and novelty of such thing, so nobody tried to do such things. It is a pity to see that the programmers’ community lost the sense of imagination and the ability to think independently and only waits for the commands from the big corporation to turn left or right and to march in new direction. Until the next big turn will be announced.

In the first half of this book I have already demonstrated how different elements can be turned into movable and resizable; now it is time to apply that knowledge to design of some real (at first simple) application.

Our program (**figure 15.1**) has two obvious groups plus a solitary control. Groups can be organized in different ways; here each of them is designed like an `ElasticGroup`, which has a single `DominantControl` element inside.[*] Each group includes a pair of controls, of which a `ListView` is obviously a dominant control and a button is subordinate. The simplest constructor of the DominantControl class needs only an array of controls, of which the first one is automatically turned into dominant.

[*] This is not the only possible solution. The first sample of the same task was demonstrated in the article [6]. There I used two `Group` objects; it was designed especially for those, who like to use the ideas of the dynamic layout. That article is accompanied by its own Demo program, of which the whole project is available, so you can download it (see section *Programs and Documents*) and look at the codes of the slightly different implementation.

```
ctrlsAll = new Control [] { listYearsAll, btnSelect };
ctrlsSelected = new Control [] { listYearsSelected, btnExclude };
```

```
groupAll = new ElasticGroup (this, new ElasticGroupElement [] {
                     new ElasticGroupElement (new DominantControl (ctrlsAll)) },
                             12, "All years");
groupSelected = new ElasticGroup (this, new ElasticGroupElement [] {
                  new ElasticGroupElement (new DominantControl (ctrlsSelected)) },
                                  12, "Selected");
```

All five controls in the program are resizable; if you want, you can resize any button even in such a way as to see its text going vertically. I do not care what any user wants to do with the view of this application. My job, as a developer, is to provide him with an instrument to select the needed data (years). User can change the view of the program in any way he wants; lists and buttons will do their work regardless of their sizes and positions.

To grant such a flexibility.

1. Declare and initialize a mover.

```
Mover mover;
mover = new Mover (this);
```

2. Organize the movable objects and register them with the mover.

```
private void RenewMover ()
{
    mover .Clear ();
    groupAll .IntoMover (mover, 0);
    groupSelected .IntoMover (mover, 0);
    mover .Insert (0, btnOK);
}
```

3. Use three standard mouse events to organize the whole moving / resizing process. Two of the methods – OnMouseDown() and OnMouseMove() – are very simple.

```
private void OnMouseDown (object sender, MouseEventArgs e)
{
    mover .Catch (e .Location, e .Button);
}
private void OnMouseMove (object sender, MouseEventArgs e)
{
    if (mover .Move (e .Location))
    {
        Invalidate ();
        GraphicalObject grobj = mover .CaughtSource;
        if (grobj is DominantControl || grobj is SubordinateControl ||
            grobj is ElasticGroup || grobj is SolitaryControl)
        {
            Update ();
        }
    }
}
```

The OnMouseUp() method has several extra lines to update the groups on possible enforced relocations of the inner elements; I have explained this, while writing about the use of the DominantControl class. (In the similar example in the program **UserDrivenApplications**, I used the Group objects to organize the groups; in that program the OnMouseUp() method consists of one line.)

```
private void OnMouseUp (object sender, MouseEventArgs e)
{
    if (mover .Release ())
    {
```

```
                if (e .Button == MouseButtons .Left)
                {
                    GraphicalObject grobj = mover .WasCaughtSource;
                    if (grobj is DominantControl)
                    {
                        (grobj as DominantControl) .CheckSubordinates ();
                    }
                    else if (grobj is SubordinateControl)
                    {
                        (grobj as SubordinateControl) .CheckLocation ();
                    }
                    groupAll .Update ();
                    groupSelected .Update ();
                    Invalidate ();
                }
            }
        }
```

The task of selection of years is a simple one; the implementation is also simple, but even this example allows to make the first acquaintance with the basic rules of the user-driven applications.

**Rule 1. All the elements are movable.**

There are no exceptions; all the objects, regardless of their shape, size or complexity must be movable. If for some object you do not see a good solution in an instant, THINK a bit more and you will find a good solution. Users have to get the full control of an application (form), which means that each and all objects must be under their control. Users are going to use an application at the best level to fulfil their work; the movability of the elements increases their chances to do this work exactly in such a way as they want, so give them this chance. If you decide to give them nearly everything, **but** this and that, then they will bump into these hillocks on the road again and again. With an adequate thought about you as a designer.

**Rule 2. All the visibility parameters must be easily controlled by users.**

Rules 1 and 2 are the projections of the full users' control over a program on the different sets of parameters. The first rule deals with the locations and sizes; the second rule - with the colors, fonts, and some auxiliary things.

The rules that are mentioned here are strongly related with each other and always used together. But because the accompanying program is a Demo application, I purposely violated the rule 2 in this particular **Form_YearsSelection.cs**, and there is no tuning of parameters in this form. It is implemented and demonstrated in details in all other forms, which are much more complicated, but not here. After looking at more complicated forms of this application and after trying the tuning possibilities in those forms, return back to this simple form and estimate your feelings in the form without such tuning of all the visibility parameters. I think that this will show you more than anything else the importance of this rule.

**Rule 3. The users' commands on moving / resizing of objects or on changing the parameters of visualization must be implemented exactly as they are; no additions or expanded interpretation by developers are allowed.**

Changing of the visualization parameters by the users is not an unknown thing and is implemented in a lot of applications. But in the standard applications, especially those that are built on the ideas of dynamic layout, users change one parameter, for example, font, and a lot of related things are changed automatically, because this is the nature of dynamic layout. With the user-driven applications you, as a designer, have to stop thinking in the way like this: "You changed the font. I am smart, I know what you really wanted to do, so I will do it for you: I will adjust the sizes of this, that, and that object. Be happy, because I save you several clicks." This is an absolutely wrong way of design for user-driven applications. The developer must not interfere with the users' commands and add anything of his own.

It can be a bit strange at the beginning of new design to control yourself and not to add anything of your own to the users' commands. You may be a designer with many years of practice, you really know what must be done to make the view of an application better in this or that situation. But this is another world; if you gave users the full control of an application, it must be really full, so you do not leave anything for yourself as a second control circuit. Whatever is gone is gone.

And eventually you will find that nobody needs your even excellent experience on adjusting the forms to their needs. Where you have to apply all your skills in design (the higher – the better) is in construction of the default view of every form. The highest credit to your skills is the big percentage of users, who would not change anything at all, but work exactly with your proposed design. And yet, the possibility of all those moving, resizing, and tuning must be there for users to try them at the first wish.

**Rule 4**. **All the parameters must be saved and restored.**

Saving the parameters for restoring them later and using the next time is definitely not a new thing and is practiced for many years. But only the passing of the full control to the users and the significant increase of the number of parameters that can be changed, turned this saving / restoring of parameters from the feature of the friendly interface into a mandatory thing. If a user spent some time on rearranging the view to whatever he prefers, then the loss of these settings is inadmissible. And as the full users' control means the possibility of changing any parameter, then the saving / restoring of all the parameters must be implemented.

It does not mean that you have to write many lines of code, if you are going to organize a form with dozens of objects inside. You can check the code of this form and of other forms with much more elements inside and see for yourself that the process of saving / restoring is not boring at all or time-consuming.

These four rules were born in different ways. Rule 1 – the main rule of user-driven applications – fixes in words the whole process of transformation from the standard applications into something totally different, based on movable elements. It took me several months to understand the entire process; the understanding came step by step, and on each stage I had no idea, what would be the next. But it turned out to be very logical; I will talk about it further on, while writing about functions and scientific / engineering applications, because this is the area, where all these things started. Rules 2, 3, and 4 became obvious to me and were formulated later, when I started to design more and more user-driven applications. With all these main rules formulated, we will see further on, how these rules work in complicated forms with a lot of elements.

## *Personal data*

File: **Form_PersonalData.cs**
Menu position: *Applications – Personal data*

Let us turn once more to the application, in which users can look through the sets of personal data (**figure 15.2**). The code of the **Form_PersonalData.cs** was already used in the chapter *Groups of elements* for the detailed explanation of the `ElasticGroup` class. This form is organized according to all the rules of the user-driven application, so let us look on the details of how those rules work in the real program.

**Rule 1**. **All the elements are movable.**

> "Is there more than anecdotal evidence to suggest that end users really want to be configuring the fine points of a user interface layout, rather than accomplishing the task that the application is intended to support? The HCI literature would seem to provide strong evidence to the contrary."
>
> Anonymous review from the UIST-2009.*

Certainly, ALL the elements are movable. I would not even try to predict the number of different users' opinions about the best view of such a form. I think that it would differ not too much from the number of users. Or it would be significantly higher than the number of users, because my experience shows that when there is a chance to rearrange the view of the form (application) at any moment, then each user has different opinions about the preferable view each day of the week and also depending on the time of the day and the weather outside. The movability of each and all parts of the form allows to rearrange its view in an arbitrary way.

It is interesting (and really funny) to watch the same reaction and hear the same critical remarks from the specialists of interface design about the implementation of movability for all the elements. "Users do not need it. They have to work with the application and not to play with moving the objects around." The funniest part of this remark that it is always pronounced before the speaker ever tried such an application, but never after. I never heard a single complaint about the uselessness of movability from anyone, who work with such applications, or even from those, who only tried them. Programmers and users prefer to try and then give out their opinion. "The biggest specialists" have different procedure; looking at a working application is never considered by them as a mandatory thing for making their conclusion. "We have out Holy books; they are always right; whatever is against them is a heresy. And whoever is pronouncing a heresy must be punished." I am not joking at all; look at the epigraph several lines prior to this. That epigraph is from the review on my proposal for one of the well known annual symposiums. Never mind that the author of that review did not bother himself with looking at an application and trying it even for a couple of minutes. Too busy and too good specialist for such a waste

* The progress of science can be based only on the exchange (and battle!) of the different ideas, proposals, theories. Hiding the name of an opponent in the scientific discussion – it is one of the most stupid and ridiculous things with which I met throughout the years of my scientific career. It is also the best illustration that the pure transfer of procedure from one area (political vote) into absolutely different area (scientific discussion) can easily turn the whole situation into the theater of the absurd. "Something is rotten in the state of Denmark."

of time. I really like this review; I admire it. I have it printed out in big letters and hanged on the wall at my work place. “The H[uman] C[omputer] I[nterface] literature provides strong evidence … that you are wrong.” Unfortunately, the review is anonymous. I would like to thank heartily the author of such an amazing review of the work, at which he never even looked.

The rules of “good design”, which can be found in many books and manuals, declare that if you have in a form several controls with the comments, then all the comments must be placed on the same side of the controls. You can see from **figure 15.2** that I have not only slightly violated, but absolutely ignored this rule. I think that with the spread of user-driven applications a lot of currently used rules of design, even those that are looked at as axioms, will be ignored, forgotten, or revised. Another world, different rules. The majority of currently used rules were declared 20 and plus years ago, when everyone was working with a single screen. Now we have much bigger screens; a lot of people use several screens. But the biggest blow to those rules comes from the movability of the screen elements.

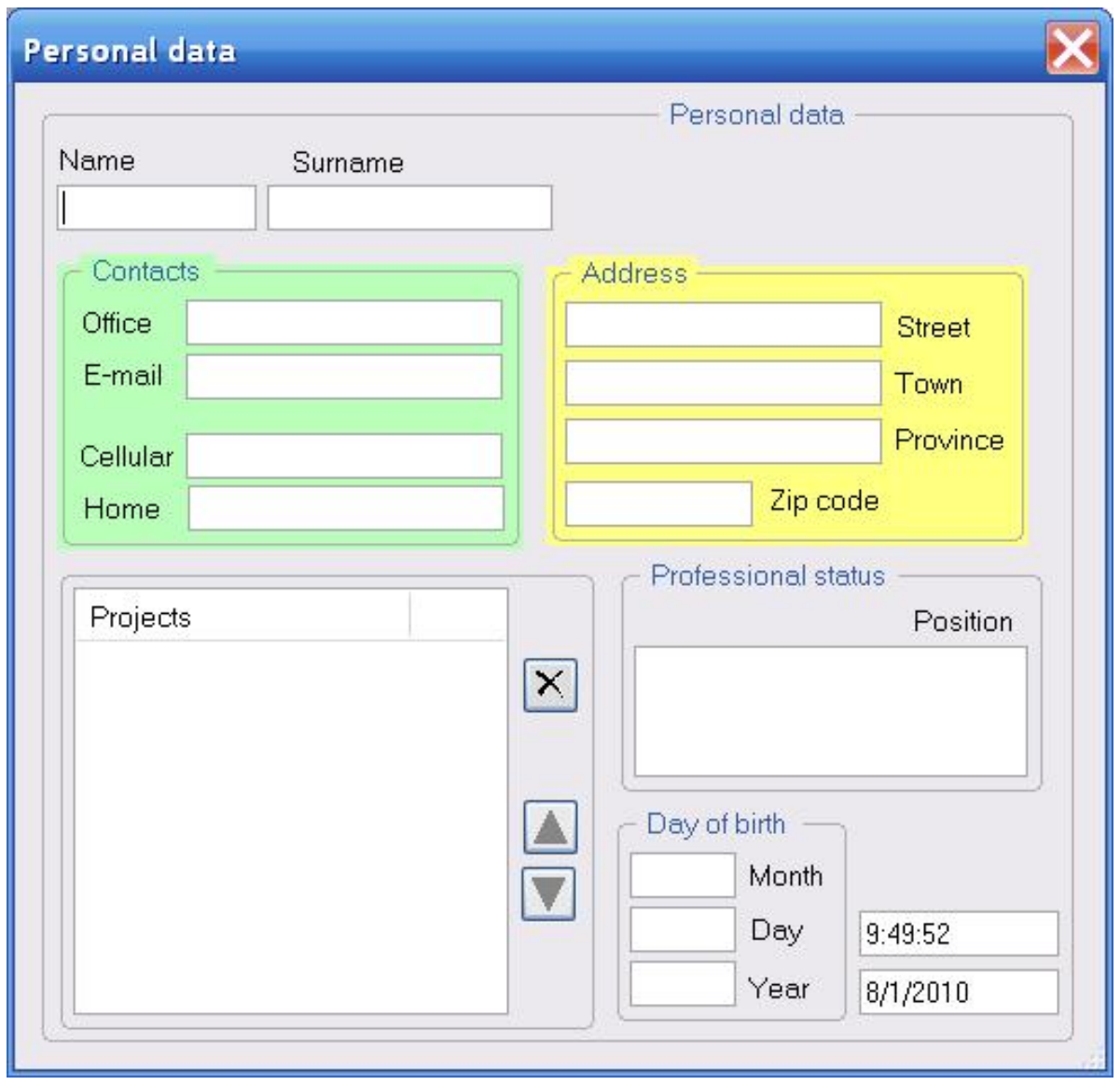

**Fig.15.2** Another view of the **Form_PersonalData.cs**

I do not see anything annoying or strange if in one group the comments are positioned on one side of the `TextBox` controls, while in another group they are positioned differently. Look at the central part of **figure 15.2**; there are two groups, which look like opposite pages of the opened book. Why not to have the comments to these groups on the outer sides of them? It is good for the books, why is it going to be wrong for the screen? By the way, in the realm of old design (fixed elements) you have no chances to check, if the announced rule is right or wrong. As a designer, you will construct the forms according to this rule; maximum that you allow yourself is to try either the left side positions or the right side positions for all the comments, but nothing more. (It is my assumption; I can be wrong.) As a user, you would have to work with whatever you have been given; in nearly all the programs this rule is implemented. Only now, with all the elements easily movable around the screen, you get the chance to check that rule for yourself and to decide about its usefulness or not.

There is another thing related to the same problem of multiple `TextBoxes` with comments: the alignment of all those comments. The preference on this item is so individual that I do not remember even reading about any rule; the alignment of comments was enforced by the designers without any second thought. (“You would have to like what I like.”) I was not arguing about the alignment of comments years ago, I would not do it now, when I give you an instrument to change this alignment easily, quickly, and in any possible way. If you have a very specific view on the good looking design, you can place those comments even in the chess order around the `TextBoxes`. I will not say a word against it. You took the application; you work with it; you change it in the way, which is the best for you.

Some people can say that these are the minor items, which can be simply ignored. I do not agree with this. And I think that the users of such programs, especially those who have to work with similar applications many hours a day (HR people), would not agree that these things can be simply ignored.

A bit more about the positioning of the elements, which is strongly related to their movability.

One of the highlighted groups at **figure 15.2** - the *Address* group - represents the standard set of controls to deal with the address information. While the order of controls in this group is natural for western countries, it is absolutely unnatural for other parts of the world, where the address is often written in the opposite order beginning from the country information. With all the elements being movable, it will take a couple of seconds to rearrange the view to whatever the user would like it to be (to whatever order he got used to throughout many years of his life). I cannot think out any other good solution for this problem of the habitual order of information. Certainly, you can hard code several variants plus some instrument for the selection of the needed one. It is definitely a lot of extra work, and are you sure that you would put into code all the needed variants? Or you can simply ignore the specific local preferences in the address information. (Let them work with what is given.) You can decide for yourself if it is a good solution or not. What would you prefer, if you have to work with such data day after day: movable elements or a fixed design?

In the chapter *Movement restrictions*, I have described the use of clipping and different levels of moving the objects across the borders. Moving the temporarily unneeded elements across the border in order to rearrange the whole view is widely practiced in the user-driven applications. There is an easy way to put the needed level of restrictions on such movement, so you never lose your objects, if you do not want to do it. The temporary relocation of the objects across the border of the form is not a trick in the user-driven applications, but a very reliable, often used, and very easy and quick way of adjusting the view of applications. This method can be used with any stand alone object (it can be a single control or a whole group with a lot of elements inside), but it is not the good idea to try this technique with any inner part of the surrounding group. If it is a part of the `ElasticGroup`, then this group's frame will go after its runaway element not looking at the distance. For the `ElasticGroup` class, there exists much better solution: any inner element can be turned into invisible and returned back to life only when it is needed again. This technique was already discussed in the section *Elastic group*.

The movability of the elements adds so many new things to the design of applications! I work on the design of really big and sophisticated programs for more than 30 years. New languages and faster computers opened new possibilities from time to time; it was never a revolutionary step, but only some improvement to the already known procedures. With the invention of the algorithm for turning elements into movable / resizable the situation is different. There are no recommendations on what can be or has to be done in one case or another; you go step by step in exploring the new continent and often you find the things, which you couldn't even imagine a month or two ago.

Complex applications have a lot of screen objects. These objects are not only individually movable, but a lot of them are complex objects, which include movable parts. To all these individual and synchronous movements users can add the moving of the arbitrary groups of objects. After series of movements user can understand that the default view was much better then the one he organized himself (you are a very good designer!). It is a good idea to give users a chance to reestablish with one click a default view of a group or of a whole form. In the standard applications you do not even think about such problem, because those applications always have a default view. For the user-driven applications, it is not another rule, but simply a sign of good design. A sign of your respect to the users of your programs.

Usually, the restoration of the default view is ordered via a context menu at any empty place. Depending on the complexity of the form, it can be done in two different ways. If the form is simple with only few inner elements, then the set of default positions, sizes, and other parameters can be saved on opening the form; this set of saved parameters can be used at any moment, when the user wants to return to the default view. With the really complex forms, I prefer to use another technique.

I have mentioned it before (rule 4) and I am going to write about it more a bit further that all the parameters of a form and all the parameters of all elements are saved in the `Registrty` at the moment when the form is closed. When the form is opened the next time, there is a checking, if the `Registry` contains the set of parameters for this form. If NOT, then the method to organize the default view is called. So, when I want to restore the default view of the already opened form, I organize it in such a way:

1. Set the flag for clearing the `Registry`, set the `DialogResult` property for reentering, and call the closing of the form.

```
private void Click_miDefaultView (object sender, EventArgs e)
{
    bClearRegistry = true;
    DialogResult = DialogResult .Retry;
    Close ();
}
```

2. Close the form; according to the previously set `bClearRegistry` flag the form's entry in the `Registry` is deleted.

```
private void OnFormClosing (object sender, FormClosingEventArgs e)
{
    if (bClearRegistry)
    {
        DeleteRegistryEntry ();
    }
    else
    {
        SaveInfoToRegistry ();
    }
}
```

3. The form is closed and the returned value from its Show`Dialog()` method is checked. As it was previously set to `DialogResult.Retry`, then the form is immediately opened again. Because the form's entry in the `Registry` does not exist, on this opening the form gets its default view.

```
private void Click_miPersonalData (object sender, EventArgs e)
{
    Form_PersonalData form =
                new Form_PersonalData (PointToScreen (new Point (200, 70)));
    while (DialogResult .Retry == form .ShowDialog ())
    {
        form = new Form_PersonalData (form .Location);
    }
}
```

User gets the default view of the form. He can either use it or make another attempt on better rearranging the form. There is no limit on repeating the above described procedure, so I am sure that user will move the elements of the form again and again.

The idea of movability penetrates all the layers of design; the movable big parts allow to change quickly the whole view of an application or a form; the same feature of the smaller or tiny parts allow to reach the aesthetic perfection in parallel with the most effective work of applications. Further on I will write how the idea of movability spreads from the main form of any application into the tuning forms and into any auxiliary forms that are used in applications.

**Rule 2**. **<u>All the visibility parameters must be easily controlled by the users.</u>**

**Form_PersonalData.cs** (**figure 15.2**) is not the simplest one, but definitely not among the most complex forms, which I have to design and work with. It is just an ordinary form with an average number of elements; each of the elements has its own parameters of visualization. The outer *Personal data* group includes five inner groups and several other elements. In total there are six groups; each group has a set of its own parameters; the number of the tunable parameters for each group can be estimated by the view of the group's tuning form. **Figure 15.3** demonstrates the tuning form of the outer group (the *Personal data* group), similar form can be called for tuning of each inner group. Thus for the **Form_PersonalData.cs** we have more then 100 visibility parameters, which <u>regulate only the view</u> of the elements on the screen. What is YOUR answer on such a simple question: "Which part of these parameters must be controlled by users?" Try to answer this question not as a developer of such program but as a USER.

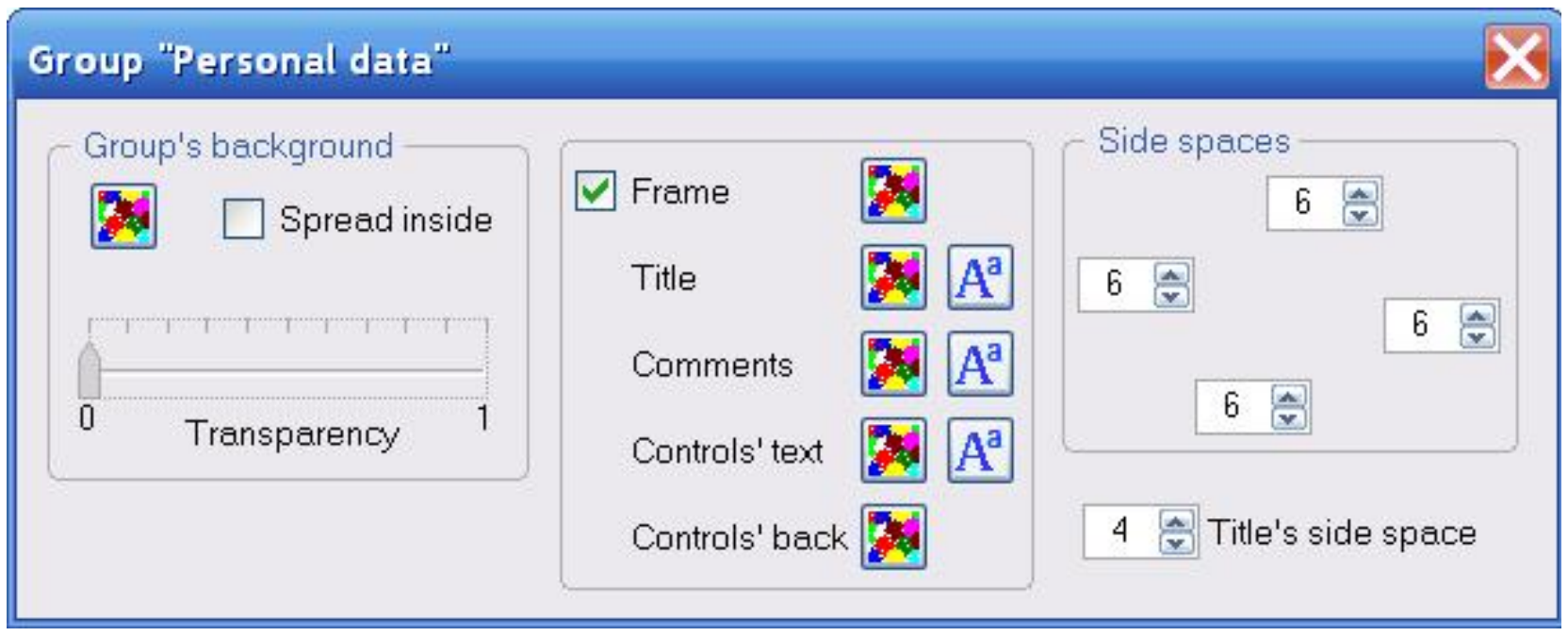

**Fig.15.3** Tunable form of the outer group

There are only three possibilities for providing the tuning of your applications:

- Giving users no possibilities for tuning. It is a funny way to look at it as some level of tuning; better to call it not providing any tuning at all.
- Giving users a chance to change some of the parameters.
- Declaring each and all parameters as tunable. Not only declaring, but giving users an instrument really to change any parameter of visualization.

I do not think that anyone can seriously consider the first possibility – no tuning at all. Certainly, if you look at yourself as the our day Malevich and produce the modern (screen) version of the *Black Square*, then such work does not need any tuning at all. On the other hand, it must be considered as a product of the contemporary art, but not as a programming product. For all other applications, some tuning is always needed.

The second option opens a wide area for discussion of what can be given to users for tuning and what not. Such discussions would never end. On the constant demands from the users of any real application you would have to increase the number of tunable parameters. So why not to avoid these long discussions and not to turn to the third option just from the beginning?

Passing the control over ALL the visualization parameters to the users looks absolutely natural for user-driven application. And again I have to mention that there are no complaints about it from the users, but there are, funny to repeat, from the "experts". Experts have two main objections.

**Objection 1** Users do not need all this setting of parameters; they need only the working program and will be happy with it.

I try to explain again and again, that the work of a program does not conflict with the possibility of tuning the parameters. All the tunings do not change the goal of the program, but only make it much more flexible and provide a lot of possibilities to rearrange the view of a program to whatever each user would like to see.

**Objection 2** Too many parameters for tuning; users will be lost in them.

I agree that the number of tunable parameters can be really big in some programs, but this is the reality of the big programs. By not allowing users to change those parameters you do not decrease their number; all the parameters are still there, but in the standard applications they are set at the developers wish, and that is all. Give users an instrument for easy changing of ALL these parameters and let users decide, when, how, and to what extent they would like to use this instrument. The decision must be given to users and not stolen from them.

Users do not change the parameters all the time. Usually they set them at the beginning of the work and then there is only an occasional additional tuning, when something needs to be changed. But at the beginning it is often the only way to get the good looking application. I spend a lot of time, trying to design the complicated forms in my scientific applications in really good view. The scientific applications are among the most complicated programs; I will write about them further on. The programs, on which I worked throughout the last two years, included the tasks, which were discussed in the laboratory for years, but were never developed because of the design problems; the movability of all the elements opened the opportunity to solve the problems and design such programs. When the applications are ready, I give these applications to colleagues and look with horror at the initial view of the same programs on other computers. I spent a lot of time on thinking out those applications up to the tiny details; the programs looked very good on my computer, but on others… Even if I would try to, I wouldn't have a chance to destroy the views of my forms more than I see, when they are used for the first time on different computers.* From my point of view, the modern PCs are like a bus, in which each passenger has his own steering wheel. Nobody knows whose wheel is the main one at one moment or another; it looks like they take this role at random, but definitely there are several of them.

With the applications constructed of movable / resizable elements, I am not worried any more about this strange effect on different computers. When a user-driven application starts at any computer for the first time, it may has not the best view, but each user gets a chance to redesign any form to a good looking one in seconds. This may happen (I am sure it will happen occasionally) with the forms from the Demo application for this book, but the figures from the book can show you what I prepared and what I would like you to see. You can rearrange each form to the identical view or you can change them in any way you want. It is a user-driven application in which the ideas are the main things and the views are under your full control.

The "experts in design" deny the idea of giving the full control to users as the biggest heresy; they want neither to discuss the possibility nor even to hear about it. I never heard a single complaint about this thing from the users, but an outcry started from those who teach for years how to design the good looking system. It looks like those people simply try to defend their monopoly on the sacred knowledge; nothing else. What are they afraid of? Are they afraid that users would organize the view of applications in such way that they really need and prefer? But what is the problem with this? Programs must have the best design (view) for each user personally, but not for the average user, as a designer imagine him. (I hope you understand the difference between these two things.)

Do not look at the users of your applications as too dull to do all the tuning. You, as a developer, write the program; the construction of a tuning part in this program is only a small part of your work, so you deal with this tuning relatively small part of your time. Users are working with your programs for a considerable amount of time (I hope you are a good programmer and users like your applications), so users would spend some time on tuning these applications according to their demands. If you exclude some possibilities of tuning from this process, users will be still looking for opportunities of tuning, but they will have to spend much more time in their efforts to go around the artificial limitations that you put there. What for? Give them the whole power of tuning. Users will be thankful to you.

There are some problems in making everything tunable. Some of these problems are born not by the number of tunable parameters, but by the way these changes are processed by the Windows system. The biggest problem may happen with the change of some font. These changes and especially some negative consequences are more related to the next rule.

* It is incredible even to imagine, what this Windows system can do to the good looking application. "We wanted to do the best, but the result was as usual…"

**Rule 3. The users' commands on moving / resizing of objects or on changing the parameters of visualization must be implemented exactly as they are; no additions or expanded interpretation by developer are allowed.**

The dynamic layout is used in applications for years and throughout these years we got used to some of its remarkable features. The dynamic layout is some kind of designing philosophy, but we, as users of different programs, deal with the implementation of this philosophy in each particular case. This can differ a lot from application to application, because it depends on the interpretation of the ideas of dynamic layout by each designer. Our expectations can be different from the developer's view; this is the cause of many problems. I still do not understand why the command to change the size of the font has to change the size of the form also. I did not ask for it, I want to see the same information at the same place, but in bigger letters; that is all. Why somebody decided, even without asking me, that I would like to enlarge the form and controls? From my point of view it is really obtrusive and I cannot understand, why it is continued to be called a friendly interface. (It is exactly like washing my car at my expense at the gas station, if somebody would decide that any filling of the gas tank must be accompanied by the total car wash. I really wonder how many of the readers would call such a gas station friendly.)

The user-driven applications give the FULL control to the users, so the designer of such applications has no rights to add anything of his own to the users' commands. Thus the command on changing the font of a single element or a group would do exactly what was ordered and not a bit more, but… the result can be slightly different from our expectations, based on all our previous experience.

- Let us return to the **Form_PersonalData.cs** (**figure 15.2**) and change the font of the comment belonging to the `CommentedControl` object. I want to remind that these are the pairs "control + comment"; such objects are widely used in the mentioned form. The comment of any `CommentedControl` object belongs to the `CommentToRect` class which is derived from the `TextMR` class. These objects are positioned by the central point; on changing the font, the anchor (central) point is not moved and the area of the text either expands in all directions or shrinks. If the comment is positioned close to its related control and you significantly increase the font of the comment, then the text expands in all directions. Chances are high that the enlarged comment will be partly covered by the associated control; the text will need some moving to reestablish the good view of this pair.
- When you increase the font of a whole group, the result depends on the type of controls in this group. Some controls will preserve their sizes and will not organize any troubles for you; others will automatically change their sizes. If several text boxes are positioned close to each other, then chances are high that after the increase of the font they will overlap; you will have to move some of the controls to restore a good view.

While working with the user-driven applications, you can occasionally observe such or similar annoying results after changing the font, but in reality such problems occur rarely enough. First, the significant increase of the font is a very rare thing; for the small changes of the size the effect is not as bad as I mentioned. Second, you do not change the font too often. Usually you change it once, make some needed movements of the elements, and after it there are no requirements on changing the font.

When you try to introduce and use something new, you have to take into consideration the features of the elements on which you and others build the applications. I want to cut the strong link between the size of controls and the size of the font, which these controls use. I also do not want to include into my applications any actions which are still out of users' control. For this reason, I never change the font for the whole form throughout its `Font` property, which, depending on the setting of the `AutoScaleMode` property, might also change the size of a form. Instead, I change the font of each control on an individual basis; the consequences depend on the type of those controls. For example, any single lined `TextBox` will change its height according to the size of the font and you have no chances to do anything with it; this reaction is fixed deep inside the system by the developers of those controls; programmers have no access to it and no control over such action. The majority of the `TextBoxes` in the **Form_PersonalData.cs** is of such a type (their `Multiline` property is set to `false`); the width of such element is under users' control, but not the height. However, you can give users the control over the height of such element, but only if you set the `Multiline` property of the `TextBox` to `true`. In the **Form_PersonalData.cs** there are two multi-line `TextBox` controls in the group *Professional status*; at **figure 15.2** only one of them is seen and another one is hidden. If the height of such control is not big, then some negative effect can be observed on the significant increase of the font for such `TextBox`: at first moment the increased symbols are seen through the same area of a control only partly, until the user resizes this control with a mouse. Everything comes with a price, but I think this is a very small price for exchange of the full control over the view.

About the violation of *rule 3.* I would prefer to have no exceptions from rule 3, but at the moment there is one, which is related to the `CommentedControl` class. (User-driven applications can use a lot of different classes and similar situation might occur with other classes later.) Any `CommentedControl` object consists of two parts: control and comment. Any moving of a control causes the synchronous movement of its comment, so there are no problems at all. Comment can be moved individually and released at any place. Controls are always shown atop all the graphical objects;

when the comment is moved across the area of any control, the text of the comment disappears from view, because it is closed by the control. What will happen if a comment is released at such a moment, when it is entirely closed from view by a control? If this is not the control of the same pair (of the same `CommentedControl` object), then the comment is left, where it was released. User can move the control aside and open an access to the comment. If the control belongs to the same `CommentedControl` object, then there would be no way to relieve the comment, because it always move synchronously with the move of the associated control. If the comment becomes invisible, it will stay invisible and inaccessible forever. To avoid this situation, the system violates *rule 3* and <u>in this particular case</u> pushes the comment slightly from underneath the control thus making the comment accessible.

The described situation happens at the moment when mover releases an object (a comment or a control), so in this case the mover is responsible for checking the possibility of such situation and avoiding it. But similar situation can happen as a result of the things not related to mover in any way; then a designer has to think about it. Consider a case of a `CommentedControl` object with the comment, using some big font and positioned mostly under its control with only a tiny part of the comment in view. Then you call a context menu on this comment and order the significant decrease of the font of the comment. As a result, the area of the text shrinks and the comment disappears under the control. This situation is especially checked in the `Click_miCommentFont()` method.

```
private void Click_miCommentFont (object sender, EventArgs e)
{
    FontDialog dlg = new FontDialog ();
    dlg .Font = cmntPressed .Font;
    if (dlg .ShowDialog () == DialogResult .OK)
    {
        long idCmntCtrl = cmntPressed .ParentID;
        GraphicalObject grobj;
        for (int i = mover .Count - 1; i >= 0; i--)
        {
            grobj = mover [i] .Source;
            if (grobj is CommentedControl && grobj .ID == idCmntCtrl)
            {
                CommentedControl cc = grobj as CommentedControl;
                cc .CommentFont = dlg .Font;
                cc .CommentEnforcedRelocation (mover);
                groupData .Update ();
                Invalidate ();
                break;
            }
        }
    }
}
```

The `CommentedControl.CommentEnforcedRelocation()` method checks the positions of comment and control; if the comment is entirerly closed by the control, then the comment is slightly moved to the side.

I mentioned the situations, when change of the font can lead to such new positioning of elements that would require some of them to be relocated by user. The standard reaction from an experienced programmer would be to think about such possibilities beforehand and to add a small bit of "improvement" from his own. <u>My advice: do not do it</u>. Such even small addition is against the full users' control of the applciations. Users are not fool or dull; you are one of them, and you are as smart, as anyone else. Users do with the applications exactly what they want to do. If you think that you are capable of organizing the good view of any program with which you work, then the users of your applications are capable of managing the programs, which you develope for them. If you think that they are too dull to do such a thing, then look into the mirror: you are one of users. We are all users; there are no exceptions.

I understand that it would be a bit strange for the experienced developers not to include into the code some "obvious imporvements", but the world of the user-driven applications is a different universe with different rules. Developers have to think about and to avoid the fatal situations like disappearance of the objects without any chance to return them back. Avoiding of such situations demonstrates your designer's level and this is a credit to you. But do not apply your skills to the automatic move of the slightly overlapping objects, which user can easily do himself. If you want, you can offer both variants and let users make their choice; I will demonstrate such things in one of the further examples.

Rules 1, 2, and 3 are formulated as separate, but in real applications they all three work together. I think you can see it even from the above text: while writing about the movability, I have to mention the tuning; while writing about the tuning, I have to mention the special and very accurate way of working with some of the elements and again have to add some words about moving.

**Rule 4**. **All the parameters must be saved and restored.**

There are many objects in the **Form_PersonalData.cs**; parameters of any element can be changed, so all these parameters must be saved until the next opening of the form. I have estimated before that there are more than 100 visibility parameters; there are also coordinates and sizes for all the elements. Each element has two parameters, which determine its hiding status and each object can be declared movable or not. For saving / restoring of all these parameters I prefer to use the `Registry`, though exactly the same thing can be organized via a binary file. All the classes, included into the **MoveGraphLibrary.dll**, have two pairs of methods to save and restore the objects of these classes; one pair of methods for `Registry`, another – for a binary file.

For saving the view of any form (to restore it in the same view the next time), I have to store the size of the form, maybe several other general parameters, and all the used objects. For the **Form_PersonalData.cs**, there are only two such objects: the outer group and the information.

```
private void SaveInfoToRegistry ()
{
    string strRegKey = Form_Main .strRegKey + strAddRegKey;
    RegistryKey regkey = null;
    try
    {
        regkey = Registry .CurrentUser .CreateSubKey (strRegKey);
        if (regkey != null)
        {
            regkey .SetValue (nameSize,
                      new string [] {version .ToString (),               // 0
                                     ClientSize .Width .ToString (),     // 1
                                     ClientSize .Height .ToString (),    // 2
                                     bShowAngle .ToString (), },         // 3
                              RegistryValueKind .MultiString);
            groupData .IntoRegistry (regkey, "Data");
            info .IntoRegistry (regkey, "Info");
        }
    }
    catch
    {
    }
    finally
    {
        if (regkey != null) regkey .Close ();
    }
}
```

Restoring of the form is as easy as saving, but you must be careful in one thing: to restore a group (the `ElasticGroup` object), the array of the inner controls must include those controls exactly in the same order as they were used for the construction of the group.

```
private void RestoreFromRegistry ()
{
    string strkey = Form_Main .strRegKey + strAddRegKey;
    RegistryKey regkey = null;
    try
    {
        bRestore = false;
        regkey = Registry .CurrentUser .OpenSubKey (strkey);
        if (regkey != null)
        {
            string [] strs = (string []) regkey .GetValue (nameSize);
            if (strs != null && strs .Length == 4 &&
                Convert .ToInt32 (strs [0]) == 608)
            {
```

```
                    ClientSize = Auxi_Convert .ToSize (strs, 1);
                    bShowAngle = Convert .ToBoolean (strs [3]);
                }
                else
                {
                    return;
                }
                groupData = ElasticGroup .FromRegistry (this, regkey, "Data",
                        new Control [] {textDate, textTime, textName, textSurname,
                                        textDay, textMonth, textYear,
                     textHomePhone, textOfficePhone, textMobilePhone, textEMail,
                     textStreet, textTown, textProvince, textCountry, textZipCode,
                     textCompany, textPosition,
                     listProjects, btnDelete, btnMoveUp, btnMoveDown });
                info = ClosableInfo .FromRegistry (this, regkey, "Info");
                if (groupData == null || info == null)
                {
                    return;
                }
                bRestore = true;
            }
        }
        catch
        {
        }
        finally
        {
            if (regkey != null) regkey .Close ();
        }
    }
```

The same request for controls to be declared in the same order as they were used during the construction is applied to all the classes, which are linked with controls. All these things are described in the **MoveGraphLibrary_Classes.doc** (see *Programs and Documents*).

# Applications for science and engineering

The rules of the user-driven applications were already formulated and discussed in the previous chapter. In this chapter I want to demonstrate the transformation under these rules of some of the most interesting and complex programs – the applications for science and engineering. This is also the area, in which all these rules were born and in which I continue to implement all the new ideas. I want to show, how the problems of this area and the logic of development brought me to design of the user-driven applications.

For many years the computers were used only as the most powerful calculators; when the graphical displays appeared later, they were immediately used for better presentation of data and results in scientific / engineering applications. The history of graphical presentation of different functions includes several decades. All programming libraries on all types of computers provide the drawing of functions, but programmers continue to write new and newer methods of their own, because the existing standard methods are too general and often not the best in each particular case.

I started to work on the programs with the presentation of functions before the PC era; in computer history it is the analogue of the Middle Ages. The hardware and software changed throughout the years, the new programming languages were born, the new ideas in programming pushed a lot of people (and me also) into rethinking of some basic principles, but for many years I work mostly on scientific / engineering applications in different areas and the core of all the applications in all these areas is still the same: "Draw me the function".

There are two milestones in design of scientific / engineering applications during the PC era. The first important moment happened when the textual mode of monitors was abandoned and the programs began to work exclusively in graphical mode. Then the increasing processor speed allowed to use more complicated (and time consuming) calculating algorithms in parallel with the real-time graphical presentation of results. For several years the hardware progress ignited the new ideas in overall design of scientific / engineering applications, making possible the things, which couldn't be implemented on slower computers.

Somewhere 10-12 years ago the design of scientific / engineering applications went into the period of stagnation. The new versions of the big and well known programs are still distributed among the clients every year or two; managers of these projects continue to declare each version as the biggest achievement in the history of mankind. "You have to buy the new product, if you want to do anything at all!" All these words do not worth a paper, on which they are printed. If there is really anything new in those programs, then it is only in the algorithmic part (the core of each program), but not in their design.

This is really a strange situation: the continuing progress in hardware; the C# language, which is much better than C++ for programmers' work on big and small scientific / engineering applications, and yet there is nothing new or helpful for users of these applications. Users do not care about the language or the programming environment that the designer used throughout his work; they care only about the result and what they get. And this is absolutely correct, because users must do their own work and the only important thing for them is: "In what way the new version of this big program is better for my work than the previous version?"

I worked on the design of big scientific / engineering applications for different areas, and it became obvious to me that the problem was not in some particular area, but it was a general problem. This general problem has its roots in the main idea: all <u>applications are designer-driven</u>.

The users of scientific / engineering applications are often better (or much better!) specialists in their particular area than the designers of the programs.* The users of these programs (scientists and engineers) try to solve some very complicated problems, and there is a paradoxical (some times even anecdotal) situation: the better specialists have to work inside the range of understanding, enforced by lesser specialists. The only way for the users to get something new from the programs is to explain to the manufacture that they really need this and that features. If they would be successful in their explanations, then maybe in a year or two they would see some additional parts in the application, but nobody knows beforehand, how those users' explanations would be interpreted by the designers.

* It is funny to watch how the developers of the scientific and engineering programs try to reject even this obvious fact. The automakers do not insist that they are better drivers than the car racers, the specialists in ballet shoes never claim that they are better dancers than the soloists of the ballet groups, but when I write that the users of the scientific applications are often much better specialists in their specific areas than the developers of the programs they have to use, then there is often an outcry from the developers of those applications: "We do not agree with this." The same point of view, demonstrated by the manufactures of applications again and again: "We know better than anyone else what is really needed. You have to like whatever is given to you."

Many years ago an adaptive interface was born and eventually became the axiom of design for all the complex applications. You can find in many books the description of the adaptive interface as a solution to "user – designer" problems, because it gives user a chance to select the best solution personally for him. What is never written and carefully hidden is the fact that user gets a chance to select only between the choices that were predetermined by a designer according to his understanding of each situation and of his vision of the best solutions for users in each particular case. Adaptive interface can soften the problem, but not solve it. When I understood that the main problem of all the scientific / engineering applications is in their being absolutely designer-controlled, I also understood that this hurdle cannot be crossed with the help of any known or will be thought out in the future form of adaptive interface. Only something different would allow to solve this main problem and move forward. Then I came to the idea of the user-driven applications.

It was obvious to me that such applications have to pass the control over the screen elements to the users, thus the algorithm of moving the screen objects became crucial and had to be invented. When I thought out such algorithm, I immediately tried to apply it to the objects, which I was using all the time in all my programs – the scientific plots. Maybe it was a mistake and I had to begin with much simpler objects (in the way I demonstrated this algorithm in the first part of the book), but an attempt to use this algorithm from the beginning on the complicated objects unveiled to me some basic features of the user-driven applications and helped to understand a lot of very important things about the whole area.

## *The iron logic of movability*

Plots, used in scientific / engineering applications, are not very simple objects. On the contrary, they are among the most complex objects used in programming; they have a lot of parameters for visualization and can be used in different situations. I work with the plots (classes for plotting) of my own design for years; the core of my plots was designed 20 years ago. At the beginning there were some significant changes, when I tried different approaches, but years of experience gave me the clear understanding of the whole set of the needed parameters and methods, so for the last several years before my switch to movable objects there were hardly any changes at all in the plots that I used in many different programs. Those plots were carefully checked, they worked in all the situations, and I did not expect any problems at all, when I started to add movability to them. Initially I thought only about turning the rectangular plotting areas into movable and did not want to change anything else, because everything worked fine. Such view on adding the movability to the reliable and working classes of objects turned out to be absolutely wrong. My understanding of the situation improved step by step. Only the first of these steps – turning the main plotting area into movable / resizable - was a voluntary one; all further steps I was enforced to do by the logic of the design.

**Step 1 – main plotting area**. I took one of my big applications with a lot of different plots, turned the main plotting areas into movable / resizable, and began to play with this new program. It was a huge improvement of my program, but the more I worked with it, the stronger became the feeling that something was incorrect. The logic of design was perfect for a fixed designer-driven application; even turning of a single part of it into movable / resizable began to corrode the whole construction. The features, which were thought up to the tiny details and were used for years without any problems, did not want to work smoothly with the new plots. The logic of movable and unmovable parts began to conflict.

Plots have different scales; usually the scales are positioned near the plotting area but from time to time you want to change their position a bit. For example, to move a scale slightly aside from the main plotting area or to switch it to the opposite side of this area. In the old programs with all the unmovable parts such actions ware organized via the tuning form: users could call the tuning form of a scale and type in some parameters, for example, the distance between the scale and the plotting area, which would change the position of this scale. With the new movable / resizable plotting areas I received a strange situation: the moving / resizing of the plotting areas could be done by a mouse; this would automatically change the positions and sizes of the related scales in an appropriate way, but the individual relocation of the scales was still possible only via the tuning forms and it looked very awkward. If a plot with all its scales can be moved around, why the scales themselves cannot be moved in the same easy way?

**Step 2 - scales**. Certainly, I designed this next change, but it was not simply a transformation of another graphical object into movable. Plots and scales are strongly linked objects with the type of "parent - children" relation. So now there were the main objects (plots), which could be moved synchronously with all their related parts, but at the same time some of those parts (scales) could be involved in individual movements. That required some type of identification for the objects, involved in moving / resizing; this identification system has to guarantee the correctness of synchronous and individual movements for the objects with any type of relations between the parts. It had to be not the new type of identification for each new class of objects, but the general solution for any classes, which will be designed in the future and involved in different types of movements. Such identification system was designed; it is described in the chapter *Complex objects*.

At the end of the second step I had movable / resizable plotting areas and the scales, which could be moved independently; at the same time the scales automatically adjusted their sizes, when the plotting area was moved or resized. Excellent? Well, yes, but…

**<u>Step 3 - comments</u>**. With the plotting areas and scales now movable in any possible way my attention turned to another huge problem for which no one knew a good solution before: good positioning of the comments along the graphs, calculated by a program. In all the numerous scientific and engineering applications the plotting areas are better or worse positioned by the designers; all these plots are unmovable, so the designers simply decide about their positions, and that is all. The results, which are shown in those areas by the graphs, often need some comments, but the exact positions of graphs are calculated throughout the work of applications, so there is no way to determine beforehand the good position for these comments. How this problem was solved before? In the same way as positioning of the old scales. When an application produces some results in the form of the plots, then users are given a chance to position some comments next to these lines by, for example, typing some positioning coefficients in the tuning form. In such a way it was done before, but now it looks very strange if the plots and scales can be moved easily around the screen, but the comments to these objects are positioned and changed in some archaic way. The comments have to be moved exactly in the same easy way as plotting areas and scales – by a mouse. In addition, the comments need to be not only moved forward but also rotated for better placement along the arbitrary lines.

I think that now it is obvious in which direction the scientific applications started to change after a single element – plotting areas – was turned into movable / resizable. There is a law that the reliability of the whole system cannot be higher than this characteristic for the lesser reliable part. The similar rule, translated into the world of programming, declares that the flexibility of the system cannot be higher than this characteristic for any part. So, to make the plots entirely movable, all the comments were turned into movable.

The comments were added into the chain of relations plots – scales – comments. The already designed system of identification was applied to the new links; it did not require any changes at all, which proved better than anything else its reliability.

**<u>Step 4 - controls</u>**. Even if you turn all the parts of the plots (areas, scales, and comments) into movable and resizable, but leave in the same form (dialogue) anything unmovable, then the users would always bump on this hillock, even on a small one. The inner area of many applications is populated both with the graphical objects and controls, so controls and groups of controls must be turned into movable / resizable. I made this change, received the new type of scientific application, and that was the new paradigm – ***user-driven applications***.

**<u>Step 5 – tuning forms</u>**. The transformation of the main part of my scientific applications from rigid construction into user-driven was definitely not the end of the road. Everything became movable / resizable in the main form of an application, and it looked fine. The positions and sizes of all the elements are now easily changed with a mouse, so each user can rearrange the view of an application to whatever he wants at any moment. But positions and sizes are not the only parameters, on which the view depends. All the plots, scales, and comments have their parameters of visualization; there are so many of these parameters that they are usually changed in the auxiliary tuning forms. After playing with the improved plots for some time, I decided to change some of the visual parameters, opened one of the tuning forms and tried to move the elements there. It was the most natural thing to do after all those moving / resizing in the main form, but it did not work: all the tuning forms were still old-fashioned. Certainly, they were old-fashioned; they were the same tuning forms, which I used all the time without any idea that there was something wrong. My amusement at the first moment, when I tried and couldn't move the elements in these tuning forms, showed me immediately what I had to do. I began to redesign all the tuning forms on the basis of movable / resizable elements.

It was not some kind of whim; there was one more thing, which required redesigning of those tuning forms according to the rules of user-driven applications. The tuning forms for all kinds of scales include a sample of a number or a text, which shows the positioning of all numbers (texts) along the scale, which is currently under tuning. This sample is movable; its movement and rotation are copied by all the numbers (texts) along the scale. When you have any movable element inside the tuning form, then the mentioned conflict between the movable and unmovable elements immediately jumps into this form, and you have to redesign it.

The next comment came from my colleagues, who began to use more and more of the user-driven applications in parallel with other programs, with which they work for years and which were designed in the old style. The colleagues caught themselves on trying to move the objects in those programs exactly in the same way, as they move them in my applications. Certainly, it does not work (all other programs are designed in an old way), but each time it happens like a mini shock to the users: why this object does not want to move, when every other is movable? It takes some part of a second to understand the reason and to remember that this is a different type of program, in which everything is still fixed…

Now I found myself in the similar situation. Several months ago I was working on some part of the program, which had to accompany another article. I was checking for many hours different parts of an application in order to find any possible mistakes; I was switching from one form to another and while I was in the main form, I tried to move it simply by some inner point, as I move any graphical object. Certainly, the form did not move in such a way, and my first thought was about some mistake, which I had to find. The next instant I remembered that that was not one of my graphical objects, but the form itself, which could be moved only by the caption bar. After 20 years of working with Windows I tried to move a window by its inner point! When it happened again, I started to think that this is not the problem of overworking, but the demonstration of the basic rule for the user-driven applications. You quickly got used to the movable / resizable objects and expect that all the objects in all the applications have to work in such a way, because it is the most natural way of dealing with all of them.

Movable / resizable objects do not want to coexist with the fixed objects in any way. **If there are movable objects in an application, they will insist, require, and demand that all other objects around them have to be turned into movable / resizable.** This requirement is not limited by the form, in which these movable objects are placed, but spreads on all the related forms, on all the forms of the same application, on other applications… The expanding universe of the movable objects. (Only in this case I know exactly, when and how the Big Bang happened. And it definitely took not the fractions of a second…)

**Step 6 - unexpected**. Whatever is mentioned several lines before is definitely not the new thing for me: colleagues mentioned with laugh their attempts to move windows by the inner points (not by their title bars!); several times I had tried to do the same. Then several months ago, when I have already finished one of the articles and was rereading the text once more before publishing it on the web, the idea came into my head that an attempt to move the forms by any inner point was not a problem at all. If the logic of applications demands the moving of the forms by any inner point, then it can be done in seconds by adding several primitive lines. It does not even require the use of mover or any special class, so my algorithm has nothing to do with this solution. Simply change the location of the form synchronously with the mouse move, if the form was pressed by a mouse at any inner point. To do such a thing, you need a very simple addition into the code of the same three mouse events: `MouseDown`, `MouseMove`, and `MouseUp`. The code is so primitive!

At the moment this moving of a form by any inner point is added to three forms of the accompanying Demo application:

- **Form_PlotAnalogue.cs** (**figure 10.3**) was discussed in the chapter *Complex objects*, but there I did not attract the attention to this specific movement of a form.
- **Form_Function.cs** is discussed in the next section of the current chapter (**figure 16.3**).
- **Form_PlotsVariety.cs** is discussed further on in the chapter *Data visualization* (**figure 17.1**).

There is one common feature between these forms: they are populated with a lot of different plots. All those plots can be moved around by pressing them at any inner point, so it is absolutely natural if the forms can be moved by pressing them at any inner point. The plots, used in these forms, are absolutely different, but the technique of moving the forms by inner points is identical; the next pieces of code are from the **Form_PlotAnalogue.cs**.

Two additional fields are used to organize such movement; one of them is the Boolean flag to inform about the form being currently in movement or not; another is the shift between the mouse location and the location of the form.

```
bool bFormInMove = false;
Size sizeMouseShift;
```

When you press the left button anywhere in the form and the mover does not catch any object, then the shift from the mouse position to the location of the form is calculated; the flag is also set to inform that the form is in move.

```
private void OnMouseDown (object sender, MouseEventArgs e)
{
    ptMouse_Down = e .Location;
    if (!mover .Catch (e .Location, e .Button, bShowAngle))
    {
        if (e .Button == MouseButtons .Left)
        {
            bFormInMove = true;
            Point ptScreen = PointToScreen (ptMouse_Down);
            sizeMouseShift = new Size (ptScreen .X - Location .X,
                                       ptScreen .Y - Location .Y);
        }
    }
    ContextMenuStrip = null;
}
```

When the mouse is moved, there can be three different situations:

- Some object was caught previously and is now moved by the mouse. In this situation the `mover.Move()` method returns `true`.

- The form is caught by the mouse and moved. The introduced flag `bFormInMove` must inform about it.
- Nothing was grabbed by the mouse.

If the mouse is moved, while no object is caught by the mover, and the flag shows that the form is in move, then that previously calculated shift `sizeMouseShift` between the mouse position and location of the form can be used to define the new location.

```
private void OnMouseMove (object sender, MouseEventArgs e)
{
    if (mover .Move (e .Location))
    {
        Invalidate ();
    }
    else
    {
        if (bFormInMove)
        {
            Location = PointToScreen (e .Location) - sizeMouseShift;
        }
    }
}
```

When the mouse is released, then there is no sense in any checks of whether the form was in move or not; simply apply the **`false`** value to the `bFormInMove` flag.

```
private void OnMouseUp (object sender, MouseEventArgs e)
{
    bFormInMove = false;
```

You can try this technique on any of your forms or in other forms of this Demo application and decide for yourself, if such an addition to moving of the standard forms can be useful. For 20 and plus years I was moving windows only by the caption bar; then the work with the user-driven applications pushed me into thinking that there can be a very useful addition even for such a well known thing. That is how user-driven applications affect all the things around.

## *The Plot class*

File: **Form_Functions.cs**
Menu position: *Applications – Functions analyser*

On my way along the steps to total movability of objects described a bit earlier, I have designed several classes of movable / resizable plots. The class, which is the last in this serious, is called `Plot`; this is the class, which I use now in all my programs. Let us look into the design of this class and how it is used in the real applications.

Applications for science and engineering can be very different with specific requirements for each particular area and very special demands from one group of scientists or another. At the same time, there are many common requirements for plotting, with which nearly all the scientist agree. The **Form_Functions.cs** includes such type of plotting, which is standard for a lot of applications

A `Plot` object consists of the main rectangular plotting area, which can be associated with an arbitrary number of vertical and horizontal scales (objects of the `Scale` class). At least one scale for each direction must exist, but the visibility of any scale is not mandatory, so you can see the plots without scales. Never mind, somewhere there is at least one horizontal and one vertical scale, because the scales determine the physical ranges of the plotting area. The main area and each of the scales can be associated with an arbitrary number of comments. All the comments belong to the `CommentToRect` class.

```
public class Plot : GraphicalObject
{
    RectCorners rectcorners;
    Underlayer underlayer;
    List<Scale> scalesHor = new List<Scale> ();
    List<Scale> scalesVer = new List<Scale> ();
    List<CommentToRect> comments = new List<CommentToRect> ();
```

It is not a problem at all to rotate the plotting area and I have demonstrated the rotation of different objects, including the rectangles (chapter *Rotation*), but throughout all the years of my work I haven't met a single scientist or engineer, who would ask to rotate the plotting area in a program, so I decided that the `Plot` class can live without it.

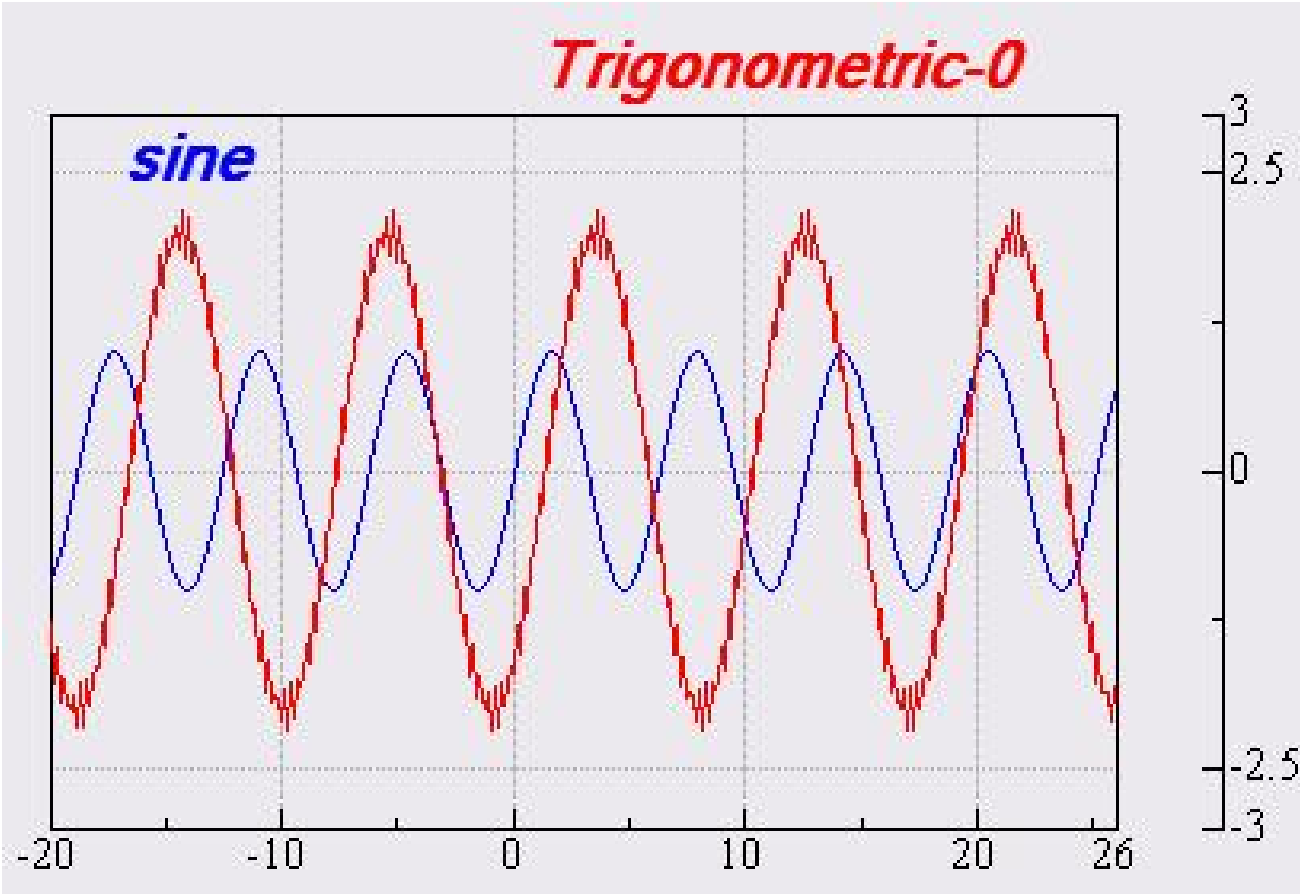

**Fig.16.1** A `Plot` object

The **Form_Functions.cs** allows to analyse a behaviour of several predefined functions and any other functions which can be defined by simply typing their text either as an Y(x) function or as a pair of functions {X(r), Y(r)}. **Figure 16.1** demonstrates a plot with two functions:

- $Y_1 = \sin(x)$
- $Y_2 = 2*\sin(0.7x – 0.9) + 0.2\sin (20x)$

This is a very simple plotting area with one horizontal scale, one vertical scale, and two comments with the names of the functions. This picture is included here not to remind you about the view of the standard sine, but only for better explanation of some of the problems and their solutions, while turning a plotting area with all possible auxiliary parts (scales and comments) into fully movable and resizable.

The main line of the horizontal scale at **figure 16.1** is placed exactly on the border of the main plotting area; such positioning of a scale is, maybe, the most common. The vertical scale is moved slightly apart from the plotting area, which is also not a rare situation. In general, scales can be positioned anywhere in relation to the main plotting area: they can be placed somewhere aside of the main rectangle, or they can overlap with it partly, if they are on the border, or they can even be placed totally atop the plotting area. **Figure 16.2** gives even better explanation of all these possibilities in positioning the scales, because it uses the scheme, which eliminates some drawing details, but shows exactly the areas, occupied by the main plotting area and the scales. This scheme is easily obtained with the help of the **Form_PlotAnalogue.cs**, which was described in the chapter *Complex objects*. The scheme gives a good understanding of some problems in design, but in some cases the situation with the real plots and scales (**figure 16.1**) is even worse, and the full problem can be better explained by comparison of these two pictures.

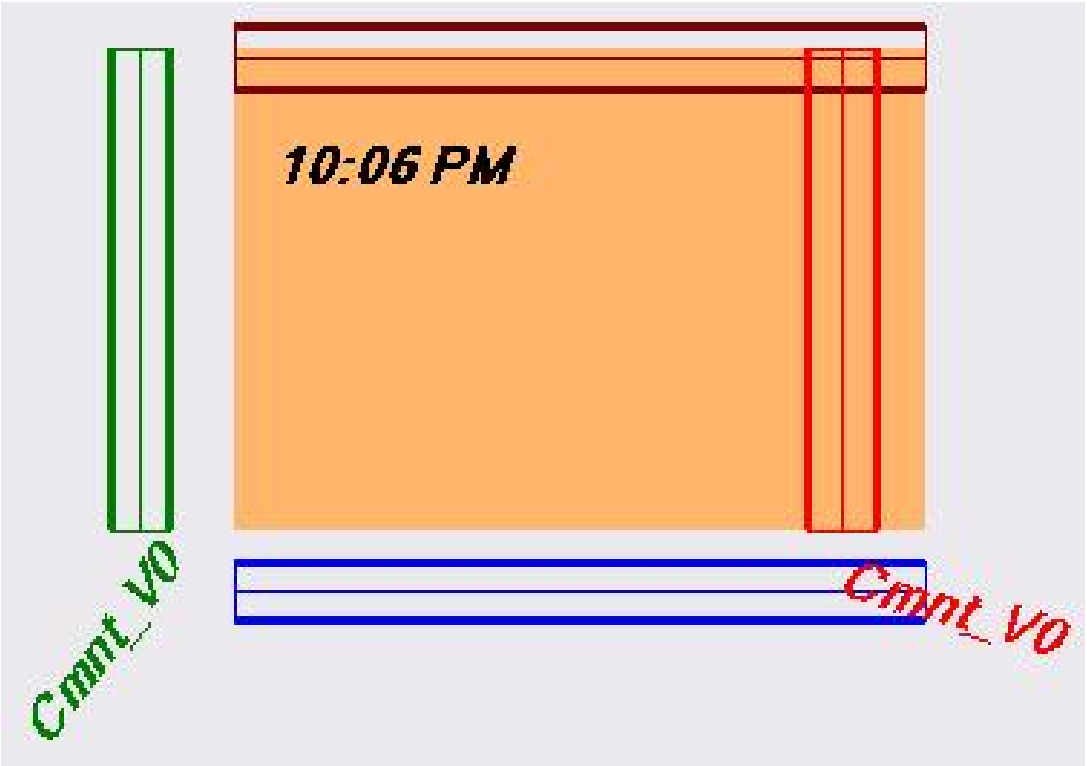

**Fig.16.2** Plotting area and scales are represented on the scheme by the movable / resizable rectangles

A `Plot` object is a complex one. As any complex object consisting of the individually movable parts, a plot must be registered in the mover's queue with the help of its `IntoMover()` method. A complex object never has a single cover for all its parts, but each individually movable part has the cover of its own; these covers must be mentioned in correct order in that method of the complex object.

The main plotting area is a simple rectangle, which can be moved by any inner point and resized by any border point; border is covered by the rectangular nodes along the sides with the additional circular nodes in the corners. These circular nodes allow to change the size of the plotting area simultaneously in two directions instead of changing width and height one after another. Such covers were described at the beginning of the book in the chapter *Rectangles*; see **Form_Rectangles_StandardCase.cs**, **figure 3.1**.

The scales are the complex objects themselves, because in addition to their main part (line, ticks, and numbers) they may have an arbitrary number of individually movable comments. A scale is movable, but its individual movement is restricted to one direction: a scale can be moved only athwart to its main line. A scale cannot be independently resized (with a mouse); its length can be changed only synchronously and forcedly with the change of the main plotting area. A cover of any scale is organized in the form of a rectangle, overlapping the main line, the ticks, and the numbers.

Any comment, associated with any scale or a main plotting area, is also represented by a non-resizable rectangle with the size, depending on the text of this comment and the used font.

The covers of all the parts are simple; the order of the components is obvious: all the scales must precede the main plotting area in the mover's queue; the comments must precede their parent (plotting area or scale) in the same queue. No problems at all? Unfortunately, there is one.

The length of the "scales" in the PlotAnalogue class is equal to the corresponding dimension of the plotting area (**figure 16.2**). Even when such a "scale" is placed along the side of a plotting area (upper scale on the figure), some part of the corner node of the plotting area still looks out from under such a "scale", so it is possible to resize the area at **figure 16.2** by the top left and top right corners. Maybe not very convenient as only the small part of the corner node is still available and can be sensed by a mouse, but at least there is a chance. The schematic scale from the **Form_PlotAnalogue.cs** always has the same size as the appropriate dimension of the plotting area. The scheme is always a simplification of the real thing; the scheme at **figure 16.2** ignores some important details of calculating a cover (a sensitive rectangle) for real scales. The details are small, but it is well known that the devil is in details.

When a real scale is placed along the border of the plotting area, it leaves nearly no chances for resizing of this area by any of the covered corners. Look at the scale at the bottom of **figure 16.1**. The sensitive area of any scale includes the main line, the ticks, and the full area of numbers along the scale. The numbers can be positioned in different ways along the scale, but the sensitive area of a scale is always a single rectangle. The upper and lower borders of this rectangle are determined by the upper coordinate of the ticks and the lower coordinate of the numbers; there are no problems with these borders. The most interesting here are the left and the right borders of the cover of a scale. The outer coordinates of this rectangle are determined by the farthest of two points: the end of the main line and the border point of the last number. Two end values (numbers) of the bottom scale at **figure 16.1** are going outside the main line of the scale, so the left and the right borders of the cover are determined by these two numbers. Unfortunately, they stay outside the main plotting area far enough to overlap, by their cover, the corner nodes of the main plotting area. In such situation the whole lower border and both lower corner nodes of the main area are overlapped by the scale; thus the plotting area cannot be enlarged down by any node.

For some time this situation looked like a deadlock in the design of plots: I had a nice working object, which looked absolutely fine in all situations, except when the scale was placed just on border. But such placement of a scale is so common (it is used more often than any other) that this inconvenience marred otherwise excellent solution. It would be nice to divide the cover of the main area into two parts (corners separately from borders and the main area) and to register them separately in the mover's queue with the scales registered in between. It would be just a solution, but… The algorithm does not allow to divide a single cover into two parts and to register them separately. Whether it is a simple cover consisting of a single node or a complex cover of many nodes, it does not matter; any cover(!) must be registered as a single entity.

Let me try to explain this problem once more, because the found solution for this problem can be useful in other cases. Suppose that I have a complex object, consisting of the rectangular area (the Underlayer class) and a set of scales (the Scale class). Each part is movable, so each part has its own cover. The whole complex object must be registered in the mover's queue; for this all the parts of the complex object must be registered in this queue.

Scales must be moved and painted atop the rectangular area, so the parts must be registered in such an order: all scales, then the rectangular area. Scales are often positioned in such a way that their covers overlap the significant (for moving) parts of the cover of the main rectangle; when it happens, the rectangle cannot be resized. It would be nice to divide the cover of rectangle into two parts and then register everything in such an order: first part of the cover for rectangle, then scales, then the second part of the cover. It would be nice, but absolutely impossible: the cover of an object must be registered as an entity.

In order to produce any even relatively good solution, I tried to use some tricks, like making holes in the cover of a scale, so that a scale would not close the corner nodes. But such solution had some other negative effects, as these holes were in the cover of a scale regardless of the position of a scale. Even with the scale staying apart from that special place on the border, the cover of a scale still had holes. It was not a catastrophe, but it was definitely a problem, which bothered me for a long time. Then I came to an excellent, elegant, and easy solution, which solved the whole problem. As often, when you think out such a solution, you cannot understand why you did not see it from the very first moment. Because it is so obvious.

A complex object consists of several parts; each part has its own cover. When a complex object is registered with the mover, those covers of the parts are placed into the mover's queue one after another. It happened so that the cover for one of the parts partly overlaps the cover of another part. Unfortunately, only the part of the overlapped cover can provide the needed movement, but this part is blocked. How to let that overlapped part of the cover work? Well, there is no way to let that part of the cover work, but what about adding something auxiliary with the identical behaviour and let this new part work ahead of everything else? What about adding into the Plot class another part, always invisible but with a cover that would duplicate those four nodes in the corners and do nothing else, but allow to resize the main rectangle at any moment? The invisible part, which is always placed ahead of all the scales in the mover's queue.

This primitive object belongs to the RectCorners class. When initiated, which is done in the Plot constructor, it gets the rectangle of the main plotting area. The cover of the RectCorners object consists of the four circular nodes in the corners of this rectangle.

```
public override void DefineCover ()
{
   int radius = 6;
   cover = new Cover (new CoverNode [4] {
     new CoverNode (0, new PointF (rc.Left, rc.Top), radius, Cursors.SizeNWSE),
     new CoverNode (1, new PointF (rc.Right, rc.Top), radius, Cursors.SizeNESW),
     new CoverNode (2, new PointF (rc.Right, rc.Bottom), radius, Cursors.SizeNWSE),
     new CoverNode (3, new PointF (rc.Left, rc.Bottom), radius, Cursors.SizeNESW)}
                         );
}
```

The constructor of the `RectCorners` class has two parameters: the first one is the rectangular area; the second one is the method, by which an object informs its "parent" (it can be a `Plot` object or any other), when any of its nodes is moved.

```
public RectCorners (Rectangle rect, Delegate_Rect inform)
```

Through this method, passed as a parameter, the `RectCorners` object sends the new rectangle to its "parent". Everything else is organized as was shown with the `PlotAnalogue` class: via the `InformRelatedObjects()` method any change of the main plotting area is reported to all the associated parts (scales and comments). In case of the `Plot` class, this change is reported also to the `RectCorners` part, so there is a two way reporting.

Now, when all the parts of the `Plot` object are organized, an object can be registered in the mover's queue. You can see from the `Plot.IntoMover()` method, that the `RectCorners` object is registered (inserted) ahead of all the scales, so there is no problem of resizing the main area "by the corners", even if the scale closes the real corners.

```
new public void IntoMover (Mover mover, int iPos)
{
    mover .Insert (iPos, this);
    foreach (CommentToRect comment in Comments)
    {
        if (comment .Visible)
        {
            mover .Insert (iPos, comment);
        }
    }
    foreach (Scale scale in HorScales)
    {
        if (scale .Visible)
        {
            scale .IntoMover (mover, iPos);
        }
    }
    foreach (Scale scale in VerScales)
    {
        if (scale .Visible)
        {
            scale .IntoMover (mover, iPos);
        }
    }
    mover.Insert (iPos, rectcorners);
}
```

There can be a lot of different `Plot` objects on the screen. **Figure 16.3** demonstrates the **Form_Functions.cs** with a lot of plots; I use this combination of plots as a reminder for myself about the predefined functions that were included into this application. It can look like an artificial example with too many plots, but I watched not once, how the scientists in our department of mathematical modeling use some of the programs, designed on the same basis of movable objects. Those scientists often have more plots on the screen, than you can see at this figure.

The plots from **figure 16.3** are simple enough: each plot (with one exception) has a single function, so each of the plotting areas has one horizontal and one vertical scale; there is also only one comment in each plot – the name of a function.

The coexistence of many plotting areas on the screen is not the only source of having a lot of different movable elements. In real applications, the use of multiple scales is not an unusual thing. It is a rare situation to see more than one horizontal scale, but the number of vertical scales can easily go up to four or five. This happens in cases, when scientists need to compare the behaviour of different functions or the distribution of some components, for example, different gases along the same time period. The values of those functions can be so far from each other that each of them requires its personal vertical scale with its specific range. The use of many scales significantly increases the number of different elements on the screen.

With several functions, shown in one plotting area, there is often a requirement for some comments to distinguish and identify these functions; as a result, there will be several comments, associated with the plotting area. The same situation happens, when you need to comment different points of even a single function. So the situation, when you have multiple comments in the plotting area, is not rare.

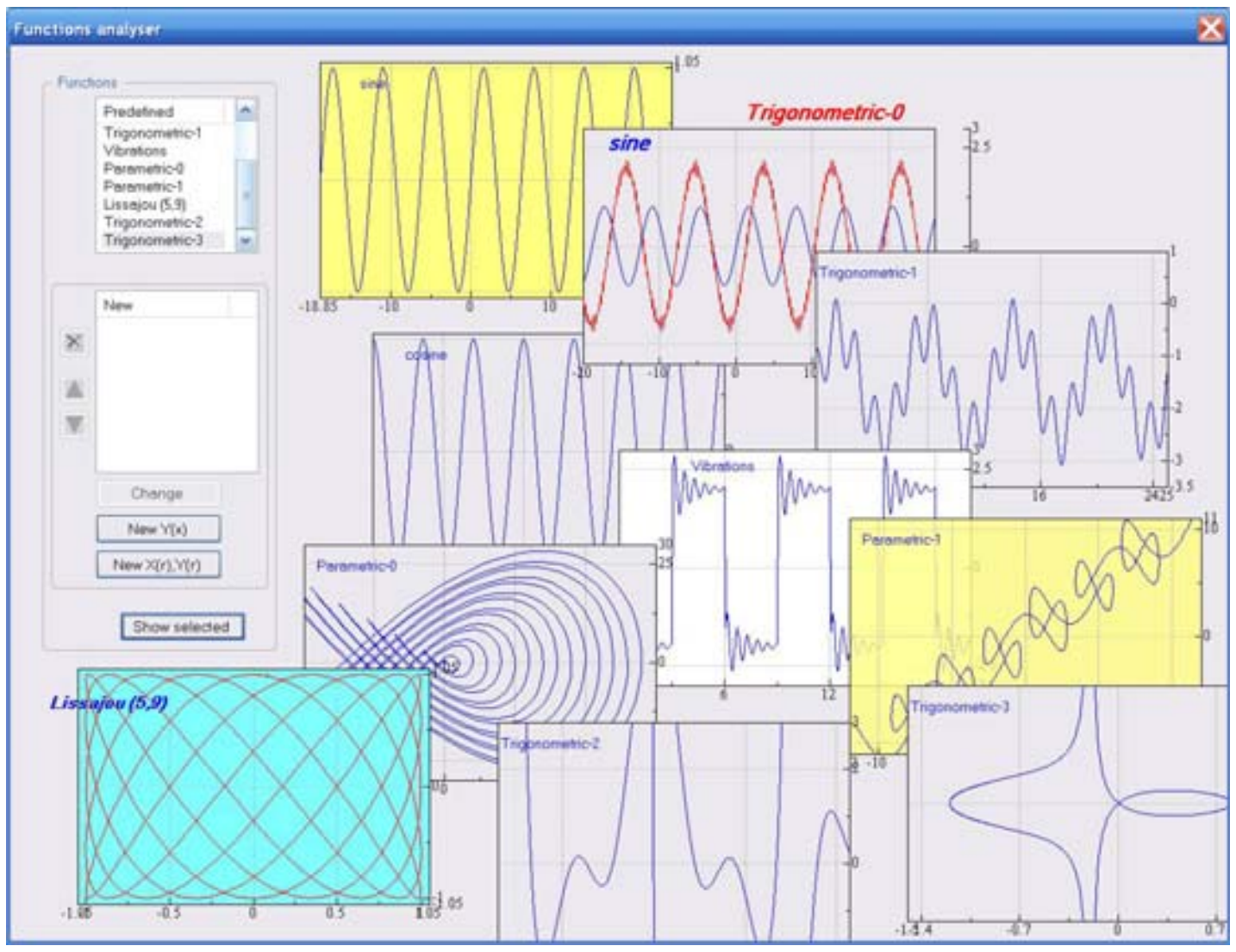

**Fig.16.3** This collection of plots reminds me about the predefined functions, which are included into application

Consider also a scale from **figure 16.4**. This is a standard view of a time scale used to show some data on geology or paleontology. There is a standard part of scale consisting of the main line, ticks, and sub-ticks along the line, plus the numbers (years). But there is also a lot of additional information, like the names of periods and eras, which make the data on a plot much more informative. This is the standard scale to be used along the area of a plot, so this scale is going to stretch or shrink synchronously with the change of the plotting area. All those words along the scale are the comments, which are repositioned automatically with the change of the length of the scale, but they can be also moved and rotated individually. I do not think that any specialist is going to change the order of periods, but inside the standard and well known order the words can be placed individually.

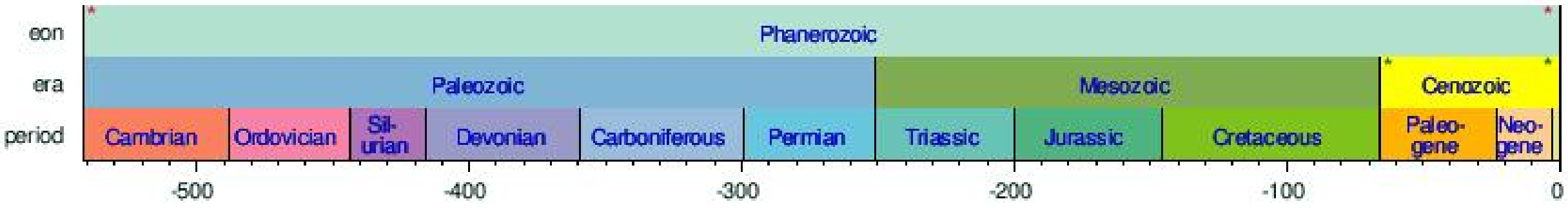

**Fig.16.4** A time scale, used in geology or paleontology, often requires a lot of different comments

I mentioned some situations, when the number of comments even for a single plot or scale can be big enough. All these examples demonstrate that the number of screen elements can vary from few to really big numbers. And this sets two big requirements for the `Plot` class:

- Changing the movability of elements.
- Changing the visibility of elements.

## Movability ON and OFF

When I started to work on the problem of movability of the screen objects, there was only one question: "How to make any object movable?" For couple of years I never paid too much attention to the problem of returning some movable objects back into unmovable, because I never considered it to be a problem. Let everything become movable; if you do not want to move one object or another, then do not do it. It is really simple: no object or its part is going to move around the screen by its own wish; if you do not want to move an object, then do not touch it. I had to change my view on the problem and to pay much more attention to it, when scientists began to use the applications with a lot of movable objects and started to bump again and again into the problem of accidental moving of the wrong object. As a result, scientists required an easy to use mechanism of switching any object from one mode of movability to another at any moment, while an application is running.

There is always a simple (I would say the easiest) way of making any object unmovable: excuse it from the mover's queue. However, this solution has some negative sides.

1. When the change of movability requires not the primitive switch between being movable and unmovable, but requires the switch between other variations of movability, then the exclusion from the mover's queue does not work at all.

2. As you can find in a lot of forms of the accompanying application, and **Form_Functions.cs** is only one of them, the decision on opening one or another menu depends on the mover's sensing of one or another type (class) of objects. If you exclude an object from the mover's queue, this possibility is over; the mover deals only with the elements that are in its queue. In such a way mover can help to exclude an object from being movable, but after it the mover will never help you to return the same object back into its queue. You can add another technique of checking the request for menu for such objects; it can be based on comparison of the clicked point and the screen areas of all the objects, but this means the significant complication of the code.

This list of possible problems must give you an idea of the solution: an object must be still kept in the mover's queue, but the cover of this object must be changed.

Before looking into two possibilities of changing a cover, I have to remind you of one example with the change of movability, which was already mentioned in this book. It was in the case of the regular polygons (the `RegPoly_Variants` class) in the **Form_RegPoly_Variants.cs** (**figure 6.1**). In that case, the free movement of any polygon around the screen can be limited to movement only along one or another direction. This is a very special case of movements limitation, which can be organized without any change of cover, but simply by a slight change of the `Move()` method of the used class. This is an easy solution in a very special case, so let us remember it as a possibility, but nothing else. In general, the demand for change of restrictions on movements requires the change of cover.

There are two main ways to change the cover of an object in such a way that it affects the movability.

- To change the design of cover.
- To keep the same cover, but change the behaviour of its nodes.

The first way can be seen in the same example of the `RegPoly_Variants` class, where three different types of resizing are provided by three different types of cover. Another example, in which the change of movability is achieved by the change of cover, is the `TrackbarC` class, which is used and demonstrated in the **Form_Trackbars.cs** (**figure 10.4**). But in the classes, which are involved in development of scientific / engineering plotting, the second way of changing covers is used.

At the beginning of the book, in the section *Algorithm*, I wrote about one of the parameters of all the nodes; this parameter is described by the `Behaviour` enumeration. When a node is used as a starting point for some movement, then this parameter of the node is set to `Behaviour.Moveable`; the translation of the movement of the invisible node into the visible movement of an object is described by the `MoveNode()` method of this object (class). If you need to turn an object into unmovable, it is enough to change the behavioral parameter of this node into the `Behaviour.Frozen`. Such node is recognized by the mover as any other, but the `MoveNode()` method is not called.

Suppose that you have placed a scale somewhere next to the plotting area in the **Form_Functions.cs** and you do not want to move this scale accidentally. In general, I would include the command of fixing any object into the menu, which is called on this object, but the procedure for the scales is slightly different. Fixing of scales was included into several of my scientific applications and I was told by the users not once that nobody wanted the individual fixing for each scale. The users prefer to place all the scales at the proper places around the plotting area and fix all the scales with one command, so this command *Fix scales* can be found in the menu, opened on the plotting area. Certainly, when you need to unfix the scales, it is done by a single command in the same menu. By clicking any of these commands, the `Movable` property of all the scales, associated with the pressed plotting area, is turned into its opposite value (`true / false`).

The `Movable` property belongs to the base class `GraphicalObject`, from which all other classes of movable objects are derived. On setting the value of the `Movable` property, the `DefineCover()` method of the involved class is called. Here is this method for the `Scale` class.

```
public override void DefineCover ()
{
    CoverNode [] nodes = new CoverNode [1];
    if (Movable)
    {
        Cursor cursor = (LineDirection == LineDir .Hor) ? Cursors .SizeNS
                                                        : Cursors .SizeWE;
        nodes [0] = new CoverNode (0, Frame, cursor);
    }
    else
    {
        nodes [0] = new CoverNode (0, Frame, Behaviour .Frozen, Cursors .Hand);
```

```
        }
        cover = new Cover (nodes);
        cover .SetClearance (false);
    }
```

The cover of a scale always consists of a single polygonal (rectangular) node. As was described before, on fixing the scale its behavioural parameter is turned into `Behaviour.Frozen`. There is also some change of the cursor. When the scale is movable, the cursor above it signals about the possibility of such movement by showing the direction, in which it can be moved. For horizontal scales it is a bi-directional arrow pointing N – S; for vertical scales it is a similar arrow pointing W – E. On turning the scale into unmovable, the cursor over this scale is switched into the `Cursors.Hand` shape.

Other objects that require the switch of movability are the comments, but for them it is done on an individual basis, so whenever you need to change the movability of any comment, you have to call menu on this particular comment. Exactly in the same way as with the scales, the change of the movability of a comment will eventually call the `TextMR.DefineCover()` method, as the `CommentToRect` class is derived from the `TextMR` class and inherits that method.

```
public override void DefineCover ()
{
   Point[] pts = Auxi_Geometry.TextGeometry (nW, nH, angle, ptMiddle, TextBasis.M);
   PointF [] ptCorners = new PointF [] {(PointF) (pts [0]), (PointF) (pts [2]),
                                        (PointF) (pts [8]), (PointF) (pts [6]) };
   cover = new Cover (new CoverNode [] { new CoverNode (0, ptCorners) });
   if (!Movable)
   {
       cover .SetNodeBehaviourCursor (0, Behaviour .Frozen, Cursors .Default);
   }
}
```

The cover for any comment consists of a single rectangular node covering the text. Knowing the sizes of the text, its angle, and middle point, I get an array of basic points for the rectangle, occupied by the text. The `Auxi_Geometry.TextGeometry()` method returns an array of nine points; four of them describe the corners of the rectangular area, which is used for a node. When a comment is fixed, the behavioural parameter of this node is turned into `Behaviour.Frozen`, the cursor above it is turned into the `Cursors.Default`. This is the difference from the unmovable scale.

Though I use this change of behaviour and cursor for a long time and in many applications, I cannot come to the final decision about the shape of a cursor over the unmovable objects. Personally, I would prefer to change it into the `Cursors.Default`. I do not need any visual indication or any other reminder about the possibility of doing something even with the temporarily unmovable objects. I know that in all my applications ALL the objects are movable; they can be temporarily turned into unmovable, but there is always a possibility of calling a context menu on them, as on any other object. When you work with the user-driven applications for some time, you come to the same understanding of the situation and do not need any kind of reminder any more. On the other hand, there are always users, who are not familiar with the user-driven applications (like the majority of readers of this book!); for them such a reminder is very helpful. Maybe this would be the best solution for everyone: it is possible (and very easy) to make this default cursor a parameter of the class, so that the default cursor over the temporarily unmovable objects can be decided by each user. Just one more tunable parameter.

## Visibility of plotting areas

Visibility of objects is another big issue in the scientific / engineering applications; the importance of it is also related to the possibility of getting too many objects and elements on the screen. In the **Form_Functions.cs** (**figure 16.3**) users have to deal with the system of plots, scales, and comments; the visibility of these objects is regulated in different ways.

In the old (or currently used !?) applications the number of plotting areas in view is usually set at the moment of the design and is never changed. To have in a user-driven application any form with a predefined number of plots would a very rare situation. For example, in the accompanying Demo application the **Form_Function.cs** has two auxiliary forms for definition of functions; each of these forms has exactly one plot, because they would never need more than one. But this is a very special and, better to say, an exceptional case. In the user-driven scientific applications the number of plotting areas is decided by each user personally. The number of such areas is never limited, so the case of 10 plots (**figure 16.3**) is neither unique nor extraordinary; I have watched many times that scientists work with a significant number of plots, while trying to analyse some difficult situations in their experiments. Each plotting area has a significant number of tunable parameters; users spent some time on tuning each plot and do not want to lose the results. What to do, if you do not want to delete a plotting area, but do not want to keep it at the screen, because it is not needed now, but will be needed later?

The simplest solution is to change the default moving restrictions for all the objects and to move some of them across the borders; this was already explained in the chapter *Movement restrictions*. The moving restrictions are regulated by the mover's `Clipping` property. By default it is set to `Clipping.Visual` and allows to move any object only inside the visible part of a form. To be absolutely correct about this situation: when any object is caught, then it is the mouse cursor that cannot move outside the visible part of the form. But an object can be grabbed for moving by any inner point (this is the best organization of the moving process), so the caught object can be positioned arbitrary in relation to the mouse cursor. The most important thing is that the mouse cannot move outside of view; so the caught point of an object cannot go outside of the view either. You can grab an object very close to its border and move nearly the whole object out of view, but regardless of where you drop this object, some, maybe tiny part of it is still visible. By grabbing this part, you can return an object into full view. This is the way in which the `Clipping.Visual` setting prevents the losing of objects. Just a reminder: this is a default setting for all the movers.

An absolutely correct and easy way to enlarge the area of existence for objects without losing them is to set the mover's `Clipping` property to `Clipping.Safe`.

```
mover = new Mover (this);
mover .Clipping = Clipping .Safe;
```

After it any object can be moved across the right or bottom border of the form and left there. If the form is resizable, then all these objects are available at any moment, as the territories across these two borders can be seen by enlarging the form.

Another solution for hiding the plotting areas is to organize an additional `List<>` of temporarily hidden plots. For example, in the **Form_Functions.cs** you can organize one more `List`.

```
List<AreaOnScreen> areasHidden = new List<AreaOnScreen> ();
```

Add *Hide area* line to the `menuOnPlot`. When this command line is clicked, the `Click_miHideArea()` method must be called.

```
private void Click_miHideArea (object sender, EventArgs e)
{
    AreaOnScreen aos = areas [iAreaTouched];
    aos .SaveAreaIntoFunctions ();
    aos .CloseTuningForms ();
    areas .RemoveAt (iAreaTouched);

    aos .Plot .Visible = false;
    areasHidden .Add (aos);
    RenewMover ();
    Invalidate ();
}
```

The touched plotting area is moved from one `List` into another; at the same time its `Visible` property is set to `false`, so this area is not shown any more. Pay attention that in parallel with hiding a plot in such a way it is excluded from the mover's queue; this is done automatically by the `RenewMover()` method, so this method must be called.

There can be some variants for restoring those hidden areas back into view. The easiest way is to restore all of them without selection. Add the *Restore areas* line to the `menuOnEmpty`; on clicking this command, the `Click_miRestoreAreas()` method has to be called. As usual, when a set of visible and movable objects is changed, the `RenewMover()` method must be called.

```
private void Click_miRestoreAreas (object sender, EventArgs e)
{
    for (int i = areasHidden .Count - 1; i >= 0; i--)
    {
        areasHidden [i] .Plot .Visible = true;
        areas .Add (areasHidden [i]);
        areasHidden .RemoveAt (i);
    }
    RenewMover ();
    Invalidate ();
}
```

If you want to organize some selection among the previously hidden areas, then it will be more complicated, but the main idea of the whole mechanism is obvious and simple: the change of a single `Visible` parameter of any `Plot` object allows to hide or restore the plotting area.

Changing the visibility of plots by switching the `Visible` property can be explored in many different ways. In one of my scientific applications, users get a chance to organize any number of different `pages` with any number of plots on each page. Each page is like a **Form_Functions.cs**, only each page has an identification – the name, which a user declares, when he wants to organize the new page. Each page contains the unique set of plots, which user has organized and placed on it. No limitation on pages; no limitations inside any page; plots are easily moved and copied from one page to another. The switch between the pages is through the dynamically changing list of their names. The classical example of the user-driven scientific instrument.

Hiding and restoring the plots can be one of the major elements of a big scientific application, but the play with hiding and unveiling the scales and comments can be even more interesting (from the programming point of view) and has more variants. This mechanism is similar to what was explained in the case of the `ElasticGroup` class and its inner elements. The hiding and unveiling of some parts of the plots will be discussed a bit later, because it is totally based on the identification mechanism, from which I want to start. The identification system was discussed in the chapter *Complex objects*; let us look at it once more.

## Identification

Any command in the **Form_Functions.cs** is started through some line of one or another context menu. If you click the right button nearly anywhere inside the form and the distance between the points of the **`MouseDown`** and **`MouseUp`** events is small, then the chances are very high that one of the context menus will be opened. The only exception is a click at the border of any control inside the group, but only because I couldn't think out any need for a menu in such situation. Any menu in my applications is opened on releasing the right button, so here is that part of the `OnMouseUp()` method, which leads to the menu opening.

```
private void OnMouseUp (object sender, MouseEventArgs e)
{
    bFormInMove = false;
    ptMouse_Up = e .Location;
    double nDist = Auxi_Geometry .Distance (ptMouse_Down, ptMouse_Up);
    … …
    else if (e .Button == MouseButtons .Right)
    {
        if (mover .Release ())
        {
            if (nDist <= 3)
            {
                MenuSelection (mover .WasCaughtObject);
            }
        }
        else
        {
            if (nDist <= 3)
            {
                iAreaTouched = -1;
                ContextMenuStrip = menuOnEmpty;
            }
        }
    }
```

If a click happened somewhere on an empty place, then the `menuOnEmpty` is called; otherwise the `MenuSelection()` method is called. Method has a single parameter – the number of the released object in the mover's queue; by this number the class of the released object is easily identified. If it was an `ElasticGroup` object, then the `menuOnGroup` is called. Other options include menus for `Plot`, `Scale`, and `CommentToRect` classes, but before any of them is called, the pressed object must be fully identified by the `Identification()` method.

```
private void MenuSelection (int iInMover)
{
    GraphicalObject grobj = mover [iInMover] .Source;
```

```
            if (grobj is ElasticGroup)
            {
                groupPressed = grobj as ElasticGroup;
                ContextMenuStrip = menuOnGroup;
            }
            else
            {
                Identification (iInMover);
                if (iAreaTouched >= 0)
                {
                    if (grobj is Plot)
                    {
                        ContextMenuStrip = menuOnPlot;
                    }
                    else if (grobj is Scale)
                    {
                        ContextMenuStrip = menuOnScale;
                    }
                    else if (grobj is TextMR)
                    {
                        ContextMenuStrip = menuOnComment;
                    }
                }
            }
        }
```

For the `Plot` objects the identification is the simplest, as it is enough to find the order of the pressed plot (`iAreaTouched`) in the `List` of plots.

```
    private void Identification (int iInMover)
    {
        GraphicalObject grobj = mover [iInMover] .Source;
        long id = grobj .ID;
        iAreaTouched = -1;
        …
        if (grobj is Plot)
        {
            for (iAreaTouched = areas.Count - 1; iAreaTouched >= 0; iAreaTouched--)
            {
                if (id == areas [iAreaTouched] .Plot .ID)
                {
                    break;
                }
            }
        }
```

For the `Scale` class it is not enough to identify the scale, but also the number of the plotting area is needed, with which this scale is associated. The **Form_Functions.cs** has a slightly simplified version for this case, as any plot in this form has only one horizontal and one vertical scale. In general, there can be more than one scale of each type. Here is a version for a case, when any plot has a single horizontal scale, but might have an arbitrary number of vertical scales; this version differs in one line from the code in the **Form_Functions.cs**. The full identification for a scale includes:

- The type of scale; either `bVerScaleTouched` or `bHorScaleTouched` parameter is set to `true`.
- The order of the touched scale in the `List` of scales of detected type (`iScaleTouched`).
- The order of the plotting area, to which this scale belongs (`iAreaTouched`).

```
    private void Identification (int iInMover)
    {
        GraphicalObject grobj = mover [iInMover] .Source;
        long id = grobj .ID;
        … …
```

```
        else if (grobj is Scale)
        {
            for (iAreaTouched = areas.Count - 1; iAreaTouched >= 0; iAreaTouched--)
            {
                if ((iScaleTouched = areas [iAreaTouched].Plot.VerScaleOrder (id))
                                                                                >= 0)
                {
                    bVerScaleTouched = true;
                    break;
                }
                else if (areas [iAreaTouched] .Plot .HorScaleOrder (id) >= 0)
                {
                    bHorScaleTouched = true;
                    break;
                }
            }
        }
```

The identification of a `CommentToRect` object requires to determine even more parameters. This identification is not more complicated, but it is longer, so I am not going to include the code, but here is the list of parameters that must be set during this identification.

- The pressed comment can belong to the main plotting area, or to the vertical scale, or to the horizontal scale, so only one of three parameters (`bMainAreaCmntTouched`, `bVerScaleCmntTouched`, or `bHorScaleCmntTouched`) must be set to `true`.
- The main plotting area or any scale may have an arbitrary number of comments, so the order of the comment in the corresponding `List` must be determined (`iCmntTouched`). The particular `List` is determined by that parameter from the previous item, which got the `true` value.
- If a comment belongs to the scale, then the order of the "parent" scale in the corresponding List of scales (`iScaleTouched`) must be determined. If by the design of application any plot has only one scale of each type, then this value is always 0; otherwise it has to be determined.
- The order of the plot (`iAreaTouched`) must be determined. The plot can be a direct "parent" of the comment, or it can be a "grandparent" through the scale.

The identification of the clicked object is important for two things:

1. To set the view of the opened menu. This can include adding the check marks, disabling some of the lines, or even deleting them.
2. To use the parameters of identification for any command that is called from the opened menu.

**Figure 16.5** shows the `menuOnPlot`, opened on one of the plots in the **Form_Functions.cs**. Two disabled menu lines in the second group of commands inform that this plot has no hidden scales or hidden comments of the main area. The existence of the *Unfix scales* line means that the scales of this plot were fixed; without this information, it would be difficult to understand why the scales are currently unmovable. One more disabled line means that no plotting area was yet declared as a sample of visibility parameters; this possibility is discussed later in the section *Faster tuning of the plots*.

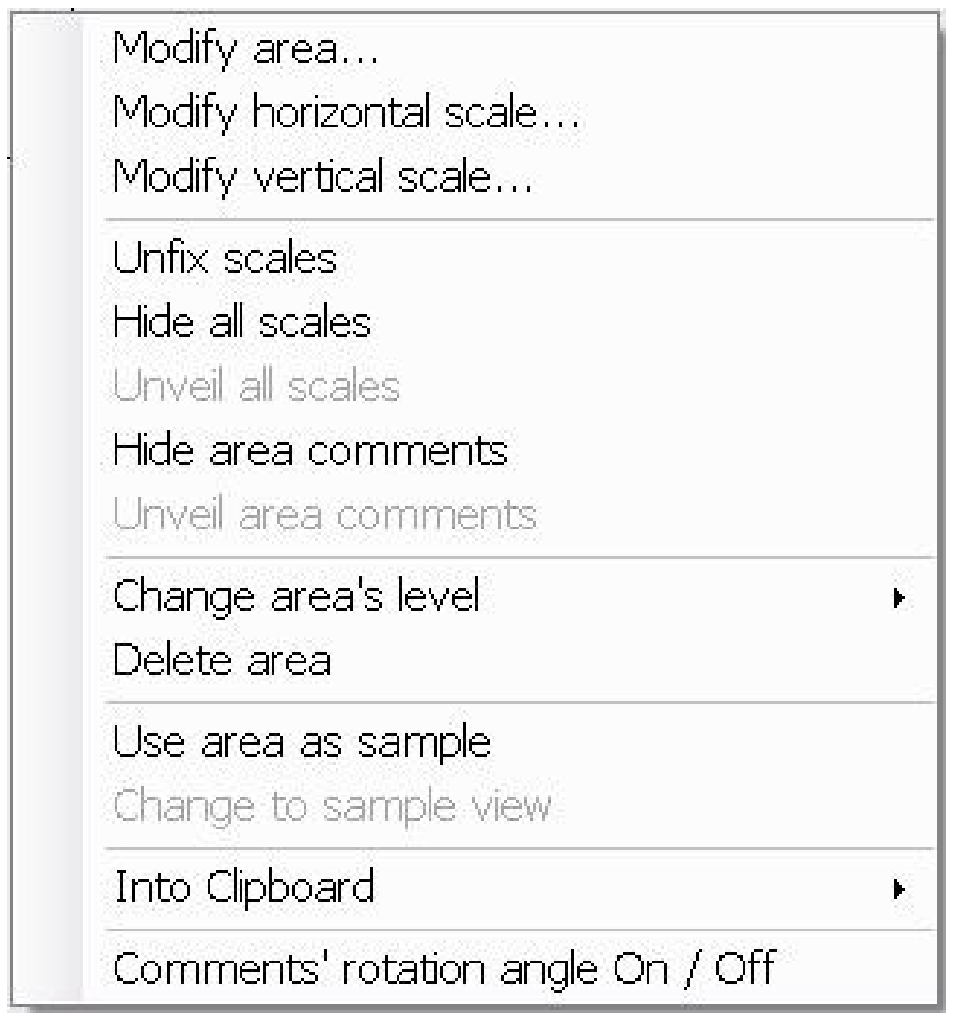

**Fig.16.5** Menu on a plot

## Visibility of scales and comments

There can be a lot of plots in a scientific application. Each plot may have an arbitrary number of scales; any scale and any plotting area can be associated with an unlimited number of comments. The screen area can be really overcrowded!

Nobody wants to keep in view the unneeded objects as they occupy the valuable screen area and distract the attention from the important data. In user-driven applications all the screen objects are movable; there are no restrictions on moving scales or comments, so it is possible to get rid of them by moving out of view across the borders, but it would be a very strange thing to do. If any comment or scale is not needed any more, then it can be deleted; the only exception is the last horizontal and the last vertical scale for each plotting area; the last scale for each direction can be only hidden, but not deleted.

Much more often is the situation, when comments, scales, and plots are not deleted, but only hidden to be restored later. These operations are based on using the `Visible` and `VisibleAsMember` properties, which any class inherits from its base `GraphicalObject` class. The use of those properties by the elements of scientific applications is very similar to what was explained in case of the groups (see subsection *Interesting sides of visibility* in the chapter *Groups of elements*).

Objects in the **Form_Functions.cs** or in any other application, using scientific plotting, organize the two level (plot – comment) and three level (plot – scale – comment) chains of relations. An element on any level can be hidden by direct command through the line of menu called on this particular element, but the hidden element cannot be restored back in exactly the same way.

I have shown a bit earlier that all the menus are called after the process of identification, which starts with clicking an object by the right button and getting the number of the clicked object in the mover's queue. But any hidden object is excluded from this queue, so it is impossible to open a menu on the hidden object. Certainly, you can continue to keep the invisible objects in the mover's queue, but it would be the biggest mess you can organize. Such thing is never done; only visible objects are registered with a mover. Thus, to restore the visibility of an element the menu for the visible object one level upper must be used.

The direct command sets the value of the `Visible` property of the touched element. If this object has a lower level (if there are some children, associated with this object), then the elements on that lower level get the new value through their `VisibleAsMember` property. The command is propagated through the `VisibleAsMember` property of all the lower levels up to the end.

I showed a bit earlier the possible command to hide the plotting area. The crucial part of the method is the changing of visibility for the touched plotting area.

```
private void Click_miHideArea (object sender, EventArgs e)
{
    … …
    aos .Plot .Visible = false;
```

The call to `Plot.Visible` property has to change the visibility of the plotting area and all the scales and comments, associated with it.

```
new public bool Visible
{
    get { return (base .Visible); }
    set
    {
        base .Visible = value;
        if (value == false)
        {
            CloseTuningForms ();
        }
        foreach (Scale scale in scalesHor)
        {
            scale .VisibleAsMember = value;
        }
        foreach (Scale scale in scalesVer)
        {
            scale .VisibleAsMember = value;
        }
        foreach (CommentToRect cmnt in comments)
        {
            cmnt .VisibleAsMember = value;
        }
    }
}
```

Comments, associated with the main plotting area, get the new value, and for them it is all. The scales have not only to change one of their visibility parameters, but to pass some value to the comments, associated with these scales. Here is the `Scale.VisisbleAsMember` property.

```
new public bool VisibleAsMember
{
    get { return (base .VisibleAsMember); }
    set
    {
        base .VisibleAsMember = value;
        bool bToMembers = base .Visible && base .VisibleAsMember;
        foreach (CommentToRect comment in comments)
        {
            comment .VisibleAsMember = bToMembers;
        }
        if (value == false)
        {
            CloseTuningForm ();
        }
    }
}
```

For the `Plot` class, the visibility of scales and comments can be changed not only through the menus, but also via the tuning forms for scales and plotting areas. This makes the whole system of changing the visibility of elements much more flexible. For example, through the menu on a scale it is possible to hide or unveil only all the comments of the scale simultaneously, but via the tuning form of the same scale it is possible to change the visibility of each comment individually. The same thing happens with the visibility of scales: it can be changed personally for each scale through the tuning form of the plot.

## Tuning of plots, scales, and comments

> The forms, discussed in this section, are the tuning forms that are used together with the plotting classes from the **MoveGraphLibrary.dll**. These forms can be semi automatically called for tuning the `Plot` and `Scale` classes, on which some scientific and engineering applications are already based and on which many other applications of such type can be based. Because of these, I want to discuss in details the work of these tuning forms and their use with the mentioned classes. Similar tuning forms can be constructed for other complex classes; for this reason I want to pay attention to some technical (programming) problems in design of such forms.

Plots, used in scientific and engineering applications, are very complex objects. When these applications are designed as user-driven, then the full control of all the visualization parameters is passed to the users of the applications. Plots and scales have a lot of tunable parameters, so passing the whole control over all these parameters to the users put a problem in front of any designer of scientific applications. The users of applications have a wide variety of opinions about their level of involvement in the tuning.

On one end are those, who do not want to do anything, but require the program always to visualize the plots in the best way. Such users are sure that the developer always has to have the same understanding of what is "the best" exactly like they have. "Do whatever you want but always show me the plots in such a way as I (user!) expect them to look. And do not bother me with any manual tuning; it is you (programmer's!) job."

On the other end are those, who prefer to have everything under their control and have an access to any parameter. I think that this view is the right one, but there is one small addition. The full and detailed control must be really easy. And this is the huge problem, when you have so many parameters to control and tune.

The list of tunable parameters is well known for years. I would say that this list is hardly changed at all in my programs during the last 15 years, but the way of tuning all the parameters changed for nearly 100 percents throughout the same period. The first significant change happened when the ideas of the user-driven applications were applied not only to the main area of applications but to all the tuning forms also. From that moment every new idea in organizing the groups and moving / resizing any set of controls and graphical objects was immediately applied to the tuning forms of the plots.*

---

* The set of tunable parameters for the plotting areas had hardly changed at all through the last 15 years, so the changes in this tuning form mostly reflect the new ideas in the form's design. **Figure 16.6** represents the tuning form for the `Plot` class, which is used throughout the **WorldOfMoveableObjects** application. The **MoveGraphLibrary.dll** also includes an `MSPlot` class, which I widely used for a couple of years. The `MSPlot` class has very similar tuning form, which has nearly the same default view, but which is constructed of different elements. The form from **figure 16.6** is based on the `ElasticGroup` and `ArbitraryGroup` objects; the tuning form of the `MSPlot` class is based on the `Group`

**Figure 16.6** demonstrates the default view of the **Form_PlotParams.cs** - tuning form for the main plotting area and related elements of the `Plot` class. This form allows:

- To change the visualization parameters of the main plotting area.
- To add, delete or modify the comments for this area.
- To switch ON / OFF the visualization of any scale.
- To change the order of lines which are used for painting of Y(x) functions and parametric functions. By double click on the area of lines or by opening a menu at the same area you can proceed to another tuning form, where the lines can be modified and added.

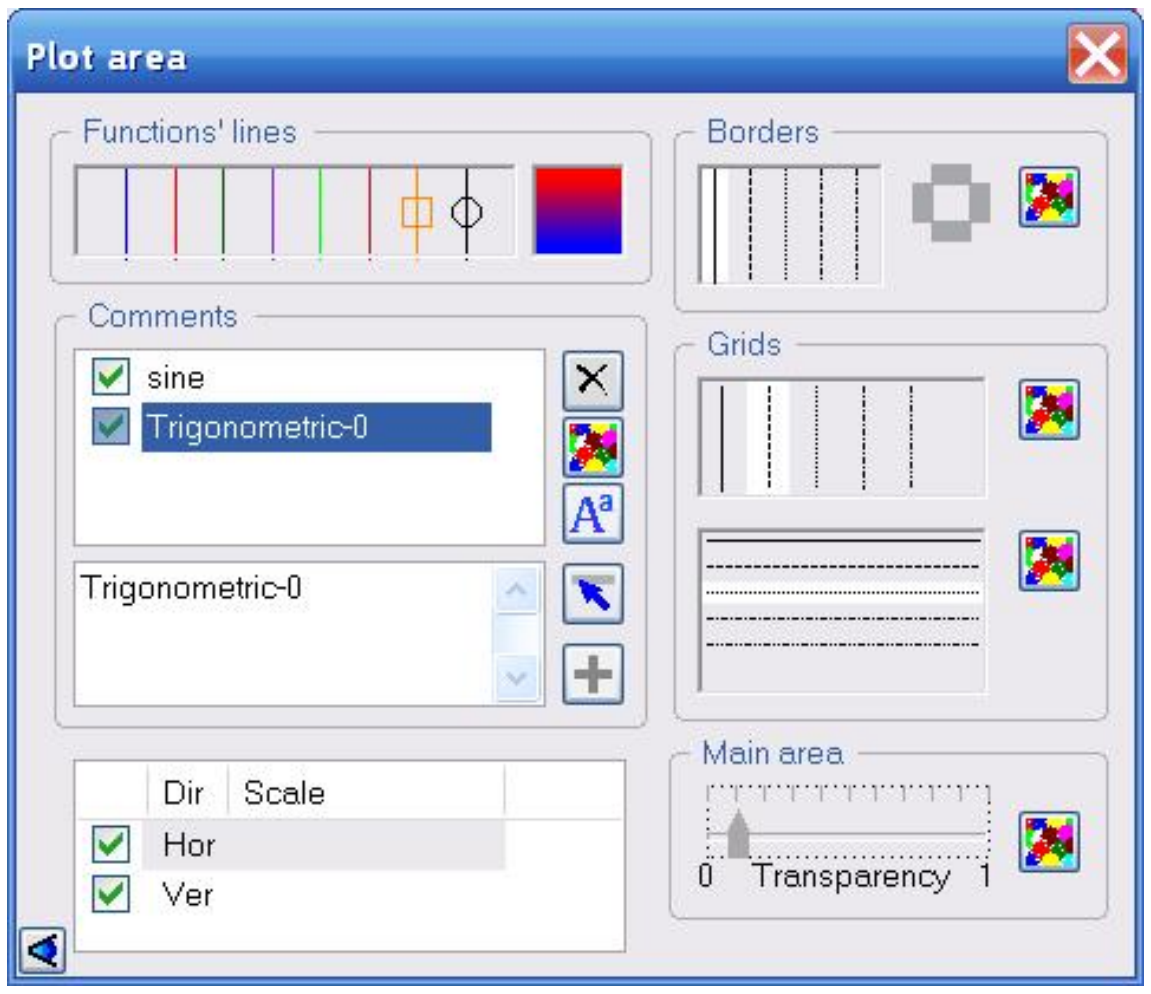

**Fig.16.6** The default view of the plot's tuning form for the `Plot` class

Before looking into some details of changing the parameters of visualization, I would like to remind the order of drawing the parts of the plotting areas. The scales are always shown atop the main area and the comments are always shown atop of their "parent", which can be a scale or a main area, but there are more details in the plots, so here is the full list of all the painted elements in the order of their drawing.

Main plotting area
  Background, if it is not transparent
  Horizontal grid
  Vertical grid
  Borders
Horizontal scales in such an order
  Main line
  Ticks and sub-ticks
  Numbers
  Comments
Vertical scales in such an order
  Main line
  Ticks and sub-ticks
  Numbers
  Comments
Comments of the main plotting area

Such order of painting allows to put any additional information on top of the main area.

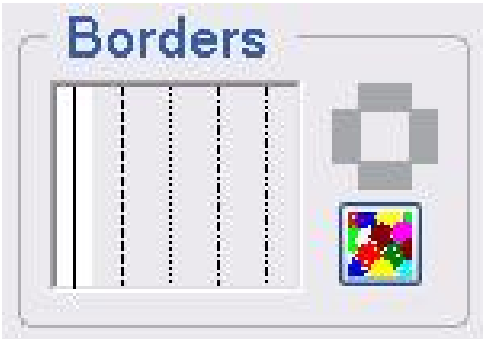

**Fig.16.7** Borders' tuning

**Figure 16.6** shows that the tuning of parameters in the form is divided between six groups; let us look what can be done in each of them.

<u>Borders</u> around the main plotting area can be shown in any color and any available standard style of lines; the width of the lines is always 1. The style of borders is selected by clicking one of the line samples. A small sketch with the four darkened rectangles (**figure 16.7**) represent four parts of the border, which have to be painted. These parts of the border can be switched ON and OFF independently. By clicking any of these rectangles you switch the flag, responsible for drawing the corresponding part of the border, between ON and OFF.

Reminder. The scales are painted after the borders, so if the line of the scale is positioned on the border, which is a common situation, then this part of the border is invisible.

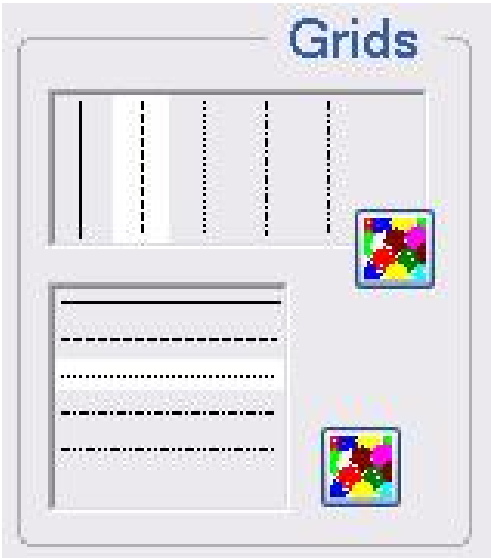

**Fig.16.8** Grids' tuning

Vertical and horizontal <u>grids</u> can be shown in any color and any available standard style of lines. If shown, the width of the lines is always 1, but it is possible to switch the grids OFF by clicking

---

elements. Though two tuning forms have the same default view, the `Group` objects implement a lot of designer's decisions, thus giving less control to the users. That was the main reason to switch to the new version (from `MSPlot` to `Plot`). It is still possible to see and try the `MSPlot` class together with its tuning form in the **TheoryOfUserDrivenApplications** program, which was designed to accompany an article with the same name. You can find there the same analyser of functions, but based on the `MSPlot` objects.

an empty strip at the end of the available samples (**figure 16.8**). The selection of styles and colors for two grids is done independently.

The pair of elements in the *Main area* group (**figure 16.9**) allows to set the <u>background color and the transparency</u> of the main plotting area. The group must look familiar as exactly in the same way these two elements are used in the tuning form for any `ElasticGroup` object.

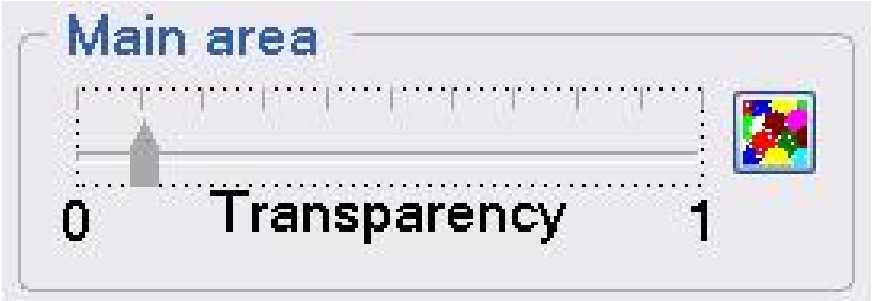

**Fig.16.9** Background color and transparency

Nearly any change of any parameter in the tuning form has an immediate effect on the original plot, but the change of transparency is an exception of this rule. When you move the slider, which determines the transparency, the areas under the samples in the tuning form (areas from **figures 16.7** and **16.8**) change their background color, but not the area of the original plot under tuning. The transparency of the original plot will change, when the slider of the track bar is finally released.

In some cases the switch to transparent mode is very useful for adding the image of one function on top of another. If you need to draw several functions in one area, this can be simply done by calling several drawing methods for the same `Plot` object. But if you want to show in the area the detailed part of some big function and at the same time you want to add somewhere in the corner of this area the small image of the whole function, then it can be done in a different way. For this, you can organize another area of much smaller size, strip it of all the auxiliary lines and additional parts, paint the same function in the small area, set the small area into transparent mode, and move it on top of the big one. (It will be like a smile of the Cat that had already vanished.) There is a limit on minimum size of any plotting area (40 by 40 pixels) to avoid its accidental disappearance, while resizing it.

Any <u>scale</u> can be shown or excluded from view by using the checkbox in the corresponding list of scales (**figure 16.10**). All horizontal and vertical scales are shown in the same list (horizontal scales first). More than often a `Plot` object is used with one horizontal and one vertical scale, but in general any number of scales can be associated with one plotting area. The scales in the list are identified by the direction and also by the first comment, associated with this scale. (Certainly, only if there are any comments at all.)

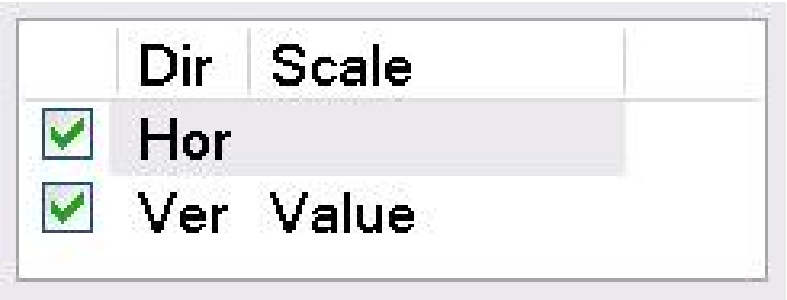

**Fig.16.10** Information about scales

The list of scales allows only to switch them ON and OFF; it is less tuning, than can be done in any other group of this tuning form, but it is very important in case of multiple scales. Any scale can be hidden from view through its own menu; all the scales of a plotting area can be hidden or restored simultaneously via the commands in the menu on that plotting area, but this list of scales is the only place to change the visualization of scales on an individual basis. This is very important in case of multiple scales, because it is really the case, when the visualization of scales needs to be applied on an individual basis.

The group to add/delete/modify <u>comments</u> (**figure 16.11**) includes more elements than any other group of the tuning form. The group consists of two main parts.

The first part contains:

- A list of all the comments associated with the main plotting area. A checkbox in each line allows to switch ON / OFF the visualization of each comment.
- 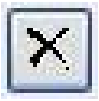 The button to delete the selected comment.
- 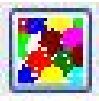 The button to change the color of the selected comment.
- 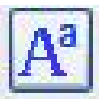 The button to change the font of the selected comment.

The second part contains:

- A `TextBox` to type in the new comment.
- 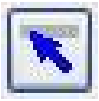 The button to replace the text of the selected comment with the new one.
- 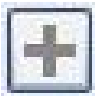 The button to add the new comment.

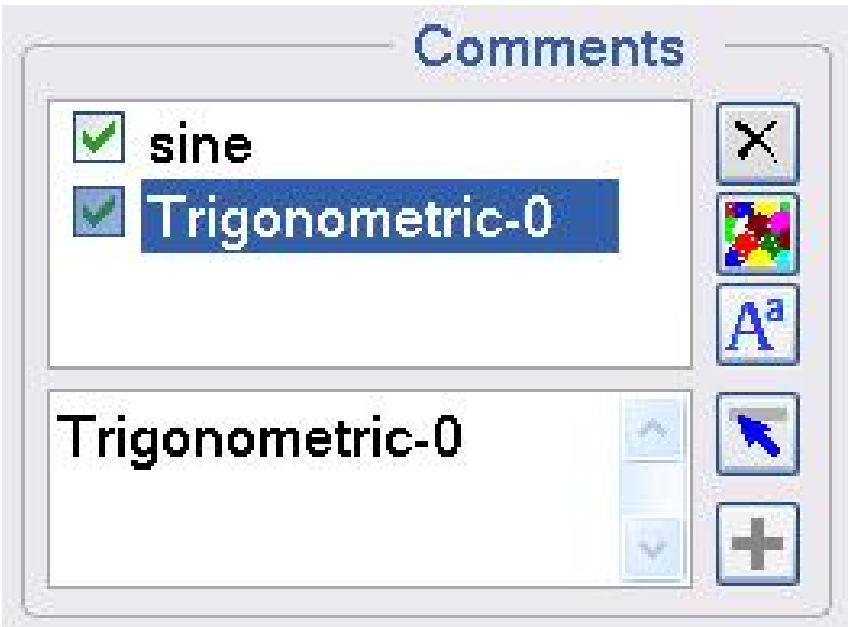

**Fig.16.11** A group to add/delete/modify comments

Several remarks on using the *Comments* group.

1. The only button in this group, which is always enabled, is the 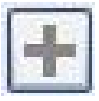 button to add the new comment. Other four buttons are enabled only when one of the comments is selected in the upper `List`.

2. When the new comment is added, two things happen: the comment itself appears in the middle of the plotting area; line with the text of the new comment appears at the end of the `List`.
3. When the color or the font for comment is changed, it is also stored as the parameter to be used for the next new comment.

The `Plot` class has a significant number of methods to draw the functions. For drawing the Y(x) or {X(r), Y(r)} functions the special lines are used. The lines belong to the `MarkedLine` class; among the parameters of such objects are color, type of line, and additional markers that can be placed along the curve. Special group (**figure 16.12**) in the tuning form allows to make some actions with these lines and opens an access to an auxiliary tuning form, in which the lines can be modified and added.

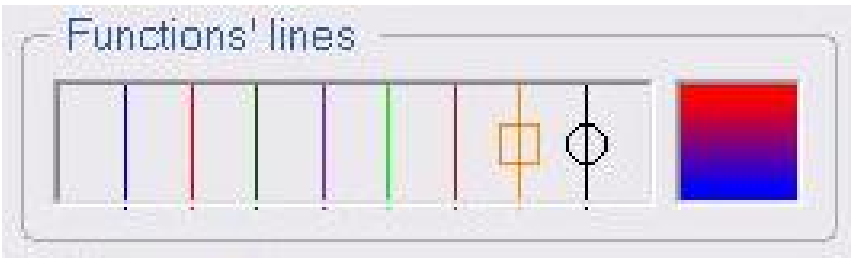

**Fig.16.12** Lines' group

Methods of the `Plot` class, which are used to draw the functions, reference the needed line by its number in the inner `List`. The samples in this group represent the lines exactly in the same order, as they are stored in the inner `List`; by moving a sample to the new position, the order of lines is changed. If the line, which is used for painting, is changed, then the graph is immediately shown in different way. (The explanation is much more complicated than the real process of changing the view of the drawn function. Move the left sample to any other position and you will see the change in your plotting area. This will immediately explain the whole process.)

Other actions with the lines can be started only via the context menu that is called at the area of samples. The minimum number of lines, associated with any plotting area, is eight. If there are more than eight samples in the area and you want to delete some of them, open the menu on the not needed one and delete it. If you want to modify some samples or add new, the same menu allows to open an auxiliary tuning form.

As I have already mentioned, the `Plot` class has a number of methods to draw Y(x) functions and also the parametric functions, described by the pair {X(r), Y(r)}. Each function is painted by a line of the `MarkedLine` class. An object of this class has two main parts – line and marker, of which at least one must be shown. So any function can be shown in three different ways:

- By a line.
- By a line with markers.
- By markers without any line.

The group from **figure 16.12** is only a part of the tuning form from **figure 16.6**. The main thing that can be done in this group is the changing of the order of lines, which are stored in the plot. This change of order immediately changes the view of the plotting area, if these lines are currently involved in the plotting. Usually, when, for example, four different functions are shown in the plotting area, then the most common situation that they are using lines with the numbers 0, 1, 2, and 3. If you reorder the samples of lines in the tuning form so that there is any change among the first lines, then you will immediately see the change in the plotting area.

To change any parameters of the existing lines or to add new lines, you have to open an additional tuning form; this can be done either by double click inside the group of samples or by opening the context menu in the area of samples. The auxiliary tuning form - **Form_PlotClrParams.cs** - is shown at **figure 16.13**.

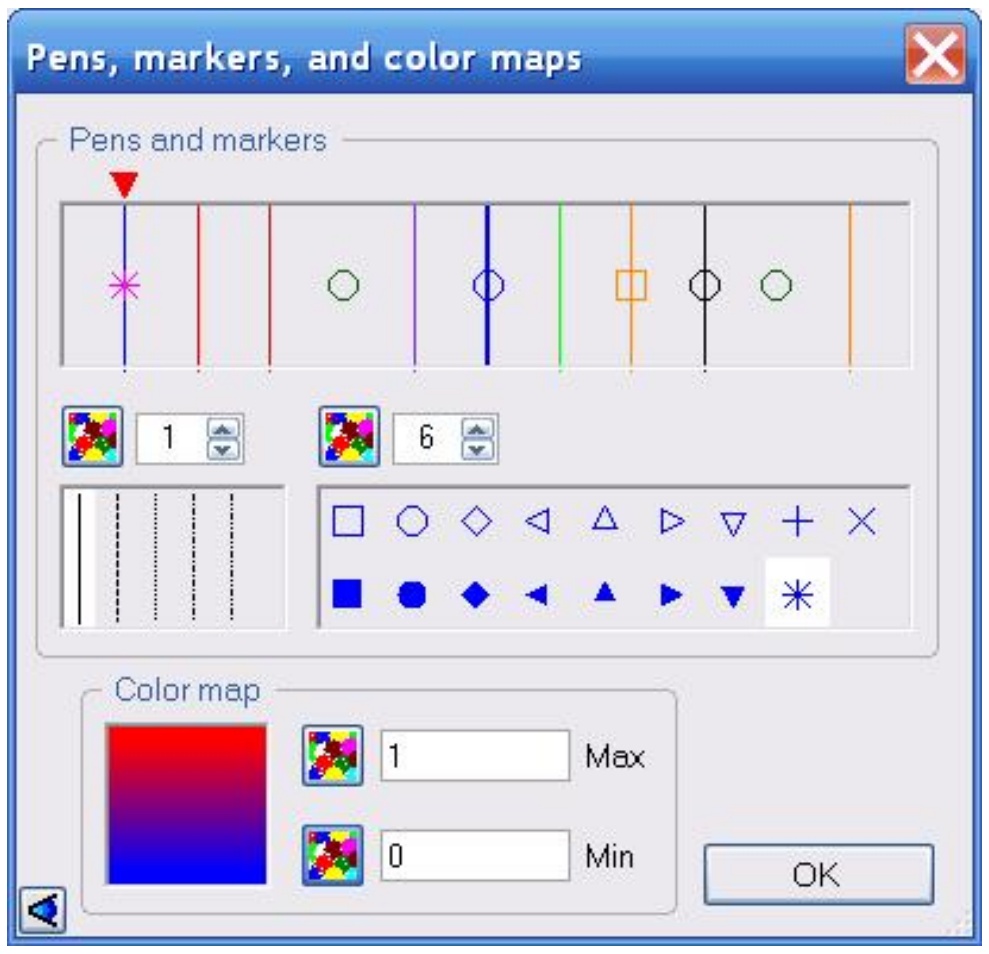

**Fig.16.13** Tuning of the lines

The order of lines can be also changed here, as in the main tuning form, but this is not the main thing. By clicking any sample line, you make it currently tunable; the small red triangle indicates such line. For this line, it is possible to set:

- Color of line.
- Width of line.
- Style of line.
- Type of markers.
- Color of markers.
- Size of markers.

All the changes, which are made here, are not going into the plotting area under tuning immediately, but only when this auxiliary form is closed with the OK button.

This tuning form has to be organized as a standard user-driven application, but I did not finish it yet in such a way (sorry…). It is possible here to change the used font (via the menu on any empty place outside the groups). On changing the font, you might want to rearrange the view of the form. All elements in the *Pens and markers group* are movable; the area of the samples and two numeric controls are resizable; so this group is organized nearly as it has to be. The *Color map* group is currently fixed.

Let us return one level back (or up) to the main tuning form of the plotting area (**figure 16.6**). This tuning form allows to do a lot of things, but you do not need all these possibilities all the time.

- The setting of the border lines or grids can be useful from time to time, but if you work with some complex scientific program day after day, then you would set the preferable parameters and after it you would need to change them extremely rare.
- It is also possible that your plotting areas do not have more than one scale of each type and you always want them to be visible; in such a case the list of scales also becomes superfluous.
- There is an easy way to change the transparency of the plotting area, but I doubt that scientists are going to use this possibility too often. They mostly do not like to change the background of the plotting area either, which makes the whole *Main area* group redundant.

Taking all these things into consideration, it leaves only two groups, which are used often enough: *Lines* and *Comments*. If only two groups are used often enough and other four very rarely, then there is no sense in using the tuning form in its default view as it is really a waste of the valuable screen space. Whenever you open the tuning form on top of the working scientific application, you overlap a lot of valuable information, which you are currently analysing. The less of the plots is closed by the tuning form, the better, and there is an easy way of shrinking the tuning form exactly to what you really need and not a pixel more. Everything is movable and resizable! Move the unneeded groups outside the tuning form and rearrange whatever is left in the way you prefer. I decided to rearrange the form according to what I have written several lines above; the results can be seen at the next figure.

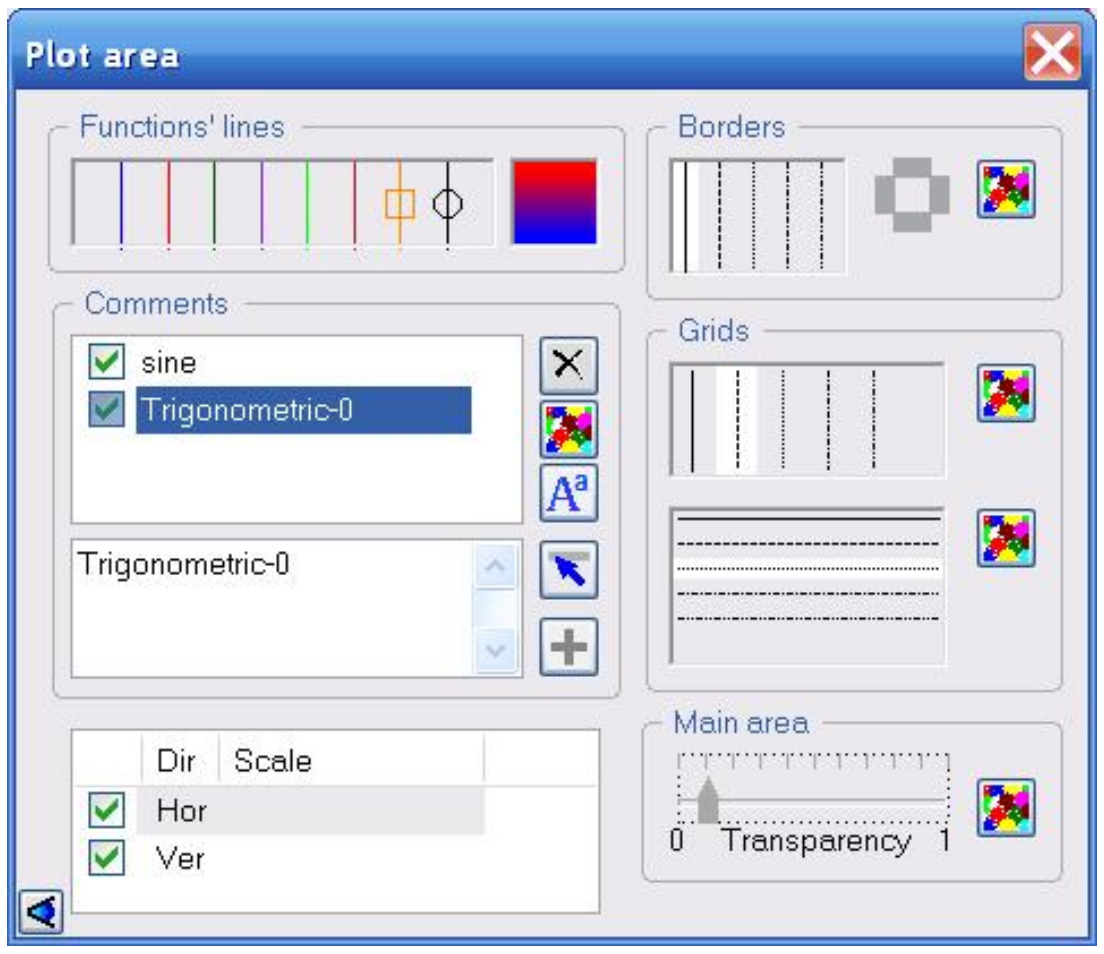

**Fig.16.14.a** Default view of the tuning form

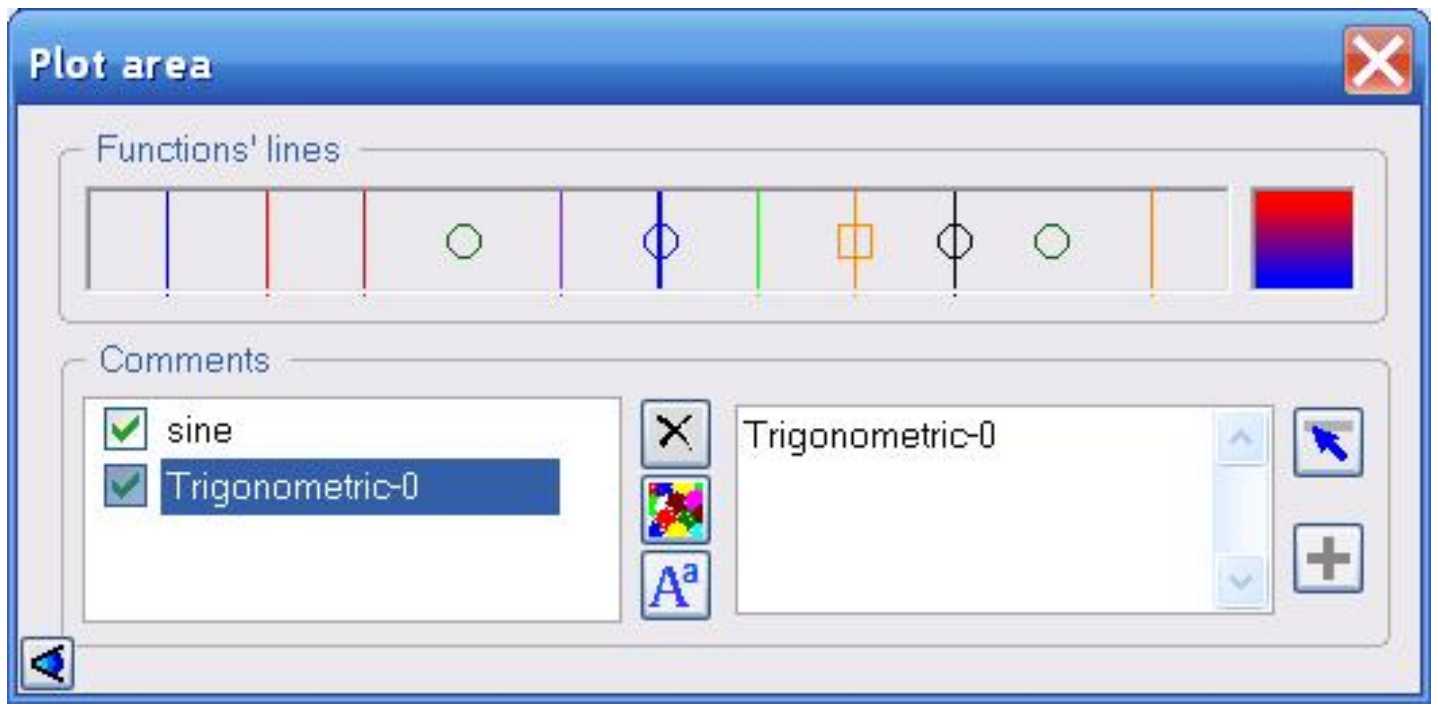

**Fig.16.14.b** One of the possible views of the same tuning form

You can see here not one, but two figures side by side; I think that it is much better for comparison than jumping between the pages back and forth. The left figure (**a**) is exactly the same that was shown before – the default view of the tuning form. The right one (**figure 16.14.b**) is **the same tuning form**. I moved four temporarily unneeded groups over the right border and dropped them there without even looking, how comfortable they feel there. Then I moved ONE control in the *Comments* group and slightly changed its size to make the two halves of the group even. Though the *Comments* group includes seven different controls, its design makes the rearranging easy to do. While writing about this group, I mentioned that it consists of two parts; these parts are organized as `DominantControl` objects, so the `TextBox` for typing the new comment moves synchronously with the two subordinate buttons. The last touch was the enlarging of the *Lines* group so that the two groups will be of the same width. All these changes took not more than two seconds. (Though the second variant of the form looks as big as the first one, but this is only because it is shown with the increased coefficient; you can see this from comparison of the same texts on both figures.)

This is a perfect demonstration of the power and effectiveness of the user-driven applications, when there is a request of rearranging a form for your particular needs. Here are four main characteristics of this process.

1. There is no fixed list of possible views of the form. The designer (in this case it is me) is not giving you the finite list of possible solutions, from which you can select one or another. Instead, you get an instrument, which gives you an infinitive number of possibilities. Any solution is determined by your demand at the current moment and by your taste of proportions, relative positions, and so on.

2. Going from one solution to another is very quick, just seconds.
3. Whatever is done is saved; none of your work on rearranging the form is going to be lost.
4. You can always return to the default view by a single command. At any moment you can call the context menu outside the groups, order the default view to be restored, and immediately return to the view, shown at **figure 16.14a**.

Let us return back to the **Form_Functions.cs** (**figure 16.3**). Any plotting area in this form has one horizontal and one vertical scale. Scales have a significant number of parameters, which can be tuned in special forms. There are different ways to open these forms:

- By double clicking the scale.
- Through the context menu on the plotting area (**figure 16.5**).
- Through the context menu on the scale itself.

The tuning forms for horizontal and vertical scales are designed in similar ways, but there are some differences, which make obvious what type of scale the opened form is related to. **Figure 16.15** shows the default view of the tuning form for horizontal scale – the **Form_HorScaleParams.cs**.

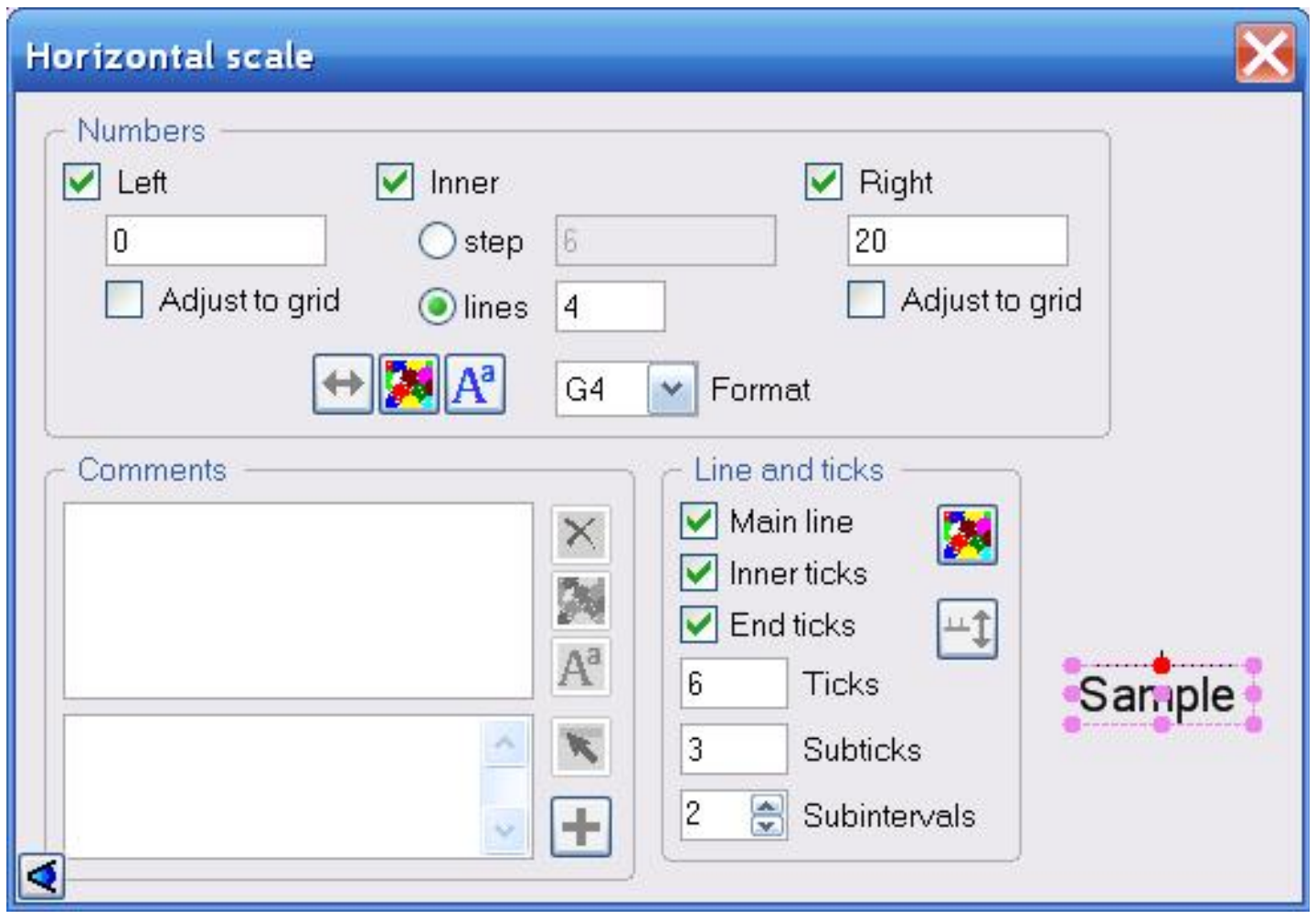

**Fig.16.15** The tuning form for horizontal scales

This tuning form has only three groups plus a sketch. The idea and the behaviour of a sketch is similar to what was demonstrated with the **Form_SpottedTexts.cs** in the chapter *Texts.*

The sketch consists of two parts: a line, representing a piece of the main line of the scale, and a *Sample* word, representing all the numbers along the scale. By moving the line, the whole sketch can be moved around the screen. (Sorry, the line cannot be seen on this picture, because it is closed from view by the upper part of the dotted frame.) The sketch in the tuning form for vertical scales has a vertical line, but the sample with the colored frame and the marks is the same.

The sketch is used for positioning of numbers along the scale. The word *Sample* is surrounded by the colored frame. There are eight color marks on this frame (four in the corners and another four in the middle of the sides) plus one more mark in the middle of the sample. The frame and eight marks are shown in one color and one mark always has different color (red). These nine colored marks are used for setting the required lining of numbers; the red mark indicates the current lining; it can be changed by clicking the needed mark with the left button. The sample can be moved by pressing with the left button on any mark or anywhere inside the frame. Sample can be also rotated by pressing inside a frame with the right button; the rotation always goes around the red mark. Whatever is done with the sample here on the sketch immediately reflects in the positioning of all the numbers along the scale.

The positioning of the numbers by moving and rotating a *Sample* is easier and more accurate, when the exact spot, to which all the movements are related, is well marked in one way or another. When there are ticks on the associated scale, then there is a tick on the sketch to mark the spot. When there are no ticks on the scale, then the spot on the sketch is marked with a small circle. The best way to position all the numbers along the scale is to move the sketch but to look at the same time not on the sketch, but on the scale itself.

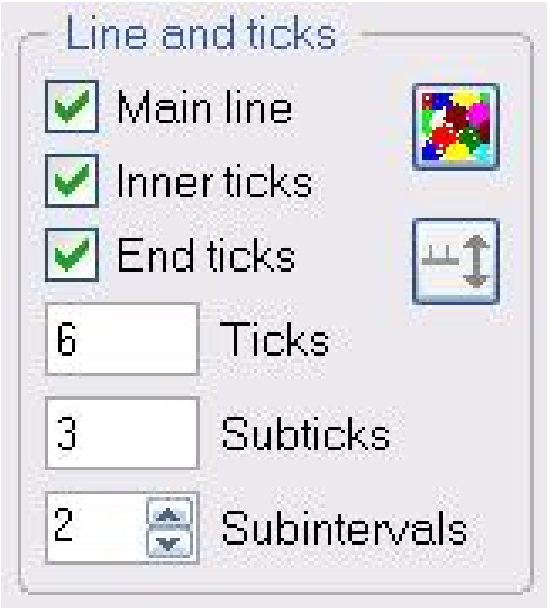

**Fig.16.16** Tuning of line, ticks, and subticks

The *Comments* group in this tuning form is identical to what was already shown in the tuning form of the main plotting area. All the comments belong to the `CommentToRect` class. The only difference is that the related rectangle ("parent") is either the main plotting area or the area of scale. The tuning of the comments, their adding, deleting, and hiding are absolutely identical to what was described a bit earlier (see **figure 16.11**).

The *Line and ticks* group (**figure 16.16**) is for tuning of all the parameters related to drawing the main line, the ticks, and the subticks. There are three check boxes, which allow users to select whether they want to see the main line or not and which of the ticks they want to see. None of these check boxes regulate the showing of the subticks, as it is done in a different way. Ticks correspond to the grid lines in the main plotting area; drawing of two ticks on the ends of the scale and of all the inner ticks is regulated independently.

An interval between the two consecutive grid lines can be divided into subintervals. The subticks are shown on the ends of subintervals, so if the number of subintervals is one, then there are no subticks in view. The number of subintervals can vary between 1 and 10. With this number set to anything but one, you will see the subticks, if their length is greater than zero. The length of ticks is set in one text box; the length of subticks - in another, but it cannot be bigger than the length of ticks.

The same color is used for the main line, all ticks and subticks; all of them are shown as solid lines with the width of 1. Ticks can be switched to any side of the main line; subticks can be only on the same side as ticks.

I have explained in the chapter *User-driven applications* how the movability of any element begins to demand the spread of the same feature on all the neighbours, on all the related tuning forms, and so on. The scales are movable, so their tuning form must be designed according to the rules of the user-driven applications. According to these rules, the group *Line and ticks* must be movable and tunable, if the users would like to rearrange it.

This group is movable and resizable automatically, as it is an object of the `ElasticGroup` class. I have already described this class in details (see chapter *Groups of elements*) and mentioned not once that this class is widely used throughout all my applications. The group can be moved around; in this way the whole tuning form is rearranged. If you do not need it; the group can be moved outside of view across the right or bottom border of the form. There are eight elements inside the group; all of them can be moved around; the frame is adjusted automatically, which is the standard feature of the `ElasticGroup` class.

Moving the inner elements of the group changes its view, but what about the content? What to do, if you never want to change, for example, the number of subintervals, the length of subticks, and the length of ticks? You can move any of the inner elements out of view in the same way, as the whole group, but this would be a very strange thing to do, as there will be a frame, starting inside the form and going somewhere beyond the borders. If you want to take out of view some of the inner elements of the *Line and ticks* group, call the context menu on this group (**figure 16.17**).

Of the eight inner elements of this group only two buttons are always shown; other six elements can be switched ON and OFF in any combination.

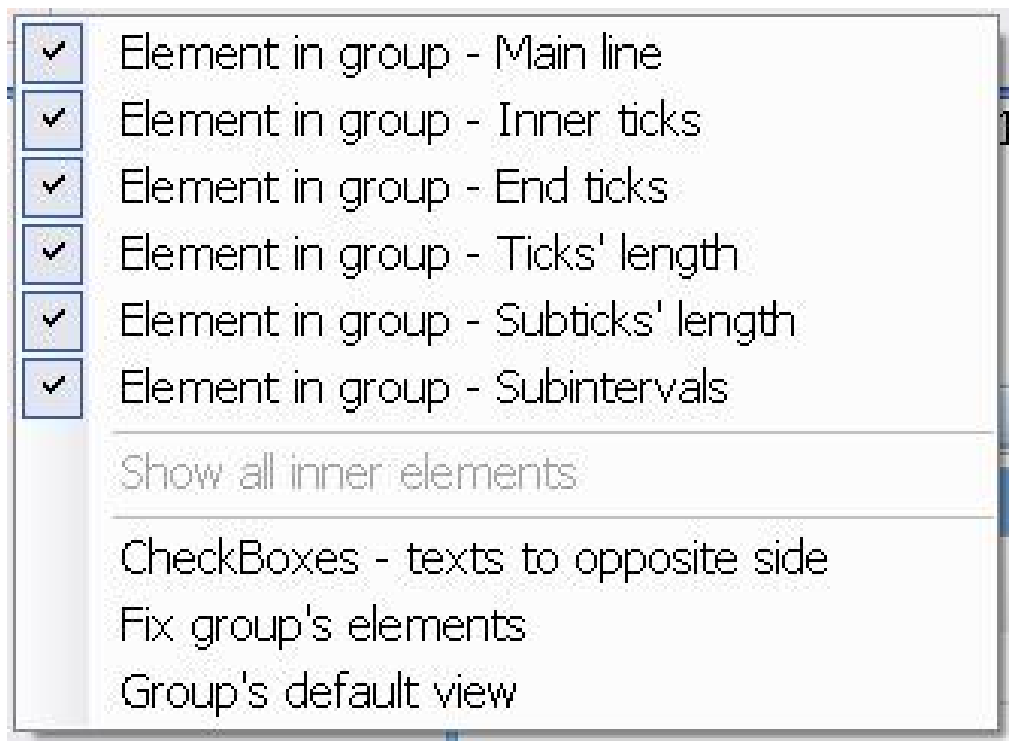

**Fig.16.17** Menu for the *Line and ticks* group

The six inner elements of the *Line and ticks* group are controls with the comments. In the chapter *Individual controls* I have introduced two different classes of the "control + text" type. The `CommentedControl` class has an individually movable comment, so a comment can be placed anywhere in relation to its associated control; the pair moves synchronously, when the control is moved (in the standard way, by its border). The comment of a `CommentedControlLTP` object has limited number of positions in relation to its associated control. There is no individual movement of the comment, so the synchronous movement of the pair can be started in two ways: by the controls border, as in the previous case, but also by the comment itself.

Objects of these two classes are visually undistinguishable, but the reaction on an attempt to move the comment is different. Because of this, I warned against using the objects of these two classes in the same form. It is exactly the thing that I organized inside the *Line and ticks* group! I know that this is not the best thing to do; it is a bad thing in design but up till now I couldn't find a better solution for this group. The moment I think out something better, I will change the group. Just now I want to write a bit more about this problem.

Big problems often arise from the small issues; the `CheckBox` control is never a big object, but up till now I cannot solve a problem, caused by this small control. If you use the standard `CheckBox` control with the text being part of it, then it can be moved around only by its invisible border somewhere around the text, but not by the text itself. Text occupies the bigger part of such control; the border is invisible and is located somewhere at the side of the textual information; I do not like this situation and never use standard `CheckBox` controls in my applications. Instead, I cut out the text of such control, leaving only the small box for a check mark. It is still a control, to which I can easily add a painted text. In such a way I can organize either a `CommentedControl` object or a `CommentedControlLTP` object. In the first case the comment can be moved individually; it provides maximum flexibility, but can you imagine a small checkbox with much bigger related text somewhere above or below this box? If you can, then this problem does not exist for you. You can call me too old-fashioned, but I cannot put a small check box and then position its much bigger comment above or below this box. The comment can be in one or two lines (I have such samples, they look normal), but the check box is always either to the left or to the right of this comment. And having a comment bigger or much bigger (this is even more often) then the check box, it is natural to move such a pair not by the border of the small box, but by any point of its comment. Which automatically means the use of the `CommentedControlLTP` class. Check boxes are always organized in my programs in such a way: it is a `CommentedControlLTP` object with the text either on the right or on the left side of a check box.

On the contrary, for all other types of controls, which are always bigger than the pure check box, I prefer to use the `CommentedControl` class. And whenever I need to put together check boxes and some other controls, there is immediately this conflict of two different classes of the "control + text" objects. These objects look similar, but are moved around differently. As I said, I know the problem very well, but do not know the good solution yet.*

Anyway, with good understanding of the problem, I put together into the *Line and ticks* group the representatives of two different "control + text" classes: three check boxes in the group are organized as the `CommentedControlLTP` objects; three other controls in the same group are turned into the `CommentedControl` objects. I want to underline again: I do not like this situation, but I do not see better solution at the moment.

The menu of this group allows to fix / unfix the inner elements and to restore the default view of the group. One command of this menu is unique throughout all the menus of all the forms in this Demo program.

In the chapter *Control + text* I wrote about these two classes of controls with comments. The `CommentedControlLTP` objects from the **Form_CommentedControlsLTP.cs** (**figure 13.2**) use the auxiliary **Form_TextsLimitedPositioning.cs** (**figure 13.3**) to position comments next to controls. In that case there are 12 different variants to position a comment; in case of a check box I think it is a bit too much and two variants - left or right - would be enough. So a command in menu (**figure 16.17**) allows just this thing: to move the comment to the opposite side of the check box.

The biggest group in the tuning form of scales is the *Numbers* group (**figure 16.18**). It allows to:

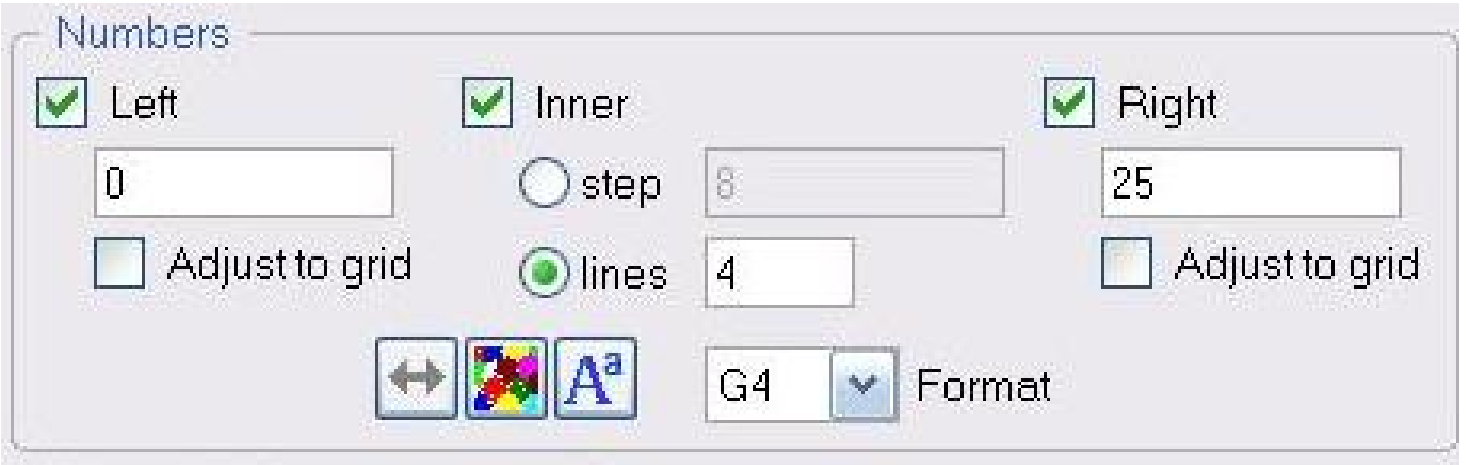

**Fig.16.18** The group for tuning the numbers along the scale

- Set the border values.
- Adjust the border values to the step of the grid.
- Define the grid either by the minimum number of grid lines or by the step of the grid.
- Set the color of numbers.
- Set the font of numbers.
- Swap the border values.
- Set the format to show the numbers.
- Independently define the drawing of the border values and inner numbers, corresponding to the grid lines.

I would say that this is the main group of tunable parameters, associated with a scale. The controls of this group are used for tuning all the time, so there is no instrument of taking out of view some of the inner elements. All the inner elements of this group are fixed, so the group can be moved around, but not rearranged by moving anything inside. Similar group is used in the tuning form of the vertical scales – the **Form_VerScaleParams.cs**, only the inner elements are regrouped to make it obvious that you have to deal with the top, inner, and bottom values along the scale (**figure 16.19**).

Changing of several parameters in the *Numbers* group is organized in a slightly different way from changing all other parameters in the tuning form. Usually, when you make any changes of parameters in the tuning form, it has an immediate effect on the object under tuning. This works both for tuning the main plotting area and the scales. The exceptions of this rule are those several parameters in the *Numbers* group, which you have to type. After typing each new value, you have to

* Please, do not declare that there is an easy and obvious solution in using in addition some special key from the keyboard. There is a wide variety of different movable objects, demonstrated in the Demo application. There were a lot of other objects, demonstrated with other applications earlier and used in other programs. There were some problems before, and I heard this advice to use the keyboard not once. Yet, up till now I always found a good solution without using the keyboard in addition to mouse. I have no prejudice against the keyboard, but this is a rude solution instead of using the brains. I prefer to think and find the good solution at the end. It worked this way up till now; I hope it will work for some time more.

press the **Enter** button, if you want to introduce the new value. Pressing of this button tells the program that the new value is ready, can be checked, and applied, if found to be correct.

The tuning forms (**figures 16.15** and **16.19**) have two more opportunities to change their view, which are available via the context menu that can be called anywhere outside the groups.

First, the font can be changed. This font is applied to all the texts in view, like comments for the controls and the titles of the groups, and to the controls in which any information is shown or typed. I have already written before that I strongly oppose the applying of the selected font to the form itself, because it changes the positions of all the elements in the form. This is done according to some inner logic, which was thought out to be correct by the developers of the Visual Studio. This logic and its application are absolutely out of users' control. If user determined to change the font, then the font must be changed, but nothing more.

Another command of the same menu allows to reinstall the default view of the form. This is done by taking into consideration the font, which user could already change.

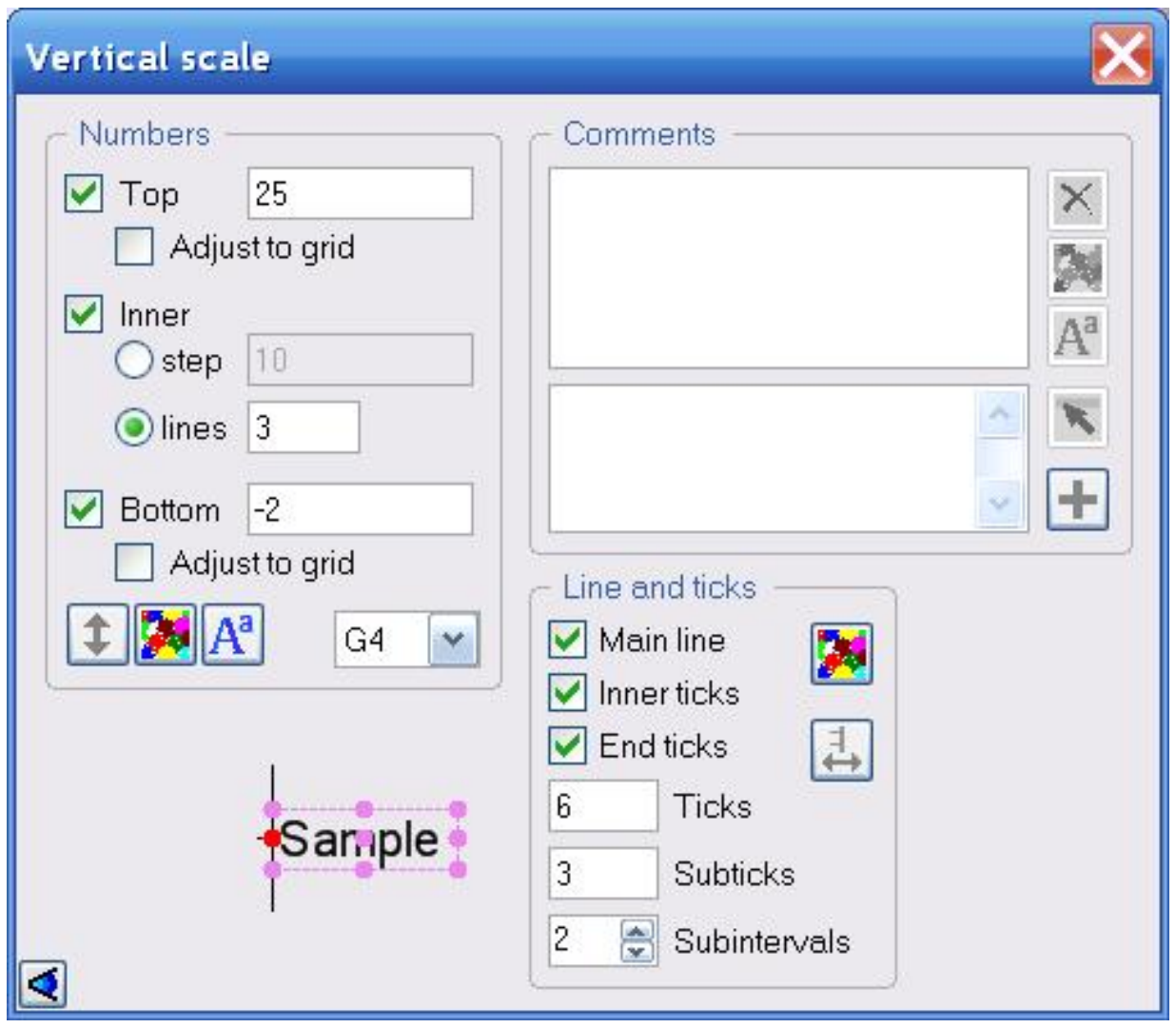

**Fig. 16.19** The default view of the tuning form for vertical scales

One more remark about tuning a scale. The *Numbers* group contains three check boxes, which allow to switch ON / OFF the visualization of different numbers along the scale; the range of possibilities is between showing all the numbers and not showing any numbers at all. The *Line and ticks* group has three check boxes, which allow to switch ON / OFF the main line and different ticks; the range of possibilities is between showing all the lines and hiding everything. If you switch OFF all six check boxes, you would receive a paradoxical situation, when a scale is not hidden, but at the same time not visible. To avoid this, the program does not allow to switch OFF all six check boxes simultaneously. If a scale must disappear from the screen, it can be hidden, for example, via the context menu, but not by switching OFF all its parts one after another.

## *Faster tuning of the plots*

File: **Form_Functions.cs**
Menu position: *Applications – Functions analyser*

I have so many years of design different scientific applications that I can think that I know about all the possible requests to such programs. In reality it turns out that I am wrong with such statement and from time to time I get the new ideas from the users. I try to put into the code any possibility that can be helpful in their work, but… First, users look at the applications with the fresh eye and have different experience than I have. Second, they look at any application as already achieved level and think about the possible additions not to the older versions, but to the newest one. Third, there are always users, who want to minimize their actions in application to the absolute minimum, but not on the cost of their research work; if they see any combination of several steps repeated one, two, or three times, then they immediately formulate a request to turn this combination into a single menu command. In such way "plots as samples" appeared in my scientific applications.

While developing the classes for plotting, I was thinking about all the needed parameters and the ways of their tuning. As the result, different tuning forms can be opened for any plotting area and its scales. Through these tuning forms all the visibility parameters can be quickly changed with an immediate effect on the associated plots. I was sure that that would be enough for tuning of all the plots on the screen. If any plot needs tuning, then its parameters are changed in the easiest and quickest way via its tuning forms. If you need to tune another plot, repeat the same procedure. What can be easier?

One day I was told what users want to do in such situation. They spent some time on rearranging one of the plots: they changed the colors and fonts, they positioned the scales, they set the border values, and they selected the types of the auxiliary lines. They turned the view of one plotting area into absolutely perfect from their point of view and they do not want to repeat this procedure again. After it they need two additional commands. The first one declares a plotting area as being a sample. The second command allows to apply the visibility parameters of a sample to any selected plot.

I agree that such pair of commands can be very helpful in a scientific application with a lot of plotting. I absolutely agree that such commands can be very useful, but there is one problem in using these commands. What exactly a user wants to copy? The majority of the visibility parameters can be copied without questions: colors, fonts, styles of lines. But there are still some questions, which have to be answered.

- Is the plotting area going to stay unchanged or has its size to be substituted by the size of the sample?
- Are the positions of the scales going to be unchanged or the new positions must be decided by the scales from the sample?
- What to do, if the number of scales in the plot under change and the sample are different?
- What to do with the comments? To leave them as they are, or to delete all the current comments and substitute them with the comments from the sample?

These are not the trivial questions. First, the answers can depend on the purpose of application, so the implementation can vary from one program to another. Second, the user's view on the problem can change (and often do change) over time. The solution according to the rules of user-driven applications would require to give users all the choices to decide, but I am not sure that in this case this is the best solution. Paradox? I am agitating all the time for the full control passed to the users, but here I oppose this thing.

Well, users have the full control over each plot. What we are discussing now is the parallel way of imposing a whole set of visibility parameters by a single command. If I am going to introduce users to another tuning form, in which I combine a lot of possibilities from several already existing tuning forms, it would be something huge and, from my point of view, absolutely unneeded. As I have already underlined, the full control is still there, it works, and nobody is going to take a bit from it. The faster transfer of the parameters from one plot to another can be looked at as an experimental parallel way, which can be constructed after consultations between the particular user and the designer.

In real life it works in such a way. I implement in the program some variant of faster tuning according to the users' request at the moment. After working with this program for some time they may come to different opinion. Then either I change that part in the first program or develop another variant for the fast tuning of plots in another program. Programs are different and users' requests can be different, so they rarely ask me to unify these tasks. But if they ask, I do. My programs are for scientists. I never work with that motto "They would have to like…"

You can see an example of the fast tuning in the **Form_Functions.cs**. **Figure 16.5** shows a typical view of context menu opened on a plotting area. Two commands of this menu allow to use any plot as a sample (*Use area as sample*) and to use the sample for changing the visibility parameters of the plot (*Change to sample view*).

The first command simply organizes an exact copy of the touched plot. The copy has the same coordinates of the plotting area and all the auxiliary parts as the touched plot, though not all of these parts are going to be used, when the sample is called by the second command.

```
private void Click_miUseAsSample (object sender, EventArgs e)
{
    Plot plotSrc = areas [iAreaTouched] .Plot;
    plotSample = plotSrc .Copy (this);
}
```

The menu line of the second command becomes enabled only after a sample was organized; at **figure 16.5** this line is disabled. When it is enabled, the `Click_miChangeToSampleView()` method can be called.

```
private void Click_miChangeToSampleView (object sender, EventArgs e)
{
    Plot plotDest = areas [iAreaTouched] .Plot;
    … …
    plotDest .CloseTuningForms ();
    Rectangle rc = new Rectangle (plotDest .PlottingArea .Location,
                                  plotSample .PlottingArea .Size);
    Plot plotNew = Plot .Copy (this, rc, plotSample);
    plotNew .CopyComments (plotDest);
    plotNew .HorScales [0] .CopyComments (plotDest .HorScales [0]);
    plotNew .VerScales [0] .CopyComments (plotDest .VerScales [0]);
    areas [iAreaTouched] .Plot = plotNew;
    RenewMover ();
    Invalidate ();
}
```

In the case of the **Form_Functions.cs** the quick tuning of the plots works under three conditions:

- The sample and the plot to be changed must have the same number of horizontal and vertical scales. In this form it is always true as each plot has exactly one scale of each direction.

- The size of the main plotting area is changed to the size of a sample.
- The comments of the area are not changed.

After setting these rules, the full procedure is going in such steps.

1. The tuning forms of the plot to be changed must be closed.

```
Plot plotDest = areas [iAreaTouched] .Plot;
plotDest .CloseTuningForms ();
```

2. The new plot is organized at the place of the changeable one, but it is the copy of the sample and has its size of the main plotting area.

```
Rectangle rc = new Rectangle (plotDest .PlottingArea .Location,
                              plotSample .PlottingArea .Size);
Plot plotNew = Plot .Copy (this, rc, plotSample);
```

3. All the comments from the plot under change are copied into the new plot.

```
plotNew .CopyComments (plotDest);
plotNew .HorScales [0] .CopyComments (plotDest .HorScales [0]);
plotNew .VerScales [0] .CopyComments (plotDest .VerScales [0]);
```

4. The plot under change is substituted by the new one.

```
areas [iAreaTouched] .Plot = plotNew;
```

5. Movable objects on the screen were changed, so the mover's queue must be reorganized; after it the form must be redrawn.

```
RenewMover ();
Invalidate ();
```

## *Analyser of functions*

All the plots at **figure 16.3** represent exclusively the predefined functions, which can be seen in the **Form_Functions.cs**. This collection of several functions was included into the program only for the purpose of easier acquaintance with this form; whenever you want to analyse any other function, it must be defined in some textual form and interpreted by a program. Before going into the details of the **Form_Functions.cs** and a couple of related forms, several words about the interpreter.

An interpreter belongs to the `FunctionInterpreter` class, which is included into the **MoveGraphLibrary.dll**. This interpreter has only few methods, but it is enough for its work. First, it analyses the text of the function and makes the decision about the correctness of this text.

```
bool Analyse (string strIn, ref List<Elem> liem,
              out int iError, out int kErrPlace)
```

The text of a function can include:

- Names of the standard functions: { sin, cos, tg, sh, ch, th, ln, lg, exp, sqrt, mod, arcsin, arccos, arctg }.
- Binary operations: { +, -, *, /, ^ }. Symbol ^ is used for degree function.
- Unary operation: `-`
- Brackets: ( and ).
- Variable: { x, p, r, X, P, R }.
- Numbers.

If there is any mistake in the text of a function, then the type of error and its place are returned. If the text is correct, then it is transformed into a special view – a collection of elements `List<Elem>`; this `List` can be used for calculations and drawing. Calculation is done by another method of the `FunctionInterpreter` class.

```
bool Calculate (List<Elem> PolishForm, double fArg, ref double fVal)
```

The return value only signals about the correctness of calculation; the result of calculation is returned via one of the parameters `(fVal).`

The results of calculations can be stored in different ways and used further on. To simplify the drawing, the `Plot` class includes a couple of methods, which use that `List`, prepared by the interpreter. The first of these methods is used to draw an ordinary Y(x) function. The function is calculated for each pixel of the horizontal axis of the plotting area.

```
    void DrawYofX (Graphics grfx, PlotAuxi auxi, List<Elem> polishY)
```

Another method is for drawing the parametric functions {X(p), Y(p)}.

```
    public void DrawParamFunc (Graphics grfx, PlotAuxi auxi, List<Elem> PolishX,
                                                            List<Elem> PolishY)
```

The use of both of these methods is demonstrated in the **Form_Functions.cs** and two auxiliary forms **Form_FuncYx.cs** and **Form_FuncXrYr.cs**, which are used to define different types of functions. Let us begin with the Y(x) functions.

In the **Form_Functions.cs** (**figure 16.3**) press the button 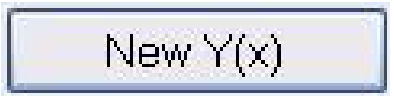

to open the **Form_FuncYx.cs** (**figure 16.20**). This is a standard form in the user-driven application, so everything is movable, resizable, and tunable.

The plotting area is of the same `Plot` class, which was just discussed. The only restriction here is the absence of the tuning form for the main plotting area. The goal of this form is to prepare the function, which is used in another place. This function will be used in the **Form_Functions.cs** in different plotting areas; all those plots are tunable, so the tuning of the main plotting area here has no effect on the demonstration of this function later. That was the main reason of not adding the tuning of this plotting area. Anyway, the scales, especially the border values, are important for preparing the needed function, so you can modify the scales via their tuning forms. As usual, those tuning forms can be called by double click on a scale, by calling the menu on a scale, or by calling the menu on the plotting area.

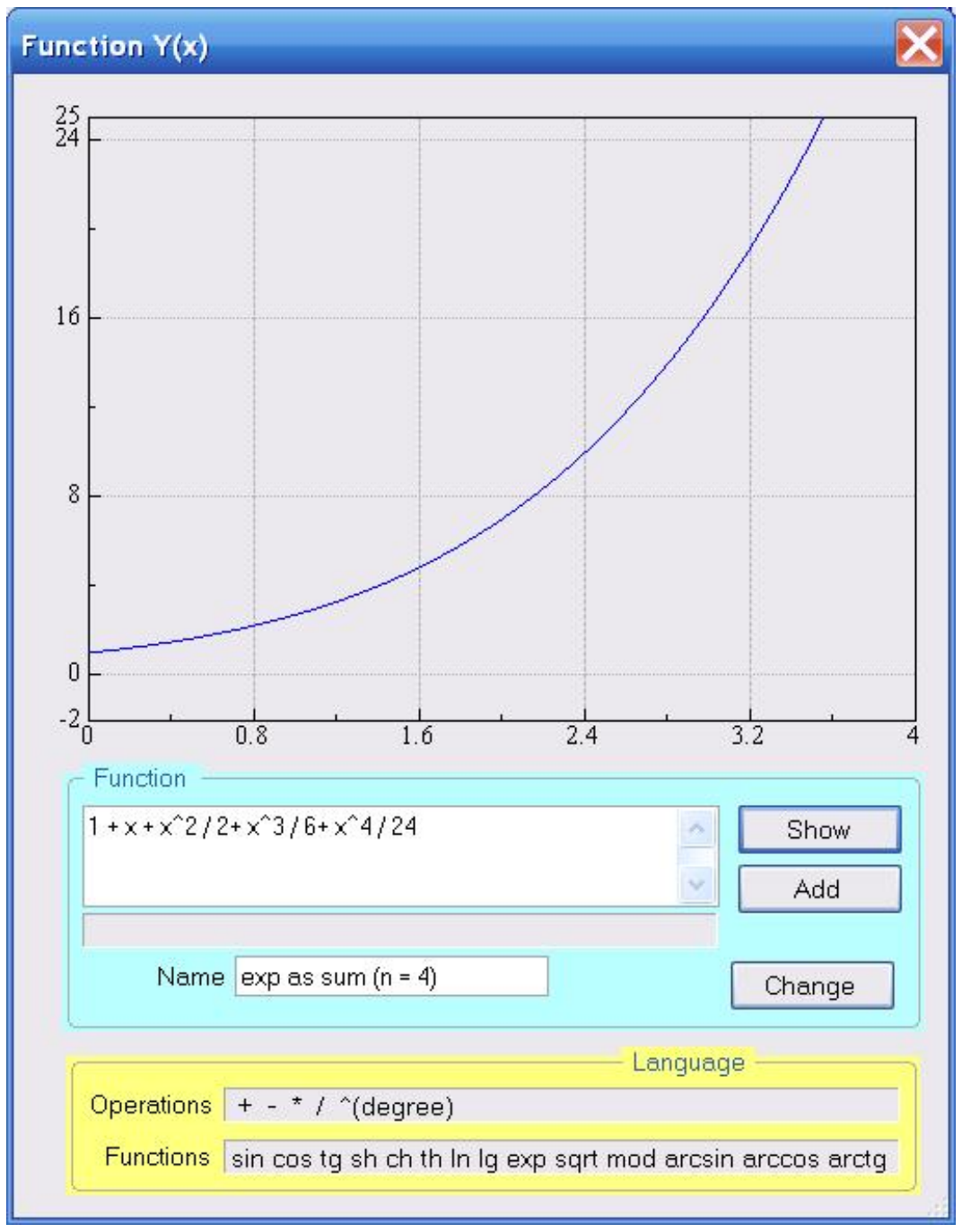

**Fig.16.20** **Form_FuncYx.cs** for editing the Y(x) functions

The **Form_FuncYx.cs** can be looked at as an example of the scientific application, but a simple one. According to its purpose, there is not going to be more than one plot; in addition there are two groups of controls. These are the objects of the `ElasticGroup` class, so they are movable, resizable, and tunable (via the context menu, which can be called on each group).

The *Language* group is an informative one: it shows the operations and functions, which can be used for typing the text of functions; the parentheses can be used for preparing more complex functions.

The *Function* group contains several controls to define any function of the Y(x) type and to start some actions on thus prepared function.

```
groupFunction = new ElasticGroup (this, new ElasticGroupElement [] {
          new ElasticGroupElement (textFunction),
          new ElasticGroupElement (labelErrorYX),
          new ElasticGroupElement (btnShowFunc),
          new ElasticGroupElement (btnAddFunc),
          new ElasticGroupElement (btnChangeFunc),
          new ElasticGroupElement (new CommentedControl (this, textName,
                                       Resizing .WE, Side .W, "Name"))},
                                  "Function");
```

The *Finction* group is the main part of this form. Type the text of the function and press button 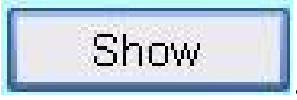

. The interpreter tries to turn the text into a special `List<Elem>`, which can be used later for calculation.

```
private bool CheckFunctionYX ()
{
    Polish_Yx = new List<Elem> ();
    bool bCorrectFunc = FunctionInterpreter .Analyse (strInput, ref Polish_Yx,
                                                      out iErr, out iErrPlace);
```

If there are any mistakes in the text, then the error in the text of a function is highlighted and the warning is shown below.

```
    if (!bCorrectFunc)
    {
        labelErrorYX .Text = FunctionInterpreter .GetErrorMessage (iErr);
        if (iErrPlace >= 0)
        {
            textFunction .Select (iErrPlace, 1);
        }
        return (false);
    }
```

If the text has no mistakes, then the function is shown in the plotting area.

```
private void OnPaint (object sender, PaintEventArgs e)
{
    Graphics grfx = e .Graphics;
    plot .Draw (grfx);
    if (Polish_Yx .Count > 0)
    {
        plot .DrawYofX (grfx, …, Polish_Yx);
    }
    groupLanguage .Draw (grfx);
    groupFunction .Draw (grfx);
}
```

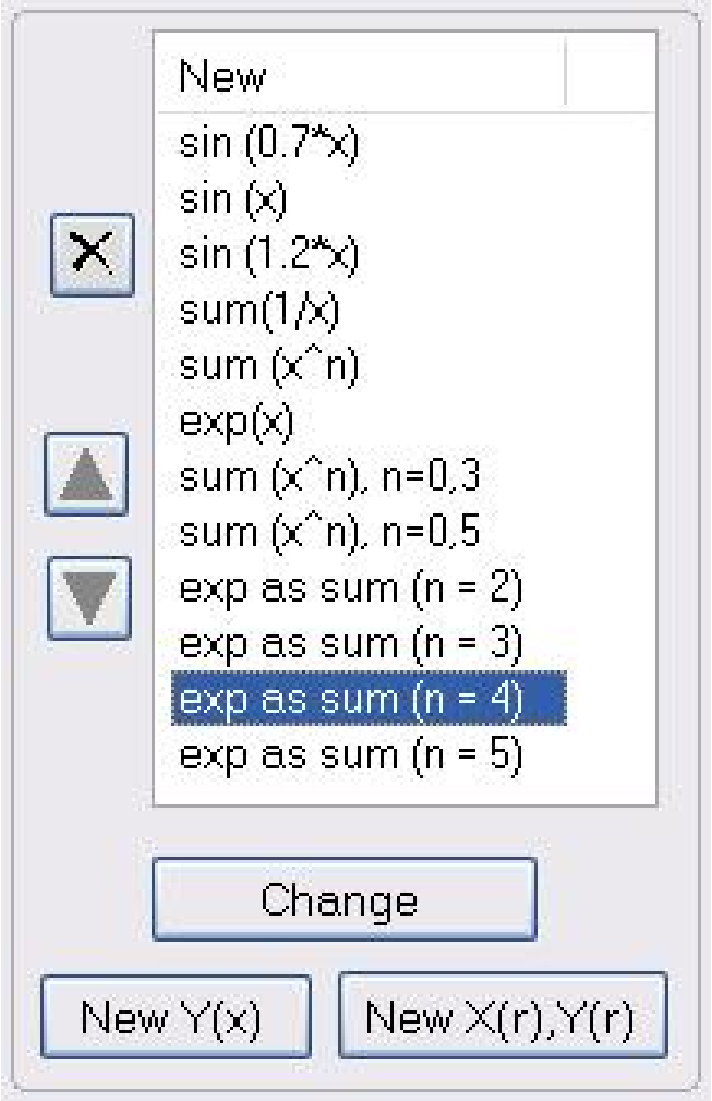

**Fig.16.21** List of functions, prepared by the user

If the function is correct, then you can either add it to the list of previously prepared functions (press Add) or replace the text of the existing function with the new one (press Change). The last option works only if some existing function was selected in the **Form_Functions.cs** before calling this **Form_FuncYx.cs**. Whether you want to add the new function or to change the old one, the name of the function must be shown.

Calling the **Form_FuncYx.cs** by the Change button can be helpful not only in cases, when you need to change the text of the existing function, but also for quicker preparation of the function, which only slightly differs from the existing one. For example, if you want to prepare several functions to demonstrate, how the exponent can be substituted by the sum of several simple functions, you may want to prepare several of them with different number of items. It is much quicker not to type each new function from the beginning, but to use the text of the previous function and add one component. As a result, you will have several new lines in the list of new functions in the **Form_Functions.cs** (**figure 16.21**).

The **Form_FuncXrYr.cs** (**figure 16.22**) is designed as close, as possible to the **Form_FuncYx.cs** (**figure 16.20**). Instead of one function, the texts of two functions must be typed in, but this is not enough. Both functions depend on one variable, so the range for this variable and its step for calculation of functions must be defined.

When any new function is prepared and the Add button is pressed, the form for editing new functions is closed and application returns back to the **Form_Functions.cs** (**figure 16.23**). The name of the new function is included into the list of all the functions prepared by the user. At the same time, the new plotting area is organized to show this function. This area will have several comments. One of them contains the name of the function; others (there will be one or two, depending on the type of the new function) – the text of the function(s). The size of such comment depends on the text of the function, so it can be small, but occasionally it can be big enough and occupy a lot of screen space. My advice: do not delete this comment. If you do not want to waste the screen space for this text, you can simply hide the comment, but it will be still possible to unveil the comment and check at any moment the text of the function that you are looking at.

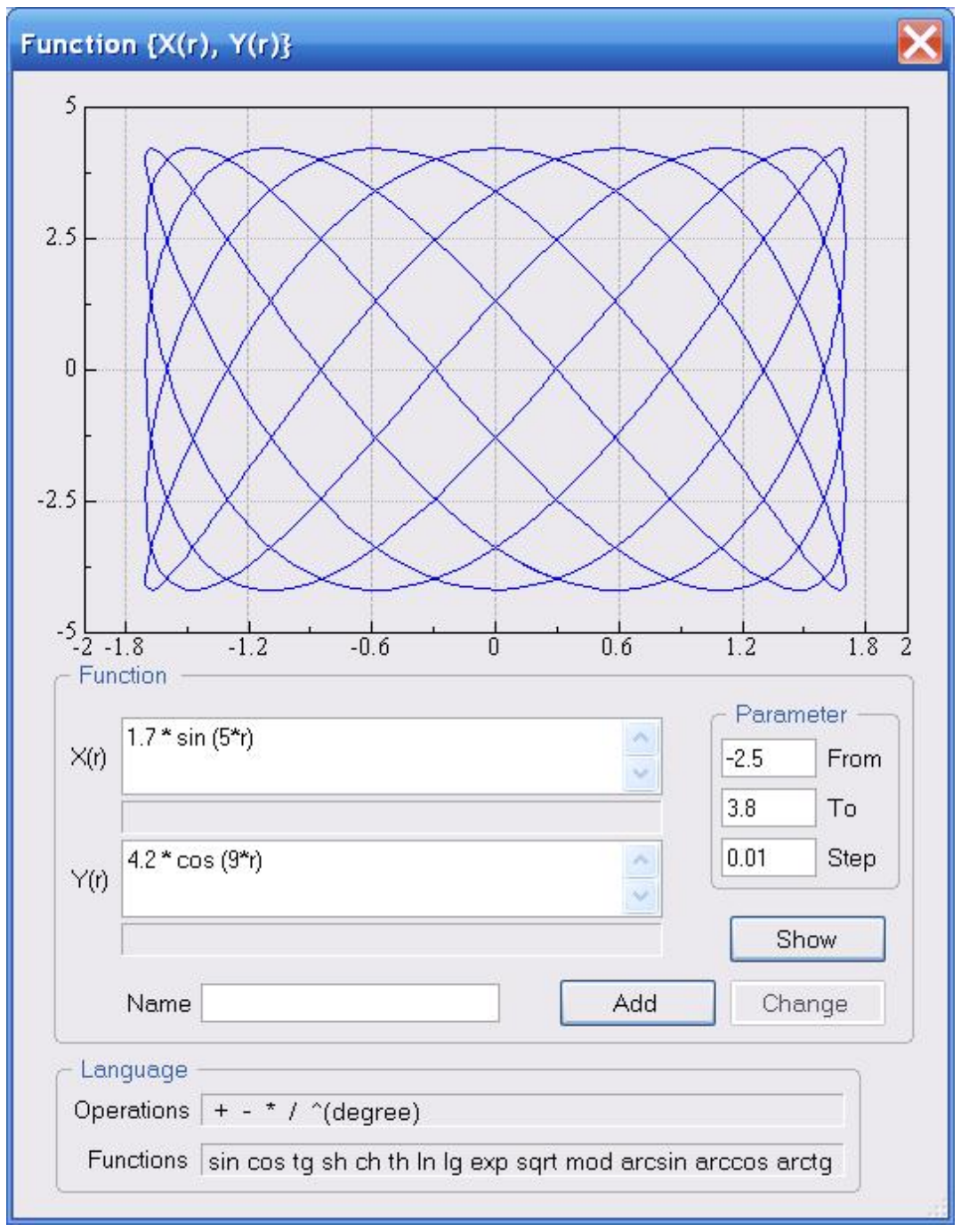

**Fig.16.22** The **Form_FuncXrYr.cs** for editing the {X(r), Y(r)} functions

The comment with the text of the function is shown only when this function is called later to be an only function in the plotting area; if you decide to look at several functions in the same area, then only their names appear in comments.

If you have deleted the text of the new function, there is still the way to reinstall it. Select the function in the list (**figure 16.23**) and press the 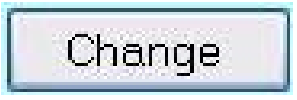

button; depending on the type of this function, either the **Form_FuncYx.cs** (**figure 16.20**) or the **Form_FuncXrYr.cs** (**figure 16.22**) will be opened. Click the 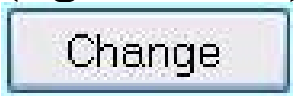

button in that form; you will return back to the analyser of functions and the comments with the text(s) of the function will be reinstalled.

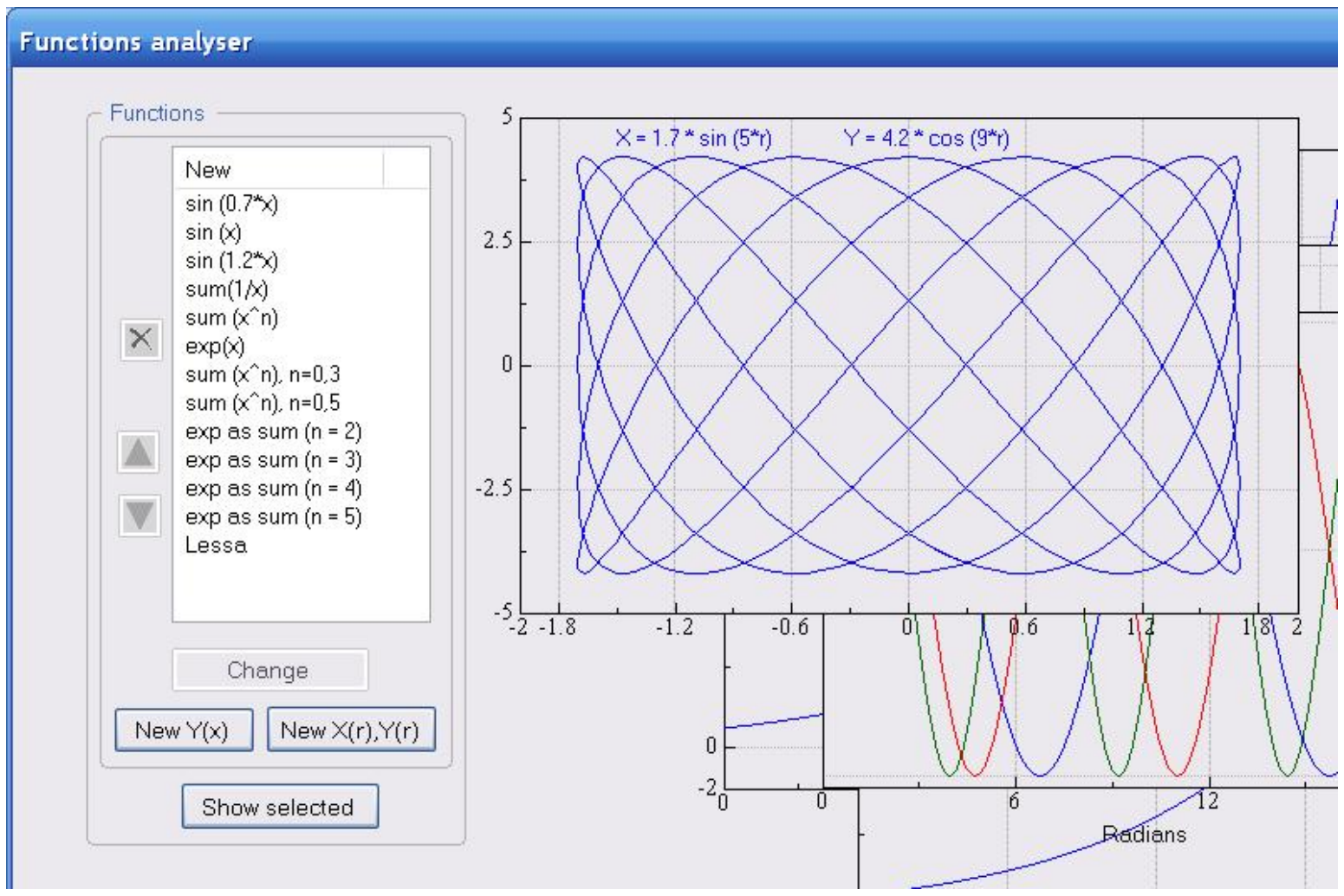

**Fig.16.23** Texts of the new functions are saved as comments

If you compare **figures 16.3** and **16.23** of the **Form_Functions.cs**, you can notice that the list of predefined functions has disappeared. This is one of the things that can be done via the context menu of the *Functions* group of that form; other options include modifying the group and fixing / unfixing its elements.

The **Form_Functions.cs** has also one more context menu that can be called at any empty spot of the form outside the *Functions* group and outside any of the plots.

## *Typical design of scientific applications*

File: **Form_Functions.cs**
Menu position: *Applications – Analyser of functions*

After looking into some details of the analyser of functions and going deep enough into the tuning forms for plots and scales, let us look into the typical design of the scientific applications on the basis of movable / resizable objects.

All the elements of the application are going to be movable and resizable, so there must be a mover to supervise the whole process.

```
Mover mover;
```

The plotting areas are supposed to be easily moved across the border of a form until the moment, when they are needed again. For this, the mover's `Clipping` property must be changed from the default one to the `Clipping.Safe`.

```
public Form_Functions ()
{
    InitializeComponent ();
    mover = new Mover (this);
    mover .Clipping = Clipping .Safe;
}
```

The plotting areas of the `Plot` class have to be associated with the functions that are shown in these areas. In the **Form_Functions.cs** this association is organized in the form of the `AreaOnScreen` class. In other cases there can be different classes with some additional properties, but their main goal is the same.

```
List<AreaOnScreen> areas = new List<AreaOnScreen> ();
```

There are a lot of ways to change the number and the order of movable objects:

- Adding / hiding / deleting plotting areas.
- Adding / hiding / modifying / deleting scales.
- Adding / hiding / modifying / deleting comments.
- Changing the order of plotting areas.
- Changing / hiding / modifying other elements, for example, groups.

Any of these changes require the renewal of the mover's queue.

```
private void RenewMover ()
{
    mover .Clear ();
    for (int iArea = areas .Count - 1; iArea >= 0; iArea--)
    {
        AreaOnScreen aos = areas [iArea];
        aos .Plot .IntoMover (mover, 0);
    }
    groupFuncs .IntoMover (mover, 0);
}
```

Moving / resizing process is organized via the standard three mouse events: **MouseDown**, **MouseMove**, and **MouseUp**. Any moving or resizing is started by the `mover.Catch()` method. The small addition in the `OnMouseDown()` method allows to organize the moving of the form itself by any inner point.

```
private void OnMouseDown (object sender, MouseEventArgs e)
{
    ptMouse_Down = e .Location;
    if (!mover .Catch (e .Location, e .Button, bShowAngle))
    {
        if (e .Button == MouseButtons .Left)
        {
            bFormInMove = true;
            Point pt = PointToScreen (ptMouse_Down);
            sizeMouseShift = new Size (pt.X - Location.X, pt.Y - Location.Y);
        }
    }
    ContextMenuStrip = null;
}
```

The moving / resizing are provided by the `mover.Move()` method. One small addition in the `OnMouseMove()` method allows to improve the view, when any control or group is moved around the screen; another provides the moving of the form by any inner point.

```
private void OnMouseMove (object sender, MouseEventArgs e)
{
    if (mover .Move (e .Location))
    {
        if (mover .CaughtSource is ElasticGroup ||
            mover .CaughtSource is SolitaryControl)
        {
            Update ();
        }
        Invalidate ();
    }
    else
    {
        if (bFormInMove)
        {
            Location = PointToScreen (e .Location) - sizeMouseShift;
        }
    }
}
```

Any moving / resizing is over, when the mouse button is released; at this moment the `mover.Release()` method must be called. The return value of this method informs if any movable object was really released; another very important parameter for making the decision on the following actions is the released mouse button.

- If an object was released by the left button, then the full identification of the released object must be done. One of the after results can be the reordering of the screen objects (plots); the drawing of objects strictly depends and must be organized opposite to their order in the mover's queue.

- If an object was released by the right button, it is often an impulse for opening a context menu. The menu depends on the type (class) of the released object; the view of the menu can depend on the particular released object, so the full identification is usually needed in this case.
- The release of the right button without any previously caught object is often a call to open a special menu at an empty spot.

```
private void OnMouseUp (object sender, MouseEventArgs e)
{
    … …
    if (e .Button == MouseButtons .Left)
    {
        if (mover .Release ())
        {
            int iInMover = mover .WasCaughtObject;
            … …
            Identification (iInMover);
            … …
        }
    }
    else if (e .Button == MouseButtons .Right)
    {
        if (mover .Release ())
        {
            … …
            MenuSelection (mover .WasCaughtObject);
        }
        else
        {
            … …
            ContextMenuStrip = menuOnEmpty;
        }
    }
}
```

The identification of an object is usually based on its number in the mover's queue, though in some cases additional information can be needed, like the number and the shape of the released node. For the main objects (plots) the identification includes only the finding of its order in the list of all the plots. For scales, the identification includes also the finding of the parental plotting area. For comments, the whole set of parameters up to the top of the parental chain must be determined; this set of parameters can differ a bit, as a comment can belong either to some scale or to the plotting area.

```
private void Identification (int iInMover)
{
    GraphicalObject grobj = mover [iInMover] .Source;
    long id = grobj .ID;
    if (grobj is Plot)
    {
        … …
    }
    else if (grobj is Scale)
    {
        … …
    }
    else if (grobj is CommentToRect)
    {
        … …
    }
}
```

Plotting areas have many parameters of visualization, which can be changed in the tuning forms. The tuning forms for the plotting areas and scales are often opened on a double click (the **MouseDoubleClick** event). During such double click, an object is caught by the mover, which is checked by the `mover.Catch()` method. The above mentioned identification is provided for the pressed object.

```
private void OnMouseDoubleClick (object sender, MouseEventArgs e)
{
    if (mover .Catch (e .Location, MouseButtons .Left))
    {
        int iInMover = mover .CaughtObject;
        Identification (iInMover);
        if (iAreaTouched >= 0)
        {
            GraphicalObject grobj = mover [iInMover] .Source;
            if (grobj is Plot)
            {
                Plot plot = areas [iAreaTouched] .Plot;
                plot .ParametersDialog (this, RenewMover, ParametersChanged,
                                        PointToScreen (e .Location));
            }
            else if (grobj is Scale)
            {
                Scale scale = grobj as Scale;
                scale .ParametersDialog (this, RenewMover, ParametersChanged,
                                         PointToScreen (e .Location));
            }
        }
    }
}
```

Another way of opening the tuning dialogs is often provided via the context menus. Different classes have their own menus, so it is not a rare situation to have around 10 different menus in a form. Menus often allow to change the visibility and movability of the elements. Hiding of an object can be organized via its own menu, but the restoration of this object back into view must be provided via the menu of the “parent” of this object. The menu of the “parent” often includes the hide / unveil and fix / unfix commands for all the “children” or some groups of “children”, which makes much easier the identical change of parameters for all the “siblings”. The same type of commands for the objects of the upper level (for all the plots in the form) can be included into the menu, which is opened on an empty place of a form. **Figures 16.24** demonstrate the context menus for plots (**a**), scales (**b**), and comments (**c**). More complex objects take the spot higher along the hierarchy of classes and have more commands in their menus to deal with their own parameters and with all the subordinate objects.

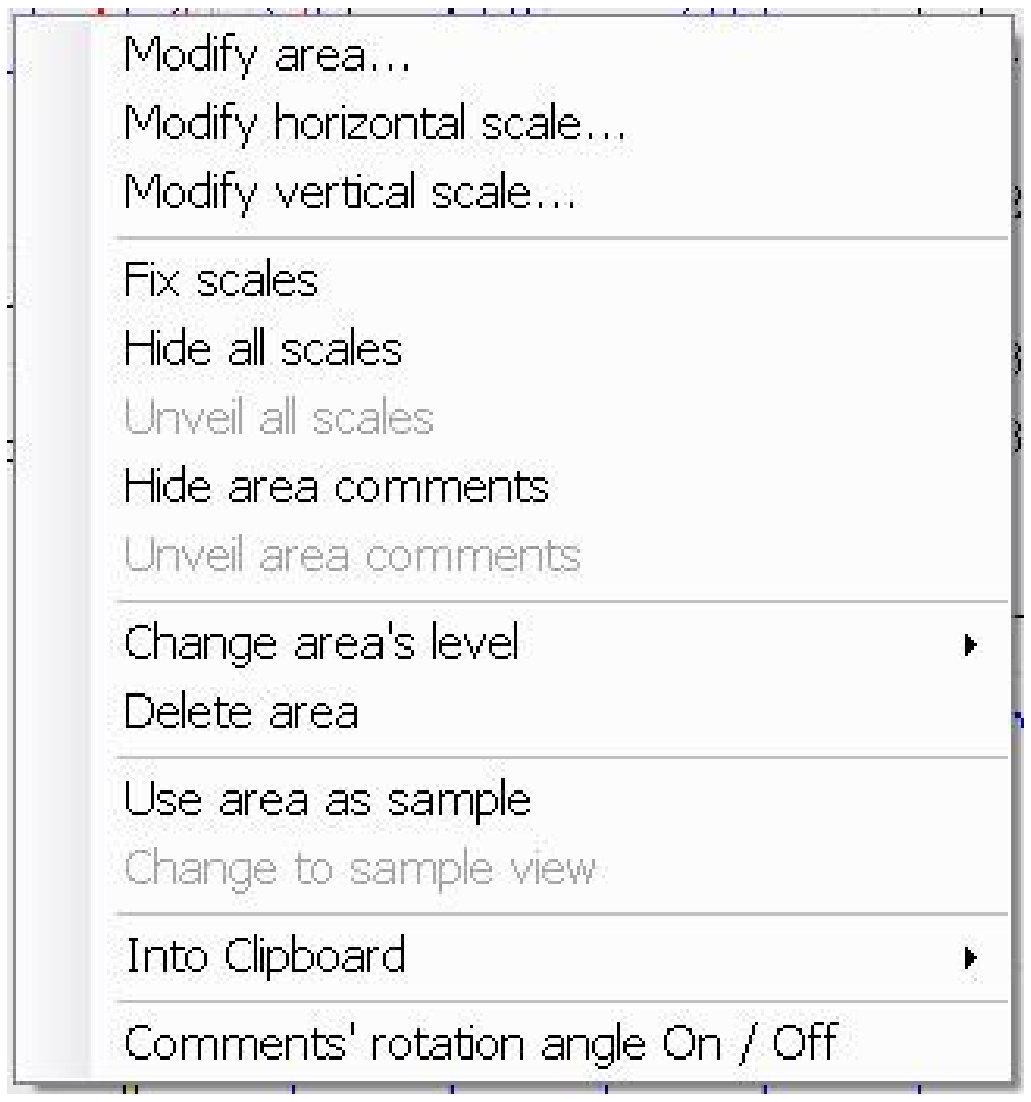

**Fig.16.24.a** Menu on plots

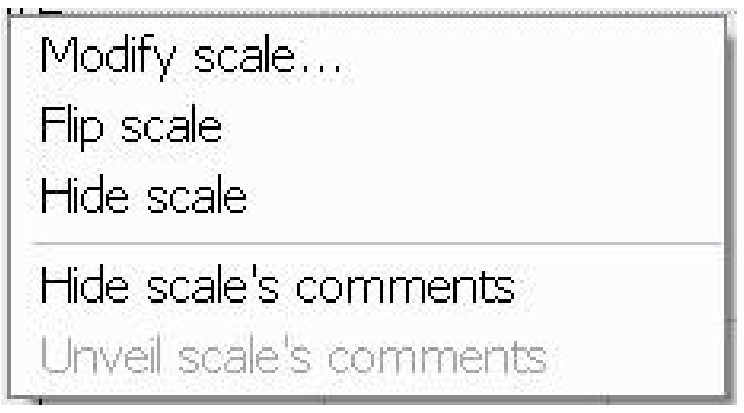

**Fig.16.24.b** Menu on scales

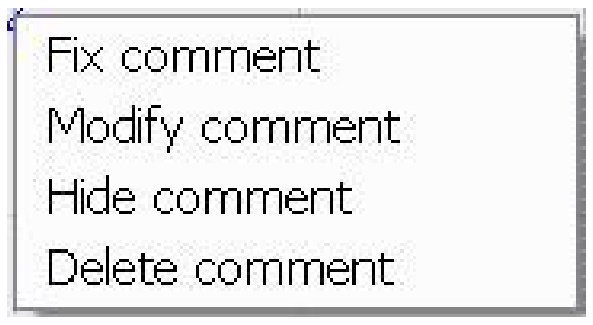

**Fig.16.24.c** Menu on comments

Scientific / engineering applications have so many tunable parameters and the usefulness of these applications so strongly depends on their ability to be tuned according to each user’s preferences that they cannot exist without the system of saving / restoring of each and all. All the widely used classes (`Plot`, `Scale`, `CommentToRect`, `ElasticGroup`, `CommentedControl`, and so on) have their methods for saving and restoring the objects on the `Registry` and in the binary files. When an application is used predominantly on the same computer, I prefer to use the `Registry` for saving / restoring. However, scientists like to take applications to different places to discuss and compare the results. Such checking and comparison of results may happen at different computers, in which case the saving / restoring via files can be preferable. It looks like the big scientific applications may need both types of saving / restoring in parallel.

## DataRefinement application

A lot of scientific branches simply cannot progress without new more and more sophisticated programs, which require huge efforts in their development. At the same time the number of users of each particular application is relatively small. Definitely not millions and rarely even thousands. I saw the information that one of the big programs, into redevelopment of which I put big efforts and which was mostly redesigned by me, is now sold in more than 130 countries. This program has thousands of users, but this is not common for scientific and engineering applications. I would call it an unusual case. The more common situation, when there are hundreds, dozens, or only several users of each program. But these applications are needed and have to be developed, because the research work cannot go on without them.

With such relatively small numbers of users, it is even more important to develop the applications in such a way that EVERY user has a chance to change it in the best way personally for him. Only the switch to the user-driven applications allow to do it; that is why I am saying all the time that user-driven applications is the way out of stagnation for the scientific and engineering programs.

The idea of the program, which I am going to demonstrate here, was discussed in the department of mathematical modeling years before I came to work there. Colleagues were thinking about such an application for a long time, but they never developed it, because they understood, how ineffective it would be, if based on the standard procedure of retyping new values again and again in some kind of a table. It took me several days to understand all the requirements and to give them the version, which works according to all their requests and expectations.

The areas of science put in front of the programmers the same tasks again and again until the new programming ideas would produce a really elegant solution. A lot of purely numerical problems were solved years ago, but the combinations of calculations and manual adjusting of the algorithms usually waits until some moment, when the new development in interface would allow to solve it easily. One of such tasks is the data refinement. A lot of researchers receive the needed data from the analog-digital converters. This data is going to be analysed by a well known mathematical methods, but the data is often obtained together with an additional noise, and the programs for calculations go crazy, while trying to solve the equations with such input values. The data has to be previously refined by a researcher, and that is where the movability of different elements is of high demand.

There are two main things that a researcher would like to do with the input data files:

- Each data file consists of a huge amount of data. When you look at the graph of any data file, you immediately see that it consists of several (or many) segments; the view of function in each segment is absolutely different from the neighbours. Those segments can be (and will be) approximated by different relatively simple functions; the first task is to set the boundaries of these segments.
- Data is received with a lot of noise. In many cases, even after division of the data file into a set of segments, it is still impossible to use the sets of (x, y) value pairs from any segment as an input for some further calculations, because this data is marred with the noise. The second goal of this application is to reduce the noise manually.

The authentic **DataRefinement** application includes much more options, than you see in this example.* I excluded nearly all the parts with storing and showing some intermediate results, because this is of interest only for the specialists, working with the data. I also excluded all the parts with any saving of produced data into the files. For the real application, this is the main and the most valuable result, but the majority of those, who are going to read this book and try the application, would be very nervous, if my application would try to write anything on their computers. I do not want the readers of this book to become nervous. The main goal of this demonstration is to show, how to design scientific applications on the basis of movable objects and how users are going to work with such applications. All these things, which I want to show, can be seen on the crippled version of the **DataRefinement** program.

I have excluded any writing of the results, but I cannot demonstrate anything without reading some data. This is the only example in the big Demo application, which cannot work without an additional input data. I put three data files into the special subdirectory (**figure 16.25**); if you want to see, how the **DataRefinement** works, you have to use one of the data files. You can also prepare your own data files in the same way and use them to check the **DataRefinement** program. Those three data files were prepared in another program, but to be sure that there will be no problems in preparation of such files, here is the short code of how it was done. In the authentic **DataRefinement** program, the original data is shown in the `ListView` control with many columns; users select the needed columns for X and Y arrays. The only limitation on the input arrays is that the values of the X array must be non-decreasing. The structure of the BIN file with data is primitive: the number of (X, Y) pairs is written as an integer, after which all the (X, Y) pairs are written as the double values.

---

* The **DataRefinement** application was commissioned by Dr.Stanislav Shabunya from the Heat and Mass Transfer Institute. Dr.Shabunya also provided the main ideas and methods of calculation, used by this program. In one of my previous articles [6], I have mentioned this application by showing one picture with several phrases of explanation.

```
bw = new BinaryWriter (fs);
int nLines = listData .Items .Count;
bw .Write (nLines);
for (int j = 0; j < nLines; j++)
{
    bw .Write (Convert .ToDouble (listData .Items [j] .SubItems [1] .Text));
    bw .Write (Convert .ToDouble (listData .Items [j] .SubItems [2] .Text));
}
```

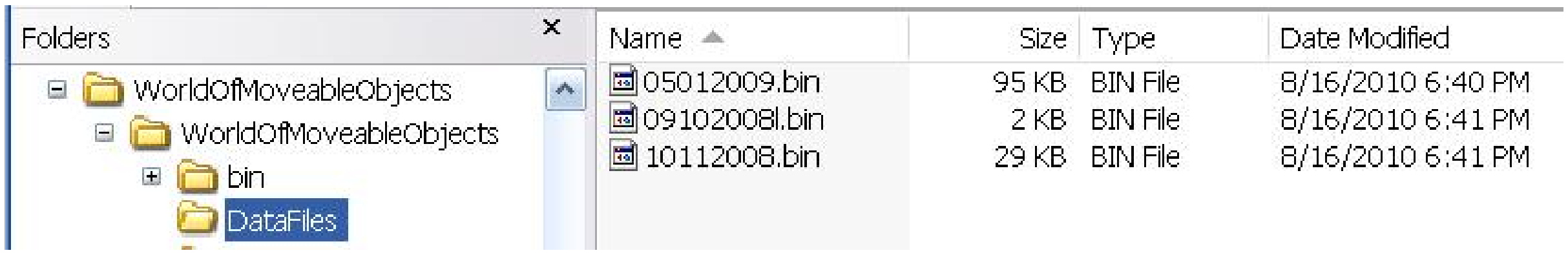

**Fig.16.25** Data files for the **DataRefinement** application

After excluding all the parts important only for the data analysis, but leaving the parts important for the discussion of movability, I have left only two forms in the example of this application:

- **Form_DataRefinement.cs** allows to read the input data file and divide it into segments.
- **Form_Zoom.cs** allows to change the data manually; new segments can be also organized in this form.

Now we can start looking at the application itself.

File: **Form_DataRefinement.cs**
Menu position: *Applications – Data refinement*

On opening the **Form_DataRefinement.cs**, you see only an empty plotting area. The data file to be analysed can be called via the standard *File – Open* menu command. **Figure 16.26** shows the **10112008.bin** file. When any new file is called (opened), there is going to be a single additional vertical line in the middle of the plotting area. This vertical line – a slider – can be moved left or right.

The sliders are used to mark the boundaries of the segments. The number of organized segments depends on the input data and on the researcher's decision. The parts of the file from **figure 16.26** are visually very well distinguishable, so I decided to divide the whole file into six segments. Sliders can be added to the view via the context menu (**figure 16.27**) that can be called anywhere on the plot. Erasing of the unneeded sliders is done via another menu that can be called directly on a slider. Each slider can be moved only up to the neighbours or the border of the plot.

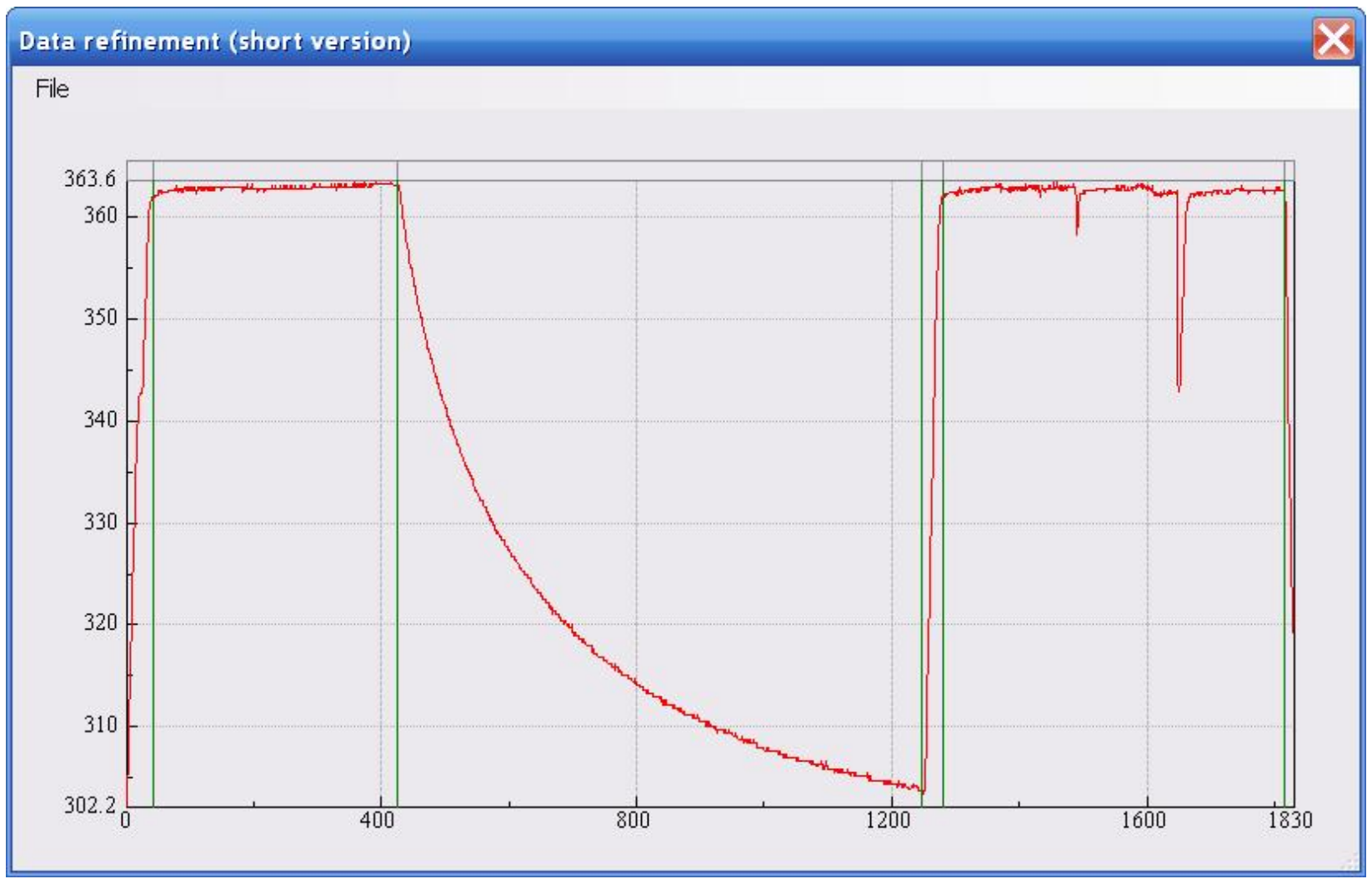

**Fig.16.26 Form_DataRefinement.cs** with the visualized data file.

The sliders in the **Form_DataRefinement.cs** allow to divide the full data file into segments, to which different methods of approximation are applied later. This form gives the full view of the data file, which may include many thousands of the (x, y) pairs. The full view makes it obvious, how to divide the big piece of data into segments, but the positioning of borders between segments is not too accurate here, as each horizontal pixel of a plot can be associated with dozens or even hundreds of real (x, y) values. The view of any data file in the plotting area in this form is used only for the rough dividing of this file into segments; more accurate partitioning and all further work on the data files is done inside the **Form_Zoom.cs**, which is called through the same context menu (**figure 16.27**). Though you see similar sliders moved across the plotting areas of both forms, the classes beyond the sliders in these two forms are different and they are organized in different ways.

All the movable lines (sliders) in the **Form_DataRefinement.cs** belong to the single object – an object of the `SegmentedData` class. When the form is opened and there is no opened data file, then an object of this class is only declared.

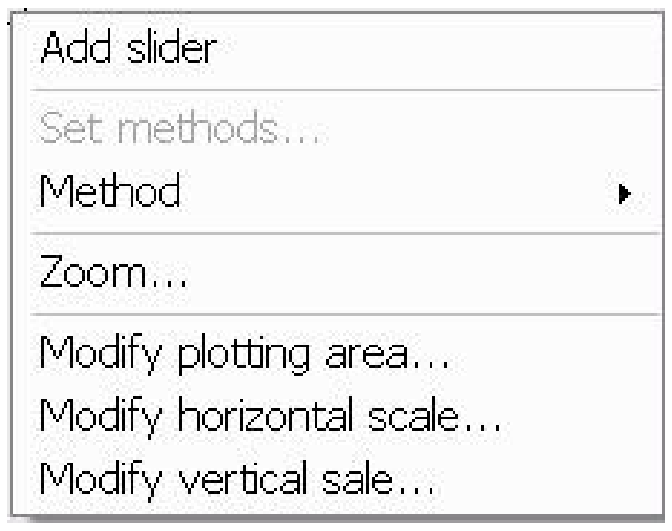

**Fig.16.27** Menu that is called inside the plot

```
        SegmentedData segmenteddata;
```

An object is initialized, when the data file is opened and the data is read.

```
    private void Click_miOpen (object sender, EventArgs e)
    {
        … …
        string filename = dlg .FileName;
        if (filename .EndsWith (".bin"))
        {
            double [] fxRead;
            double [] fyRead;
            if (ReadBinaryData (filename, out fxRead, out fyRead))
            {
                fxOriginal = fxRead;
                fyOriginal = fyRead;
                segmenteddata = new SegmentedData (plot .PlottingArea,
                                                   fxOriginal, fyOriginal);
                AdjustPlots ();
            }
        }
        RenewMover ();
        Invalidate ();
    }
```

The newly constructed `SegmentedData` object has two `Lists` of data, a `List` of segments, and a `List` of coordinates, on which the lines (sliders) must appear.

```
    public class SegmentedData : GraphicalObject
    {
        Rectangle rc;
        Pen pen;
        List<double> args = new List<double> ();
        List<double> vals = new List<double> ();
        List&lt;Segment&gt; segments = new List&lt;Segment&gt; ();  // minimum number will be 2
        List<int> cxLine = new List<int> ();
```

The first and the last lines are always positioned on the first and the last (x, y) pairs, so they are always on the borders of the plotting areas. These two lines are never shown and never moved, so they do not need to be covered by any nodes. The number of the sliders inside the plotting area is variable; these sliders can be added and deleted, but at least one slider inside the area always exists. Each movable line is covered by a simple thin rectangular node which can be moved only left or right. Movable sliders are shown atop the plotting area and included into the mover's queue before this area. The area itself is resizable by any border point. I do not want the cover of any slider to block any part of the sensitive border of the plot, so I make the cover for the sliders a bit shorter and move the borders of these covers several pixels inside the plotting area.

```
    public override void DefineCover ()
    {
        CoverNode [] nodes = new CoverNode [cxLine .Count - 2];
        for (int i = 0; i < nodes .Length; i++)
        {
            int cx = cxLine [i + 1];
            Rectangle rcNode = new Rectangle (cx - halfsense, rc .Top + halfsense,
                                       2 * halfsense, rc .Height - 2 * halfsense);
            nodes [i] = new CoverNode (i, rcNode, Cursors .SizeWE);
        }
        cover = new Cover (nodes);
    }
```

All the sliders together, regardless of their number, represent a SINGLE object of the `SegmentedData` class. This is seen very well from the `IntoMover()` method, where the sole object of this class is registered as a simple one.

```
        private void RenewMover ()
        {
            mover .Clear ();
            plot .IntoMover (mover, 0);
            if (segmenteddata != null)
            {
                mover .Insert (0, segmenteddata);
            }
        }
```

The sliders can be moved only left or right and their movements are restricted by the borders of the plotting area and the neighbouring sliders. These movements can be restricted in different ways. Usually, and I have written about it in the chapter *Movement restrictions*, the limits of movements for nodes are used inside the `MoveNode()` method. But the code of the `SegmentedData.MoveNode()` method does not show any restrictions at all, so there must be something different.

```
public override bool MoveNode (int i, int dx, int dy, Point ptM, MouseButtons btn)
{
    bool bRet = false;

    if (btn == MouseButtons .Left)
    {
        cxLine [i + 1] += dx;     // +1, because no node on the left line
        bRet = true;
    }
    return (bRet);
}
```

Sliders exist only inside a rectangular plotting area. If there is a single slider, then its area of existence is equal to this rectangular plotting area. If there are more sliders, then area of existence for each of them is some smaller part of the plotting area; it is also a rectangular area, but with lesser width defined by the neighbouring sliders. Each slider is covered by a thin node of several pixels width, so I even skip the possible difference of one or two pixels between the position of the cursor and the position of the node. As each node can be moved only inside some rectangular area, then it is possible to set an ordinary `Cursor.Clip` restriction on the mouse movement between the **`MouseDown`** and **`MouseUp`** events.

```
    private void OnMouseDown (object sender, MouseEventArgs e)
    {
        ptMouse_Down = e .Location;
        if (mover .Catch (e .Location, e .Button))
        {
            int iInMover = mover .CaughtObject;
            if (mover [iInMover] .Source is SegmentedData)
            {
                int iNode = mover .CaughtNode;
                int iLine = iNode + 1;
                int cxL = segmenteddata .LineCoordinate (iLine - 1) + 1;
                int cxR = segmenteddata .LineCoordinate (iLine + 1) - 1;
                Rectangle rcClip = new Rectangle (cxL, segmenteddata .Area .Top,
                                         cxR - cxL, segmenteddata .Area .Height);
                Cursor .Clip = RectangleToScreen (rcClip);
            }
        }
        ContextMenuStrip = null;
    }
```

When the mover in the **Form_DataRefinement.cs** is initialized, then its standard clipping inside the visual part of the form is organized by default. The smaller clipping rectangle is enforced, only when the `SegmentedData` object is caught by the mover, but not the plot or any of its movable parts; for them the clipping is unchanged. When any object is released, you do not need even to check the class of the released object; simply return the standard clipping of the cursor; the mover's clipping is unaffected by all these things.

```
        private void OnMouseUp (object sender, MouseEventArgs e)
        {
            Cursor .Clip = Rectangle .Empty;
```

Was there any other way to organize the restrictions of the movements for sliders? Certainly, it can be done in a classical way. In the working version there are lines on the left and right borders of the plotting areas. By "*there are lines*" I mean that there are corresponding elements in the `List<int> cxLine,` but these first and last members of this `List` are not covered by the nodes and never painted. You can cover them with the similar rectangular nodes, but make them even thinner (for example, 2 pixels wide) and set their behavioural parameter to `Behaviour.Transparent`. In such a way they will be not movable, but, being transparent, they would not block the resizing of the plotting area by those borders. Then you can delete the special clipping from the `OnMouseDown()` method of the form, but include a simple check for movement of nodes into the `SegmentedData.MoveNode()` method. The two transparent nodes at the ends will be not considered in the `MoveNode()` method, because they are transparent; all other nodes will use the positions of two neighbours to set the restrictions on their movements. It will be a more standard way of organizing the movement restrictions, but I do not think that it would be better in this particular case. The sliders would be moving correctly, but cursor would easily leave a caught slider and move farther on, when a slider would be stopped at the border. For this reason I decided to demonstrate in the **Form_DataRefinement.cs** a slightly different way of organizing the movement restrictions. I want to underline that it is possible because of the simplicity of the case and rectangular area of the movements.

Several interesting situations may occur, when the caught object is released. The first is the release of a slider after movement, which means the release by the left button. As a rule, the data in the input file consists of a huge number of (x, y) pairs, so each pixel along the horizontal scale of the plotting area corresponds to dozens, hundreds or even more (x, y) pairs. A slider has to be associated with one particular (x, y) pair, so in such a case it is only the question of selection of one of these pairs, which correspond to the particular horizontal coordinate (pixel). But there are situations, when there are gaps (in horizontal coordinate) between the neighbouring (x, y) pairs; the released slider cannot be left somewhere in between, but must be moved to the horizontal coordinate of the nearest (x, y) pair. This is done by the `SegmentedData.AdjustLineToValue()` method.

```
    private void OnMouseUp (object sender, MouseEventArgs e)
    {
        Cursor .Clip = Rectangle .Empty;
        if (mover .Release ())
        {
            int iInMover = mover .WasCaughtObject;
            GraphicalObject grobj = mover [iInMover] .Source;
            if (e .Button == MouseButtons .Left)
            {
                if (grobj is SegmentedData)
                {
                    (grobj as SegmentedData) .AdjustLineToValue (e .Location,
                                                                 mover .CaughtNode);
                    Invalidate ();
                }
            }
```

There are two movable objects in the **Form_DataRefinement.cs**: a plot of the `Plot` class and sliders, which all belong to the same `SegmentedData` object. Each of the objects is associated with its own context menu. When the plot is released by the right button and prior to this some data file was already opened, then there are at least two segments in the plotting area. The segment corresponding to the cursor position must be determined, as this parameter is important for the opened menu (**figure 16.27**). The number of segment is important for adding the new slider, for setting the method of approximation for particular segment, and for opening the **Form_Zoom.cs**.

```
private void OnMouseUp (object sender, MouseEventArgs e)
{
    if (mover .Release ())
    {
        … …
        else if (e .Button == MouseButtons .Right)
        {
            ptMouse_Up = e .Location;
            double nDist = Auxi_Geometry .Distance (ptMouse_Down, ptMouse_Up);
            if (nDist <= 3)
```

```
            {
                if (grobj is Plot && segmenteddata != null)
                {
                    for (iClickedSegment = segmenteddata .Segments .Count - 1;
                         iClickedSegment >= 0; iClickedSegment--)
                    {
                        if (segmenteddata .LineCoordinate (iClickedSegment) < e .X
                            && e .X <
                              segmenteddata .LineCoordinate (iClickedSegment + 1))
                        {
                            break;
                        }
                    }
                    if (iClickedSegment >= 0)
                    {
                        ContextMenuStrip = menuOnPlot;
                    }
                }
```

If there is an attempt to open the menu on a slider, then there is a check of another type. This menu contains only one comment – *Delete slider*. The only slider in view cannot be deleted, so the menu is not opened, if there are only two segments.

```
private void OnMouseUp (object sender, MouseEventArgs e)
{
    if (mover .Release ())
    {
        … …
        else if (e .Button == MouseButtons .Right)
        {
            … …
                else if (grobj is SegmentedData && segmenteddata.Segments.Count >2)
                {
                    iClickedSliderNode = mover .CaughtNode;
                    ContextMenuStrip = menuOnSlider;
                }
```

The sliders live and move only inside the plotting area. The individual movements of sliders were already discussed, but there are also the related movements; when the plot is moved or resized, sliders have to change their positions in the proper way.

```
    private void OnMouseMove (object sender, MouseEventArgs e)
    {
        if (mover .Move (e .Location))
        {
            if (segmenteddata != null)
            {
                if (plot .PlottingArea != segmenteddata .Area)
                {
                    if (plot .PlottingArea .Width == segmenteddata .Area .Width &&
                        plot .PlottingArea .Height == segmenteddata .Area .Height)
                    {
                        // plot is simply moved, so sliders must be moved also
                        int dx = plot.PlottingArea.Left - segmenteddata.Area .Left;
                        int dy = plot.PlottingArea.Top - segmenteddata .Area .Top;
                        segmenteddata .Move (dx, dy);
                    }
                    else
                    {
                        segmenteddata .Area = plot .PlottingArea;
                    }
                }
```

```
            }
            Invalidate ();
        }
    }
```

When the plotting area is moved, then the sliders are moved synchronously by the `SegmentedData.Move()` method.

```
        public override void Move (int dx, int dy)
        {
            rc .X += dx;
            rc .Y += dy;
            for (int i = 0; i < cxLine .Count; i++)
            {
                cxLine [i] += dx;
            }
        }
```

When the plotting area is resized, then the positions of all the sliders must be recalculated, which is done by the `SegmentedData.Area` property.

```
        public Rectangle Area
        {
            get { return (rc); }
            set
            {
                rc = value;
                SetLineCoordinates ();
                DefineCover ();
            }
        }
```

After looking into the most interesting corners of the **Form_DataRefinement.cs**, it is time to look at the **Form_Zoom.cs**, which is even more interesting. This form allows to work with the same data files, but with higher accuracy, because it shows only one segment at a time (**figure 16.28**). When the **Form_Zoom.cs** is opened, it shows the data only from that segment, at which the menu in the previous form (**Form_DataRefinement.cs**) was called, but there is no need to go back and force between two forms to jump from one segment to another. There are two `ListView` controls in the form: the first one includes the list of segments and highlights the line of the currently selected segment; the second one shows all the (x, y) pairs of data for the current segment. **Form_Zoom.cs** also allows to add the new sliders thus increasing the number of segments.

As in all the user-driven applications, everything in the **Form_Zoom.cs** is movable and resizable. The number of segments, on which the data file is divided, can vary from few to dozens. When the number of segments is small, there is no sense to make the list of segments big, but if there are a lot of segments, then it makes sense to enlarge this list. Anyway, it is the users' choice to place the list anywhere in the form and set its size. The same thing happens to the list of dots; this list shows the values of (x, y) pairs for the currently viewed segment.

Selection of the approximation method for each segment and declaring some parameters for the selected method are done in the *Method and settings* group. This is an object of the `ElasticGroup` class. This class was described in the chapter *Groups of elements*; I have mentioned not once that this class became the most useful for organizing groups in different kinds of programs. The flexibility of such groups allows users to rearrange them in absolutely different ways according to the preferences and views of each user. Plus, there is an auxiliary form for tuning all the visualization parameters of the group; for this call the context menu anywhere inside the group.

Several of the previous examples (and some examples further on) demonstrate that the real applications can use a lot of different context menus with a significant number of available commands.

- The **Form_Functions.cs** (**figure 16.3**) uses five different context menus. Each of those menus has a lot of command lines (**figures 16.24**).
- The **Form_PersonalData.cs** (**figure 15.2**) uses nine different context menus.
- The **Form_PlotsVariety.cs** (**figure 17.1** in the next chapter) uses 10 different context menus.

The number of context menus and the number of commands, available through them, can significantly vary. There are some applications with a low number of used menus and each of them can include only one or two commands. The **Form_Zoom.cs** is just such a case: recently there were three menus with only a single command in each of them; not long ago I have added several commands into one of those menus.

Previously I have demonstrated the groups, in which the currently unneeded inner elements were temporarily hidden; in other cases the inner elements can be fixed to prevent their incidental move. But these actions are needed not always. Some of the inner elements of the *Method and settings* group are not used with all the methods, but there is no sense of hiding couple of controls and then restoring them back, if another method is selected. The group is also not too densely populated; it can be always moved around by the inner empty spots, so there is no sense in adding the commands for fixing and unfixing the inner elements. The only thing which is really needed in this case is the calling of the tuning form for the group; this is the only command which is included into the menu of the group.

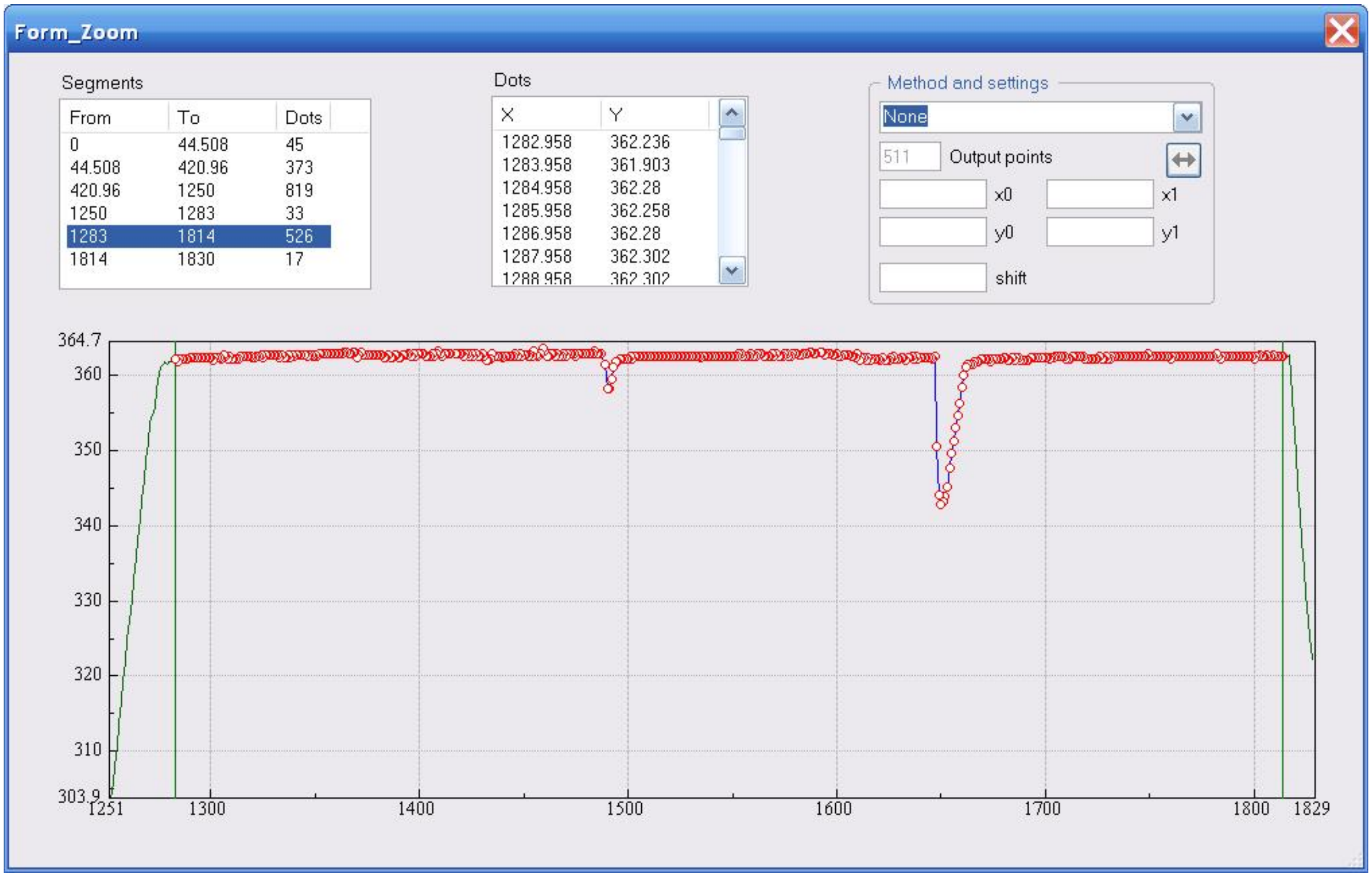

**Fig.16.28** **Form_Zoom.cs** is used to zoom any part of the data file to such a level that makes it easy to work even with the individual (x, y) pairs.

Two `ListView` controls do not need commands for hiding and restoring. The list of segments is in use all the time; if anyone does not want to keep the list of dots in view, then this list can be simply moved out of view across the border. The really needed thing is the possibility of changing the font for these controls; this is done via the only command, which is included into another menu that can be called at any empty place of the form.

The main part of the work in this form is done inside the plotting area. There is another context menu, which includes a command to add a new slider into view. There are also three commands in this menu to call the tuning forms of the plotting area and scales.

The tuning forms for scales of this plotting area have one significant limitation. Usually, the tuning of the ranges is one of the main things that these forms allow to do. But in the **Form_Zoom.cs** this is not allowed! The plotting area has to show one segment at a time; so in each case the X range is strictly defined by positioning of the sliders. In reality slightly more than a single segment is shown at a time, but this is purposely organized for better visualization and making easier the decision on moving the sliders. For these purposes the parts of the function beyond the sliders are also shown but in a different way. The sliders are often placed at the places, where the shape of the function changes, so it is very useful to see the function on both sides of a slider. Dots, corresponding to the (x, y) pairs inside the segment in view, can be moved up and down, so the vertical range of the plot is made slightly wider than the range [minimum, maximum] defined by the values on this segment. To prevent the manual (through the tuning forms) change of the range for the plot, some of the parameters are declared immutable, which makes them unchangeable through the tuning forms.

```
plot .HorScales [0] .ImmutableParams = Immutable .ValueLT | Immutable .ValueRB |
                                       Immutable .ValuesSwap;
plot .VerScales [0] .ImmutableParams = Immutable .ValueLT | Immutable .ValueRB |
                                       Immutable .ValuesSwap;
```

The biggest opportunities in the **Form_Zoom.cs** come not with the wide variety of available commands, but with what users can do manually with the mouse. And all these possibilities are based on using two classes of movable objects.

The first is the `UpDownDots` class. Such object, as it is obvious from the name of the class, is a collection of dots, which can be moved up and down. The **Form_Zoom.cs** always shows only one segment in the plotting area. On initialization, an object of this class gets the area of the plot, the boundary values, and the arrays of X and Y values, belonging to segment.

```
public class UpDownDots : GraphicalObject
{
    Rectangle rc;
    ValuesOnBorders borderValues;
    int nRad;
    Color clrDots;
    Pen penLines;
    double [] fX;
    double [] fY;
    PointF [] pts;
```

Based on the plotting area and the values on borders, the positions of all the (x, y) pairs, belonging to segment, are calculated; these pairs are shown as the small circles, but their size is enough to make their grabbing by the mouse and moving easy. The number of dots in segment can vary from few to several thousands. When the number of dots is huge, then the neighbouring dots can overlap (**figure 16.28**), but even in such a case you can see some areas of signal, where the pattern is obviously broken. This is the clear indication of some noise being added to the signal. There can be other situations with the individual dots staying far away from each other and seen as a very rough terrain (**figure 16.29**). In this case it is obvious, which of the dots must be moved up or down to make the line along the dots smoother and the calculations easier and more accurate.

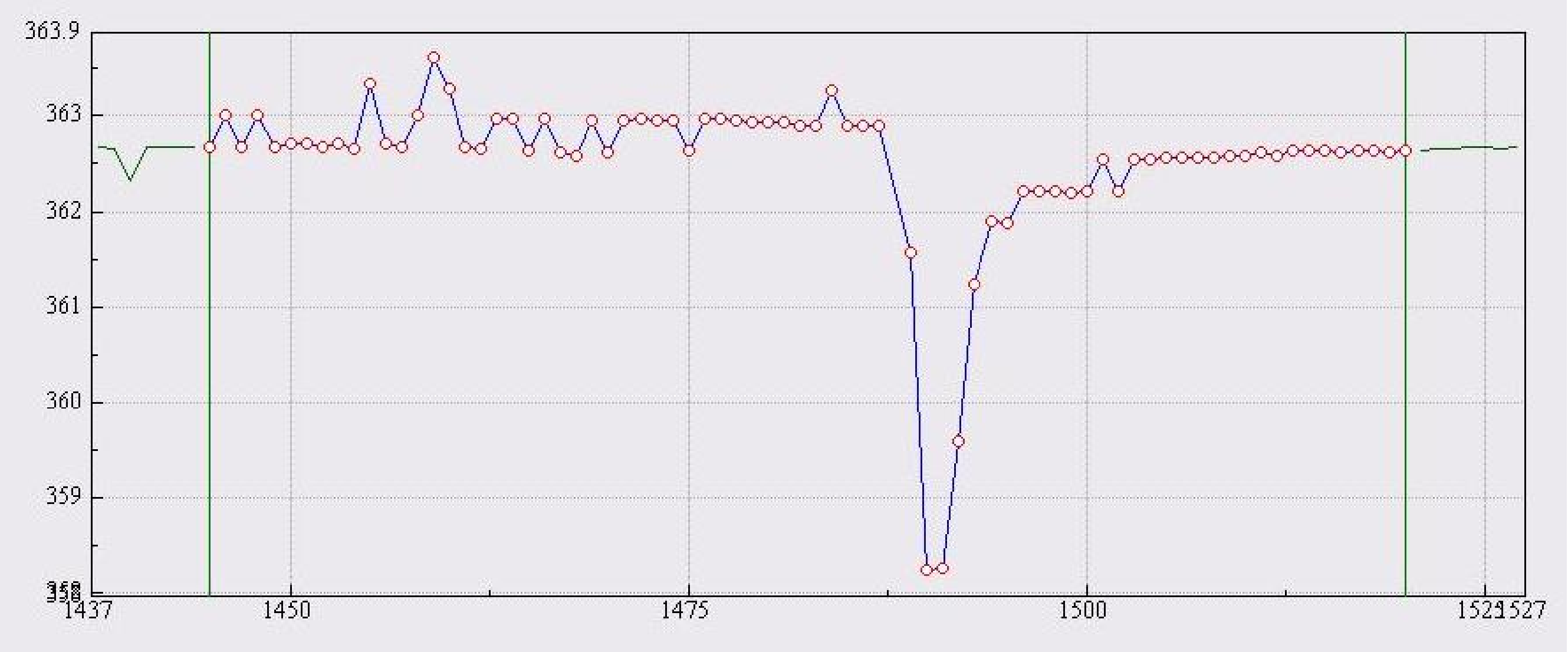

**Fig.16.29** This segment contains not too many dots; they do not overlap and it is more obvious, which of them have to be moved up or down to make the function smoother.

The design of a cover for such an object is easy enough. Each dot is covered by a small circular node, which allows to move this particular dot up and down. There is no need to move the whole collection of dots by itself, so no other nodes are needed.*

* The cover of any object is determined by its view and the requirements for its movements. I am slightly jumping ahead, but in the next example I will demonstrate a similar looking collection of dots, which are involved in different movements; this difference in movements will be the cause of different cover.

```
    public override void DefineCover ()
    {
        CoverNode [] nodes = new CoverNode [pts .Length];
        for (int i = 0; i < pts .Length; i++)
        {
            nodes [i] = new CoverNode (i, pts [i], radius, Cursors .SizeNS);
        }
        cover = new Cover (nodes);
    }
```

Moving of any node up or down is aimed on making the whole function smoother, so there is no sense in moving any dot out of view; this would be the only restriction on individual movements of dots.

```
public override bool MoveNode (int i, int dx, int dy, Point ptM, MouseButtons btn)
{
    bool bRet = false;
    if (btn == MouseButtons .Left)
    {
        float cyNew = pts [i] .Y + dy;
        if (rc .Top <= cyNew && cyNew <= rc .Bottom)
        {
            pts [i] .Y = cyNew;
            fY [i] = Auxi_Geometry .CoorToVal_Linear (rc .Top, rc .Bottom,
                                 borderValues .Top, borderValues .Bottom,
                                 Convert .ToInt32 (pts [i] .Y));
            bRet = true;
        }
    }
    return (bRet);
}
```

There is no need to move the whole collection of dots by itself, but this collection of dots is shown atop the plotting area. If the plotting area is moved, then all the dots must move synchronously; the `Move()` method is very useful in such a case.

```
        public override void Move (int dx, int dy)
        {
            rc .X += dx;
            rc .Y += dy;
            Size size = new Size (dx, dy);
            for (int i = 0; i < pts .Length; i++)
            {
                pts [i] += size;
            }
        }
```

The second interesting class, which is used in the **Form_Zoom.cs**, is the `LocalSlider` class. An object of this class is represented by a vertical line, which can be moved only left or right between the defined boundary positions.

```
    public class LocalSlider : GraphicalObject
    {
        int iArg;
        int cx, cyT, cyB;
        Pen penLine;
        int cxLeftBound, cxRightBound;
```

The cover for such a movable object is primitive and consists of a single rectangular node around the line; the width of the sensitive strip along the visible line of a slider is fixed to six pixels.

```
    public override void DefineCover ()
    {
        Rectangle rcNode = new Rectangle (cx - 3, cyT, 6, cyB - cyT);
        cover = new Cover (new CoverNode [] {
                             new CoverNode (0, rcNode, Cursors .SizeWE) });
    }
```

The possible individual movements of a slider to one side or another are restricted by the `cxLeftBound` and `cxRightBound` values, which are defined at the moment of initialization.

```
public override bool MoveNode (int i, int dx, int dy, Point ptM, MouseButtons btn)
{
    bool bRet = false;
    int cxNew = cx + dx;
    if (cxLeftBound <= cxNew && cxNew <= cxRightBound)
    {
        cx = cxNew;
        bRet = true;
    }
    return (bRet);
}
```

The sliders are supposed to live only inside the plotting area, so when the plot is moved, the existing sliders (there can be either one or two) have to move synchronously. This is the same related movement, which was described in some of the previous examples by the `InformRelatedElements()` method. A slider is described by its two end points, so the `Move()` method of such an object is really primitive.

```
    public override void Move (int dx, int dy)
    {
        cx += dx;
        cyT += dy;
        cyB += dy;
    }
```

As you can see, the **Form_Zoom.cs** contains a couple of simple movable objects, but the team-work of these objects produces very interesting results. Is it not amazing that one or two rectangles, which can be moved left and right, plus a set of small circles, which are moved up and down, can produce together a very efficient instrument for data analysis?

In nearly all the complex examples of this Demo application the movable objects are already prepared before calling the `RenewMover()` method of a form; in such a way this method becomes very simple and easy to understand. The situation with the **Form_Zoom.cs** is slightly different. Some of the movable objects are constructed inside the `RenewMover()` method; the mover's queue depends on the result of this construction. This peculiarity depends on the fact that the number of the selected segment determines the amount of sliders to be included into the mover's queue: if it is the first or the last segment of the data file, then there is only one slider in view (and in work); otherwise there are two sliders. It is not enough to determine the number of sliders at the moment, when the form is called for the first time; the number of sliders can be changed throughout the work by adding or deleting the sliders and by changing the current segment. For all these reasons, I found it more convenient to initiate the needed sliders inside the `RenewMover()` method.

The slider can move only inside the main plotting area, so the constructer of a slider uses the coordinates of this area. It also uses the coordinates of dots, because the sliders have to be positioned exactly on the dots, but never between them.

```
    private void RenewMover ()
    {
        mover .Clear ();
        plot .IntoMover (mover, 0);
        Rectangle rc = plot .PlottingArea;
        int iSegment = listSegments .SelectedIndices [0];
        Segment segment = srcData .Segments [iSegment];
        if (iSegment < srcData .Segments .Count - 1)
        {
            rightSlider = new LocalSlider (segment .RightNumber,
                              Convert .ToInt32 (dots [dots .DotNumber - 1] .X),
                                          rc .Top, rc .Bottom, Color .Green);
            mover .Insert (0, rightSlider);
        }
        else
        {
            rightSlider = null;
        }
        if (iSegment > 0)
```

```
        {
            leftSlider = new LocalSlider (segment .LeftNumber,
                                          Convert .ToInt32 (dots [0] .X),
                                          rc .Top, rc .Bottom, Color .Green);
            mover .Insert (0, leftSlider);
        }
        else
        {
            leftSlider = null;
        }
        mover .Insert (0, dots);
        group .IntoMover (mover, 0);
        ccSegments .IntoMover (mover, 0);
        ccDots .IntoMover (mover, 0);
    }
```

Sliders and dots, which are the movable elements discussed here in details, are visualized atop the plotting area, so they must be placed ahead of the plot in the mover's queue. But is there any difference in the order of dots and sliders? It looks like it does not matter, but it is not so. A slider is always placed at exactly the same **x** coordinate as some dot (see **figure 16.29**). The width of the node in the cover of a slider is exactly the same as the diameter of the node in the cover for a dot. If the slider would stay in the mover's queue ahead of the dots, then the node for this particular dot would be entirely closed by the node of the slider; there would be no chance to move this dot. Because of this, the dots have to precede the sliders in the mover's queue; you can see this from the code, which is shown above.

**Figures 16.28** and **16.29** demonstrate that the selected segment of data is shown in more details, but some additional parts of data are also shown beyond the sliders. (Certainly, these parts are shown only if it is not the beginning or the end of the data and there is anything beyond the sliders.) The dots, which happen to be inside those extra ranges, are saved separately and their values are used for two purposes.

- The **y** values of these (x, y) pairs are considered, while determining the vertical range of the plotting area.
- The **x** values of these (x, y) pairs are used to calculate the boundaries of the possible movements for sliders.

For example, if there are some values beyond the left slider, then it can be moved to the left up to the border of the plotting area. If one slider is moved in direction of another, then the limit of its movement is determined so that they would be never placed at the same dot.

```
private void OnMouseDown (object sender, MouseEventArgs e)
{
    ptMouse_Down = e .Location;
    if (mover .Catch (e .Location, e .Button))
    {
        if (e .Button == MouseButtons .Left)
        {
            GraphicalObject grobj = mover .CaughtSource;
            if (grobj is LocalSlider)
            {
                LocalSlider lsCaught = grobj as LocalSlider;
                int cxLeftLimit, cxRightLimit;
                if (leftSlider != null && leftSlider .ID == lsCaught .ID)
                {
                    if (fxLefter .Count > 0)
                    {
                        cxLeftLimit = plot .PlottingArea .Left;
                    }
                    else
                    {
                        cxLeftLimit = Convert .ToInt32 (dots [0] .X);
                    }
                    leftSlider .StartMovement (cxLeftLimit,
                                 Convert .ToInt32 (dots [dots .DotNumber - 2] .X));
                }
```

This call to the `LocalSlider.StartMovement()` is similar to the `StartResizing()` method, which was demonstrated many times in the examples of the first part of this book. At the moment, when an object is caught and the movement is going to start, the boundaries of the possible movement are calculated and set to an object as parameters of this method.

```
public void StartMovement (int cxL, int cxR)
{
    cxLeftBound = cxL;
    cxRightBound = cxR;
}
```

These boundaries are saved and used throughout the movement, while an object, in this case it is a slider, is moved left or right. These saved values are used inside the `LocalSlider.MoveNode()` method to determine the possibility of proposed movement, this method was shown two pages back.

With the movement of any dot, it is much easier. It can be moved only up or down; the boundaries of the possible movement are determined beforehand (these are the borders of the plotting area), so, when any dot is caught for moving, only the corresponding line in the `List` of dots must be highlighted.

```
private void OnMouseDown (object sender, MouseEventArgs e)
{
    ptMouse_Down = e .Location;
    if (mover .Catch (e .Location, e .Button))
    {
        if (e .Button == MouseButtons .Left)
        {
            GraphicalObject grobj = mover .CaughtSource;
            … …
            else if (grobj is UpDownDots)
            {
                int jNode = mover .CaughtNode;
                listDots .Items [jNode] .Selected = true;
                listDots .EnsureVisible (jNode);
            }
        }
    }
```

The `OnMouseDown()` method is often used to determine the boundaries of the possible movement; the significant changes are often done, when an object is released; for this we have to look into the `OnMouseUp()` method.

- If any dot is released, then the **y** value of the associated (x, y) pair is calculated, based on the new position of the dot. But this is not all, as the vertical range of the plotting area depends on the values from the shown segment. If needed, the range is changed.
- The release of a slider causes bigger changes. A slider can be released at any **x** coordinate, but it has to be placed exactly at some dot. This new dot, nearest to the position of the release, must be determined; the position of the slider must be adjusted. Any move of a slider to another dot means the redistribution of dots between two segments. This changes the information in both `ListViews` and the whole view of the plotting area, as it has to be changed according to the data of the shown segment.

There are enough movable objects in the **Form_Zoom.cs**: two `ListView` objects, the group with all its inner elements, plotting area, one or two sliders, and a number of dots. Movements of all these elements are described in the `MoveNode()` method of each particular class, so nothing else is needed. The only exception is the moving / resizing of the plotting area, because this area has strongly related objects: sliders and dots. Whenever the plotting area is changed, some other changes must be made, which is done inside the `OnMouseMove()` method. You can be surprised not to see the check of the class inside this method. Instead, there is a comparison of two rectangular areas from the `Plot` and `UpDownDots` objects; in this case it works fine. (If you would decide to change this part of code to the classical style with checking the class of the object under move, do not forget that the change of the plotting area can be caused both by the `Plot` and `RectCorners` objects.)

There are different reactions on moving and resizing of the plotting area. In the first case all the associated objects must be moved synchronously, which means calling their `Move()` methods. Those methods for the `UpDownDots` and `LocalSlider` classes were shown a bit earlier; both are simple.

```
private void OnMouseMove (object sender, MouseEventArgs e)
{
    mover .Move (e .Location);
    Rectangle rc = plot .PlottingArea;
    if (dots != null && rc != dots .Area)
    {
        if (rc .Width == dots .Area .Width && rc .Height == dots .Area .Height)
        {
            // area is simply moved, so dots must be moved also
            int dx = rc .Left - dots .Area .Left;
            int dy = rc .Top - dots .Area .Top;
            dots .Move (dx, dy);
            if (leftSlider != null)
            {
                leftSlider .Move (dx, dy);
            }
            if (rightSlider != null)
            {
                rightSlider .Move (dx, dy);
            }
        }
```

The resizing of the plotting area causes the recalculation of positions for dots and sliders. This has to start with the dots! When all the dots get their new positions, then the positions of sliders can be determined, because they have to reside exactly on the same **x** coordinates as the end dots of the segment.

```
private void OnMouseMove (object sender, MouseEventArgs e)
{
    mover .Move (e .Location);
    Rectangle rc = plot .PlottingArea;
    if (dots != null && rc != dots .Area)
    {
        … …
        else
        {
            // area is changed, so dots must be recalculated
            dots .Area = rc;
            if (leftSlider != null)
            {
                leftSlider .AdjustXCoor (leftSlider .ArgNumber,
                         Convert .ToInt32 (dots [0] .X), rc .Top, rc .Bottom);
            }
            if (rightSlider != null)
            {
                rightSlider .AdjustXCoor (rightSlider .ArgNumber,
                         Convert .ToInt32 (dots [dots .DotNumber - 1] .X),
                                       rc .Top, rc .Bottom);
            }
        }
```

One feature of the user-driven applications is not realized in the **Form_Zoom.cs**: not all the parameters of visualization can be tuned by users. Well, everything is prepared for this thing, but there is no access to some of the parameters. Sliders and dots with the connecting segments are drawn with the predefined colors, which cannot be changed in this working version. You can easily improve the design by adding a menu for the sliders with one command to change their color. Another menu can be added for dots to change their radius and color and to change the color of the line. If you do not want to add menus, you can organize another group of the `ElasticGroup` class; several controls inside the new group will do the job of setting the visualization parameters. All the needed properties to change the colors can be found in the `UpDownDots` and `LocalSlider` classes.

At the moment the drawing of sliders and dots uses the predefined colors.

```
private void OnPaint (object sender, PaintEventArgs e)
{
    Graphics grfx = e .Graphics;
    … …
    if (dots != null)
    {
        dots .Draw (grfx);
    }
    … …
    if (leftSlider != null)
    {
        leftSlider .Draw (grfx);
    }
    if (rightSlider != null)
    {
        rightSlider .Draw (grfx);
    }
```

There are no questions about the drawing of sliders; these straight lines are drawn with a predefined pen. If you look into the `UpDownDots.Draw()` method, it also looks simple; there is no problem in drawing a set of segments between the known points and to mark these points with the small circles.

```
public void Draw (Graphics grfx)
{
    for (int i = 0; i < pts .Length - 1; i++)
    {
        grfx .DrawLine (penLines, pts [i], pts [i + 1]);
    }
    if (bShowEnlargedDots)
    {
        for (int i = 0; i < pts .Length; i++)
        {
            Auxi_Drawing .FillEllipse (grfx, pts [i], radius, clrDots);
        }
    }
}
```

But if you check the value of the `bShowEnlargedDots`, you will find that it is **`false`**, so those small circles on **figure 16.29** are not drawn by the `UpDownDots.Draw()` method. Then who is responsible for drawing those circles?

Does not their view look familiar to you? Such small red circles filled with white inside did not appear in the last big and complex examples, but they were used frequently enough in the first part of the book, when I was explaining the design of covers. Yes, this is simply the visualization of cover for the `UpDownDots` class! I have mentioned that covers are never shown in the real applications, but here is a refutation of this statement.

```
private void OnPaint (object sender, PaintEventArgs e)
{
    Graphics grfx = e .Graphics;
    … …
    for (int i = 0; i < mover .Count; i++)
    {
        if (mover [i] .Source is UpDownDots)
        {
            mover [i] .DrawCover (grfx);
        }
    }
```

A strange and maybe funny (or amazing) example of using the `MovableObject.DrawCover()` method, but it works well enough in this case. Certainly, if you add an instrument to change the visualization of dots, then you have to take this part out and rely on drawing the small circles inside the `UpDownDots.Draw()` method.

The scientists, who work with this application (the original one; not the truncated version, which is shown here), use both types of movable objects (sliders and dots) to organize the needed segments and to exclude the noise from data for further calculations. Then one of the methods is selected for approximation; the parameters for each method can be also changed. The calculated results are shown at the same plot together with some additional results of calculations, which help to estimate the accuracy of approximation and to compare the used methods. The discussion of the used methods is definitely out of the scope of this book, but the use of the movable objects for design of such applications is demonstrated.

Nearly every serious task could be solved with some older methods. The question is how effective the new method is in comparison with the old one. Tycho Brahe was extremely accurate in all his measures, but when the calculus was invented, the astronomers received absolutely new methods for their work.

The data refinement can be done without the movable objects, but it would be so ineffective that there are big doubts if it can be used in such a way. On the contrary, the design of this application on the basis of movable objects turned it into very effective and useful program. (The original **DataRefinement** application was significantly changed in parallel with writing this book. The most significant changes were made in the **Form_Zoom.cs**; I have to keep one of my hands with another in order not to start rewriting the whole section of the book about this application.)

You can look through the long lists of data and by comparison of the values decide, which of them must be adjusted, but it is a very laborious work. Even then it would be a huge problem to decide, for how much this or that value must be changed. By moving the dots up or down, the same thing can be done easily. The human eye is a perfect instrument to estimate the smoothness of a curve. The calculations, which are done in parallel with such movement of dots, immediately show the accuracy of your manual adjustment.

The short additions of the data beyond the boundaries of the shown segment help to decide about the best positions of the borders for segments. Again, the calculations, going in parallel with such movement, and the visualization of these results immediately signal about the wrong or right movement of the borders.

Both things allow to refine the input data quickly and accurately. Users do not need to adjust manually the positions of hundreds or thousands of original dots. The methods of approximation will do the job perfectly, but some refinement of the noisiest parts of the input files must be done. That is the main purpose of the **DataRefinement** application.

## *Manual definition of graphs*

Different branches of science and engineering require and use a huge variety of specialized programs. The majority of these programs use graphical presentation of data and results, because it is much easier to understand the sets of numbers in the form of some plots than as columns in a table. From time to time very special types of plots are needed, like sonograms used in voice recognition and speech analysis, but the most often used is the case of Y(x) functions. The input data for drawing Y(x) graphs can be provided in different ways.

- Analytical equations transformed into the precoded methods.
- Experimental results stored in files or in the shared memory.
- Textual notation, which need an interpreter.
- Interactive application.

The first three variants were already demonstrated in the previous examples; now it is time to play with an interactive application, though it maybe not the best term to use. Anyway, the request for such an application sounds simple enough: it must be an easy to use program to enter the (x, y) pairs of values; an ordered collection of such pairs can be looked at as a Y(x) function. In such a way a profile of some device or surface can be defined; this is a well known and widely required task. The **MoveGraphLibrary.dll** includes a `Profile` class, which was designed to be used in such tasks. In one of the previous articles I have demonstrated the use of this class; see **Moveable_Resizable_Objects.doc** and **Test_MoveGraphLibrary.zip**, which are mentioned in the section *Programs and Documents*. But I do not want to copy that example into the book. Instead, I am going to demonstrate absolutely different example with several interesting classes.

An idea of a program, which gives you an opportunity to define a Y(x) function as a set of (x, y) pairs is very old. You can skip those decades, when every computer user was a programmer, and start from the moment, when people became divided on programmers and users. From the moment, when edit boxes and later tables gave users one or another way of entering new values into the programs, this task was implemented many, many times. It was already done in the textual mode; it became much easier for users, when the graphical mode allowed to visualize the results of typing the new values; in such a way all mistakes are more obvious. When you have side by side a table of values and its graphical presentation, then the only problem is to find in the table exactly that value, which you need to change. This minor problem also has not bad and obvious solution: each (x, y) pair in the table is associated with some dot on the graph; change the color of the dot that is associated with the pair currently selected in the table. In such a way many programs implemented this task of the definition of a Y(x) function and nearly all users thought that nothing else was needed. Well, as I have already mentioned, it looks OK, when you do not know about the possible movability of each and all elements. But when you know how easy it is to make any object movable, you quickly find out that this movability is needed and very helpful in all the corners of all the programs. The need of movability for elements penetrates the scientific applications at all levels and through all the corners.

File: **Form_GraphManualDefinition.cs**
Menu position: *Applications – Manual definition of graphs*

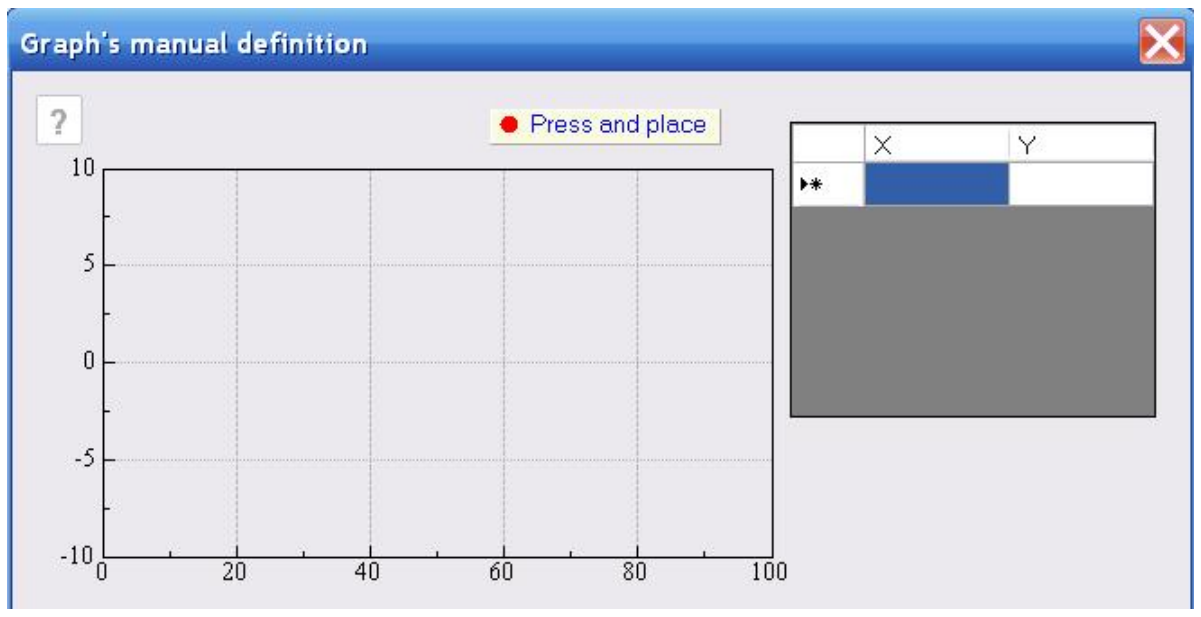

**Fig.16.30** At the beginning the **Form_GraphManualDefinition.cs** does not show any function

When you open the **Form_GraphManualDefinition.cs** for the first time, it shows two big empty objects and the small one (**figure 16.30**). The big objects are the plotting area and the table; the small one is a rectangle with a bright circular spot and a short information next to it. You can ignore this small object and define your function in an old way by typing the new values into the table. Or you can use this small object of the `DotNest` class and make the definition of a function much easier. Just do, what the information tells you to do (**figure 16.31**). Press the colored spot and move it to the new location inside the plotting area, where you want to place a new dot, which represents the new (x, y) pair. If you release the spot anywhere outside the plotting area, then nothing would happen, as it is a wrong place. If you release a spot anywhere inside the plot, then the new (x, y) pair will appear in the table and the new dot in the plotting area.

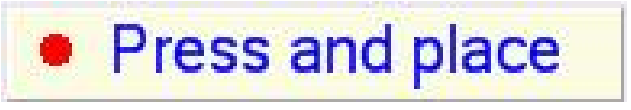

**Fig.16.31** `DotNest` object

The `DotNest` is relatively simple, but interesting class. Parameters of its constructor define the place, size, and view of an object.

```
public DotNest (Form form,
                PointF ptLT,        // top left corner of rectangle
                Font font,          // font of the text
                Color clr,          // color of the text
                bool show_frame,    // flag to decide the showing of a frame
                Color clrBack)      // background color
```

The specified font defines the size of the text; an extra space is added to the left of the text; this is the place for the colored patch. The whole rectangular object can be moved around by any inner point; the only exception is the patch, which can be moved individually. The patch must be shown atop the rectangular area and must be caught ahead of the area, on which it resides. The cover for such an object is obvious: two nodes, of which the small circular one must be the first.

```
public override void DefineCover ()
{
    CoverNode [] nodes = new CoverNode [] {new CoverNode (0, ptDot, nDotRadius),
                                           new CoverNode (1, rc)};
    cover = new Cover (nodes);
}
```

The patch has an interesting behaviour: it can be moved anywhere around the screen, but whenever it is released, it must be returned home to its “nest”. The `MoveNode()` method is easy, as there are no restrictions on movement for both nodes.

```
public override bool MoveNode (int i, int dx, int dy, Point ptM, MouseButtons btn)
{
    bool bRet = false;
    if (btn == MouseButtons .Left)
    {
        switch (i)
        {
            case 0:
                ptDot = ptM;
                bRet = true;
                break;
            case 1:
                Move (dx, dy);
                break;
```

```
        }
    }
    return (bRet);
}
```

The patch - the colored circle – is a relatively small circle. In the case of the patch I can ignore the difference between the cursor and the center of a circle at the moment, when the circle is pressed by the mouse, and use the changing mouse position (`ptM`) as the real place of the patch (`ptDot`). But, as you will see shortly, not all even small objects allow such a substitution; some of the small objects may demand to take into consideration any tiny discrepancy between the mouse position and the real position of an object.

The returning of the patch to the "nest" works in such a way. Two `Point` fields are defined in constructor:

- `ptDotNest` describes the position at home (the "nest"); its relative position inside the `DotNest` object never changes.
- `ptDot` describes the current position of the patch and it can be anywhere.

When a patch is moved around by a mouse, it can be at any location, so the `ptDot` value is used for drawing a patch.

```
public void Draw (Graphics grfx)
{
    grfx .FillRectangle (brushBack, rc);
    if (bFrame == true)
    {
        ControlPaint .DrawBorder3D (grfx, Rectangle .Round (rc), borderstyle);
    }
    Auxi_Drawing .DrawText (grfx, text, font, 0, clrText,
                            Point .Round (rcText .Location), TextBasis .NW);
    Auxi_Drawing .FillEllipse (grfx, ptDot, nDotRadius, clrDot);
}
```

When the patch is released, then some of the actions depend on the point of its release, for example, a new dot must be added to the graph, if the patch was released inside the plotting area. But the patch itself must be returned home immediately.

```
private void OnMouseUp (object sender, MouseEventArgs e)
{
    … …
    if (e .Button == MouseButtons .Left)
    {
        if (mover .Release (out iWasObject, out iWasNode, out shapeWasCaught))
        {
            GraphicalObject grobj = mover [iWasObject] .Source;
            … …
            else if (grobj is DotNest && shapeWasCaught == NodeShape .Circle)
            {
                dotNest .DotReturnHome ();
```

This method `DotNest.DotReturnHome()` is primitive and simply assigns the `ptDot` the "nest" place. The patch is back and ready to be used again.

```
public void DotReturnHome ()
{
    ptDot = ptDotNest;
}
```

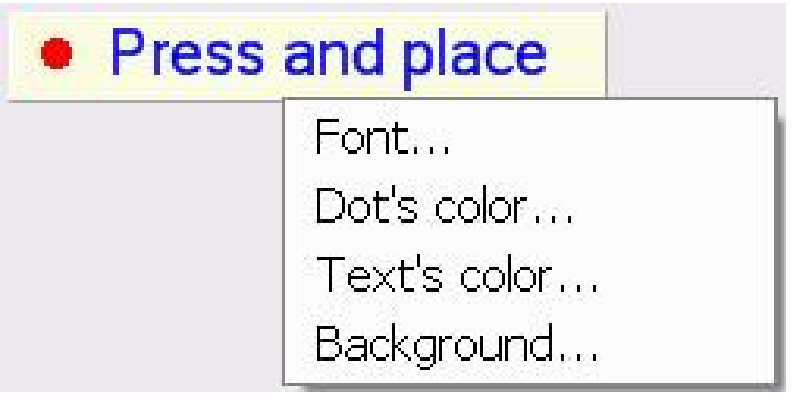

**Fig.16.32** A menu on the `DotNest` object

Though the `DotNest` object is a small one, but it is an object in the normal user-driven application, so all its parameters of visibility must be controlled by users. A menu can be called on this object to change all four possible parameters (**figure 16.32**). Only one parameter is excluded from the tuning in the current version – the transparency of an area. This `DotNest` object can be placed not only outside the plotting area (see **figure 16.30**), but also atop the plotting area (**figure 16.33**); especially for such a case some transparency of this object can be helpful. Because of this, the `DotNest` object is initialized with some transparency; this coefficient is not changed, when the background color is changed. You can easily add the tuning of transparency for the `DotNest` object; this is demonstrated, for example, in tuning the `ClosableInfo` object in the **Form_PersonalData.cs**.

The `DotNest` object is used to add the new dots on the graph. The first time you drop the patch inside the plotting area, a small dot will appear at the spot. Beginning from the second dot and further on all of them will be linked in order by segments (**figure 16.33**). All the (x, y) pairs are ordered according to the increase of their **x** component; the same **x** value for the consecutive dots is allowed. In some cases the positioning of the dots on the plotting area is just the best way to define the function. In other cases more accurate definition of values may be required; for this purpose you can use editing of numbers in the table. The new (x, y) pairs can be also added through the table; the new lines will be inserted according to the increase of their **x** components. There is one more way of adding new dots to the graph: press the graph with the left button and the new dot will appear at this spot; I will write about it a bit further.

Usually the tuning form of the horizontal scale allows to change both end values and in a real application I would certainly allow such a thing. However, if you change the end values in such a way that the X numbers would decrease from left to right, then I would have also to change the logic of entering the new values into the table. This would only make the code more complicated, but would have nothing to do with the discussion of the movability of the objects and the design of such application, so I blocked this feature by declaring some of the parameters for the scale immutable.

```
plot .HorScales [0] .ImmutableParams = Immutable .ValueLT | Immutable .ValueRB |
                                        Immutable .ValuesSwap;
```

The next interesting object in the **Form_GraphManualDefinition.cs** is the graph (class `DotsOnPlot`), which is painted as a set of colored dots connected with the segments. Well, this is how it looks at **figure 16.33**, but there is a way to change the graphs view. I will return to this a bit later, but now let us look at the cover of the `DotsOnPlot` object.

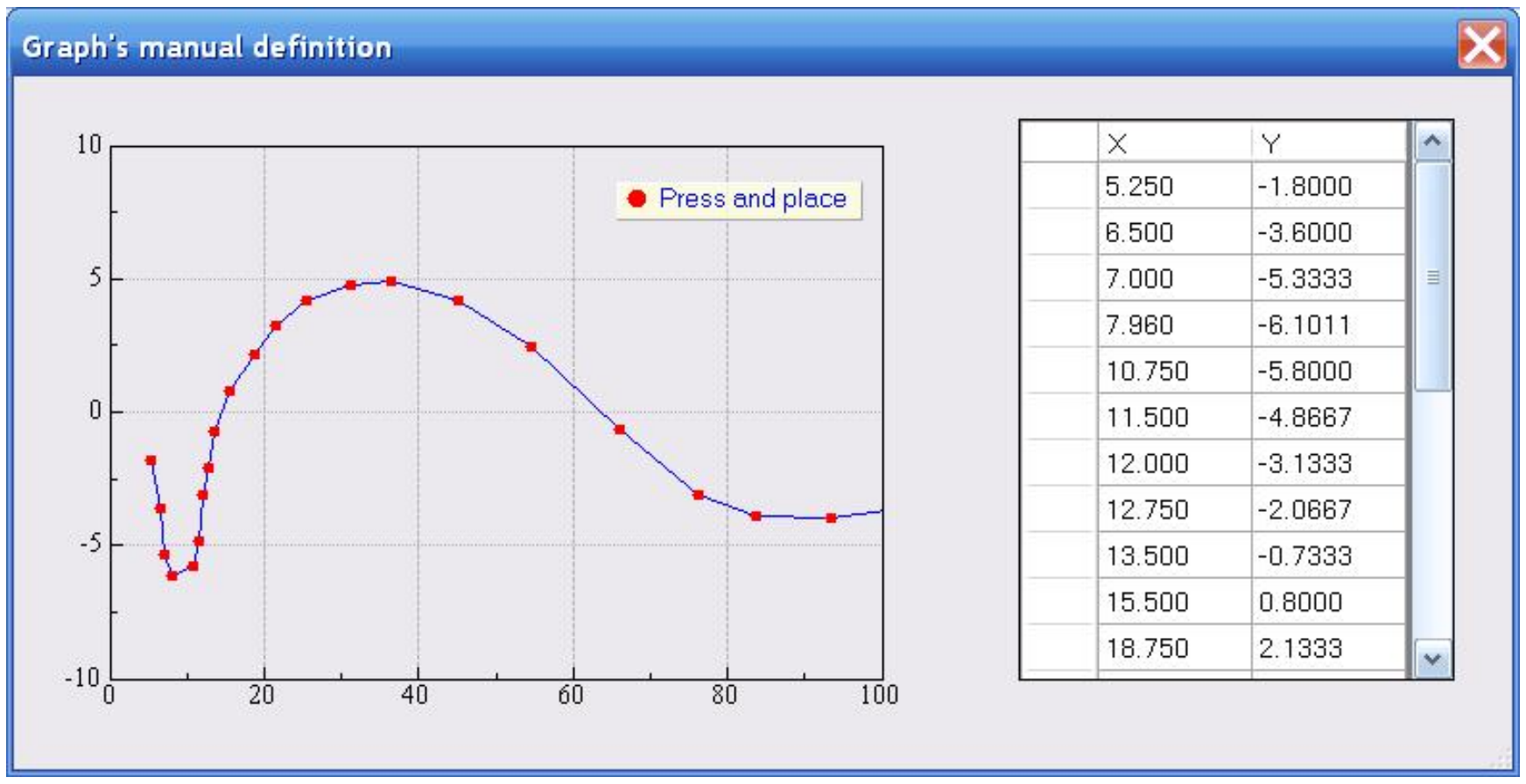

**Fig.16.33** A graph can be determined in parallel by adding/moving/deleting the dots and by editing/adding the values in the table

The main part of the `DotsOnPlot` class is a collection of points, at which the dots must be placed. This collection is linked with the `Plot` object, which has its boundary values, so each dot of the `DotsOnPlot` object gets the associated (x, y) values.

```
public class DotsOnPlot : GraphicalObject
{
    Plot plot;
    List<Point> pts = new List<Point> ();
    List<double> args = new List<double> ();
    List<double> vals = new List<double> ();
```

The design of cover for the `DotsOnPlot` class is determined by several ideas of how this class is going to be used.

- All the dots must be moved individually, so each of them must be covered by a small circular node. No other movement is needed for this object, so it looks like no other nodes are needed either, but there are other things that have to be considered.
- Moving the patch of the `DotNest` object and dropping it somewhere inside the plotting area is not the only way of adding new dots to the graph. I want also to add the new dots by pressing any segment at any needed place. This can be organized by analysing the geometry of this set of dots, but can I use mover's help for such a task? If each segment would be covered by a thin strip node, then mover can inform about the mouse pressing at the needed place. These strip nodes are not going to be used for any movement of an object(!); they are artificial, but very important pieces of the whole design.
- Dots can be moved outside the plotting area, though the graph is not shown outside this area. The problem of painting up to the borders of rectangular area (see the right end of a graph at **figure 16.33**) is easily solved by using some standard clipping throughout the time of drawing. Much more important is to organize it so that the circular and strip nodes outside the plotting area would be not seen by mover. The obvious solution to this is to cover everything outside the plotting area by a transparent node; it is impossible to do it with one node, but four transparent nodes close everything around the plotting area.

Thus we have such order of nodes in the cover of the `DotsOnPlot` class: four transparent nodes to cover everything outside the plotting area, then circular nodes on the dots, and then the strip nodes between each pair of consecutive dots.

```
public override void DefineCover ()
{
    int nNodes = 4 + pts .Count;
    if (pts .Count > 1)             // bShowLine &&
    {
        nNodes += pts .Count - 1;
    }
    CoverNode [] nodes = new CoverNode [nNodes];
    Rectangle rc = plot .Underlayer .Area;
    int cxL_out = -10;
    int cxR_out = rc .Right + 4000;
    int cyT_out = -10;
    int cyB_out = rc .Bottom + 4000;
    nodes [0] = new CoverNode (0, Rectangle .FromLTRB (cxL_out, cyT_out,
                                    rc .Left, cyB_out), Behaviour .Transparent);
    nodes [1] = new CoverNode (1, new Rectangle (rc .Right, cyT_out, cxR_out,
                                            cyB_out), Behaviour .Transparent);
    nodes [2] = new CoverNode (2, new Rectangle (cxL_out, cyT_out, cxR_out,
                                            rc .Top), Behaviour .Transparent);
    nodes [3] = new CoverNode (3, new Rectangle (cxL_out, rc .Bottom, cxR_out,
                                            cyB_out), Behaviour .Transparent);
    for (int i = 0; i < pts .Count; i++)
    {
        nodes [4 + i] = new CoverNode (4 + i, pts [i], halfsize, Cursors.Hand);
    }
    if (pts .Count > 1)         // bShowLine &&
    {
        int j0 = 4 + pts .Count;
        for (int i = 0; i < pts .Count - 1; i++)
        {
            nodes [j0 + i] = new CoverNode (j0 + i, pts [i], pts [i + 1],
                                            Behaviour .Frozen, Cursors .Hand);
        }
    }
    cover = new Cover (nodes);
}
```

There can be one more condition for using the strip nodes between the consecutive dots. The dots, representing the (x, y) pairs, can be shown in three different ways: special marks with lines between them, lines without marks, and marks without lines. When the lines are not visualized, then it can be decided that in this case users are not allowed to add a new dot (mark) by pressing somewhere on the invisible line. In this case you can use an additional flag `bShowLine`. I checked such a variant, but then moved this flag into comments; in the current version it is possible to add the dots by pressing the cursor on the non-visible, but existing strip nodes.

There are six movable objects in the **Form_GraphManualDefinition.cs**. Two of them - the information about this form, organized as a `ClosableInfo` object, and the small button to show this information, if it is hidden at the moment – are not seen at **figure 16.33**; four others are in view. Any controls must precede all the graphical objects in the mover's queue. I always prefer to put the small objects ahead of the big ones. This rule is used independently for graphical objects and controls. According to this rule, the small button precedes the table in the mover's queue. The order of graphical objects is partly obvious, as graph must be shown atop (and thus must stay ahead) of the plotting area; about others the designer has to make some decision. I decided to put all the objects in such an order: small button, table, information, `DotNest` object, graph, and plotting area. The `RenewMover()` method has to register all objects according to the declared order.

```
        void RenewMover ()
        {
            mover .Clear ();
            plot .IntoMover (mover, 0);
            mover .Insert (0, dots);
            mover .Insert (0, dotNest);
            if (info .Visible)
            {
                mover .Insert (0, info);
            }
            mover .Insert (0, scGrid);    // table turned into SolitaryControl object
            mover .Insert (0, btnHelp);
        }
```

Everything is declared movable and registered with the mover, now is the time to look into the details of this moving. The most interesting moment is adding the new dot by pressing any segment of the graph. It starts by pressing the strip node but ends with the caught circular node; all this without leaving the `OnMouseDown()` method. To make such a trick, the mover would have to release the first node and catch another one. I want to remind about one of the previous examples – the **Form_SpottedTexts.cs** (**figures 5.2** and **5.3**) from the chapter *Texts*. In that form there is a text sample, which can be caught either by the nodes on special spots of the text or by any inner point; in the last case the sample is immediately released. The **Form_GraphManualDefinition.cs** uses similar technique, but in a more interesting way.

I have described the cover of the graph (class `DotsOnPlot`): transparent nodes around the plotting area, circular nodes on the dots, and strip nodes on segments connecting the consecutive dots. When the graph is caught by the circular node, it is the beginning of movement for this particular dot. The number of the dot is easily obtained from the number of the caught node; the restrictions on the movement of any particular dot are easily calculated. The special case happens, when the graph is caught by the strip node.

```
private void OnMouseDown (object sender, MouseEventArgs e)
{
    ptMouse_Down = e .Location;
    if (mover .Catch (e .Location, e .Button))
    {
        if (e .Button == MouseButtons .Left)
        {
            if (mover .CaughtSource is DotsOnPlot)
            {
                if (mover .CaughtNodeShape == NodeShape .Strip)
                {
                    int iStrip = mover .CaughtNode - (4 + dots .DotNumber);
                    PointF ptBase;
                    double fDist = Auxi_Geometry .DistanceToLine (e .Location,
                                                  dots .Points [iStrip],
                                                  dots .Points [iStrip + 1], out ptBase);
                    mover .Release ();
                    int iDot = iStrip + 1;
                    dots .InsertNewDot (iDot, Point .Round (ptBase));
                    InsertTableLine (iDot);
                    NewPairIntoABArray (iDot,
                          new Data_ABPair (dots .Args [iDot], dots .Vals [iDot]));
                    Invalidate ();
                    mover .Catch (e .Location, e .Button);
                }
```

The graph is caught by the strip node, which covers one of the segments. The button is pressed at the place, where the new dot is supposed to appear. Unfortunately, the point of cursor cannot be used as the point for the new dot, but more accurate calculations are required. In many previous examples, when some vertex was caught via its small circular node or a line was caught by its thin strip node, I easily ignored the difference of one or two pixels and declared the position of a cursor as the new position of an object. Why is it not allowed here?

Any strip node has some width; to make the catch easier, it has the width of six pixels (it can be changed), so, as a rule, the cursor is pressed not exactly on the line but somewhere in vicinity. It is not a problem to declare the point of the cursor as a new dot for horizontal lines or the lines with not a big inclination, but it can be a disaster for vertical lines. The pair of consecutive dots may have the same **x** coordinates, but the next dot in array cannot have lesser **x** coordinate than the previous dot. Suppose that you have a vertical segment and you press the mouse somewhere in the middle but one or two pixels to the side of the line. The strip node is caught, but this point cannot be declared as coordinate for the new dot, because it violates the rule for **x** coordinates either with the previous or with the next dot. That is why some calculations are needed.

Taking into consideration the order of nodes in the `DotsOnPlot` cover and knowing the number of the caught node, it is easy to obtain the number of a strip.

```
    int iStrip = mover .CaughtNode - (4 + dots .DotNumber);
```

This number gives the numbers of two dots at the ends of this strip (segment of the graph). One of the methods from the **MoveGraphLibirary.dll** calculates the distance between a point and a line. I do not need this distance at the moment, but the same method returns the coordinates of the closest point on the line (`ptBase`); this point is going to be the point of the new dot. Thus calculated point lies on the line and guarantees that the rule for **x** coordinates will be not broken.

```
    PointF ptBase;
    double fDist = Auxi_Geometry .DistanceToLine (e .Location,
                         dots.Points [iStrip], dots.Points [iStrip + 1], out ptBase);
```

The number for the new dot is also determined by the number of the caught strip.

```
    int iDot = iStrip + 1;
```

After the number of the new dot and its coordinates are known, the information about the new dot must be included at all the needed places.

```
    dots .InsertNewDot (iDot, Point .Round (ptBase));
    InsertTableLine (iDot);
    NewPairIntoABArray (iDot, new Data_ABPair (dots.Args[iDot], dots.Vals [iDot]));
```

All the information is renewed, the graph with the new dot is already repainted, and the new dot is under the cursor, so it is natural to catch it.

```
    mover .Catch (e .Location, e .Button);
```

I would assume that the reason to press any segment anywhere inside is to add some curve to the segment by moving this new dot to another location; that is why I wrote that catching of the new dot immediately after its initialization would be natural. If somebody decided simply to add a new dot without moving it anywhere, then he would have to release the button. I think that that would be a rare case, but for this person it would be also natural.

What other interesting details can be found in the **Form_GraphManualDefinition.cs**? Some of them are in the `OnMouseMove()` method.

```
private void OnMouseMove (object sender, MouseEventArgs e)
{
    if (mover .Move (e .Location))
    {
        GraphicalObject grobj = mover .CaughtSource;
        if (grobj is Plot || grobj is RectCorners)
        {
            dots .PlotChanged ();
        }
        else if (grobj is DotsOnPlot && mover.CaughtNodeShape == NodeShape .Circle)
        {
            int iDot = mover .CaughtNode - 4;
            RenewTableLine (iDot);
        }
        else if (grobj is SolitaryControl)
        {
            Update ();
        }
        Invalidate ();
    }
}
```

When the plotting area is moved or resized, the dots of the graph must be also moved. The resizing of the plotting area can be done by sides or by corners. I have already explained why a special `RectCorners` object was included into the `Plot` class; both cases must be taken into consideration for recalculating the dots.

```
if (grobj is Plot || grobj is RectCorners)
{
    dots .PlotChanged ();
}
```

The `DotsOnPlot` class keeps two types of parameters for each dot: the coordinates and the physical (x, y) values.

- When the values are changed through the table, then the physical border values of the plotting area are used for calculation of the new coordinates. This is done in the `CellEndEdit_datagridXY()` method.
- When any dot is moved individually, its new coordinates are used to calculate the new (x, y) pair.

```
else if (grobj is DotsOnPlot && mover.CaughtNodeShape == NodeShape.Circle)
{
    int iDot = mover .CaughtNode - 4;
    RenewTableLine (iDot);
}
```

- When the plotting area is moved or resized, the unchanged (x, y) pair is used to calculate the new coordinates.

```
public void PlotChanged ()
{
    for (int i = 0; i < args .Count; i++)
    {
        pts [i] = plot .PhysToPoint (args [i], vals [i]);
    }
    DefineCover ();
}
```

Several interesting actions take place, when an object is released by the mover; a lot depends on the class of the released object.

- The `ClosableInfo` object demonstrates some needed information, but it can be closed when you are familiar with all the commands and do not need such a reminder any more. The cover of an object consists of two rectangular nodes; the small cross in the corner, by clicking which the panel can be closed, is covered by the first node. If the `ClosableInfo` object is released and the number of the caught node was zero, then the panel must be hidden and the mover's queue must be changed.

```
private void OnMouseUp (object sender, MouseEventArgs e)
{
    ptMouse_Up = e .Location;
    double nDist = Auxi_Geometry .Distance (ptMouse_Down, ptMouse_Up);
    if (e .Button == MouseButtons .Left)
    {
        if (mover .Release (out iWasObject, out iWasNode, out shapeWasCaught))
        {
            GraphicalObject grobj = mover [iWasObject] .Source;
            if (grobj is ClosableInfo && iWasNode == 0)
            {
                (grobj as ClosableInfo) .Visible = false;
                btnHelp .Enabled = true;
                RenewMover ();
                Invalidate ();
            }
```

- When the patch of the `DotNest` object is released, it has to return into the "nest". But if it is released inside the plotting area, then the new dot must be added to the graph and the new line with the (x, y) values of this dot must appear in the table. In the code below you can see that the conditions for this case include not only the release of the patch inside the plotting area, but also this point has to be outside the `DotNest` object. The second condition is not crucial and can be excluded; it is up to you to decide about it. The `DotNest` object is movable, so it can be positioned outside the plotting area or over it; in such way it is shown at **figure 16.33**. The object is made slightly transparent; the graph can be seen through this rectangular area. Anyway, this check of the new position for a dot to be not covered by the `DotNest` object is not crucial.

```
private void OnMouseUp (object sender, MouseEventArgs e)
{
    ptMouse_Up = e .Location;
    double nDist = Auxi_Geometry .Distance (ptMouse_Down, ptMouse_Up);
    if (e .Button == MouseButtons .Left)
    {
        if (mover .Release (out iWasObject, out iWasNode, out shapeWasCaught))
        {
            GraphicalObject grobj = mover [iWasObject] .Source;
            … …
            else if (grobj is DotNest && shapeWasCaught == NodeShape .Circle)
            {
                dotNest .DotReturnHome ();
                if (plot .PlottingArea .Contains (e .Location) &&
                    !dotNest .RectAround .Contains (e .Location))
                {
                    int iDot = dots .InsertNewDotInXOrder (e .Location);
                    InsertTableLine (iDot);
                    NewPairIntoABArray (iDot,
                      new Data_ABPair (dots .Args [iDot], dots .Vals [iDot]));
                }
                Invalidate ();
            }
```

The **Form_GraphManualDefinition.cs** has three different context menus; the menu selection is done in an ordinary way inside the `OnMouseUp()` method. Previously there were four menus, but the one, which was called on the `ClosableInfo` object, had a single line in it, so I changed it for direct calling of the appropriate tuning form.

```
private void OnMouseUp (object sender, MouseEventArgs e)
{
    … …
    else if (e .Button == MouseButtons .Right)
    {
        if (mover .Release (out iWasObject, out iWasNode, out shapeWasCaught))
        {
            if (nDist <= 3)
            {
                GraphicalObject grobj = mover [iWasObject] .Source;
                if (mover .WasCaughtSource is ClosableInfo)
                {
                    info .ParametersDialog (this, RenewMover, ParametersChanged,
                                            null, PointToScreen (ptMouse_Up));
                }
                    else if (grobj is DotsOnPlot)
                {
                    ContextMenuStrip = menuOnGraph;
                }
                else if (grobj is DotNest)
                {
                    ContextMenuStrip = menuOnDotNest;
                }
                else if (grobj is Plot)
                {
                    ContextMenuStrip = menuOnPlot;
                }
            }
        }
    }
```

As you can see from this piece of code, three different menus are used inside the **Form_GraphManualDefinition.cs.** When the number of menus is big (around 10 or more), I prefer to organize the menu selection in a separate method; with a small number of menus the code becomes easier to understand if the selection is organized inside the `OnMouseUp()` method.

`menuOnDotNest` is used for tuning the `DotNest` object and was already shown at **figure 16.32**.

`menuOnPlot` can be called anywhere inside the main plotting area, but not on the line of the graph; menu allows to call three auxiliary forms for tuning the plotting area and its scales; it also allows to delete the graph (**figure 16.34**).

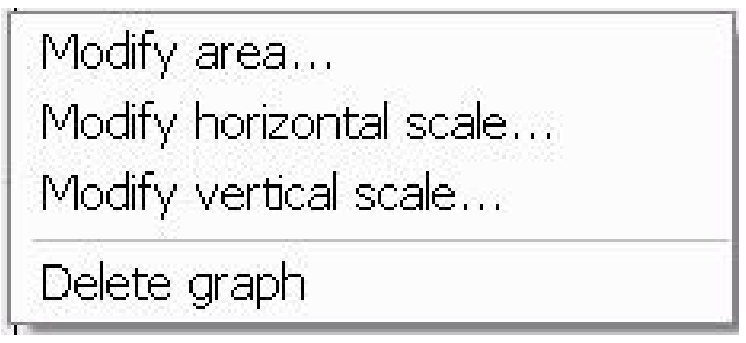

**Fig.16.34** Menu on plotting area

`menuOnGraph` can be called on the dots of the graph and on the… Here is a small problem in correct definition. The `DotsOnPlot` object is shown at **figure 16.33** as a set of dots connected by a line; when the graph has such a view, the `menuOnGraph` can be called on any dot and any segment between the dots. But the same menu can be called at the same places between the dots even when the segments are not shown. The cover of the `DotsOnPlot` object does not depend on the current view of the graph and the calling of menu is initialized by the mover, which senses the cover of this object. There is a small difference in the view of the called menu, which depends on whether the menu is called directly on the dot or somewhere between; in the last case the first command of this menu is disabled (**figure 16.35**).

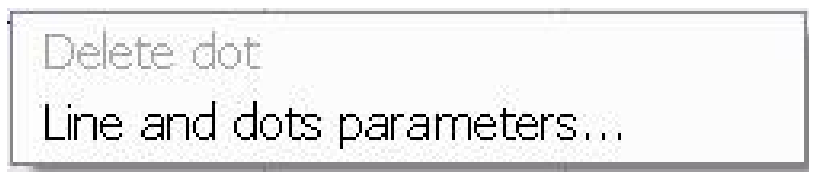

**Fig.16.35** Menu on the graph; the first command is available only when the menu is called directly at some dot.

The second command of menu calls the auxiliary tuning **Form_MarkedLineParams.cs**, which allows to change all the visibility parameters of the graph (**figure 16.36**). The graph can be shown in three different views: line with the markers, line without markers, or markers without line. For segments you can change the color, width, and style. For the markers you can change the view, color, and size; pay attention that this is half of the real size, so for the circles you are declaring radius in such a way. (This tuning form is a bit unfinished: there is a menu outside the groups, but there is no menu to change the groups.)

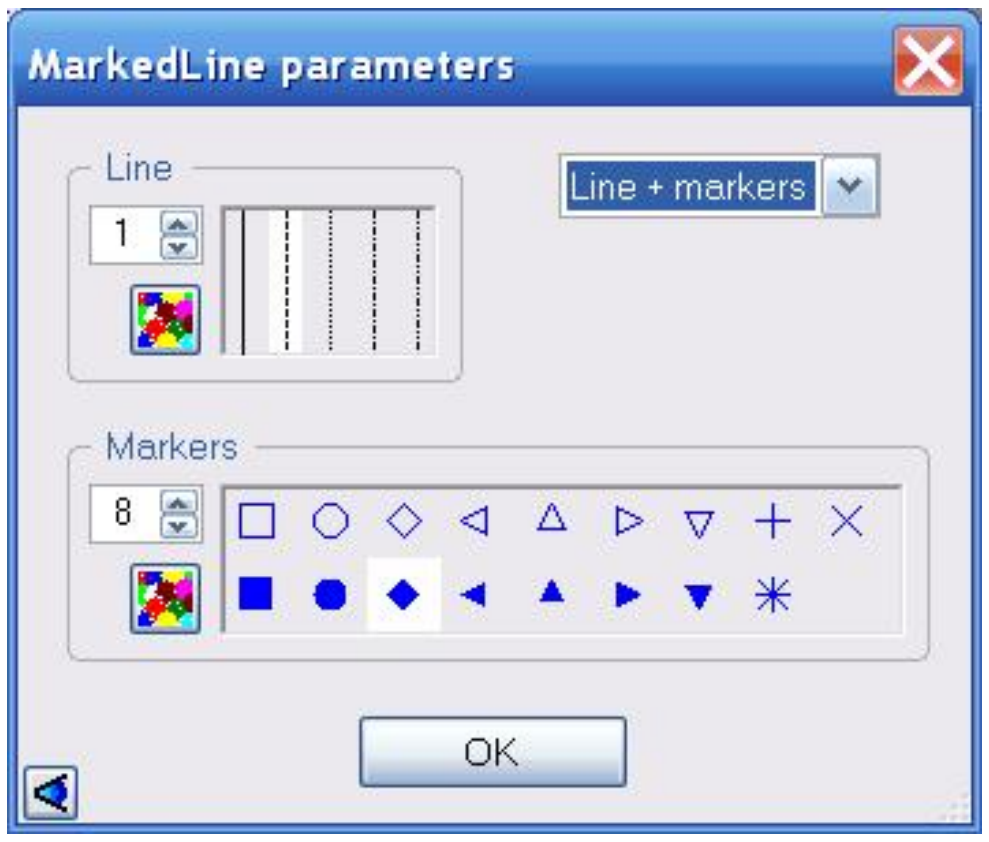

**Fig.16.36** **Form_MarkedLineParams.cs** can be used to change the view of the graph.

<u>Short conclusion to this chapter.</u>

Not surprisingly at all that this chapter about the applications for science and engineering became the biggest in the book. The whole idea of total movability of each and all screen elements was born on demand from this area and the majority of new improvements are either aimed directly at this area or at some related tasks. I have demonstrated several examples in this chapter; overall they show how the movability of the objects can be used in different kinds of scientific / engineering applications. I think that this type of applications demand the highest level of movability for all the involved objects in each and all tasks. There is no end of the new more and more interesting tasks in the realm of scientific applications, which require the design of new, interesting, and unusual movable objects. Implementation of these movable objects in different programs opens the new possibilities not only in design, but, which is more important, in the research work, for which those programs are only instruments. With more sophisticated instruments the research work becomes more exciting. This is the best result and the best evaluation of the importance of applying the new programming technique to the area of scientific applications.

# Data visualization

Data can be visualized through many different types of plots. Graphs of Y(x) functions were demonstrated in the previous chapter. This chapter shows some other very popular plots, turned into movable and resizable.

The world of data is infinitive. There are many forms of data visualization; there are many systems and libraries, which allow to show data in different ways. Throughout my professional life I worked on different scientific / engineering problems, which use mostly the standard plots of the Y(x) type or something similar. There were also special plots, like sonograms, which are among the most interesting things I was working on. With the invention of an algorithm for turning objects into movable / resizable, I was especially looking for unfamiliar objects to check the algorithm with the unusual forms or in strange situations. The variety of plots, widely used outside the familiar to me scientific areas, turned out to be a good range for my ideas. The plots, which I am going to discuss in this chapter, are more often used for economical and financial analysis than anywhere else. For some time I was using the term "financial plotting" for the objects, which will appear in this chapter, but this is not a correct term, though it can give a main idea. The area of financial analysis, the discussions of stock market use some types of plots, which you will never see anywhere else. From my point of view, all these plots are shouting to be turned into movable / resizable, because they perfectly fit with the ideas of user-driven applications and can be changed in such a way to the huge benefit of their users. And there are millions and millions of people, who daily analyse the huge amount of economical and financial data. For such users the turn of all the used plots into objects, which anyone can rearrange according to his personal preferences, means a lot in understanding this huge amount of data and in the outcome of their analysis.

I do not want to introduce you to the whole variety of plots for financial and economic analysis, which can be turned into movable and resizable. This chapter is limited to only three types of plots: bar charts, pie charts, and ring sets.

File: **Form_PlotsVariety.cs**
Menu position: *Applications – Variety of plots*

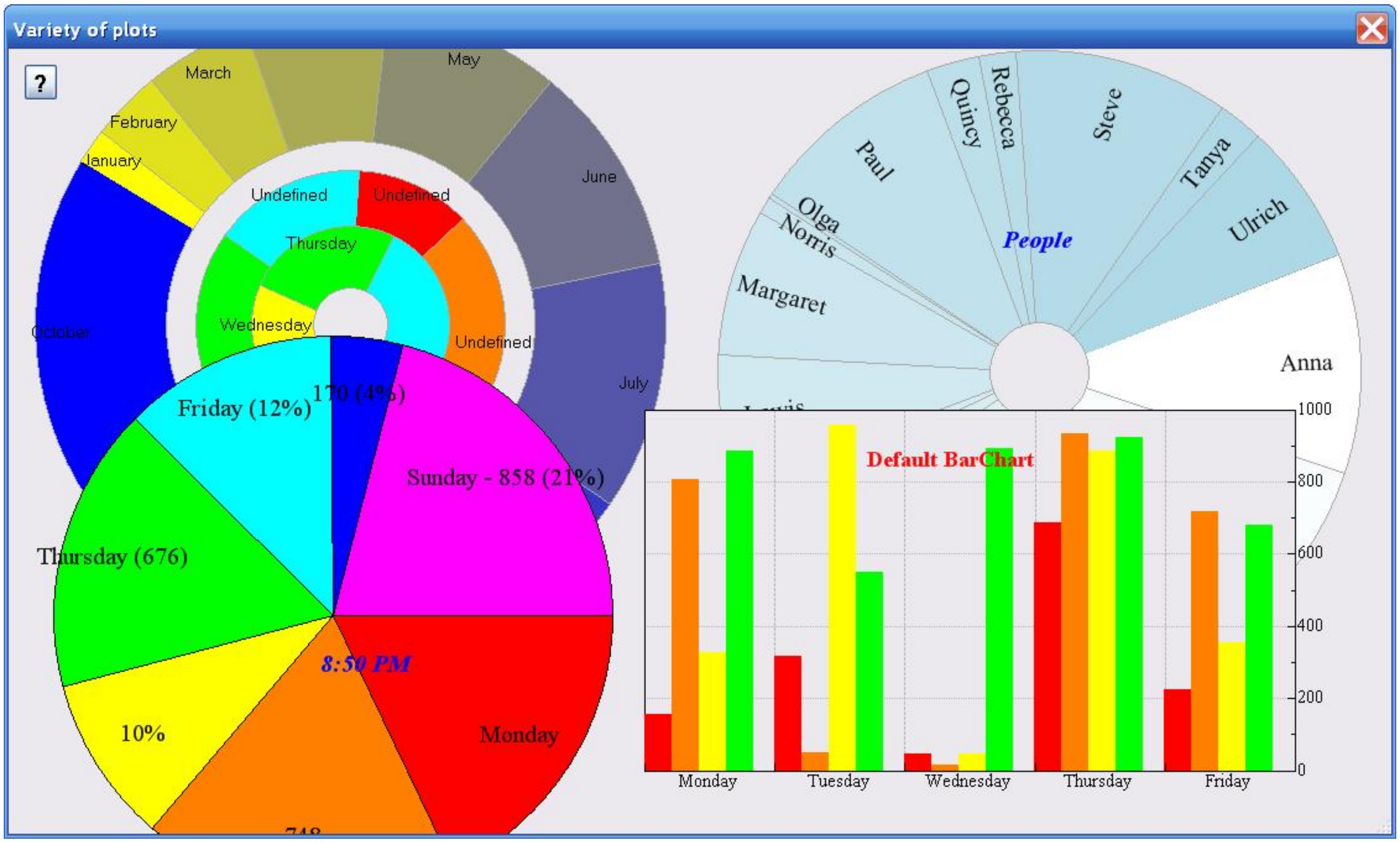

**Fig.17.1** Variety of plots

**Figure 17.1** demonstrates the default view of the **Form_PlotsVariety.cs**. The `BarChart` and the `PieChart` classes are represented at this view by one object each; the `RingSet` class is represented by two objects to show that an object of this class can consist of any number of rings. Before discussing the **Form_PlotsVariety.cs**, let us look into the details of the three mentioned classes.

## Bar charts

A bar chart consists of a set of rectangular bars; the length of bars is proportional to the values, which they represent. There are many different types of bar charts, but I am going to demonstrate and discuss only the "classical" one (**figure 17.2**).

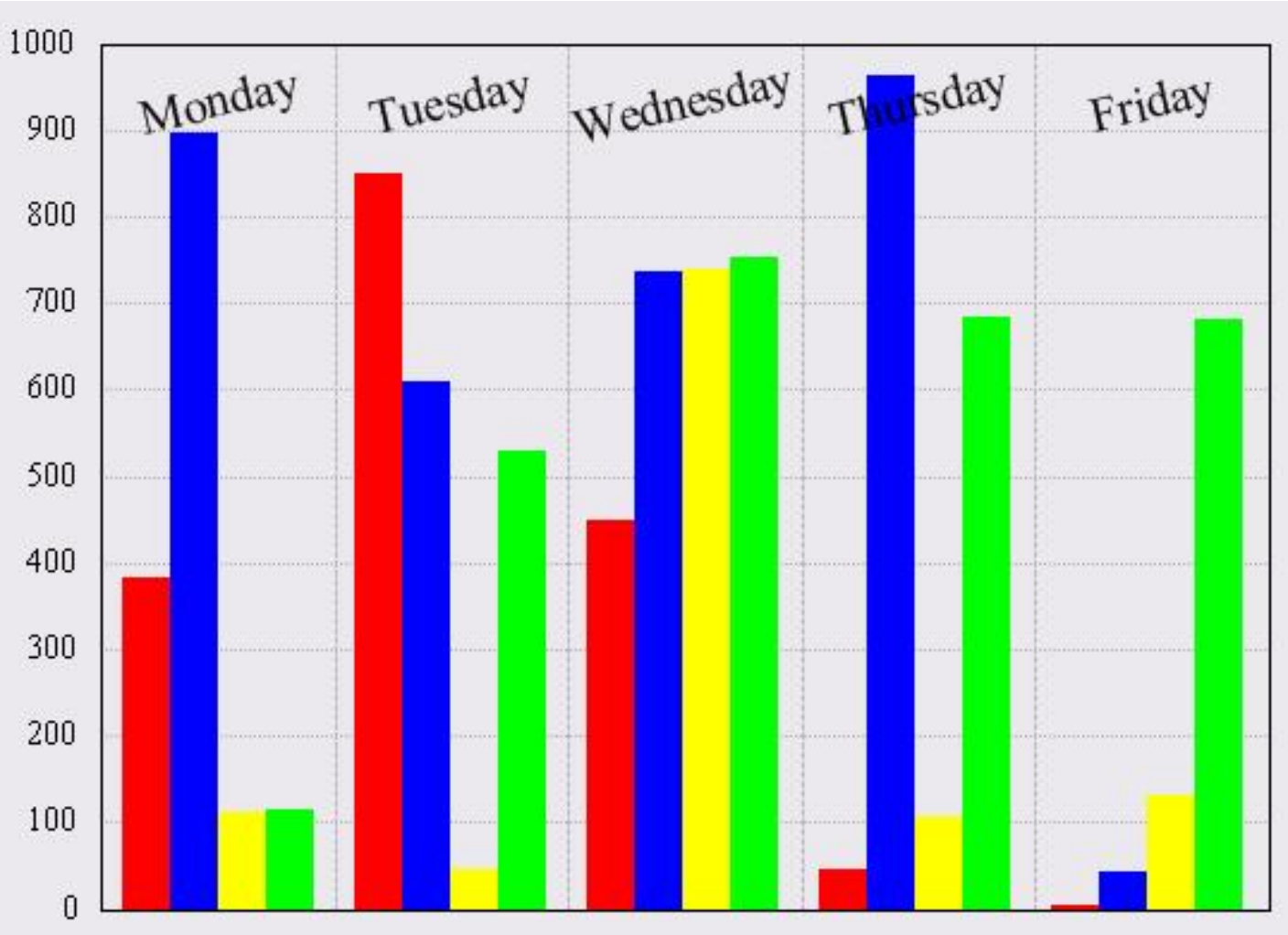

**Fig.17.2** Bar chart

From many aspects the `BarChart` class is similar to the `Plot` class, which was described in the previous chapter. Because of these similarities, I try to keep them in behaviour and in use as close as possible, so that users and programmers would not need to learn a lot of new things, if they are familiar with one of them. But just a bit later you will see that the tuning forms of these two classes are slightly different at the moment. This happened, because I have already changed the tuning form for the `Plot` class to newer version, while the `BarChart` class continues to work with the older version. In a short period of time I will change it, so they will be very close again.

A `BarChart` object consists of the main rectangular plotting area (of the `RectArea` class), one horizontal scale, and one vertical scale. One of these scales, the one which goes along the bars and shows the numeric values, belongs to the `Scale` class. Another scale, orthogonal to the first one and showing the textual information, belongs to the `TextScale` class. You cannot determine beforehand the types of horizontal and vertical scales, as the bar chart can be easily turned around at any moment. The angle of each turn is fixed to 90 degrees, but can be made in any direction. The main area and each of the scales can be associated with an arbitrary number of comments. All the comments belong to the `CommentToRect` class.

```
public class BarChart : GraphicalObject
{
    RectCorners rectcorners;
    RectArea rarea;
    TextScale scaleText;
    Scale scaleNum;
    List<CommentToRect> comments = new List<CommentToRect> ();
```

A `BarChart` object can be moved and resized in exactly the same way as any `Plot` object: move it by any inner point of the main plotting area and resize by the borders of the same rectangle. The scales of a bar chart can be positioned anywhere in relation to the main plotting area, so there was the same problem with the resizing in case, when a scale is placed over the border. The same problem was solved in exactly the same way: you can see a `RectCorners` field in the `BarChart` class, so a bar chart can be resized by any corner regardless of the positions of scales.

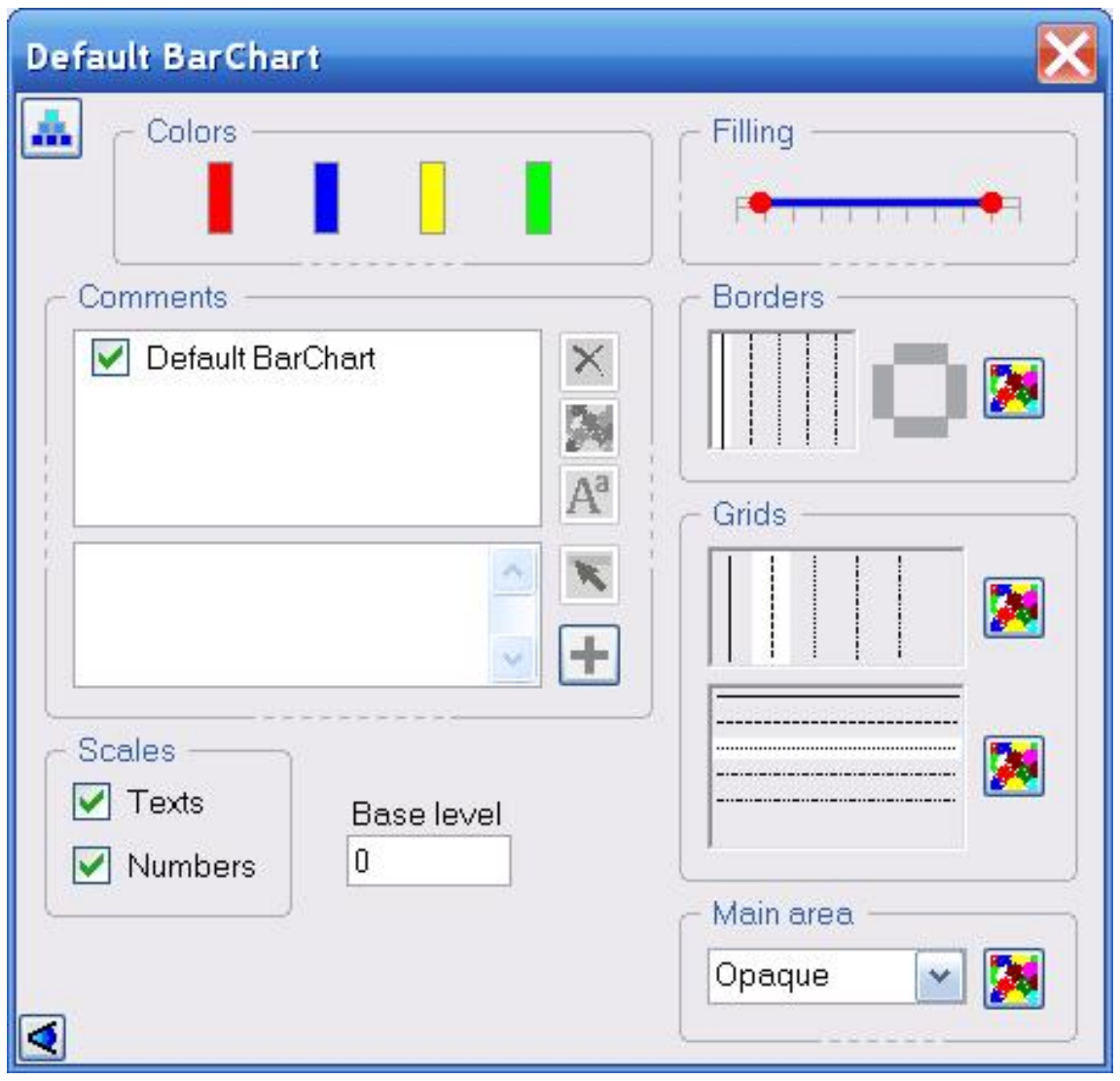

**Fig.17.3** The tuning form for the main area of the bar chart

Whatever can be done with the bar chart is best seen in comparison of the bar chart from **figure 17.2** and the tuning form for its main area (**Form_BarChartParams.cs**, **figure 17.3**). Three groups in this tuning form - *Comments*, *Borders*, and *Grids* - look nearly the same, as in the tuning form for the `Plot` class, but there are some differences.

*Borders* and *Grids* groups are movable, but not resizable, and the positions and sizes of their inner elements are fixed.

The stretches of the dashed lines in the frame of the *Comments* group indicate that the group is resizable. This is an object of the `Group` class, so the sizes of the inner elements are changed, when the frame of the group is resized (press the frame at any point and move), but the rules of resizing are decided by the developer (me…) and not by the users. The next version will be more friendly and work as in the tuning form for the `Plot` class (**figure 16.6**). As for now, you still have a chance to compare two tuning forms and decide for yourself, what you would prefer as a USER.

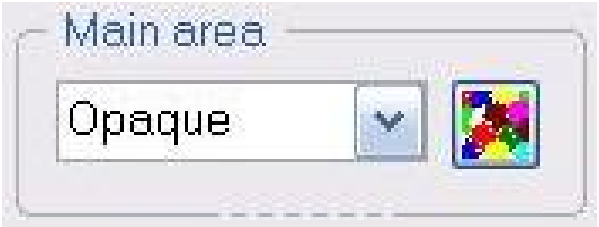

**Fig.17.4** Setting the main area color

The current version of the `BarChart` class allows to declare the main plotting area either opaque and select the color for the background (**figure 17.4**) or make it transparent, in which case the background color has no sense. You can compare it with the flexible transparency of the main area for the `Plot` class (**figure 16.9**).

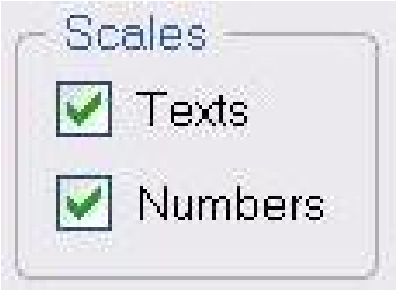

**Fig.17.5** Scales can be shown or hidden

Any `BarChart` object has exactly two scales, of which one is numerical and another is textual. In the *Scales* group, it is possible to decide about their visibility (**figure 17.5**). The scales can be hidden from view via their context menu, which is easier and faster to do, but this group in the tuning form is the only way to reinstall the visibility of scales.

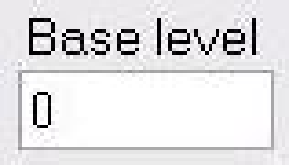

**Fig.17.6** Setting the base level

With the base level set to the value on one of the borders, you see the rectangular bars going in one direction only from that border. By changing the base level to something between the two border values (**figure 17.6**), you can receive the rectangular bars going from that level in both directions. In some cases such view can be more informative.

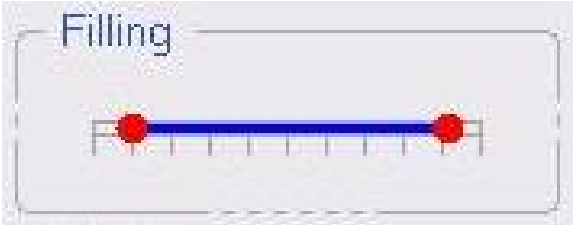

**Fig.17.7** Changing the filling of the bar chart

The *Filling* group allows to change the filling of the main area by the colored bars. Two red balls at the ends of the wide blue line (**figure 17.7**) can be moved by a mouse left and right.

The rectangular plotting area of the `BarChart` is divided into segments according to the structure of the associated data; the bar chart from **figure 17.2** has five segments (data for five days). Each segment has the same width. A segment includes several colored bars of equal width and may have an empty space at both ends. The number of sets is the same for all segments and is also determined by the structure of the associated data; the bar chart from **figure 17.2** has four sets. The graphical object from **figure 17.7** is used to regulate the filling of each segment. All the sets (all the colored bars) have equal width, so it is possible that several pixels at the border of each segment will be not used, even if the filling is set (at least, demanded) from one end to another. Human eye is a perfect instrument, which can see even a couple of pixels gap between colored bars. To avoid this, the width of the whole main area can be changed.

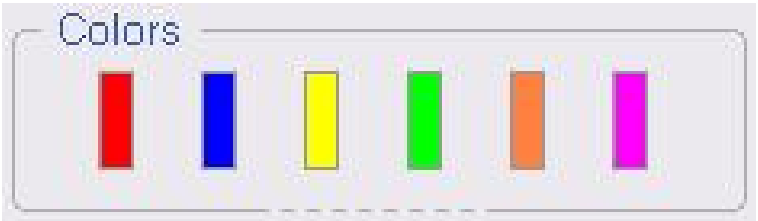

**Fig.17.8** *Colors* group

A bar chart is represented by a series of the colored rectangular bars; the *Colors* group is the place to change the colors and their order (**figure 17.8**). This group may include more colored samples, than you see in the bar chart (**figure 17.2**). The samples inside the group can be moved by a mouse from place to place, thus changing their order. The change of any parameter in the tuning form is immediately reflected in the view of the bar chart under tuning, so a change in the order of samples immediately changes the view of the original bar chart. Thus you can include extra samples into the *Color* group and select the best combination of colored bars in the bar chart by changing the order of samples. Only the first colors equal to the number of sets is used at any moment. The order of colors is changed by the mouse; all other operations are done via the context menu, which can be called inside the group (**figure 17.9**).

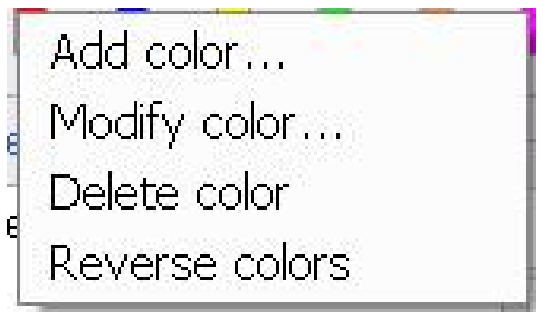

**Fig.17.9** Menu inside the *Colors* group

The view of this menu partly depends on whether it was called on one of the colored samples or anywhere else inside the group between the samples. The *Modify color* and *Delete color* commands are available only if the menu was called exactly on one of the samples. Two other commands allow to add a color to the set and to reverse the whole set of colors.

Couple of additional things can be still seen and used in the tuning form of bar charts. They are present in all the tuning forms discussed in this chapter, but these things are the remnants of the older versions and will disappear in the nearest future (in the next version). But while they are present, I want to mention them.

The first one is the 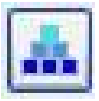 button in the top left corner of the form. Clicking this button restores the default view of the form. The same restoration of the default view of the form can be done now via the context menu that can be called anywhere outside the groups, so this button is not needed any more. It quietly waits in the corner until the moment of the next redesign.

One more "dinosaur" is another command in the same menu: it allows to switch the covers ON / OFF! In the first chapters of this book, in all the examples, where I explained the design of covers for different simple objects, I included the special button for visualization of those covers. But you never see anything like this in any real application. Couple of years ago I included the same "cover visualization" button into all the tuning forms, because users did not know that the tuning forms were designed under the same rules as the applications themselves. By including the "cover visualization" button into the tuning form I attracted some attention to it; after clicking such button, users became aware that all the groups in the forms were also movable. Later I took those buttons out, but left the possibility of switching the covers ON / OFF through the menu command. Now I think that the users of such applications know that everything is movable, so the command for visualization of covers can be taken out. As with the "default view" button, this command is going to disappear with the next redesign of this tuning form.

`BarChart` objects have two scales of two different types. The numeric scale of the `Scale` class was discussed in the previous chapter together with its tuning form. The textual scale of the `TextScale` class was not used anywhere before we came to the bar charts, so it is time to look at this scale and its tuning form (**Form_TextScaleParams.cs**, **figure 17.10**).

A textual scale is simpler than the numerical; not surprisingly that the tuning form of such scale is also simpler. A textual scale may have any number of comments, which can be positioned arbitrary to the scale and moved and rotated individually. The tuning of such comments was already described in the previous chapter in connection with the numeric scales, there are absolutely identical *Comments* groups in the tuning forms of both types of scales (compare **figures 16.11** and **17.10**) so there is no need to discuss the tuning of comments again.

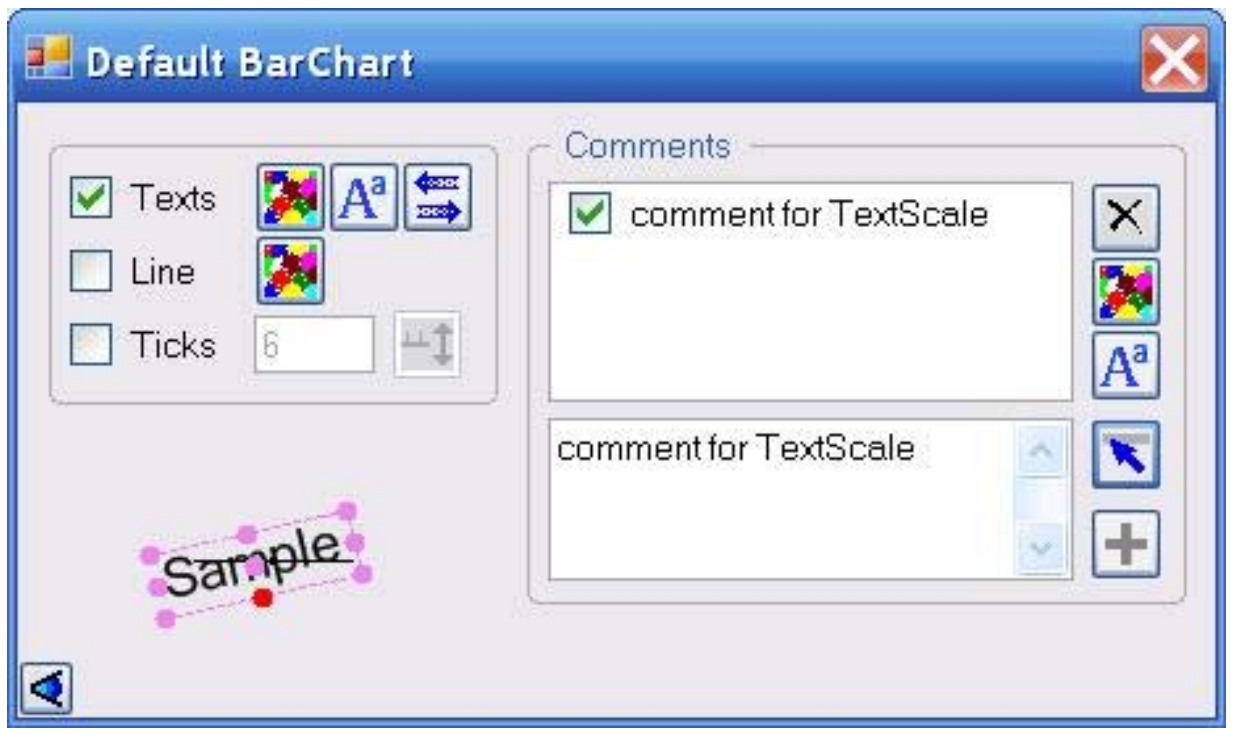

**Fig.17.10** Tuning form for the `TextScale` objects

If you exclude for some time from discussion the comments of the scale, then the remainder of a textual scale consists of three main parts:

- The main line.
- The ticks.
- The words (texts) along the line; one text per segment.

These three parts can be switched ON / OFF in different combinations, but it is forbidden to switch OFF all three of them. Switching of all three parts would mean the disappearance of the scale; this can be achieved much easier by hiding the scale via the context menu.

The <u>main line</u> has only one parameter – color.

The <u>ticks</u> have the same color, as the main line. The ticks can be flipped to another side of the main line by clicking the 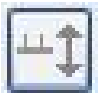 button. This is the view of the button for the horizontal scales; for vertical scales the picture on the button is different, so it is never confusing. The length of the ticks can be changed, but the minimum allowed length is set to two pixels. It is not the limitation of the users' right to do whatever they want, as ticks can be easily switched OFF. This limitation allows to avoid an awkward situation, when the ticks are left as the only part of the scale in view and then their length is set to zero. In this case the scale theoretically exists in view (it is not declared as hidden), but the area, occupied by the visible parts, is zero, so the mover does not sense such an object. Thus the tuning form of a scale cannot be opened, the parameters of a scale cannot be changed; the scale becomes a ghost without any chances to return back to life. I am not sure that any program needs such ghosts.

The color and the font of the <u>texts</u> can be changed in an ordinary way via the standard dialogs. Texts have as many parts, as there are segments in the associated data: one part of the texts per segment. The 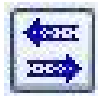 button allows to mirror the image of the bar chart; the texts are associated with the segments, so the order of texts is also reversed. This is the view of the button for the horizontal scales; for vertical scales the picture on the button is different, so it is never confusing.

The texts and the comments of the textual scale may look the same, but these are the elements with the different purposes. They behave differently in some situations; the understanding of such differences is important. The texts are always positioned along the line of a scale; the line can be invisible, but the lining for all the texts and their angle (the same for all) is still regulated by the single *Sample* from the tuning form (**figure 17.10**). The textual scale can be placed outside the main plotting area or atop this area. Each text has its own length and they can be lined in different ways (by a corner, by the middle point, etc.), but their anchor points are always lined in the same way.

Five words near the top of **figure 17.2** are the texts of the textual scale. The main line and the ticks were made invisible; then the scale was moved to the top of the plotting area. The relative sizes of the colored bars inside the bar chart are determined by the data; in this case the highest bar partly covers one of the texts. Certainly, the whole scale can be moved even higher, so that the bar will not cover the text, but there is also another solution.

**Figure 17.11** shows the upper part of the same bar chart, but in addition to the textual scale, which I moved slightly down, there are comments, which nearly duplicate the original view of the textual scale from **figure 17.2**. By putting these two collections of words next to each other I want to show that visually they are indistinguishable; they use the same font, they are rotated on the same angle, and I could easily set them the same color. One word of the upper set is moved out of the line; this can be done only with the comments, which can be moved individually.

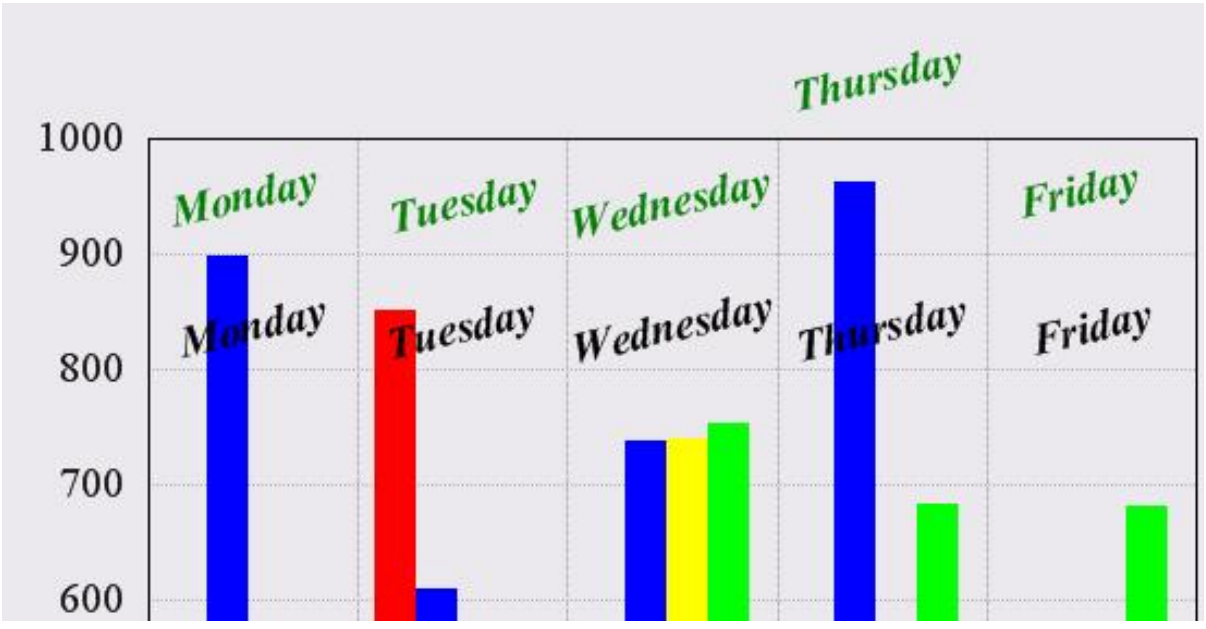

**Fig.17.11** Comments and texts of the `TextScale` object can look similar, but they behave differently

Everything comes with a price. The individualism of the comments requires more efforts in their placement and rotation; the easiness of placement and rotation of all the texts strips them of any chance to go out of line. Users have the full control over everything in user-driven application (I will return to discussion of these things again a bit later), so there are really interesting things that users can do here. Data for a bar chart arrives from some outside source; it can be a result of calculations or information from the database. The texts for the scale in this case were proposed by the author of an application, which turned to be me. If for any reason users do not like these texts, they can turn OFF the scale and substitute it with their own comments. The behaviour of texts and comments in response to any movement, resizing or rotation of the bar chart is identical. There is no way to find out that the textual scale was substituted by some comments, which were not there originally!

## Pie charts

Pie charts were among the first really complex nonrectangular objects, to which I tried to apply the ideas of movability. The moving / resizing of the old versions of pie charts and rings [11] can look odd in comparison with the currently used, but the awkwardness of those old variants was exactly the thing that pushed me to think about different solutions and eventually resulted in development of the N-node covers. This type of covers allowed to include into normally movable and resizable a whole variety of non-rectangular objects. Now the pie charts can be moved by any inner point and resized by any border point.

A `PieChart` object is initialized by a central point, the radius of a circle, and an array of values. The number of values determines the number of sectors; the ratio of values determines the angles of the sectors, as their sum must be equal to 360 degrees.

An object of the `PieChart` class consists of three main parts:

- A circle, divided into sectors.
- Comments, associated with the whole circle.
- Comments, associated with the sectors.

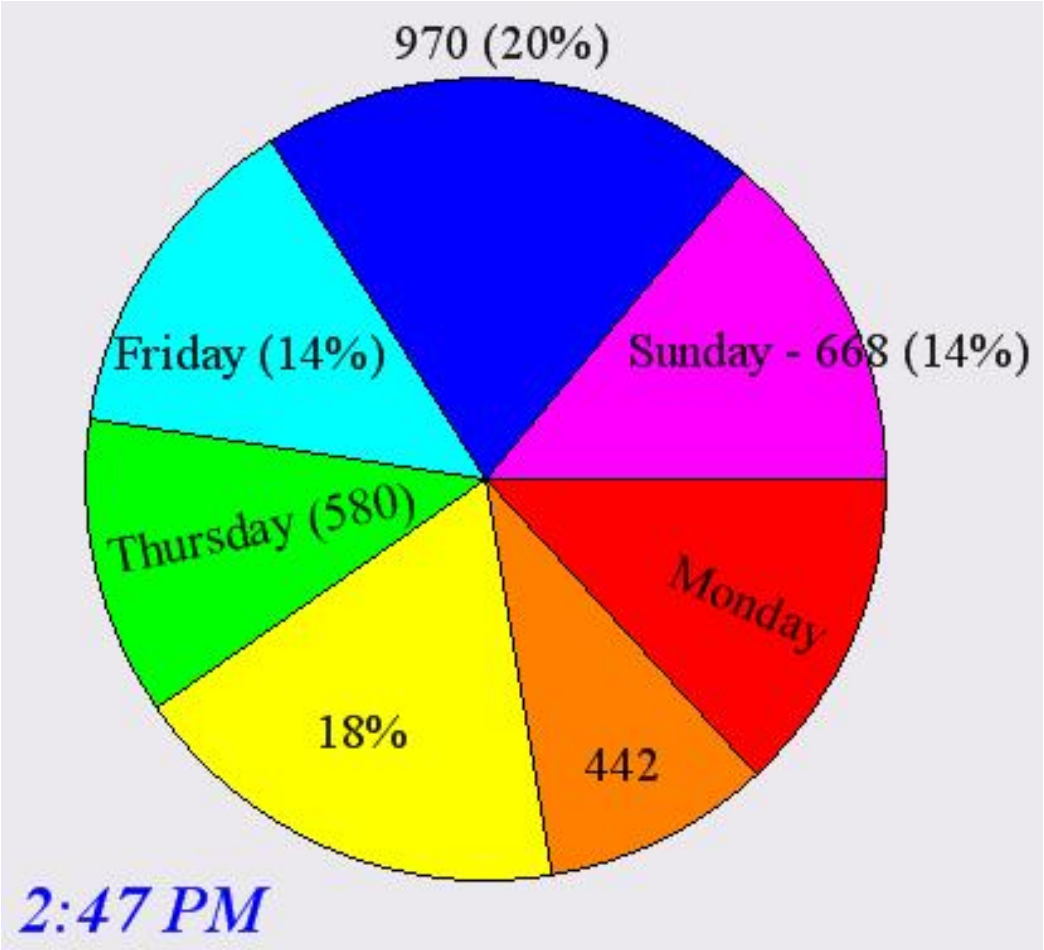

**Fig.17.12** Pie chart

```
public class PieChart : GraphicalObject
{
    Point ptCenter;
    int nRadius;
    double [] vals;
    List<CommentToCircle> circlecomments = new List<CommentToCircle> ();
    List<CommentToCircleSector>sectorcomments = new List<CommentToCircleSector> ();
```

**Figure 17.12** shows a `PieChart` object, consisting of seven sectors. Though all the comments, associated with the sectors, are positioned in different ways, it is obvious, with which of the sectors each of them is associated. The only comment, associated with the whole pie chart, can be seen in the bottom left corner of the picture.

The cover, which makes a circle movable by any inner point and resizable by any border point, was already explained with the `CircleNR` class in the chapter *N-node covers*; such cover is shown at **figure 7.2**. The `PieChart` class uses exactly the same idea for cover design.

- The whole area of a pie chart is covered with a single big circular node, which provides the movement in any direction.

- The curved border is covered by a set of small circular nodes that provide the resizing. The overlapping of these nodes guarantees that a pie chart can be resized by any border point. The combined area of all these small nodes looks like a strip with the real border line in the middle. The size of the nodes is selected in such a way as to make this sensitive strip not too wide (not to cover too much of the screen space), but wide enough to make the resizing easy. The distance between the neghbouring circles is determined by their radius and the request to organize an easy to grab strip without gaps inside. With the radius of small nodes equal to 5 pixels (`nrSmall`) and the neighbours on the border placed not farther than 8 pixels from each other (`distanceNeighbours`), the cover works perfectly. It produces a sensitive border without any gaps and minimum 6 pixels width.

```
public override void DefineCover ()
{
    int nOnPerimeter = Convert.ToInt32 ((2*Math.PI *nRadius) / distanceNeighbours);
    CoverNode [] nodes = new CoverNode [nOnPerimeter + 1];
    nodes [0] = new CoverNode (0, ptCenter, nRadius -nrSmall + 1, Cursors.SizeAll);
    nodes [0] .Clearance = false;
    for (int i = 1; i <= nOnPerimeter; i++)
    {
        nodes [i] = new CoverNode (i, Auxi_Geometry .PointToPoint (ptCenter,
                                  2 * Math.PI * (i - 1) / nOnPerimeter, nRadius), nrSmall);
        nodes [i] .Clearance = false;
    }
    cover = new Cover (nodes);
}
```

As for any other complex object, the most important things are the individual and related movements of the parts of an object. Comments, associated with the whole circle, belong to the `CommentToCircle` class. The number of such comments, associated with a pie chart, is not limited in any way.

```
List<CommentToCircle> circlecomments = new List<CommentToCircle> ();
```

The `CommentToCircle` class is derived from the `TextMR` class, so any such comment can be moved anywhere around the screen and rotated around its own central point without any mentioning in the code. The synchronous and related movements of the circle and such comments are described by several rules.

1. When a circle is moved forward, all the comments to the whole circle move synchronously.
2. If a circle is resized, then each comment moves along the line of the radius, but the exact movement depends on whether a comment was inside or outside the border of circle when the resizing was started. The position of any comment is determined by its center. If this point was outside the border of a circle at the beginning of resizing, then it has to stay outside and at exactly the same distance from the border throughout the whole process of resizing. If the comment was inside, then the ratio between its distance from the center of a circle and the radius of a circle is going to stay unchanged.
3. A comment to the circle does not react at all to the rotation of a circle.

To fulfil these rules, a `CommentToCircle` object has to include and use four fields

```
public class CommentToCircle : TextMR
{
    double coef;                 // positioning coefficien
    double angleToComment;       // angle from the center of a circle to the center of commen
    int radiusCircle;            // radius of a circle
    Point ptCenterCircle;        // center of a circle
```

The value of the positioning coefficient `coef` gives a clear understanding of whether this comment is inside or outside the border. For the inside positioning, this coefficient belongs to the [0, 1] range and equals to the ratio between the distance from the center of comment to center of circle and the radius of circle. Any coefficient greater than one means the distance from the border to the comment placed somewhere outside the circle.

Of the mentioned four parameters, the last two describe the circle ("parent"), while the first two describe the relative position of the comment. When the comment is moved individually, then the first two parameters have to be recalculated to describe the new position. Here is the code of the `CommentToCircle.Move()` method.

```
    public override void Move (int dx, int dy)
    {
        base .Move (dx, dy);
        double distance = Auxi_Geometry .Distance (ptCenterCircle, Location);
        if (distance <= radiusCircle)
        {
            coef = distance / radiusCircle;
        }
        else
        {
            coef = Math .Max (1.0, distance - radiusCircle);
        }
        angleToComment = -Math .Atan2 (Location .Y – ptCenterCircle .Y,
                                       Location .X – ptCenterCircle .X);
    }
```

When the position or the radius of a circle are changed, all the comments of this circle are informed about the changes via the `PieChart.InformRealtedElements()` method.

```
        private void InformRelatedElements ()
        {
            … …
            foreach (CommentToCircle comment in circlecomments)
            {
                comment .NewParentSizes (ptCenter, nRadius);
            }
        }
```

On receiving the new size of a circle, the `CommentToCircle.NewParentSizes()` method uses the unchanged coefficient and angle to calculate the new position of comment.

```
    public void NewParentSizes (Point ptC, int nR)
    {
        ptCenterCircle = ptC;
        radiusCircle = nR;
        Location = Auxi_Geometry .CirclePointBySpecialCoefficient (ptC, nR,
                                                         angleToComment, coef);
    }
```

<u>Comments, associated with the sectors of a pie chart</u>, belong to the `CommentToCircleSector` class. As with many other classes of comments, this one is derived from the `TextMR` class, so it is a comment, which can be moved anywhere around the screen and rotated without any mentioning in the code. A pie chart has a `List` of comments, associated with the sectors.

```
  List<CommentToCircleSector> sectorcomments = new List<CommentToCircleSector> ();
```

It would be not a problem to link any number of comments with a sector, but I couldn't think out any case, when more than one comment per sector would be needed, so a `PieChart` class has one comment per a sector. The comments to sectors have similar system of rules for synchronous and related movements. The first two rules are exactly the same, but the third one is different.

3. Comment of a sector rotates synchronously with the rotation of a circle.

The rule 3 defines only the movement (rotation) of the central point of comment. There is more about the changing of orientation (angle) of the text involved into the rotation of the whole pie chart; I will write about it a bit further.

The angles of all the sectors are determined by the array of values, associated with the pie chart. Each sector has its starting angle and the angle of the circle, which it occupies.

A `CommentToCircleSector` object has four fields to describe its position; three fields are the same as in the case of a `CommentToCircle` object, but one is different.

```
    public class CommentToCircleSector : TextMR
    {
        double coef;                         // positioning coefficien
```

```
        double angleToSectorStart;          // angle from the side of a sector
        int radiusCircle;                   // radius of a circle
        Point ptCenterCircle;               // center of a circle
```

The movement of the comments to sectors as a reaction to the moving / resizing of a circle are organized via the same `PieChart.InformRealtedElements()` method. The difference with the previous case is that these comments get three parameters from the "parent".

```
    private void InformRelatedElements ()
    {
        … …
        for (int i = 0; i < vals .Length; i++)
        {
            sectorcomments [i] .NewSectorParams (ptCenter, nRadius, startSector);
            … …
        }
```

The `PieChart` class has much more tunable parameters, than the `BarChart` class, so it has more complex tuning form (**figure 17.13**). The form consists of three main groups; two of them are used to work with two different types of comments; the third one is to deal with the colors. There is also a small addition of two controls to decide about drawing or not the borders and to select the color of borders, if the borders have to be shown.

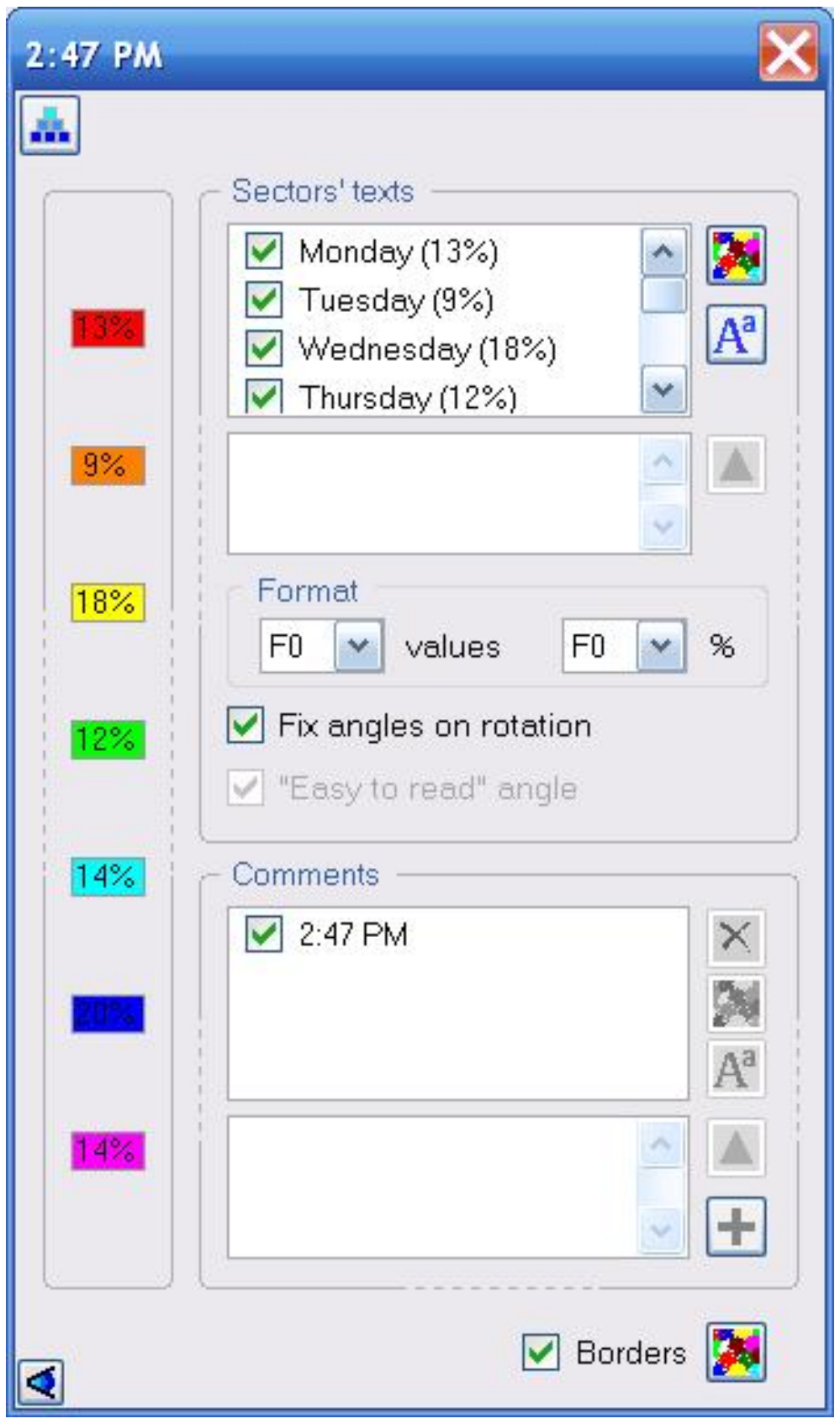

**Fig.17.13** Tuning form for the `PieChart` class

The narrow group without a title but with a frame includes the samples of colored rectangles. The colors and their order are the same as the colors of the sectors in the pie chart. Different sectors may have the same colors, so in addition there is information inside the samples about the percent of the pie chart, occupied by the corresponding sector.

Samples can be moved by a mouse to other positions inside the group. If a sample is released at another position, then the order of colors in the pie chart is changed. It is not the reordering of the sectors! The sectors are defined by the array of values at the moment when the chart was initialized. Each sector gets its angle according to the associated value from the array. Each sector also gets a comment; the comments can be moved and rotated. Nothing of these is going to change, when the colored samples are moved. Sectors are painted with different colors; only the order of colors is changed on such movement.

Colors can be changed. Double click in the sample you want to change; this will open the standard dialog to select the color.

The *Comments* group is used for tuning the comments, associated with the whole circle. The work of this group was described in details in the previous chapter (**figure 16.11**), only here you see the previous version of the same group, but the work is the same.

The *Sectors' texts* group is the biggest in the form. As its title tells, the group allows to tune the comments, associated with the sectors. Inside the group, there is nearly the standard, but slightly simplified set of controls to deal with the comments. Select any line with comment; then you can change its color, font or text in the standard way. But the simplified version does not allow to delete the comments or add the new. These actions are not allowed with the comments to sectors because they make no sense. Each sector is associated with some comment. It can be shown or hidden (one way to do this is to use the check box in the line with comment), but it always exists, so you cannot delete it. Each sector is always associated exactly with one comment, so there is no sense in adding more comments of this type.

The inner *Format* group allows to change the format of the numbers' presentation. The formats can be declared independently for showing the numbers and the percents.

All the comments (of any type) can be rotated individually, but the comments to sectors are also involved in synchronous rotation with the pie chart. When the circle is rotated, the centers of all the comments to sectors rotate synchronously, but what has to happen with the angles of all the comments? Texts to sectors at **figure 17.12** are shown at different angles. How these angles have to change, if you start turning the pie chart around? Do they have to keep the same angles regardless of the turn of a circle? Or do they have to keep the same angle in relation to their sectors? Does it mean that the turn of a circle for 180 degrees has to turn the text of each sector for the same 180 degrees? If you turn the pie chart from the **figure 17.12** for 180 degrees and all the texts to sectors for the same angle, then all of them will be shown upside down. Do you expect such a result? Different reactions of the texts to sectors on rotation of circle can be organized; two additional check boxes in the group allow to regulate this process.

If the check box *Fix angles on rotation* is switched ON, then the angles of all texts to sectors are not changed during the rotation of the pie chart, and you are looking at the graphical big dipper. If it is switched OFF, then each text does not change its angle to the "parent" sector, and you can see some of the texts upside-down or at very interesting angles.

To avoid looking at the texts upside-down, there is an additional "*Easy to read*" option, but it can be switched ON only when the angles are not fixed.

I would recommend switching ON the tuning form and turning a pie chart around with different combinations of these check boxes. This will explain to you all the possibilities much better and in shorter time, than reading of the previous sentences.

There are other possibilities for tuning the pie charts; they are available not through the tuning form, but via the context menu. I will return to these possibilities a bit later.

## Ring sets

Rings and ring sets have a lot of common with the pie charts; I often worked in parallel on these two types of plotting. Whenever anything new was introduced with one of these types, the same changes had to be done immediately on the other, because they are very much alike. Pie charts have the minimum allowed radius to prevent their accidental disappearance on squeezing. For the same reason, there is a limit on minimum size of the inner radius of a ring and the minimum allowed width of the rings. The covers for two objects are also of the same type. They were similar two years ago [11]; they were changed in parallel, when I came to the idea of the N-node covers for the objects with the curved borders.

A `RingSet` object may have any number of rings, but at the moment of construction it has only one ring, so it is initialized by a central point, two radii (outer and inner), and an array of values. The number of values determines the number of sectors; the ratio of values determines the angles of the sectors, as their sum must be equal to 360 degrees. Thus initialized `RingSet` object contains a single ring of the `RingArea` class (**figure 17.14**). Later the rings can be added or deleted. The rings are added only on the outside, but any ring can be deleted. No overlap of the rings is allowed, but the gaps between the rings are (**figure 17.15**). All the rings of the `RingSet` object are concentric; whenever any ring is moved forward, all the rings move synchronously to continue being concentric. The rotation of each ring is independent of all others; rotation of any ring does not affect all other rings at all.

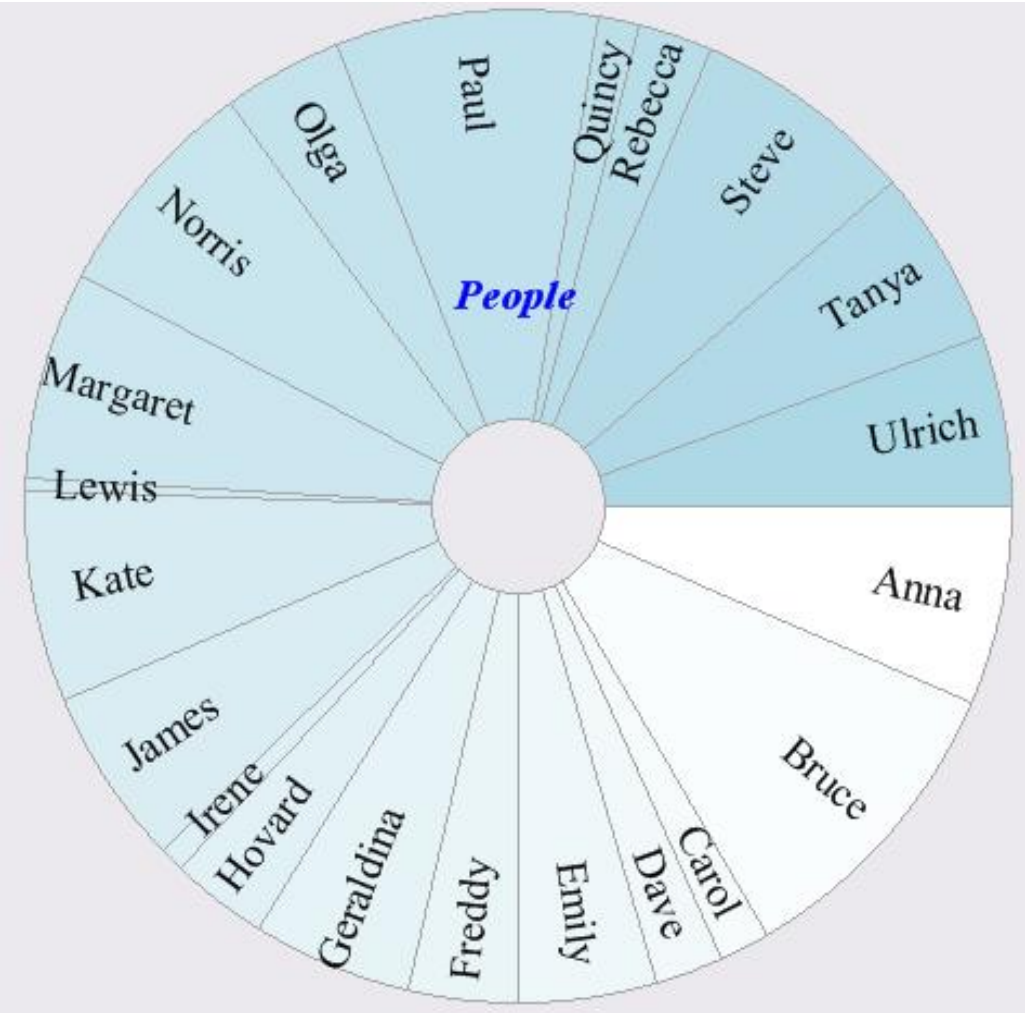

**Fig.17.14** Any ring set starts with one ring; then other rings can be added.

```
public class RingSet : GraphicalObject
{
    Point ptCenter;
    List<RingArea> rings = new List<RingArea> ();
    List<CommentToRing> areacomments;
```

The covers that make possible the resizing of the rings by any border point were already demonstrated at **figures 7.2** and **8.1**. A set of small nodes cover each circular border in such a way that it is covered by a bended sensitive strip without any gaps. As a result, any ring can be resized by any point of their inner or outer borders. Borders of the rings are covered by the nodes beginning from the inner ring and going outside, so the nodes over the inner borders precede the nodes of the outer borders. Thus, when two rings stay side by side without any gap, then their common border can be moved only inside, but not outside. When you move such border inside, the inner ring shrinks a bit, the nodes of two borders do not overlap any more, and then any border can be moved.

An object of the `RingSet` class consists of three main parts:

- A `List` of rings; every ring is divided into sectors.
- Comments, associated with the whole set of rings.

- Comments, associated with the sectors.

The positioning of the comments is similar to how it is done in the pie charts, but there are some differences.

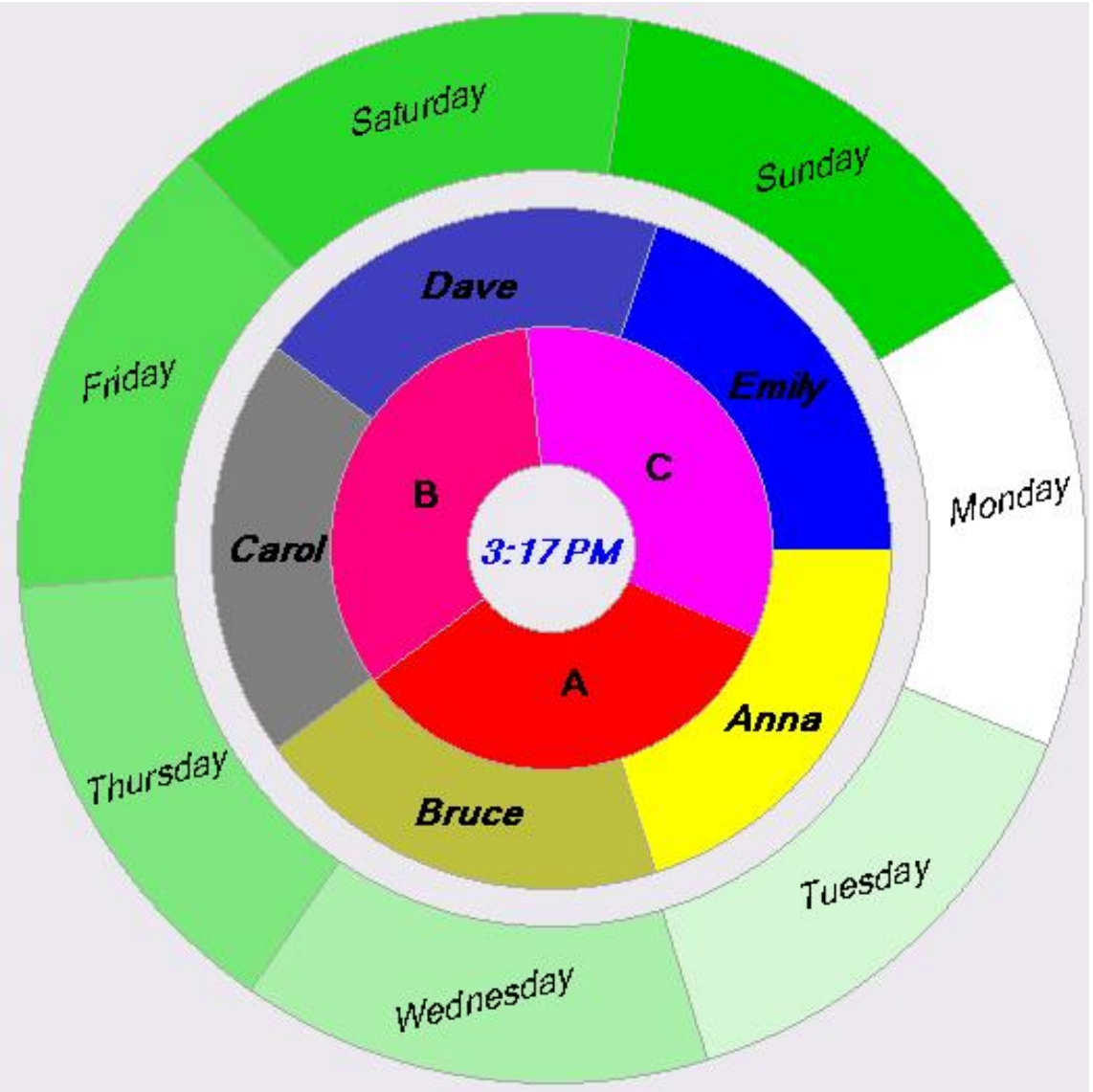

**Fig.17.15** A `RingSet` object may contain any number of rings

Comments, associated with the whole set of rings, belong to the `CommentToRing` class. For positioning of such comments two radiuses of the ring are needed: the inner and the outer. When the `RingSet` object consists of a single ring, these two parameters are obviously the parameters of this ring; when there are several rings united into an object, then the most inner and outer of all the radii are used. The synchronous and related movements of the `CommentToRing` objects are described by such rules.

1. When the rings are moved forward (this move always involves all the rings), the general comments move synchronously.
2. Rotation of any ring does not affect the general comments.
3. Not every resizing of a ring affects the general comments, but only the change in the most inner and outer borders do. The change of the location of comment depends on its initial position in relation to these two borders.

The positioning of any `CommentToRing` object is determined by five values

```
public class CommentToRing : TextMR
{
    double coef;                    // positioning coefficien
    double angleToComment;          // angle from the center of the ring to the center of comment
    int radiusInner;                // the inner radius of a set of rings
    int radiusOuter;                // the outer radius of a set of rings
    Point ptCenter;                 // center of a ring
```

The special positioning coefficient is calculated in different ways for three different situations.

- If a comment is inside the inner border, then this is a coefficient from the [-1, 0] range with -1 corresponding to the central point and 0 – to the inner border.
- If a comment is between the inner border and the outer, then the coefficient belongs to the [0, 1] range with 0 corresponding to the inner border and 1 – to the outer border.
- If a comment is outside the outer border, then the coefficient is greater then 1 and equal to the distance (in pixels) between the comment and the outer border.

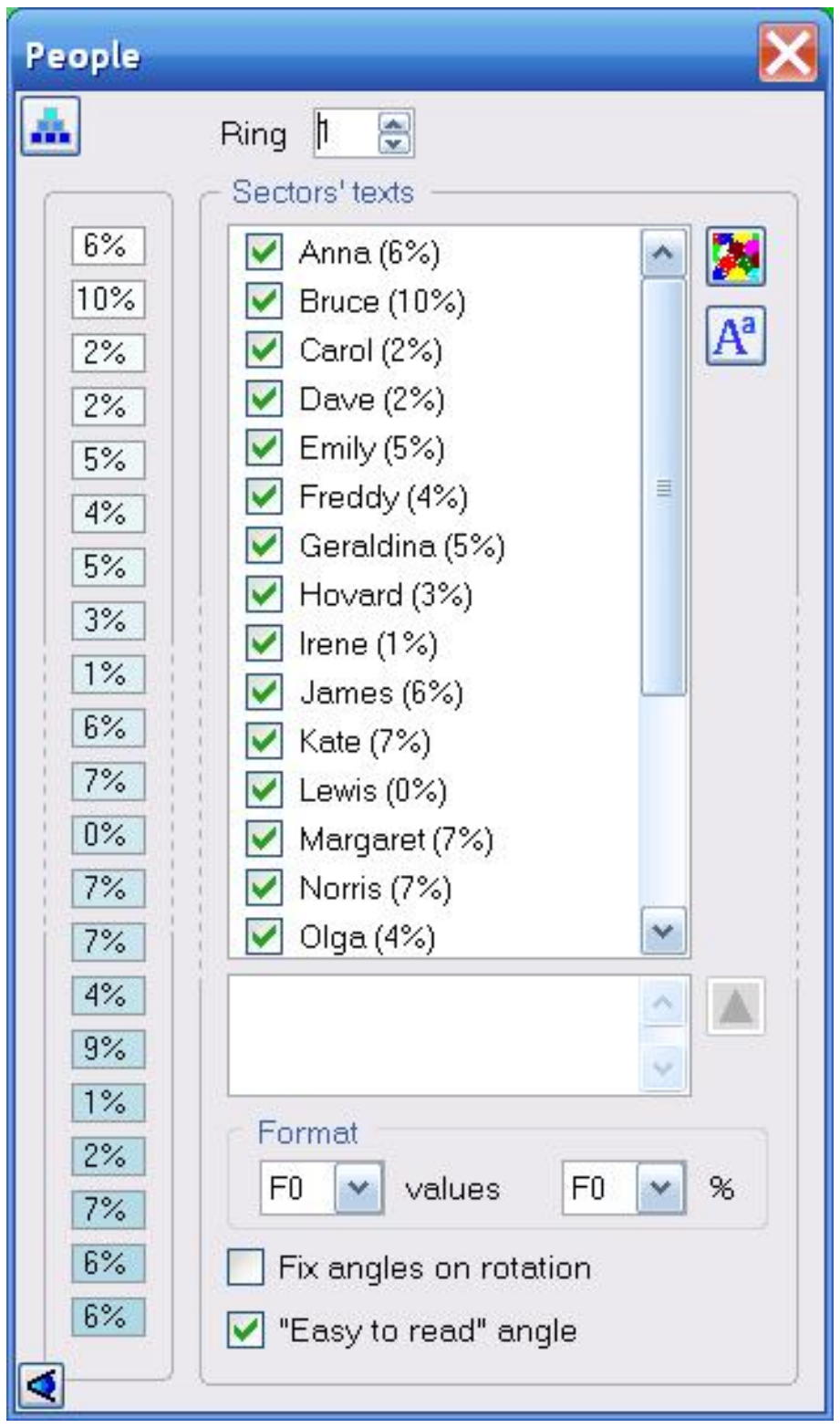

**Fig.17.16** Tuning form for the `RingSet` objects

When the `CommentToRing` object is resized, it tries to keep the positioning coefficients of all the general comments unchanged. In this way, the comments outside all the rings are kept at the constant distance from the outer border. The comments inside are moved to the new positions in such a way that the coefficient, calculated with the new sizes, is unchanged.

Comments, associated with the sectors, belong to the `CommentToRingSector` class.

1. When the rings are moved forward, all these comments move synchronously.
2. They rotate synchronously with their ring.
3. They react to the resizing of their ring exactly in the same way, as the general comments to the resizing of the whole set of rings.

The tuning form for the RingSet objects (**figure 17.16**) is nearly identical with the tuning form of the PieChart class. Some differences in view at **figures 17.13** and **17.16** are the result of my rearranging of the last form, which is the standard procedure in the user-driven applications. I moved out couple of pieces, which I do not need, and enlarged the remaining two groups, so that it would be easier to work with the big amounts of data. The only new part in this form is the small selection of the ring for tuning at the top of **figure 17.16**.

Figures **17.14** and **17.15** demonstrate some interesting examples of the comments for the sectors.

At **figure 17.15** all the comments of the outer ring are turned on the same angle. Certainly, each comment can be turned around individually, but that would be a long and tiresome work, if you have many sectors and want to turn all the comments on exactly the same angle. Originally, all the comments for the sectors are shown horizontally (zero angle). You can unfix the angle on rotation (in the tuning form) and turn the ring. All the comments will turn exactly on the angle of the turn. What if you want these angles to be different? What if you want to turn the ring for 150 degrees (sectors look better in such a way) and the comments only for 15 degrees? There must be some way to do it.

**Figure 17.14** shows even stranger view. All the comments of the sectors are positioned along the radii. I hope you did not expect me to do it manually in order to produce such a view?

Both figures **17.14** and **17.15** were prepared in the **Form_PlotsVariety.cs;** a lot of things can be done with the objects of this form not via tuning forms of these objects, but via different context menus. All these things are included into application in accordance with the well known rule that users can do in the user-driven applications whatever they want. Let us look at the work of this rule in the **Form_PlotsVariety.cs**.

## *Variety of possibilities among the variety of plots*

The **Form_PlotsVariety.cs** demonstrates the use of only three different types of plots: bar charts, pie charts, and the sets of rings. I could easily include more plotting classes there, but this would not change the main purpose of this form: to demonstrate the design of applications for economical and financial analysis according to the ideas of the user-driven applications. I will remind the rules of such applications.

**Rule 1**. All the elements are movable.

**Rule 2**. All the visual parameters must be easily controlled by the users.

**Rule 3**. The users' commands on moving / resizing of objects or on changing the parameters of visibility must be implemented exactly as they are; no additions or expanded interpretation by developer are allowed.

**Rule 4**. All the parameters must be saved and restored.

The first rule works as an axiom for all of my programs, so there is nothing to discuss. Everything is movable.

Rule 4 is implemented in the **Form_PlotsVariety.cs** in the standard way for all my forms and applications. I turned these procedures into standard, so it would be easier for anyone to understand them in any new form, if he looked into their implementation even once. Every class from the **MoveGraphLibrary.dll**, which may need the saving / restoring, has two methods to store the objects of this class in Registry and in binary file, plus two other methods to restore the same objects from those two sources. If I use anywhere a new class, not from that DLL, such class has to have similar four methods. In this way I can store and restore objects of any class I work with. I have already mentioned in the previous chapter, why the saving / restoring through the binary files is more popular with the scientific / engineering applications; for all the examples or this book I prefer to use Registry.

Saving of all the objects is done at the closing of the form; at this moment the SaveInfoToRegistry() method is called. The only unique parameter, which is usually needed for saving, is the string that allows to distinguish the objects of the same class at the time of restoring. Objects of four classes have to be saved in the **Form_PlotsVariety.cs**: BarChart, PieChart, RingSet, and ClosableInfo. All these objects use their IntoRegistry() methods.

```
private void SaveInfoToRegistry ()
{
    … …
    info .IntoRegistry (regkey, "Info");
    for (int i = 0; i < elems .Count; i++)
    {
        switch (elems [i] .ElementType)
        {
            case MedleyElem .BarChart:
                strTypes [i] = "Bar_";
```

```
                elems [i] .BarChart .IntoRegistry (regkey,
                                                   strTypes [i] + i .ToString ());
                break;
            case MedleyElem .PieChart:
                strTypes [i] = "Pie_";
                elems [i] .PieChart .IntoRegistry (regkey,
                                                   strTypes [i] + i .ToString ());
                break;
            case MedleyElem .RingSet:
                strTypes [i] = "Rings_";
                elems [i] .RingSet .IntoRegistry (regkey,
                                                  strTypes [i] + i .ToString ());
                break;
        }
    }
    regkey .SetValue (nameElemTypes, strTypes, RegistryValueKind .MultiString);
```

The `IntoRegistry()` method of any complex object relies on the same methods of its parts. For example, in the subsection *Bar charts* the design of the `BarChart` class was described and its constituents were mentioned.

```
        RectArea rarsea;
        TextScale scaleText;
        Scale scaleNum;
        List<CommentToRect> comments = new List<CommentToRect> ();
```

Not surprisingly, the `BarChart.IntoRegistry()` is based on the similar methods of those inner parts.

```
    public void IntoRegistry (RegistryKey regkey, string strA)
    {
        … …
        rarea .IntoRegistry (regkey, "BcRa_" + strA);
        scaleText .IntoRegistry (regkey, "BcTs_" + strA);
        scaleNum .IntoRegistry (regkey, "BcNs_" + strA);
        for (int i = 0; i < comments .Count; i++)
        {
            comments [i] .IntoRegistry (regkey, strA + "BcCmnt_" + i .ToString ());
        }
```

The restoration of the previously saved view and all the objects is done inside the `OnLoad()` method of the form. If it is the first start of the form or for any reason there is a problem with the restoration, then the default view is organized.

```
    private void OnLoad (object sender, EventArgs e)
    {
        RestoreFromRegistry ();
        if (!bRestore)
        {
            DefaultView ();
        }
        RenewMover ();
        btnHelp .Enabled = !info .Visible;
    }
```

Though it is not forbidden, I wouldn't recommend editing the `Registry` manually in order to change the parameters of visibility between two sessions. Changing of each and all visibility parameters in the **Form_PlotsVariety.cs** can be much better done via the turning forms, which were already discussed, or via the context menus.

Rule 2 is implemented partly by the tuning forms and partly through the context menus. Individual changes of the objects are mostly done via their tuning forms. Some of these changes are duplicated in menus, but with the increase of the number and variety of objects the bigger percentage of the changes is done through the menu commands. A lot of these commands are spread on several objects simultaneously, which cannot be done with the tuning forms. Rule 3 is closely related to rule 2 and controls the implementation of all the numerous tuning commands.

The **Form_PlotsVariety.cs** has 10 different context menus! The users do not care, how many different menus are there, but they do care about one thing, or at least they find such a thing very quickly: the right click anywhere in the form opens a context menu. It happens at each point inside the form! It can be at an empty spot, or it can be at one or another object. In some real applications, used for financial analysis, you can use lesser different menus, but the **Form_PlotsVariety.cs** is to some extent a demo program with the idea of demonstration as many possibilities as I could think out. The tuning of the bar charts was discussed a bit earlier; now let us check how the tuning through the tuning forms and the commands, available via menus, work together.

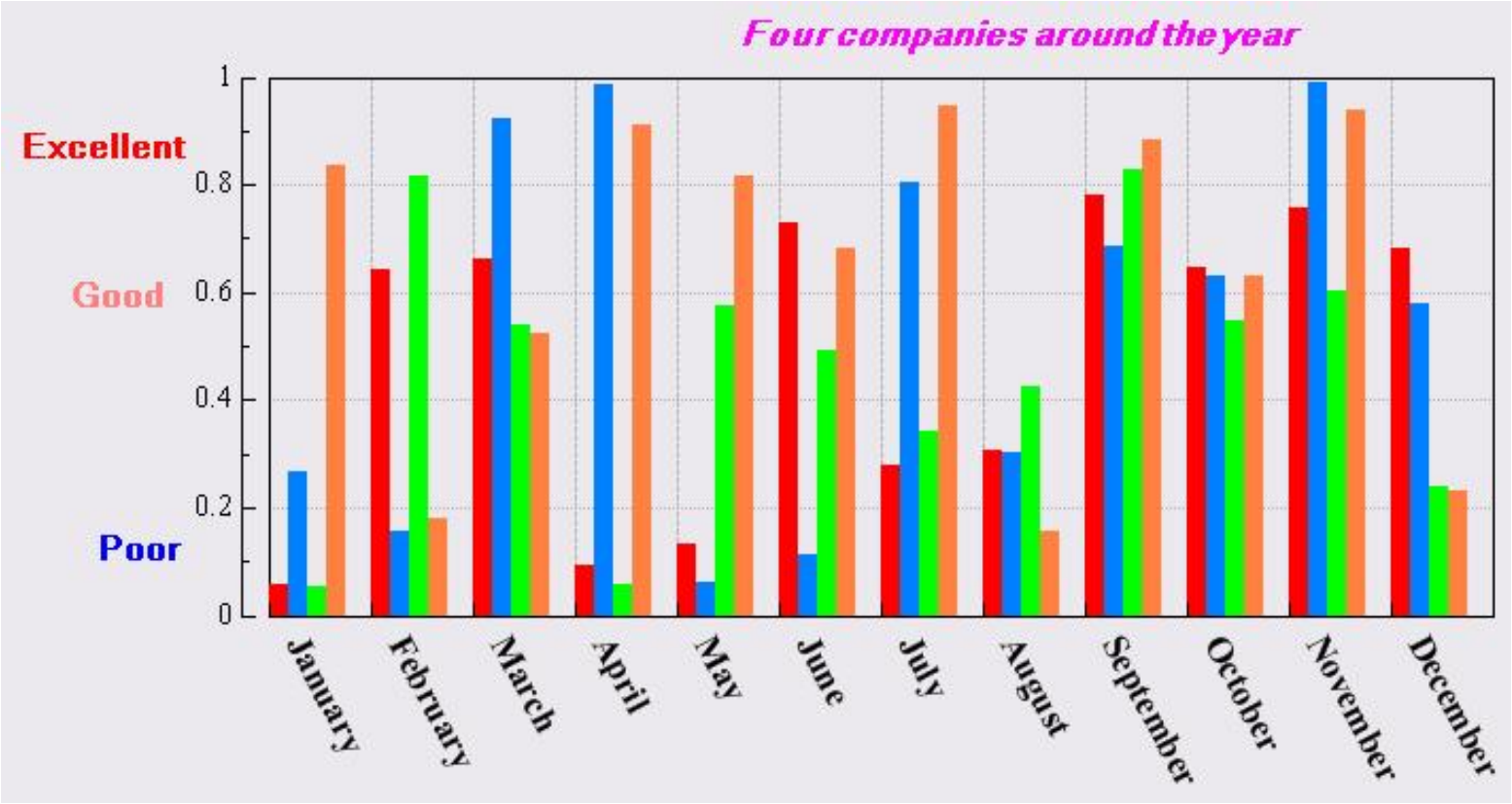

Fig.17.17 A typical bar chart.

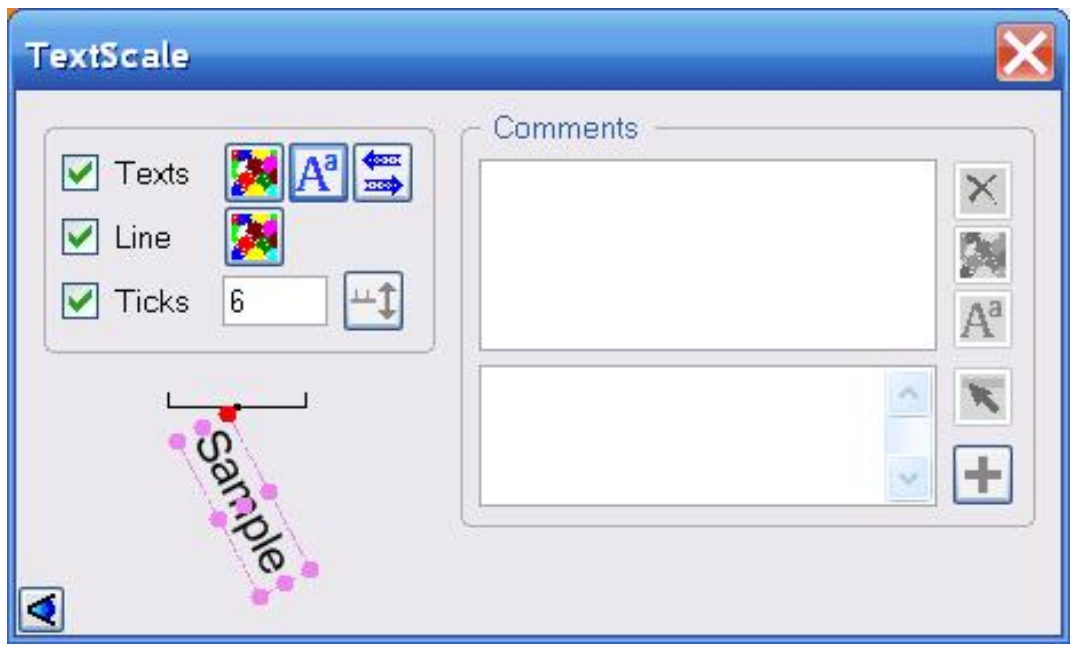

Fig.17.18 Tuning of the textual scale

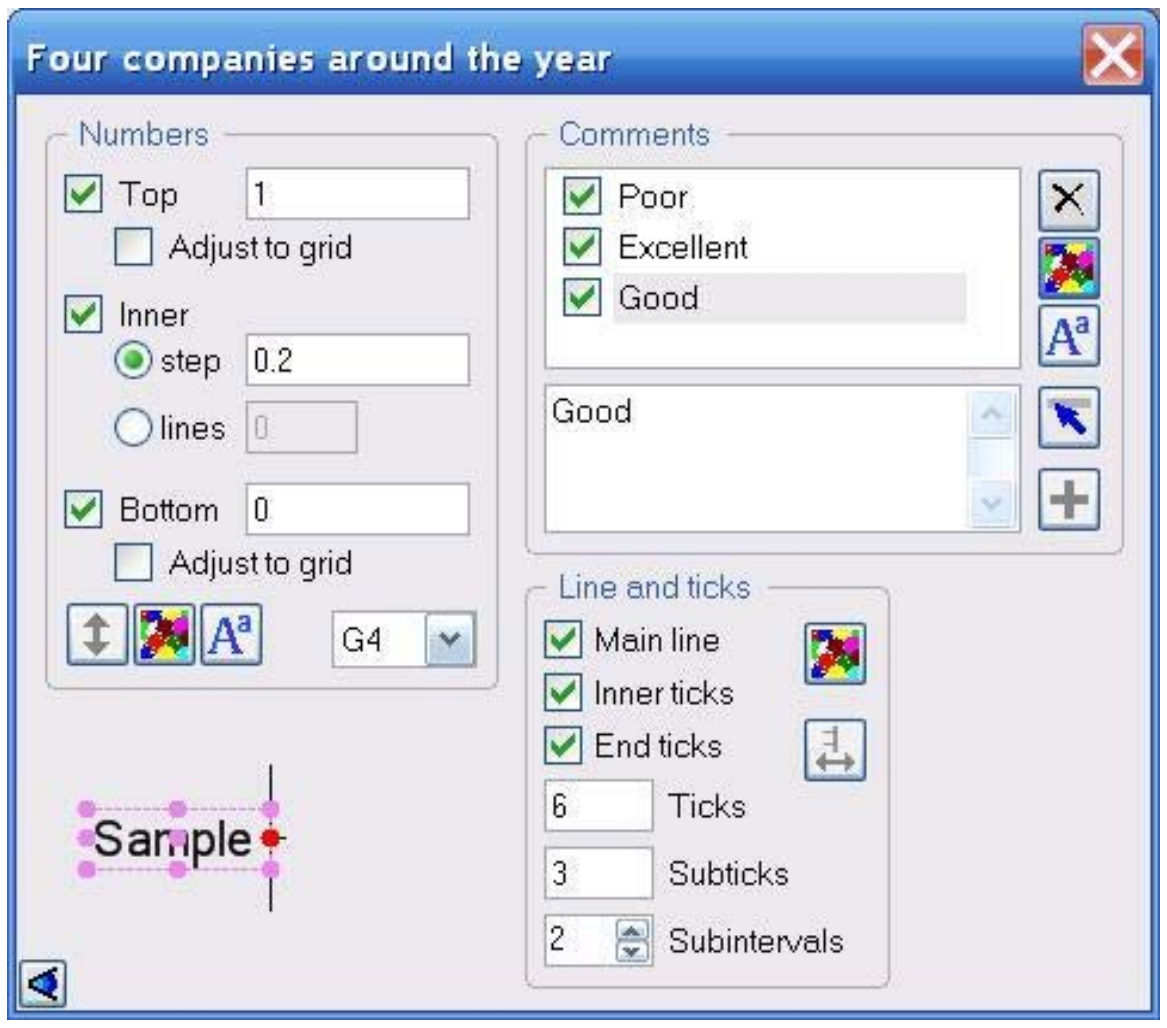

Fig.17.19 Tuning of the numeric scale

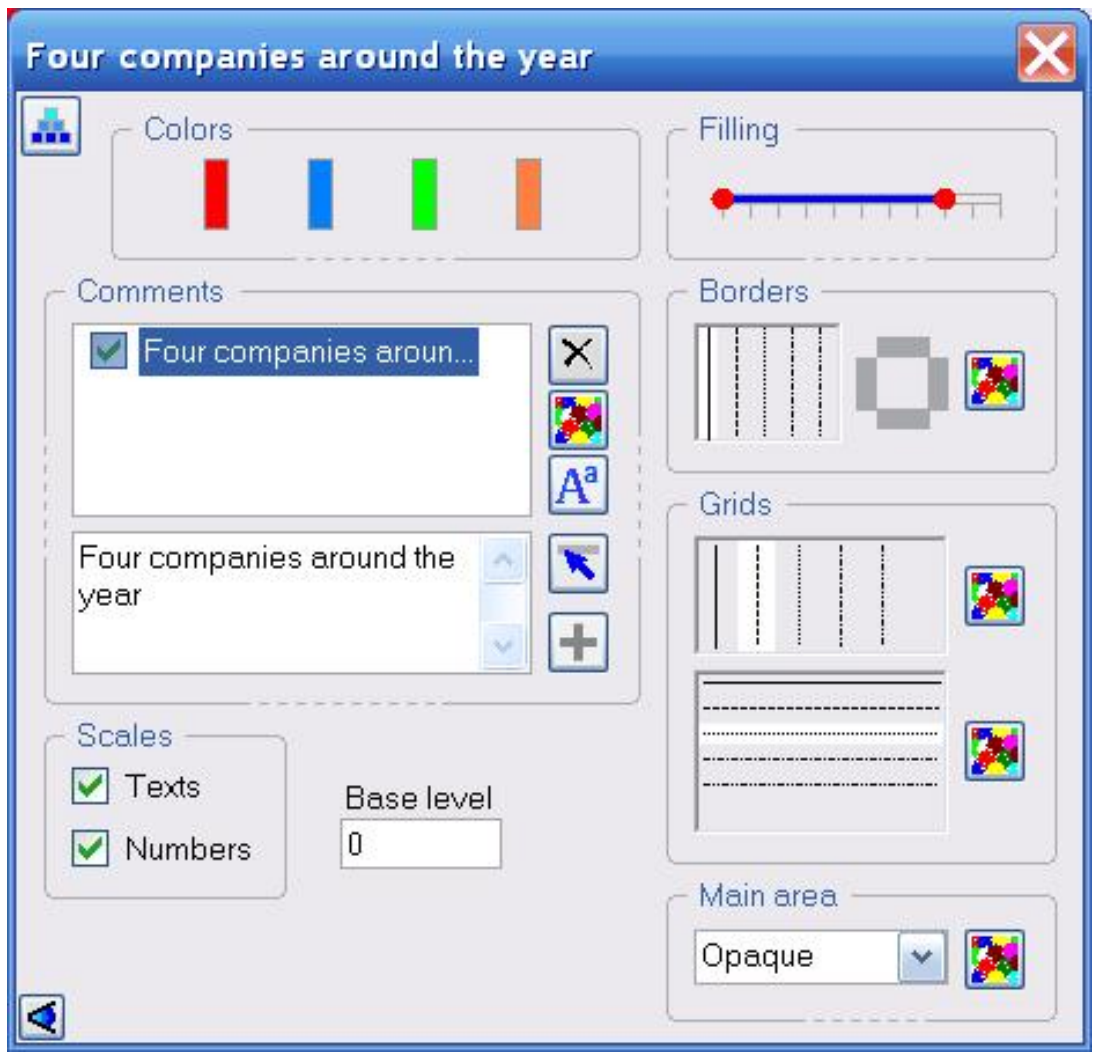

Fig.17.20 Tuning of the main area

**Figure 17.17** demonstrates the typical bar chart; **figures 17.18 – 17.20** show three tuning forms exactly for this bar chart. It is a standard procedure that for the fastest and best tuning of the plot all the associated tuning forms are called to the screen at the same time and positioned somewhere around the tunable chart. Thus the combination of such four figures can be often seen at the screen throughout the real work. (I have placed these four figures together in the text; I only hope that they will not move around and not make a mess of the whole text.)

Bar charts consist of several parts:

- The main plotting area.
- Numeric scale.
- Textual scale.
- Comments, which can be associated with any of the previous parts.

On any of these objects a context menu can be called. Menu on the main plotting area of the bar chart has two submenus, of which I decided to show one (**figure 17.21**). Menus for scales (**figure 17.22**) and comments (**figure 17.23**) are significantly simpler.

Menus often include several commands, which can be achieved in some other ways. This does not mean that one of these ways must be excluded; it is not the question of redundancy, but the question of providing customary paths for different users.

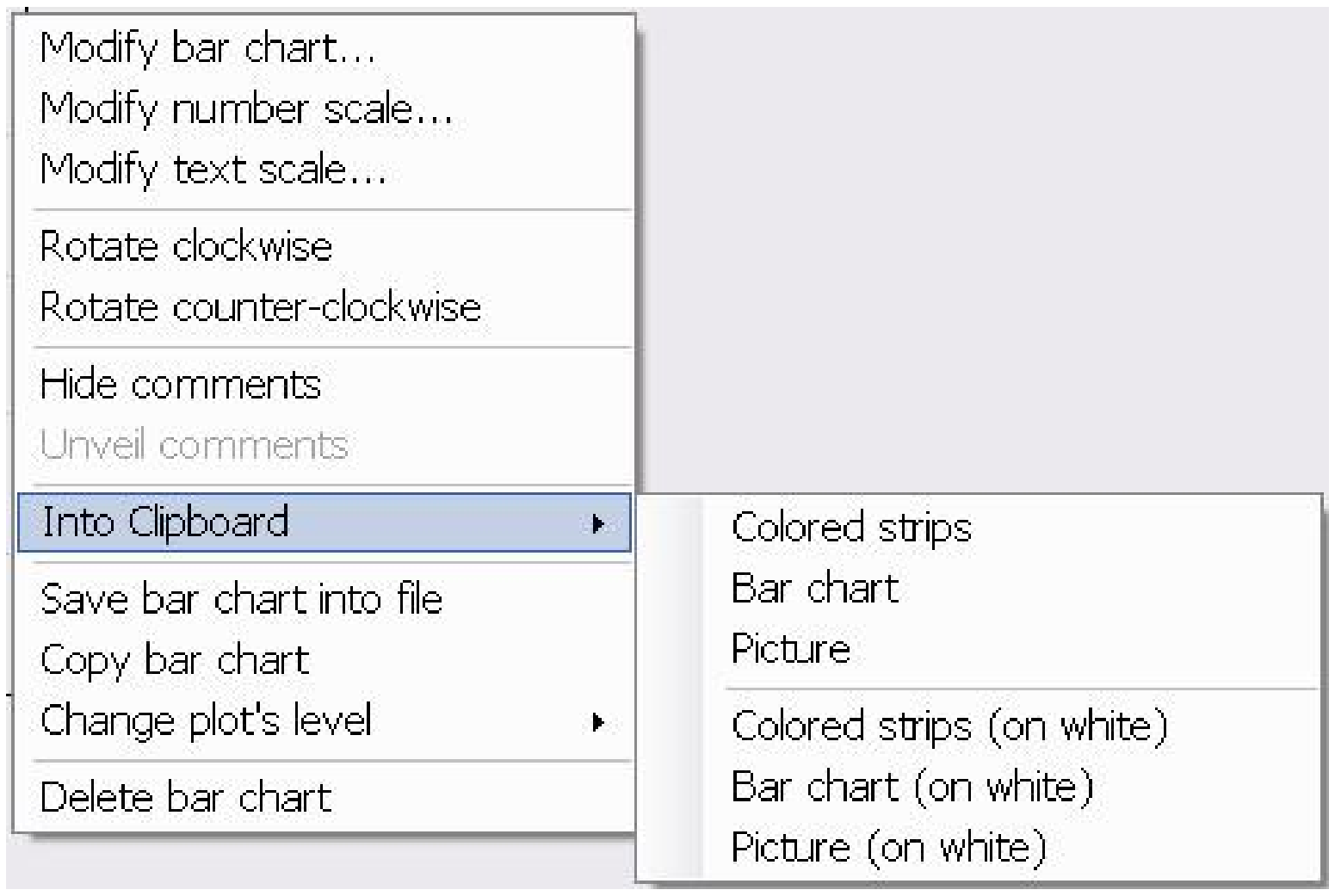

**Fig.17.21** Menu on the bar chart's area with one of the submenus

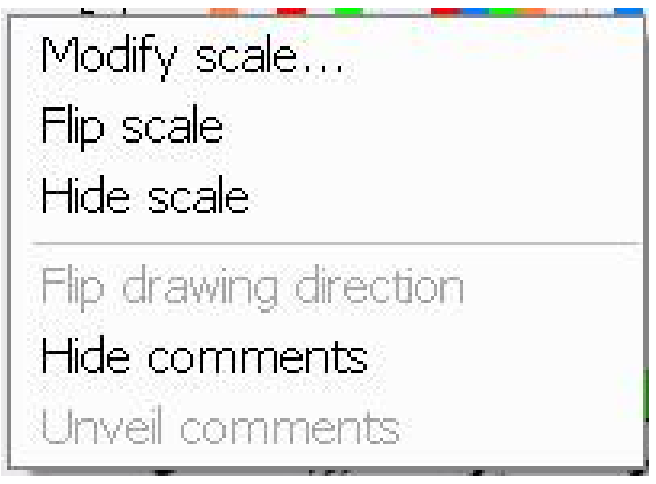

**Fig.17.22** Menu on scale

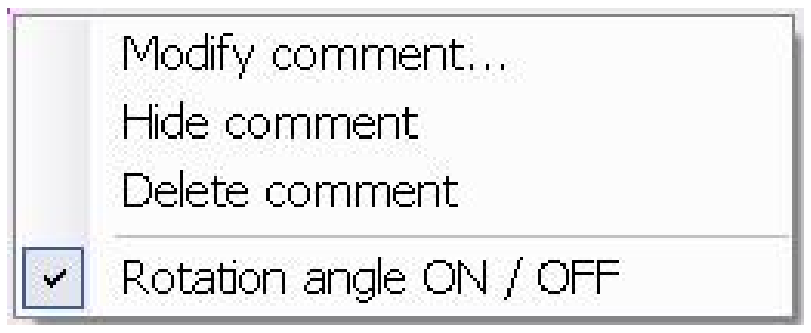

**Fig.17.23** Menu on comment

The first three commands in the menu on the main area of bar chart (**figure 17.21**) open one or another required tuning form; the same form can be opened by a double click on the main area or the scale. If you need to modify the comment, then there are not two, but three different paths. Comment is modified inside the tuning form of its parent, so there are the same two ways; plus you can go through the menu of comment itself (**figure 17.23**). I would say that the last way through the menu of commen is the most natural and at least the most reliable, as it is impossible to distinguish the belongings of the comment by its view or position. In a lot of cases the parent of comment is obvious from the position of comment, but not always. The three comments on the left side of **figure 17.17** more likely belong to the vertical scale, than to anything else, but they can also belong to the main area. The comment at the top of the same **figure 17.17** can belong to the main area, but what would you say if I move the same comment and position it below the scale at the bottom? Would you continue to insist that this comment belongs to the main area or the nearest scale (textual) would look like an obvious "parent"?

Rotation of the bar charts is organized only through the main menu (second group of commands at **figure 17.21**). The only unchangeable thing throughout the rotation is the size and position of the main plotting area, but everything else goes around according to the selected direction of rotation.

Hiding and unveiling the comments is again about different ways in different situations. The same menu of comment (**figure 17.23**) is opened on any comment regardless of its association with the main area or with one of the scales. But with the command from this menu a comment can be only hidden. There is no way to call a menu on the hidden object, so a hidden object can be unveiled only through the parental tuning form or through the menu of the parent. In the first case (via the tuning form) the comments can be hidden / unveiled on an individual basis by checking or unchecking each line in the list of comments (**figures 17.19** and **17.20**). Through the menus for the main area or a scale (**figures 17.21** and **17.22**) the commands to hide or unveil comments can be applied only to the whole set of associated comments.

All the complex applications can be (and often are) used for grabbing the view of the results from the screen into some kind of document. The wide variety of tuning of all the screen objects makes it easier to get the screen image exactly in the view, which is needed for the documents. For this reason I include into all the complex applications different commands for taking the information from the screen into the `Clipboard`; submenu from **figure 17.21** shows the possibilities. All the commands for taking any part of the screen into the `Clipboard` are available only via the menus.

Several more interesting commands from the menu on the main area of a bar chart.

<u>Copy bar chart.</u> Do you need a second copy of the data you are analysing now? Possibly NOT, but this command may help you in one interesting situation. Tuning of the parameters for visualization allows to change the view, but it is difficult to predict, whether the changed view will be better or not. If the final result of several changes comes with the realization that the original view was better, this does not improve the user's spirit. Maybe the better way is first to copy the whole bar chart and start changing one of the copies. In this way it is much easier to estimate, if the changes are making the view better or worse. The number of copies is unlimited; the unneeded charts can be deleted at any moment.

Another command allows to save the bar chart into the file, but I will return to this possibility later.

There can be a lot of different plots on the screen; they can stay apart, or they can overlap. The whole set of charts configure a multi level system with each chart occupying its own level. A small submenu, not shown at **figure 17.21**, but available via the *Change plot's level* menu line, allows to move the charts up or down through the levels. I saw many times that scientists often changed the order of charts, while working with a big number of plots. If such commands are very helpful in analysing a lot of plots in the scientific applications, then maybe the similar commands would be useful in programs for economical and financial analysis.

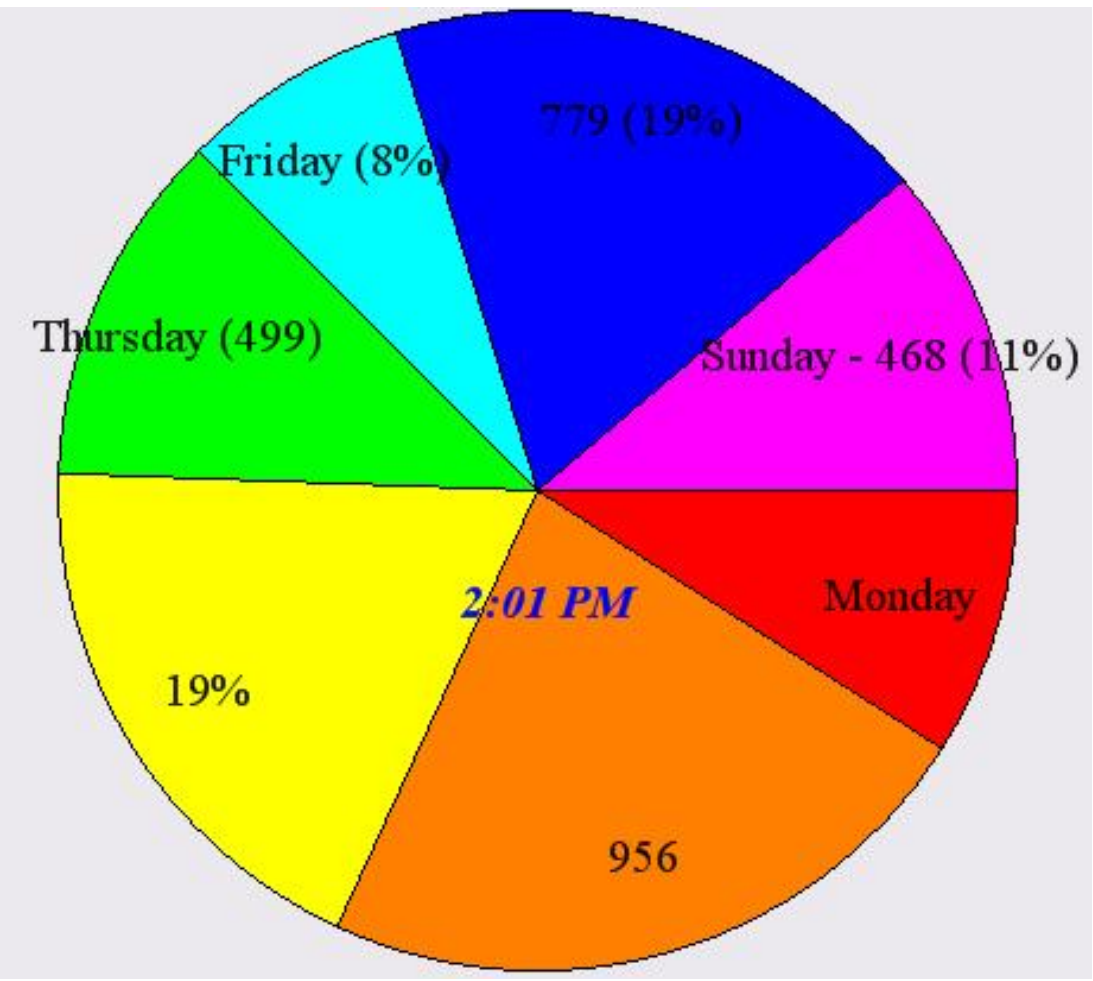

**Fig.17.24** Pie chart sample

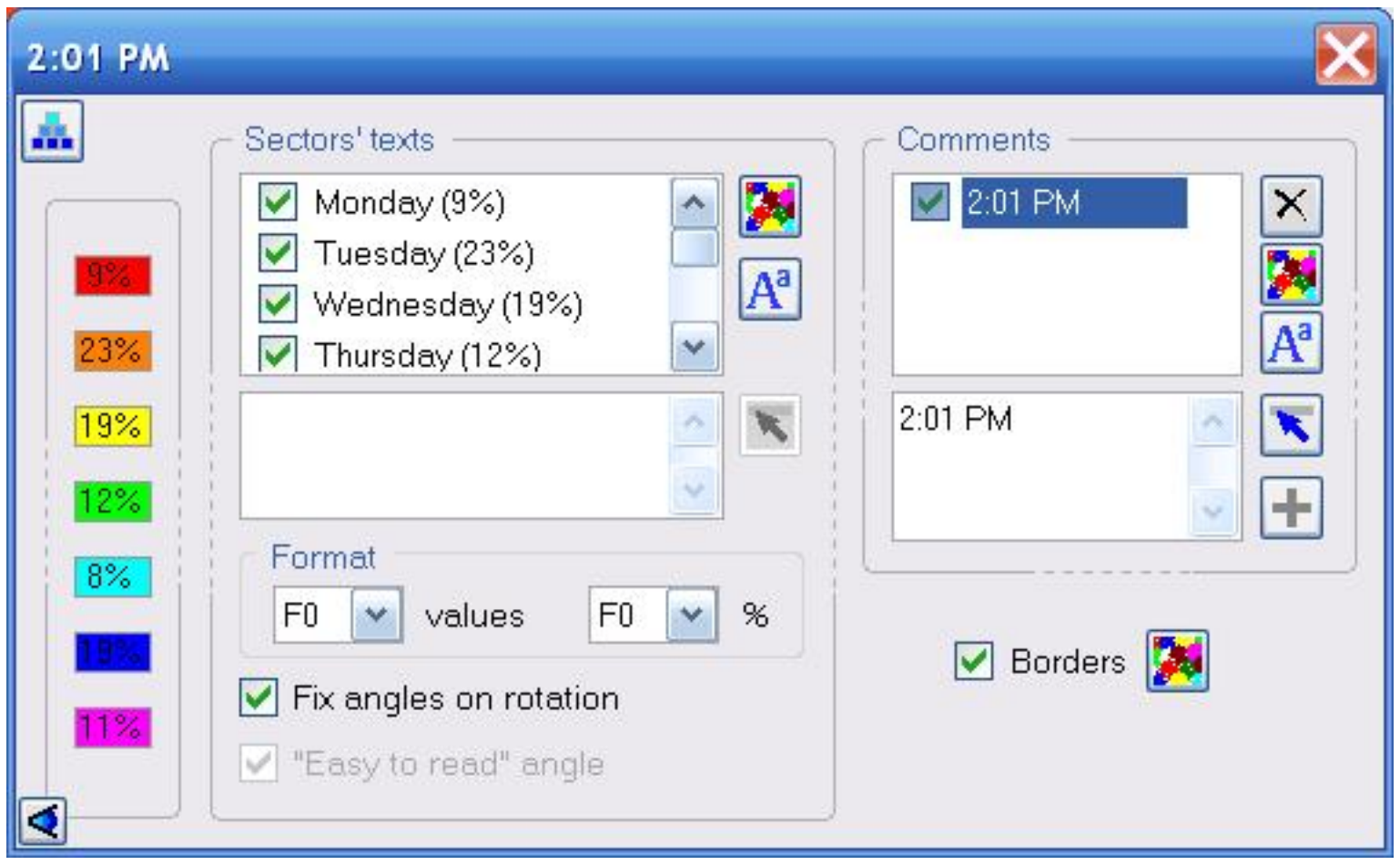

**Fig.17.25** Tuning form for the pie chart

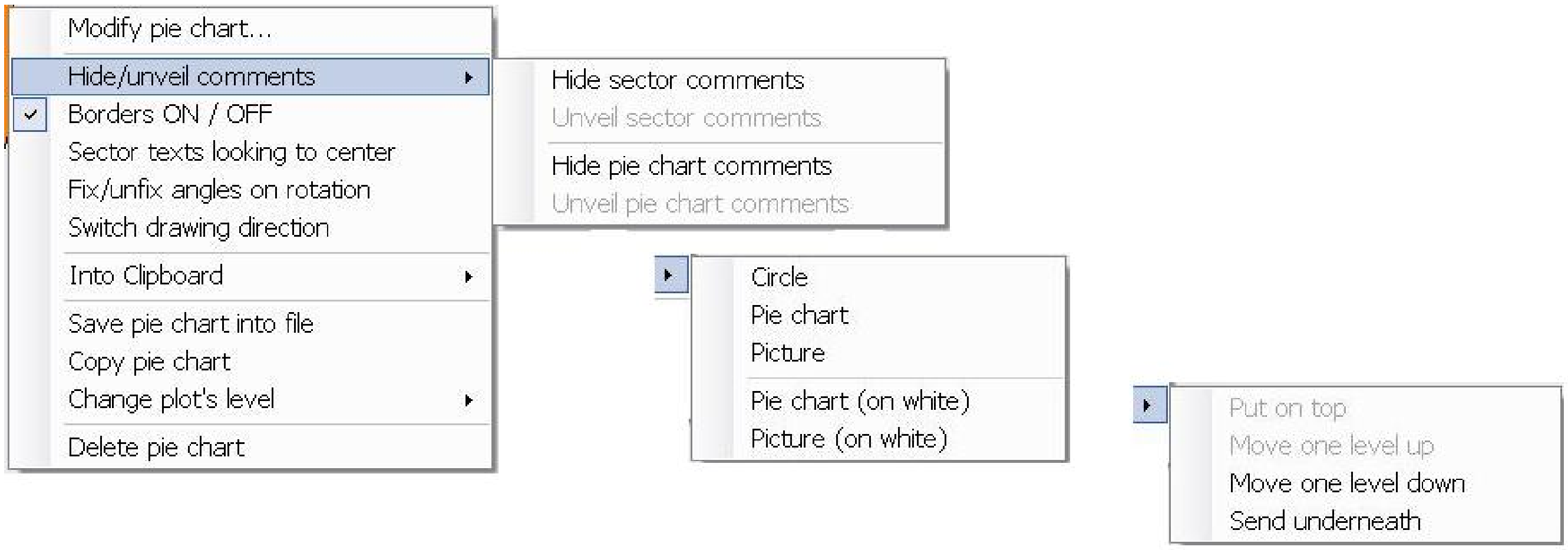

**Fig.17.26** Menu on the pie chart with all the available submenus

Pie charts and rings have less tuning forms than bar charts. There are no scales in these round plots, so any object of the `PieChart` or `RingSet` class is associated with exactly one tuning form. But several plots from **figure 17.1** make it obvious that those round plots have more variations in their views than the bar chart. If there are more variants and all these possibilities have to be tuned somewhere (this is a user-driven application!), then there must be another way of changing the parameters. One obvious way to do this is through the context menus, and the menus for round plots provide all the needed possibilities of tuning. In order to avoid multiple jumping between different pages, I want to show next to each other the `PieChart` object (**figure 17.24**), its tuning form (**figure 17.25**), and the menu on the pie chart together with its submenus (**figure 17.26**). It is impossible to see all submenus simultaneously in the program, but I think that such a view in the book gives the best understanding of all the available commands.

Some of the commands in this menu look identical to those that were already shown with the bar charts, but others are definitely new. One of them is the "Sector texts looking to center"; this command allowed to prepare the view of the pie chart at **figure 17.14** in an instant. If the *Fix angles on rotation* in the tuning form is not checked, but *"Easy to read" angle* is checked, then throughout the rotation of the pie chart the texts of sectors are always positioned along the radii and never turn upside down.

I have explained earlier that changing the order of the colored samples in the tuning form does not change the order of values, but only the order of colors, associated with the values. From comparison of **figures 17.24** and **17.25** you can see that the values are shown from the *Red* sector in the clockwise direction. The "Switch drawing direction" command does not change the order of values and their association with the colors, but changes the direction of drawing.

For the bar charts, the menu on comments is the simplest; the same primitive menu is called on any type of comments whether it is associated with the main plotting area or any type of scale. With the pie charts the situation is absolutely different. For the comments associated with the whole pie chart, the same primitive menu is called with the same few opportunities (**figure 17.23**). But for text of any sector there are a lot of possibilities (**figure 17.27**).

At this figure you can see the same menu with both of its submenus; I decided to show it in such a way because each of the submenus is bigger than the menu itself. The menu is small and nearly the same as for all other types of comments; only one command line disappeared and is substituted by two others. No comment of a sector can be deleted; that is why that line disappeared. But the information of any commentfor a sector can be shown in different views; these possibilities are available through the first submenu. The parameters of any comment for a sector can be used as a sample for all the siblings; these things are available through the second submenu.

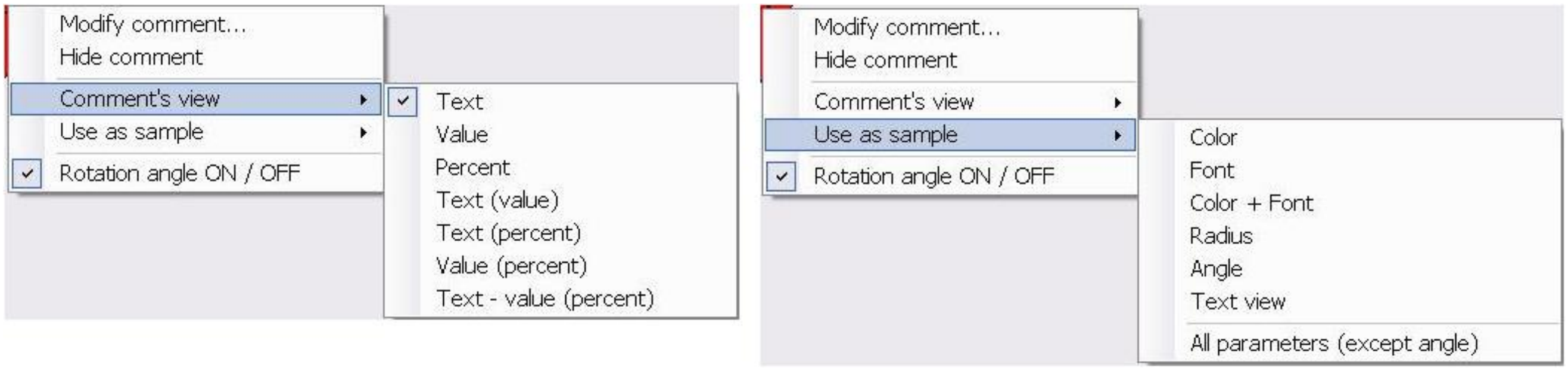

**Fig.17.27** Menu on the sector's comment with two available submenus

The sectors are associated with some values. An array of values is used for initialization of the pie chart; the number of sectors is equal to the number of values in that array. Values cannot be negative; the situation of all zeros is also forbidden; the sum of the array must be positive. Together all the sectors fill the whole circle; the angle of each sector corresponds to the percentage of its value in the whole sum. Sectors can be also associated with some texts. The comments to sectors can show in different combinations those texts, real values, and percents. Each comment for a sector can be shown in its own way; comments in sectors at **figure 17.24** demonstrate all seven possibilities of different views. The first submenu from **figure 17.27** allows to select the view of the comment. The order of views for the comments in sectors at **figure 17.24** is the same as the order of views in the first submenu at **figure 17.27**; everything starts from the red sector and goes clockwise. The comment for Red sector shows only the associated text; the comment for Orange sector – only the value, and so on.

Each comment for a sector has a lot of individual parameters and the easy ways to change them.

- Color. Select the needed line in the list of comments in the tuning form (**figure 17.25**) and call the standard dialog with the 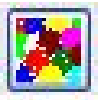 button.
- Font. Select the needed line in the list of comments in the tuning form (**figure 17.25**) and call the standard dialog with the 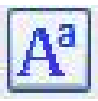 button.
- Text. Select the needed line in the list of comments in the tuning form (**figure 17.25**), type the new text underneath and press the 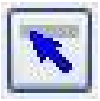 button to substitute the old text with the new.
- Radius. A comment can be moved at any moment.
- Angle. A comment can be rotated at any moment.
- Text view. Change of the view is done through the menu and submenu (**figure 17.27**).

Suppose that you have really a big number of sectors; **figure 17.14** represents such a case for the rings, but the tuning of the pie charts and rings are organized identically. You changed the view of one comment and decided that it would be nice to spread the same view on all the siblings. You do not need to repeat the same procedure again and again; instead you can use the commands from the second submenu (**figure 17.27**). When any command of this submenu is selected, the corresponding parameter from the pressed comment is spread on all the siblings (on the comments to all other sectors).

The combination of individual and group tuning is very powerful. For example, you set the same color for comment of all the sectors, but then you saw that one of them is poorly visible on its sector (the case of the black text on the blue sector at **figure 17.24**). In this case you can change the color of this particular comment; in the mentioned pie chart the white comment on blue looks much better. Or you need to attract the attention to one particular comment; you can set the same font to all the comments and then enlarge the needed one.

If you have an idea that the users of the complex financial applications are too dull to get and to use all these possibilities, you better ask THEM about it, but not decide instead of them. I do not understand why a TV remote control needs anything except the single red button. I can never understand, what the need in all other buttons is, but I do not demand that all the remote controls have to contain only one button and nothing else. I have a feeling that other people have a better understanding of such devices and use all (or bigger part) of all those buttons at one moment or another. The same thing with the variety of visibility parameters for all the screen objects. Give the choice to the users; do not think that you are cleverer than anyone and that you have to decide for others, what they can understand and what they can do. Give them the control over everything and an easy way of performing this control. The users will decide (individually!) what they really need and how are they going to use all those choices.

I am not finished yet with the context menus in the **Form_PlotsVariety.cs**; **figure 17.28** shows the menu that is opened at any empty place.

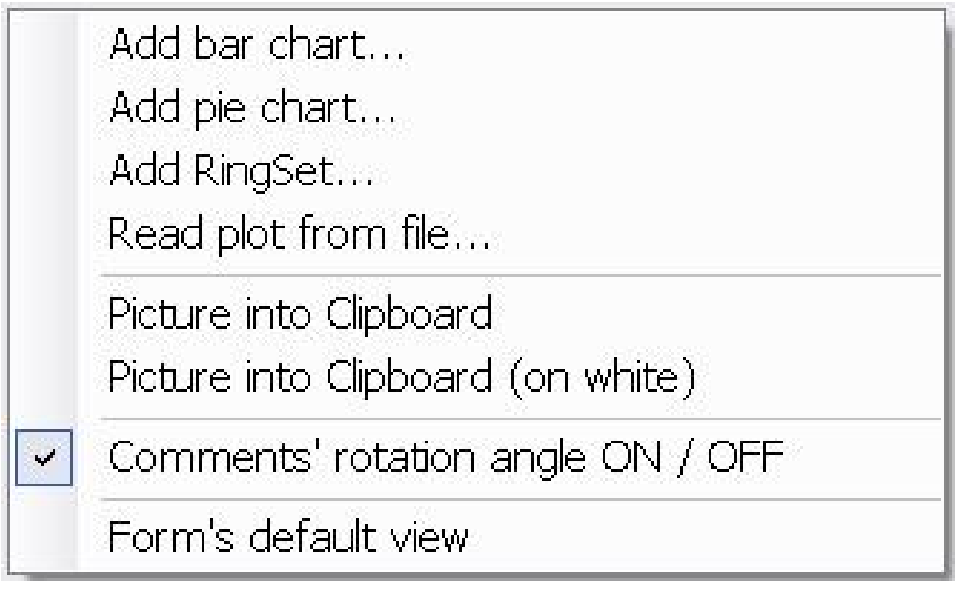

**Fig.17.28** Menu at any empty place in the **Form_PlotsVariety.cs**

The last command from this menu can be of some interest. If there are too many plots in the form, you can get rid of them one by one (via their menus), or you can select the last line in this menu and return to the default view of the form with only four plots (**figure 17.1**). Only do not expect the new pie chart and rings to be exactly the same as on that picture, as their associated values are picked at random at the moment of the initialization.

The "Form's default view" is a standard command, which can be found in menus of many forms of this application, but in the **Form_PlotsVariety.cs** it was the reason to add several more commands to the form and organize one situation, which is not repeated anywhere else in the whole Demo program.

In the section *DataRefinement application* of the previous chapter, I have already mentioned that I do not use writing to the files anywhere in this program. Too many readers are absolutely sure that whenever the program is writing anything on their computers, this is a virus and nothing else. I prepared the program, which accompanies this book, to show the readers all the possibilities of the user-driven applications, but I do not want those readers-users to become nervous; for this reason the saving of information (objects) into the files is not used anywhere throughout the program. Better to say that up till this moment I couldn't even demonstrate the saving of the objects into the files, because I purposely avoided such actions. However, the more I worked with the **Form_PlotsVariety.cs**, the more I felt the need of the possibility to save and restore the objects of this form via the files. It is not done somewhere behind the curtains, but at least it can be done, if the user wants.

There are so many chances to add the new objects and modify them in this form, but then you have (or had before) only two choices: either to keep these objects permanently or to delete them. With only these two chances, I had a big problem even in preparation of the book with the parallel checking of the program. I prepared some objects for demonstration, I made some pictures of the views, and I couldn't use the "Form's default view" command without losing those objects.

Now there is a chance to save the objects, delete them from view, but have an opportunity to restore them later. Commands for saving the bar charts and pie charts into file can be seen in menus at **figures 17.21** and **17.26**; saving of the `RingSet` objects is done in exactly the same way. It is a standard procedure, where you can select the location and the name of the file, but the default name includes the type of an object, for example, "PieChart", and the long number, which is nothing else, but the original **id** of this object. You can use any names of your own; the file includes a short indication of the object, which is stored inside, so the restoration is not going to produce an elephant instead of the monkey, even if you change the name.

```
private void Click_miPiechartSave (object sender, EventArgs e)
{
    … …
    fs = new FileStream (binFile, FileMode .Create, …);
    bw = new BinaryWriter (fs);
    bw .Write ("PieChart");
    elems [iElement] .PieChart .IntoFile (bw);
```

I have mentioned before that each object has a pair of methods for saving / restoring via the binary files; this code uses the `PieChart.IntoFile()` method. The "Read plot from file" command (**figure 17.28**) opens the standard dialog to select the file for reading. When the file is selected, then its content is first checked for the quick identification of the type of the object; after it the method of the appropriate class is called for the restoration of an object. This piece of code shows the restoration of a pie chart; there are similar pieces for bar charts and rings.

```
private void Click_miReadPlotFromFile (object sender, EventArgs e)
{
    OpenFileDialog dlg = new OpenFileDialog ();
    …
    fs = new FileStream (filename, FileMode .Open, FileAccess .Read);
    br = new BinaryReader (fs);
    string str = br .ReadString ();
    … …
    else if (str == "PieChart")
    {
        PieChart piechart = PieChart .FromFile (this, br);
        if (piechart != null)
        {
            piechart .Move (ptMouse_Up .X - piechart .Center .X,
                            ptMouse_Up .Y - piechart .Center .Y);
            elems .Insert (0, new SingleElement (piechart));
            bRenew = true;
        }
    }
    … …
    RenewMover ();
    Invalidate ();
```

Two important remarks to the restoration of the plots from files.

1. The original plot was stored in the file with the whole set of its parameters, which includes the unique identification numbers for all its parts. The `FromFile()` method of any class restores an object, but gives it the new **id**; it happens at all the levels, so all the identification numbers of the restored complex object and its inner parts will be different. Thus, you can copy the objects from the file without producing a mess in the identification system.

2. The plots are restored with the same coordinates that they had on the moment of saving into the files. It is possible that the original object was moved around the screen between the moments of storing and restoration; it is also possible that it was not moved at all. In the last case you get the copy exactly at the same place and there will be no indication of two different objects at the same spot. To avoid this, the restored object is positioned at the place, where the menu with the command for restoration was called. The top left corner of the menu is used for positioning the restored plot; for the bar chart it will be the top left corner of the main area; for the pie charts and rings it will be the central point. It is possible that the original object was moved and the menu was called in such a way that the position of the restored object will put them on top of one another, but the probability of such a thing is negligibly small.

It would be an unusual thing to have in my big application some form, in which users would be allowed to do everything with the existing objects, but not to add objects of their own. The first three commands of the menu (**figure 17.28**) close this possible flaw in design and allow to add the new objects to the form; the description of these possibilities is in the next subsection.

## *The same design ideas at all the levels*

The **Form_PlotsVariety.cs** is an example of an application for financial analysis. From my point of view any type of financial plotting can be developed in two major modifications. One of them works with the fixed array of data (values), which is passed as a parameter at the moment when the plot is initialized. There can be numerous ways of showing this data, but the values are not changed throughout the life of the plot. Another type of plots may have all the same variations of visualization, but also includes an instrument of changing the associated data. The first type of plots is needed for analysis of already collected data; the second type is often needed for data preparation. As both types of plots are needed for different purposes, I decided to demonstrate both, but slightly divided the areas of their use. All three classes of plots used in the **Form_PlotsVariety.cs** - `BarChart`, `PieChart`, and `RingSet` - belong to the first type. The similar plots of the second type are used in the forms, where the new objects are prepared.

The newly prepared objects are included into the **Form_PlotsVariety.cs**, where the whole lot of possibilities for their tuning exists. For this reason, I do not see any sense in duplicating all those possibilities at the moment and at the place, where the new objects are organized. There are only basic things plus an instrument for tuning all the initial data (values). Because the corresponding three classes for organizing the new bar charts, pie charts, and rings are lacking in tuning, their names include the word *primitive*. Though from the point of working with the associated values they are much more interesting than similar objects in the **Form_PlotsVariety.cs**. Let us start with the preparation of the new bar charts.

**Figure 17.29** demonstrates the **Form_DefineNewBarChart.cs**, in which the new bar charts are prepared. The form consists of two major parts – a group and a bar chart - plus a couple of buttons. Any big application includes some small auxiliary forms, in which one, two or several parameters are prepared. The change of each parameter needs only one small control, so such a form usually contains several controls. Throughout the years I have developed many forms of such a type without any serious thought about their design. Usually I put those controls in the way I liked them to be and that was all.

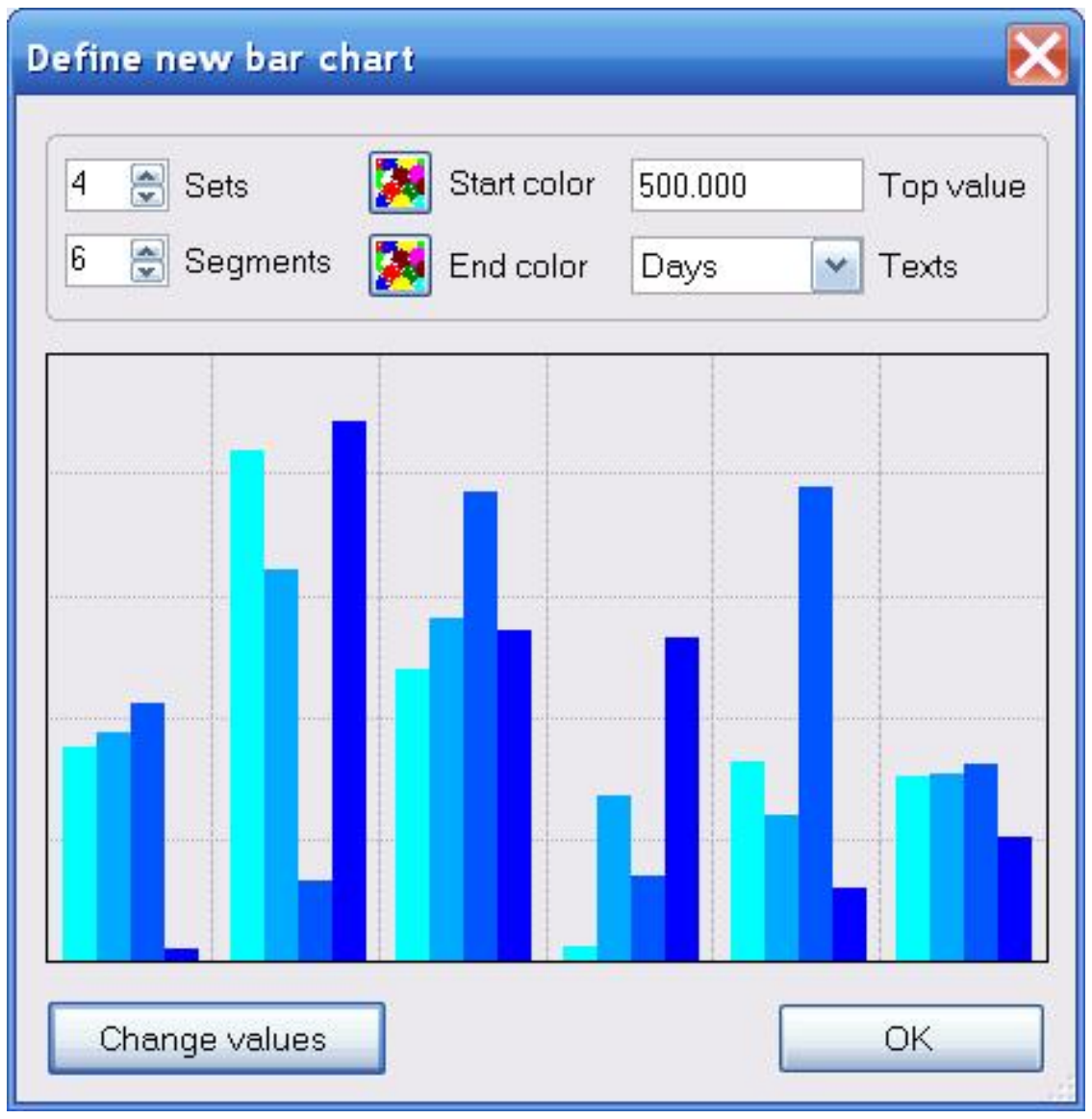

**Fig.17.29** The form to define new bar charts

The switch to the user-driven applications ended this practice. If everything is movable and resizable in the main forms of an application, then no auxiliary forms can be designed in an old style. The logic of the application itself would require all these even small, less important, and rarely used forms to be designed according to the now standard rules: everything must be movable and tunable. It is not an imposed burden; in a short time you begin to do it automatically. If by chance you missed it (it can happen only at the beginning), you bump into this strange behaviour of the unmovable and unchangeable objects long before any of the users. You stumble over it the very first time you open this small form for checking. You automatically try to move one or another object a bit and you are shocked that they do not obey your mouse. The code for making everything movable in such small forms is so simple that there is no sense even in trying not to implement such thing. And very shortly the implementation of movability becomes an automatic process.

Everything in the **Form_DefineNewBarChart.cs** is movable. The needed controls are united into a group of the `ElasticGroup` class; the bar chart belongs to the `PrimitiveBarChart` class.

```
    public class PrimitiveBarChart : GraphicalObject
    {
        int nSets;                // each set has its own colour
        int nSegments;            // each segment includes a strip of every colour
        List<SingleBar> bars = new List<SingleBar> ();
        List<Color> clrs = new List<Color> ();      // one color per each set
```

What makes this bar chart interesting is the use of the manually resizable columns (bars), which belong to the `SingleBar` class. In the first part of the book, while discussing many different graphical primitives turned into movable, I have demonstrated a whole set of rectangles with different types of movability and resizability. Among those cases were the rectangles with a single moving border (**Form_Rectangles_SingleSideResizing.cs, figure 3.2**). This is the type of figure which I need here for manual changing of values for a bar chart.

To exclude any unneeded code, I organized the `SingleBar` class exactly for this example: I fixed the lower end of each bar at the bottom of the rectangular area, while the upper side of the bar can be moved inside the area up and down between the two borders of rectangular area. The width of each bar is determined by the width of the area (it is resizable), the number of segments, and the number of sets. A small part of each segment is left empty.

Here is the cover for the `SingleBar` class; the `rcMax` in this code is the biggest rectangle that a bar can fill.

```
public override void DefineCover ()
{
    CoverNode [] nodes = new CoverNode [] {
           new CoverNode (0, new Rectangle (rcMax .Left, cyTop - halfsense,
                                  rcMax .Width, 2 * halfsense), Cursors .SizeNS),
           new CoverNode (1, new Rectangle (rcMax .Left, cyTop, rcMax .Width,
                                             Math .Max (2, rcMax .Bottom - cyTop)),
                                  Behaviour .Transparent, Cursors .Default) };
    cover = new Cover (nodes);
}
```

The cover for each resizable bar consists of two nodes: the first node covers the movable upper end of the bar and allows to change its height; the second node covers the whole area of the bar… and is not used at all!

The last strange statement is obvious from the `StripInRectangle.MoveNode()` method. It shows from the beginning that only the node with the zero number is considered in this method.

```
public override bool MoveNode (int i, int dx, int dy, Point ptM, MouseButtons btn)
{
    bool bRet = false;
    if (btn == MouseButtons .Left && i == 0)
    {
        int cyNew = cyTop + dy;
        if (rcMax .Top <= cyNew && cyNew <= rcMax .Bottom)
        {
            cyTop = cyNew;
            bRet = true;
        }
    }
    return (bRet);
}
```

The movable side of the bar is not allowed to go outside the rectangular area; the method moves the top side, if the `nodes[0]` was caught; the `nodes[1]` is not even mentioned in this method. To explain the strange appearance of the not usable node in the cover of those bars, I will first return to the old case of the `OneSideRectangle` class and then back to our tunable bar chart.

In that old case I have demonstrated several rectangles (**figure 3.2**). Each of those rectangles can be moved around the screen by any inner point, plus one side of each rectangle is movable and allows to change the area of rectangle. To fulfil these two purposes, I have to use the cover consisting of two nodes: one node for moving the sliding side, another for moving the whole rectangle. Two different purposes require two different nodes.

In the case of the **Form_DefineNewBarChart.cs** we have more objects and more different possibilities. The resizable bars are not the independent rectangles that can move anywhere by themselves. On the contrary, each of these bars is anchored at the particular place inside the rectangular area and cannot be positioned anywhere else. The position of each bar and its width is determined by a couple of other parameters and depend only on these parameters. The bottom of each bar is fixed; only its top can be moved.

The bars are the parts of the complex object; being the parts of the `PrimitiveBarChart` object, they are registered in the mover's queue by the `IntoMover()` method of this class.

```
new public void IntoMover (Mover mover, int iPos)
{
    mover .Insert (iPos, this);
    foreach (SingleBar bar in bars)
    {
        mover .Insert (iPos, strip);
    }
}
```

The `PrimitiveBarChart` class has a very simple standard cover for a rectangle, resizable by its sides and corners, so the code of this `IntoMover()` method indicates that, when the mover analyses its queue for moving anything, it first checks through all the resizable bars and then comes the resizable area of the bar chart.

As you can see from the `SingleBar.DefineCover()` method at the previous page, the second node of the cover is declared with the `Behaviour.Transparent` parameter. This means that if the mouse is pressed on such a node, then mover ignores it, looks through this node, and finds the next node at the same spot. This node can be only the rectangular node of the area of a bar chart. Mover grabs it and moves. The `PrimitiveBarChart.Move()` method is simple.

```
public override void Move (int dx, int dy)
{
    area .Move (dx, dy);
    InformRelatedElements ();
}
```

This is the standard procedure for the complex objects: when the main area is moved, all the related parts are informed about the new position of the parent and move synchronously. In the case of the PrimitiveBarChart and SingleBar classes, the bar chart calculates the new allowed area for each bar and sends it to each of them.

```
private void InformRelatedElements ()
{
    for (int i = 0; i < bars .Count; i++)
    {
        bars [i] .MaxRectangle = BarMaxRect (i);
    }
}
```

Each bar is recalculated according to the new rectangle, allowed for it.

```
public Rectangle MaxRectangle
{
    get { return (rcMax); }
    set
    {
        double fill = FillCoef;
        rcMax = value;
        cyTop = Auxi_Geometry .ValToCoor_Linear (rcMax .Top, rcMax .Bottom,
                                                  1.0, 0.0, fill);
        DefineCover ();
    }
}
```

If you press inside the area of a bar chart not at the sensitive top of any bar, then it does not matter whether you press on an empty place or on a colored strip; in any of these cases the bar chart area moves together with all the bars.

If you use for the second node of a bar the Behaviour.Frozen or Behaviour.Nonmoveable parameter, then any of these nodes would block the mover from looking further on; such bar chart would be movable only by the empty spots, but not by the colored strips. I think that such a decision would be worse than the used one.

There is another possible solution which allows to move the bar chart by any inner point: get rid of the second node in the SingleBar class. It is not needed at all, so just leave only one node over the sliding side. It will work, but I simply do not like the situations when part of an object is not visible by mover. In this case it is a big part; nearly the whole area of a bar will be ignored. I want to underline: everything will work exactly as it works now, there will be no side effects. I simply do not like to exclude the big parts of the objects from the attention of mover. I prefer to use the transparent nodes for such cases.

One more remark about the covers of the SingleBar objects. The second node is determined by the area of a bar, but this area can diminish to zero when the upper side of the bar is moved down to its lowest position. Nodes with the zero area are not allowed, so there can be two solutions to this problem. The first one is to limit the minimum allowed area of the bar by some value, greater than zero, for example, let the height of the bar never be less than two or three pixels. It is a good solution, but what to do, if some values need to go to zero? The second solution is shown in the code: the value associated with the bar can diminish to zero, but the area of node does not! The height of any node cannot become less than two pixels.

```
new CoverNode (1, new Rectangle (rcMax .Left, cyTop, rcMax .Width,
                                 Math .Max (2, rcMax .Bottom - cyTop)),
```

The drawing procedure shows the real bar; it can squeeze to nothing and the bar disappears. But the node still exists. The first node, which moves a bar up and down, is wider than 2 pixels, so in such situation it will totally cover the second node. I have already explained this technique, while discussing the disappearing objects in the first part of the book, but decided to remind about this small, but very important detail in design of such objects.

The rules of user-driven applications are applied not only to the main form of any program, but they spread deep inside into the farthest corners. From the **Form_Main.cs** of this application we went to the **Form_PlotsVariety.cs**; from there we called an auxiliary **Form_DefineNewBarChart.cs**. Even here all the declared rules must be applied. Everything is movable, everything is tunable, and everything is saved and restored.

The **Form_DefineNewBarChart.cs** has a group of the ElasticGroup class, so there is a menu to call the standard tuning of this group (**figure 17.30**). Other commands of the same menu allow to fix / unfix the inner elements of the group and give some other options. Another menu (**figure 17.31**) can be called anywhere outside the group.

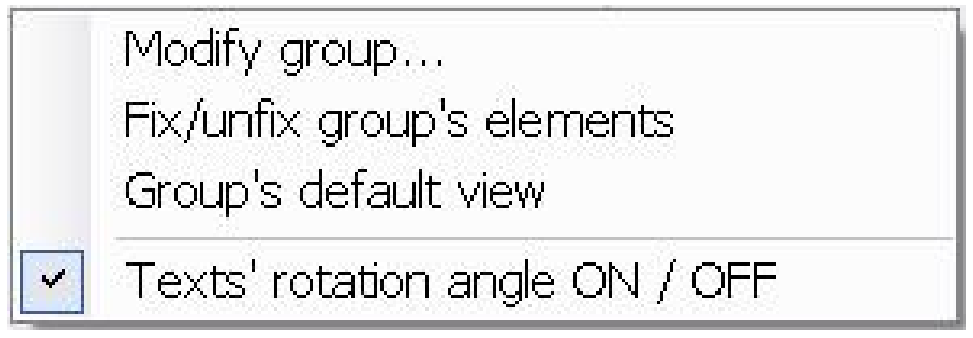

**Fig.17.30** Menu on the group in the **Form_DefineNewBarChart.cs**

You can see that the first menu (**figure 17.30**) contains a command to reinstall the default view of a group; the second menu (**figure 17.31**) includes similar command to recover the default view of the whole form. Similar commands are used throughout all the complex groups and forms of all the demonstrated examples. In the user-driven applications it is possible to change the views in any possible way. Some modifications can significantly improve the view (there are users, who are much better designers than the authors of those applications); some of the changes may have an opposite effect. With these menu commands users receive an opportunity to recover the initial view; whether they would like to do it or not, it is up to them.

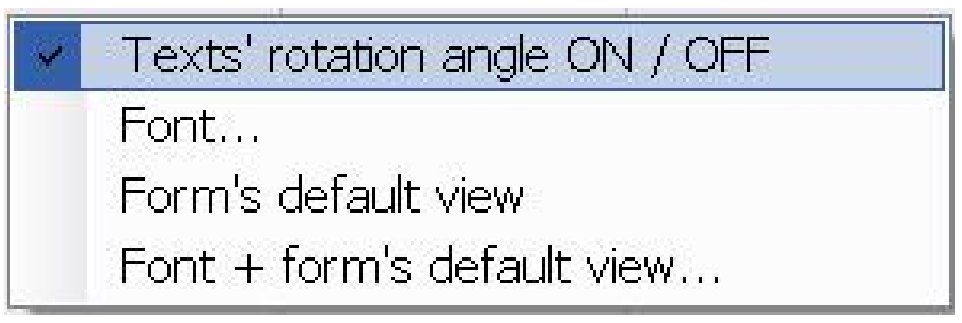

**Fig.17.31** Menu outside the group in the **Form_DefineNewBarChart.cs**

The possibility to change the font becomes one of the most important menu commands and one of the most valuable for a lot of users. The increasing percentage of the middle aged and older users simply cannot use the applications designed predominantly by the younger developers, because of the trend in eyesight change, of which the younger people rarely think at all. You can find the commands for changing of the fonts in nearly every corner of all of my examples, so there is no surprise in having such a command for this form (**figure 17.31**). But the last command in the same menu can be a bit surprising. I continue to insist, and I have mentioned it several times, that the change of a font must be only the change of a font and not a bit more. A designer is not allowed to adjust the view according to the new size of the font. This would be against the law of user-driven applications: developers are not allowed to interpret the users' commands but have only to provide the reaction on whatever was asked.

If somebody asked another font, then the font would be changed. It can partly corrupt the whole view, but there is an opportunity to move and resize all the objects, so the view can be improved by some moving / resizing. At the same time there can be a request for the same general view, but with the changed font. This can be (and must be) a separate command; that is what the last command in the menu from **figure 17.31** allows to do. I did not spread the use of such commands throughout the Demo application, but it is implemented in the **Form_DefineNewBarChart.cs** and a couple of similar forms.

Let us close the small form to add the new bar chart and return back to the **Form_PlotsVariety.cs**, in which other types of plots can be also added. First three commands from menu at **figure 17.28** open the ways to three different forms to define new bar charts, pie charts, and rings. Pie charts and rings are so close in design that it is enough to demonstrate and discuss one of them; this time I prefer to show the work with the rings. By the way, the command from menu (**figure 17.28**) is not the only way to come to the **Form_DefineNewRing.cs** (**figure 17.32**). If you need to add another ring to already existing object of the `RingSet` class, you call the context menu on this object and the command *Add new ring* in that menu will bring you to the same form.

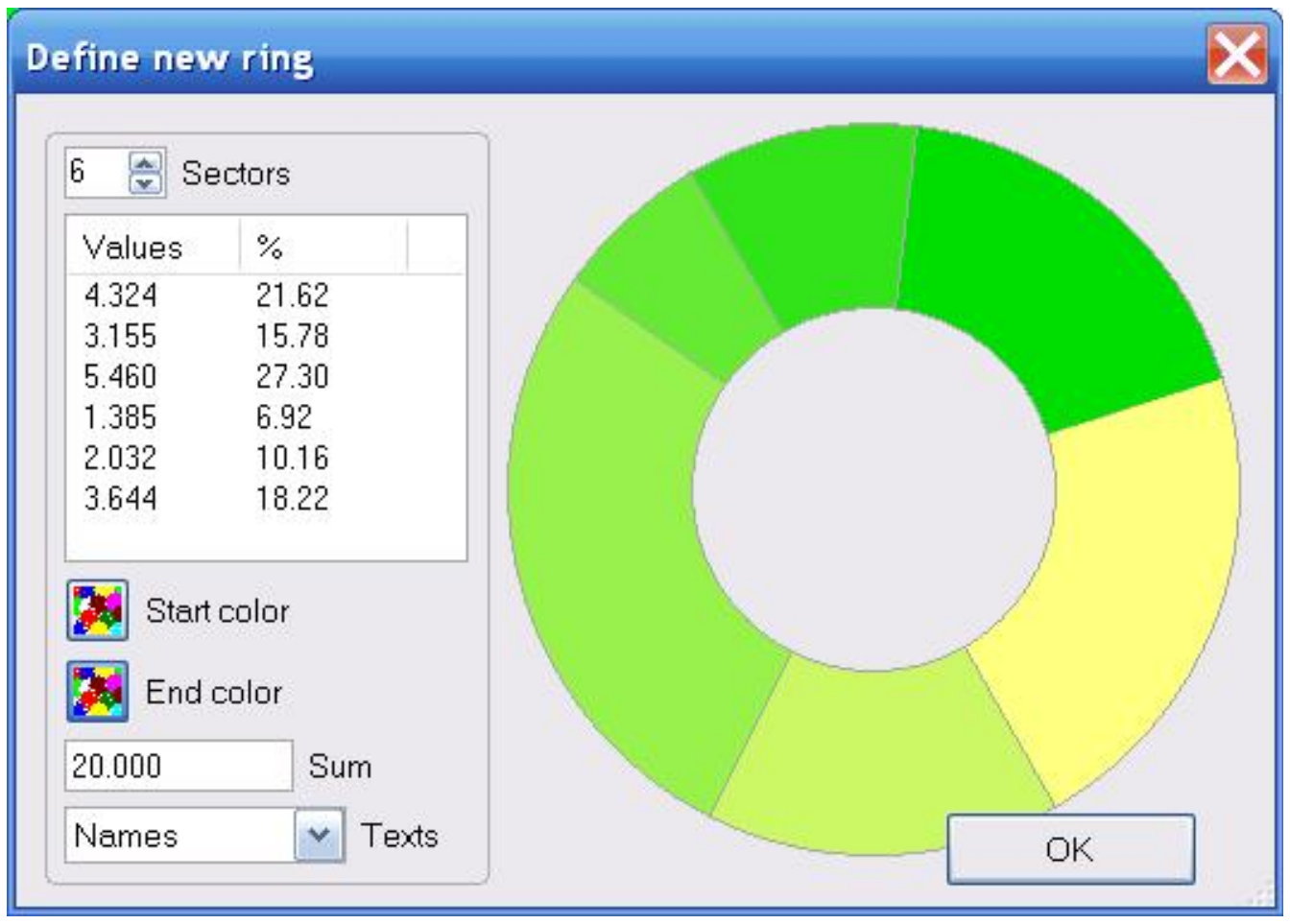

**Fig.17.32** The form to define the new rings

The forms to define the new bar charts (**figure 17.29**) and rings (**figure 17.32**) are similar; the minor discrepancies are rooted in the differences of rectangular and round plots. The interesting element in the new form is the ring with the changeable sectors.

In the first part of the book, while demonstrating different graphical objects, I used an example of a circle with the sliding partitions (chapter *Sliding partitions*, section *Circles with the sliding partitions*, **figure 9.3**). There is a similar example of the rings with the sliding partitions (**Form_Rings_SlidingPartitions.cs**), which can be started through the menu *Graphical objects – Basic elements – Rings – Sliding partitions*, but that example was not included into the text. This is one of the reasons, why I decided to demonstrate here not the circle, but the ring.

The ring, which you see at **figure 17.32**, belongs to the `PrimitiveRing` class. As I explained a bit earlier, the word *primitive* corresponds only to the lack of tuning for many of the possible parameters, but from the point of the cover design it is definitely not a primitive object. Well, it is not far away from other rings, which were discussed in the first part of this book, but with each new feature the number of nodes is growing, so I would prefer not to include the full code of the `PrimitiveRing.DefineCover()` method here, but only to give some explanation.

The ring is resizable by both borders (inner and outer). For resizing of the objects with the curved borders we have the standard technique of the N-node covers; in the case of the `PrimitiveRing` class I use the small polygonal (trapezoid) nodes, which stay side by side to each other and produce a sensitive strip without any gaps. The ring is movable; this is organized by using two big circular nodes, of which the first and the smaller one, closing the hole, is transparent. The borders between the neighbouring sectors must be movable. These borders are straight, so it is enough to cover them with the narrow strip nodes. The obvious solution is to construct each strip node on two points where the border of a sector crosses the outer and inner circles.

The number of the small nodes to cover the outer and inner borders of the ring is determined by the length of those borders and the width of the small neighbouring nodes (`wNode`).

```
private void NodesOnBorders ()
{
    nNodesOnOuter = Convert .ToInt32 ((2 * Math.PI * rOuter) / wNode);
    nNodesOnInner = Convert .ToInt32 ((2 * Math.PI * rInner) / wNode);
}
```

Cover of a ring includes the nodes in such an order:

1. Polygonal (trapezoid) nodes on the outer circle.
2. Polygonal (trapezoid) nodes on the inner circle.
3. Strip nodes on the borders between the sectors.
4. The circular node on the hole.
5. Bigger circular node to cover everything inside the outer circle.

```
nodes = new CoverNode [nNodesOnOuter + nNodesOnInner + vals .Length + 2];
```

The sensitive strips, which are organized along the circular borders, go for `hHalf` pixels on both sides of the covered border. This strip is divided into small trapezoids; four corners of each trapezoid are positioned on two radiuses. Trapezoids stay side by side, so the two corners of the next are the same, as they were for the previous trapezoid. Going around the circle and calculating these nodes one after another, I have to calculate only two new points for each one. Here is the part of the `DefineCover()` method for the nodes on the outer circle.

```
public override void DefineCover ()
{
    PointF [] pts = new PointF [4];
    … …
    rBelow = rOuter - hHalf;
    rAbove = rOuter + hHalf;
    pts [0] = Auxi_Geometry .PointToPoint (center, 0, rBelow);
    pts [1] = Auxi_Geometry .PointToPoint (center, 0, rAbove);
    for (int i = 0; i < nNodesOnOuter; i++)        // nodes on outer border
    {
        pts [2] = Auxi_Geometry .PointToPoint (center,
                            2 * Math .PI * (i + 1) / nNodesOnOuter, rAbove);
        pts [3] = Auxi_Geometry .PointToPoint (center,
                            2 * Math .PI * (i + 1) / nNodesOnOuter, rBelow);
        nodes [i] = new CoverNode (i, pts, Cursors .Hand);
        pts [0] = pts [3];
        pts [1] = pts [2];
    }
```

Nodes on the inner border are calculated in exactly the same way.

The strip nodes over the partitions between the sectors are positioned along the radiuses. The position of the first one is determined by the `angle` parameter of a ring. The angles for all the sectors are stored in the `sweep[]` array; these values allow to calculate the angle for the next partition, when the current one is known.

```
public override void DefineCover ()
{
    … …
    PointF ptIn, ptOut;
```

```
        double angleForLine = angle;
        for (int i = 0; i < vals .Length; i++)  // nodes on borders between sectors
        {
            ptIn = Auxi_Geometry .PointToPoint (center, angleForLine, rInner);
            ptOut = Auxi_Geometry .PointToPoint (center, angleForLine, rOuter);
            nodes [nSmallNodes + i] = new CoverNode (nSmallNodes + i, ptIn, ptOut);
            angleForLine += sweep [i];
        }
```

To understand the resizing of sectors by moving their sliding partition you have to look into the `OnMouseDown()` and `OnMouseMove()` methods of the **Form_DefineNewRing.cs** and into the `PrimitiveRing.OnMoveNode()` method. Everything starts when a partition is caught by the mover. The nodes over the partitions are the only strip nodes in the cover of a ring, so both the class of the caught object and the form of the caught node are checked for decision about starting such movement.

```
        private void OnMouseDown (object sender, MouseEventArgs e)
        {
            ptMouse_Down = e .Location;
            if (mover .Catch (e .Location, e .Button, bShowAngle))
            {
                GraphicalObject grobj = mover .CaughtSource;
                if (grobj is PrimitiveRing)
                {
                    if (e .Button == MouseButtons .Left)
                    {
                        if (mover .CaughtNodeShape == NodeShape .Strip)
                        {
                            ring .StartResectoring (mover .CaughtNode);
                        }
```

If the node over the partition is caught, then the number of this node is passed as a parameter to the `PrimitiveRing.StartResectoring ()` method.

- From the number of the node the number of the partition ( `iBorderToMove` ) is calculated. The calculation is primitive and based on the known number of nodes along the two circular borders and the known order of nodes in the cover.
- From the number of the partition its angle ( `angleCaughtBorder` ) is calculated. The angle of the first border is equal to the angle of a ring; the angles for all the sectors are also known.
- From the same number of the partition the numbers of the sectors on both sides are determined. Usually these are the two consecutive numbers, but the special case of the border between the first sector and the last must be considered. The calculations also depend on the direction of the currently used drawing of a ring, as it can be either clockwise or counterclockwise.
- The numbers of the sectors on two sides of the sliding partition allow to calculate the limiting angles for turning the partition clockwise ( `min_angle_Resectoring` ) or counterclockwise ( `max_angle_Resectoring` ). I use the angles in the normal mathematical way with the angles increasing, when you go counterclockwise; this explains the use of *min* and *max* in the names of these two angles.
- When the partition is moved one way or another, it only affects the redistribution of values between two sectors, but their sum is not changed; this sum is calculated for further use.

```
public void StartResectoring (int iNode)
{
    iBorderToMove = iNode - (nNodesOnOuter + nNodesOnInner);
    double angleCaughtBorder = angle;
    for (int i = 0; i < iBorderToMove; i++)
    {
        angleCaughtBorder += sweep [i];
    }
    if (dirDrawing == Rotation .Clockwise)
    {
        iSector_Clockwise = iBorderToMove;
```

```
        min_angle_Resectoring = angleCaughtBorder + sweep [iSector_Clockwise];
        iSector_Counterclock = (iSector_Clockwise == 0) ? (vals .Length - 1)
                                                        : (iSector_Clockwise - 1);
        max_angle_Resectoring = angleCaughtBorder - sweep [iSector_Counterclock];
    }
    … …
    two_sectors_sum_values = vals[iSector_Clockwise] + vals [iSector_Counterclock];
```

Some of the above calculations depend on the direction of the drawing of a ring, because the numbering of sectors can go one way or another. The **Form_DefineNewRing.cs** does not allow to change the drawing direction of the `PrimitiveRing` object, which is used there. However, objects of this class can be used in other places (programs), so I decided to keep in this class the code for both variants of drawing.

The real moving of any object is described in its `MoveNode()` method; this is the place to look further at the details of moving the partition. Two remarks on the next piece of code.

- Previously the limiting angles for the movement of partition were calculated, but in this case I do not want to diminish any sector up to total disappearance, so there is an allowed minimum angle of 0.05 radian for any sector.
- The united angle for two sectors on the sides of the sliding partition is not going to change; this united angle is strictly linked with the sum of two values, which was calculated beforehand. On every change of the angle of partition I calculate the part of the sum that is now associated with the sector counterclockwise from the partition. By knowing this part and the sum of two values, I can calculate the value for each of sectors. After the two values are changed, the standard `SweepAngles()` method is used to calculate all the needed angles.

```
public override bool MoveNode (int i, int dx, int dy, Point ptM, MouseButtons btn)
{
    … …
    // border between two sectors
    double angleMouse = Auxi_Geometry .Line_Angle (center, ptM);
    … …
    if (min_angle_Resectoring + 0.05 < angleMouse &&
                                   angleMouse < max_angle_Resectoring - 0.05)
    {
        double part_Counterclock = (max_angle_Resectoring - angleMouse) /
                              (max_angle_Resectoring - min_angle_Resectoring);
        if (iBorderToMove == 0)
        {
            angle = angleMouse;
        }
        vals [iSector_Counterclock] = two_sectors_sum_values * part_Counterclock;
        vals [iSector_Clockwise] =
                          two_sectors_sum_values - vals [iSector_Counterclock];
        SweepAngles ();
    }
```

Everything is movable and tunable in any corner of an application regardless of how big it is. At the moment we are discussing the **Form_DefineNewRing.cs** which is the third level of the main Demo application. The purpose of this form is to prepare the new ring; the main visibility parameter of such an object is an array of colors, associated with the sectors. When the new ring is ready and appears in the **Form_PlotVariety.cs**, there is an easy way to change all the colors via the tuning form of the ring. In order to make the preparation of a new ring faster, users are not required to declare the colors for all the sectors in the **Form_DefineNewRing.cs**; instead all the colors are organized as a smooth palette between the two end colors. However, the color of any sector for the ring under construction can be changed by calling the menu on this sector and using one of the commands. This menu (**figure 17.33**) can be called anywhere outside the group, but the first command is enabled only when the menu is called on the ring.

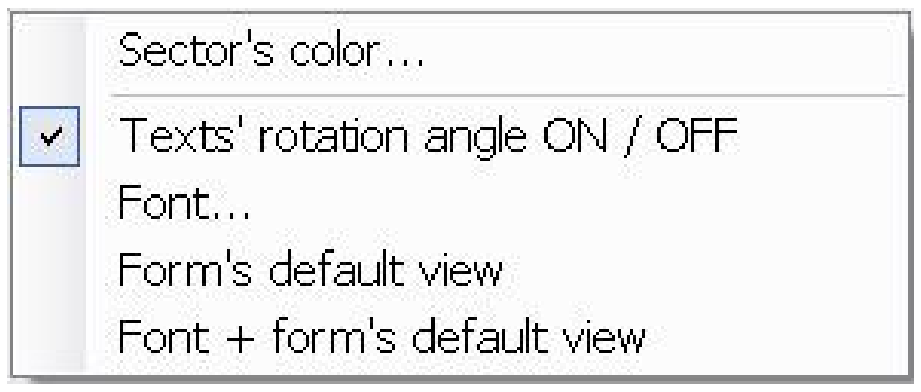

**Fig.17.33** Menu outside the group in the **Form_DefineNewRing.cs**

There is also an interesting group of the last three commands in this menu. The *Font* and *Form's default view* commands are used in many places of this application and represent the standard actions. The last command is the combination of the two previous commands. This additional command appears not everywhere, and here are several words for an addition of such command.

I say and write all the time that I strongly oppose the idea of an automatic change of the sizes, when a user asked only to change the font. Flies and meat must go separately. The change of a font must change the font only and nothing else. (**Rule 3** of the user-driven applications. *The users' commands on moving / resizing of objects or on changing the parameters of visualization must be implemented exactly as they are; no additions or expanded interpretation by developer are allowed.*) However, the `CommentedControl` objects can be vulnerable by this rule when the size of the font is increased. The vulnerability of these objects depends on the relative positioning of a control and its comments; there can be an infinitive number of variants of such positioning, so the results cannot be predicted. If the user agrees with the default view of the form but wants to increase the font without damaging this view, then he can use the last command. If the user wants to change the font without anything else, or wants to restore the default view without any other changes, then both things are also available.

This subsection is about the applying the same ideas of the user-driven applications at all levels of the complex applications with a lot of data presentation. Let us return one level up to the **Form_PlotVariety.cs** and look at one more graphical object and, not surprisingly at all, at its tuning.

The very first figure of this chapter shows the big view of the form without one existing object. But in the top left corner of that **figure 17.1** there is a small button. When this button is enabled, it can be clicked; then the information panel appears (**figure 17.34**). This is an object of the `ClosableInfo` class.

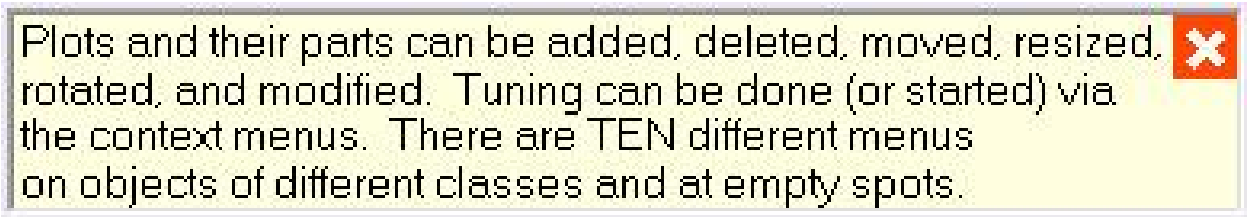

**Figure 17.34** Information in the **Form_PlotsVariety.cs**

Throughout the last several years I have developed several of the demonstration programs to accompany the articles on movable / resizable objects and the aspects of their design. As there were no applications before, in which ALL the objects were movable and resizable, then the readers of the articles (who are also the users of these programs) have no idea beforehand, how to move and resize all the objects they are looking at. In order to give users some idea of this process, the majority of the forms in those applications have one or another type of the help information. A short one, just a sentence or two. I change the design of those info panels from time to time, but mostly they are simply the objects of the `TextM` class. As any other object of the applications, these information panels have to be movable, so this class works perfectly for my purpose. The panels are easy to organize; they are movable by any point; they give the needed help information.

In the previous applications such panels were nearly in every form; in this Demo application only a couple of them are left (**figures 6.1** and **6.8**). These are the two forms for explaining some details in the beginning. After the users get the main idea that everything is movable by any inner point and resizable by any border point, such information is not needed any more. However, in the complex forms, which are the applications themselves, some information is still needed, but not all the time. It is enough to show it at the beginning; later users can hide or unveil it again at their wish, but this needs some easy to use and very compact (in screen space) method to do it. The result was the design of the `ClosableInfo` class and its use in combination with the 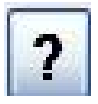 button. This type of a short Help, based on the `ClosableInfo` class is used in several forms of this application, like the **Form_PersonalData.cs** (**figure 14.16**), the **Form_GraphManualDefinition.cs** (**figure 16.30**), and the **Form_PlotsVariety.cs**. As I said, the **figure 17.1** of the last form does not show this information panel, but you can see it at **figure 17.34**. The `ClosableInfo` class is derived from the `TextM` class.

```
public class ClosableInfo : TextM
```

There is nearly nothing new in this class, which was not in the base class. There is a small cross in the corner, which allows to close the information and which gave the name for the class. While discussing the **Form_GraphManualDefinition.cs**, I wrote about the cover of the `ClosableInfo` class and the way it is hidden and restored back in view. Here I want to mention one more aspect of using this class; this is related to that part of the main rules "*everything is tunable*", which demands some extra work on this class.

The information panel is small, but it uses several parameters for visualization:

- Color of the text.
- Font.
- Frame, which can be shown or not.
- Background color.
- Transparency of the whole area.

This information panel is moved around the screen as any other object of the form; if all other objects are totally tunable, then the information panel must be also tunable. Just to stay in line with everything else.

If an object has two or three tunable parameters, then it is easier to organize the tuning through several lines in context menu. If there are ten or more tunable parameters, the design of an auxiliary tunable form is preferable; the cases of the `Scale`, `TextScale`, `Plot`, `BarChart`, and other complex classes are the best demonstrations for this variant. The `ClosableInfo` class with its five tunable parameters is just the boundary situation. You can decide either way and I demonstrate both ways of tuning such object. The information panel in the **Form_PersonalData.cs** has a menu for tuning (**figure 17.35**), while the similar information panel in the **Form_PlotsVariety.cs** is changed with the auxiliary tuning form (**figure 17.36**).

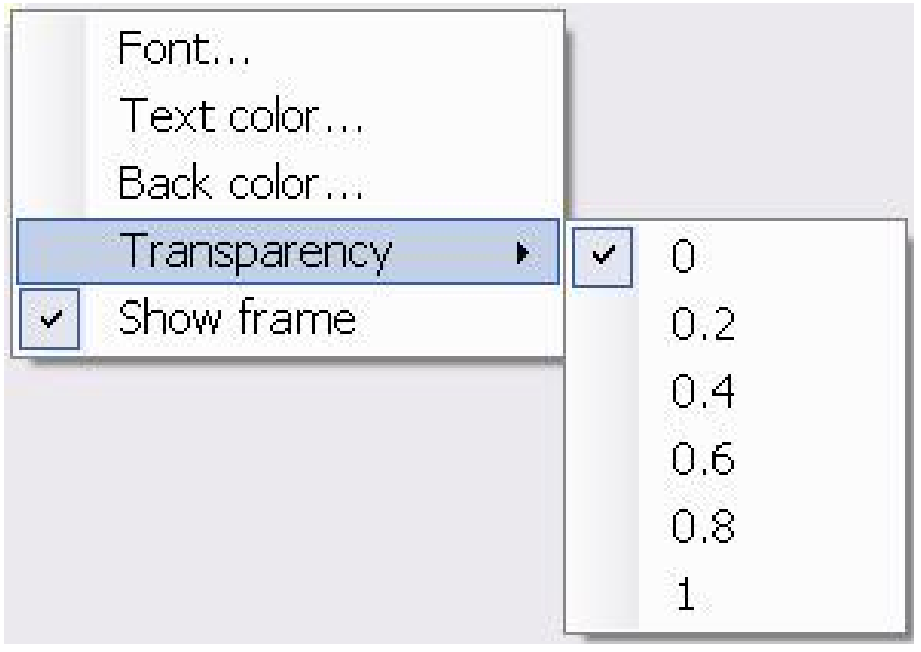

**Fig.17.35** Menu for tuning the parameters of the `ClosableInfo` objects

You can compare and decide for yourself, which way of tuning you prefer, but try to distinguish two different cases:

- What do you prefer as a developer?
- What do you prefer as a user?

The case of menu is slightly easier for the development, but it does not allow the smooth change of the transparency. Maybe it is not important at all; I am not sure. I am very skeptical about the need for transparent areas for scientific plotting (though I have implemented this feature in the `Plot` class), but I have a feeling that the smooth change of the transparency can be useful in information panels.

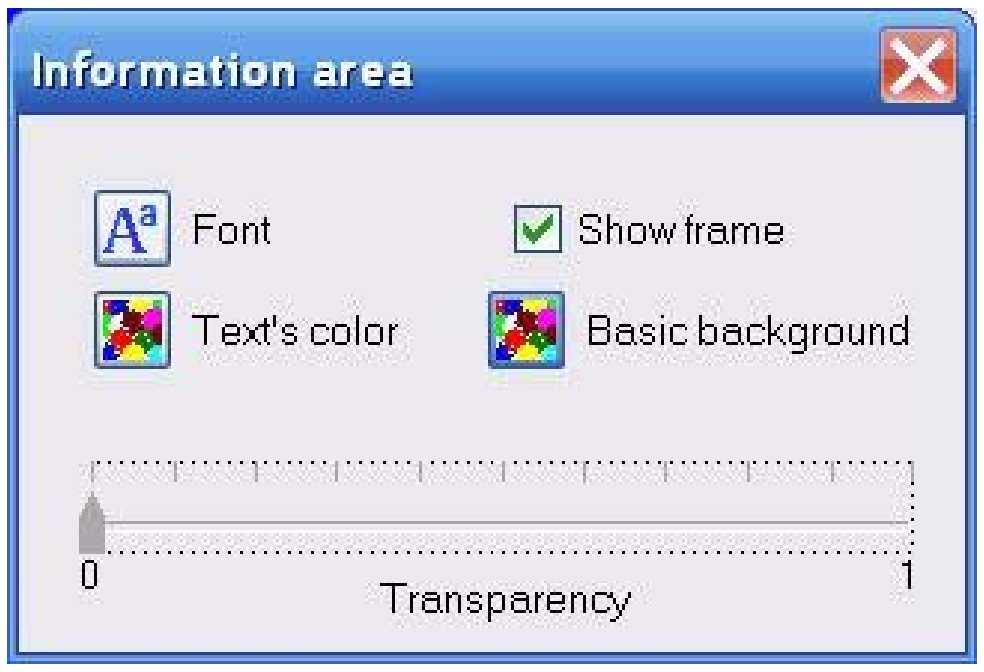

**Fig.17.36** Tunable form for the `ClosableInfo` objects

The design of an additional tuning form does not take a lot of time, but it gives users a full view of all the possibilities. The view of all the things together gives users a better chance to estimate the choices and decide what they really want to change.

I purposely designed two different ways of organizing the same tuning. I do not want to insist on the preferences of one or another. Both things are developed according to the requirements of the user-driven applications, but there are variants inside those rules.

A final remark on this case. The **Form_ClosableInfoParams.cs** (**figure 17.36**) is a simple one, but even such a simple form needs (requires!) some tuning. The rules of the user-driven applications work up to the deepest corners, so there is a small (tiny) menu inside this form (**figure 17.37**). Not surprisingly that the only commands in this menu allow to restore the default view of the form and to change the font; exactly what I was writing two pages back. All the objects in the tunable form are movable; the view of the form is saved and restored. Did I forget anything? Well, maybe it is not worth even mentioning, but there is no way to change the color of comments in this form. If you want, you can consider it as a small exercise. If you add such a command into this menu and several lines into the code of the **Form_ClosableInfoParams.cs** for the identification of the pressed comment, then it would be a final touch for this small form.

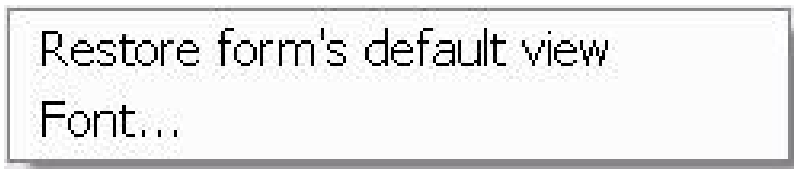

**Fig.17.37** Menu inside the **Form_ClosableInfoParams.cs**

The most amazing thing about this chapter I understood only when I was rereading this text for the second time: different types of plots are discussed from all the sides and up to the farthest corners of their design including the tuning forms, but there was not a word about the big form itself, where an unlimited number of complex objects and elements can be moved, rotated, hidden, unveiled, tuned, resized, and so on. This is really a complex application and there is nothing special in it? Surprisingly, it is so! When all these types of plots (and there can be much more different classes involved) are designed according to the rules of user-driven applications, then all these classes can be used in many places without any tricks or special procedures for one or another case.

At the very beginning of the book I explained that three mouse events – **`MouseDown`**, **`MouseMove`**, and **`MouseUp`** – are used in all the possible situations. These events have to use one mover's method each; nothing else is needed. If there are complex objects for which some tuning forms are called, then the **`MouseDoubleClick`** event is also used, which relies on one of the already used mover's methods. Just to be sure that nothing special is missed in discussion of the **Form_PlotsVariety.cs**, let us make a small tour of this form. It is worth doing it, because this is one of the complex forms and a lot of real applications are designed in exactly the same way. Just to remind you about the field of discussion, here is a view of this form with an increased number of plots (**figure 17.38**). Do not try to find any sense behind the shown objects; I simply added several plots more and moved them around the screen; several of the smaller elements were also moved around, when they were unlucky to be found under the mouse cursor. As we liked to say many years ago, while playing in the yard: “Do not blame me if you couldn’t hide”. If anything happened to be under the cursor, why not to move it or rotate?

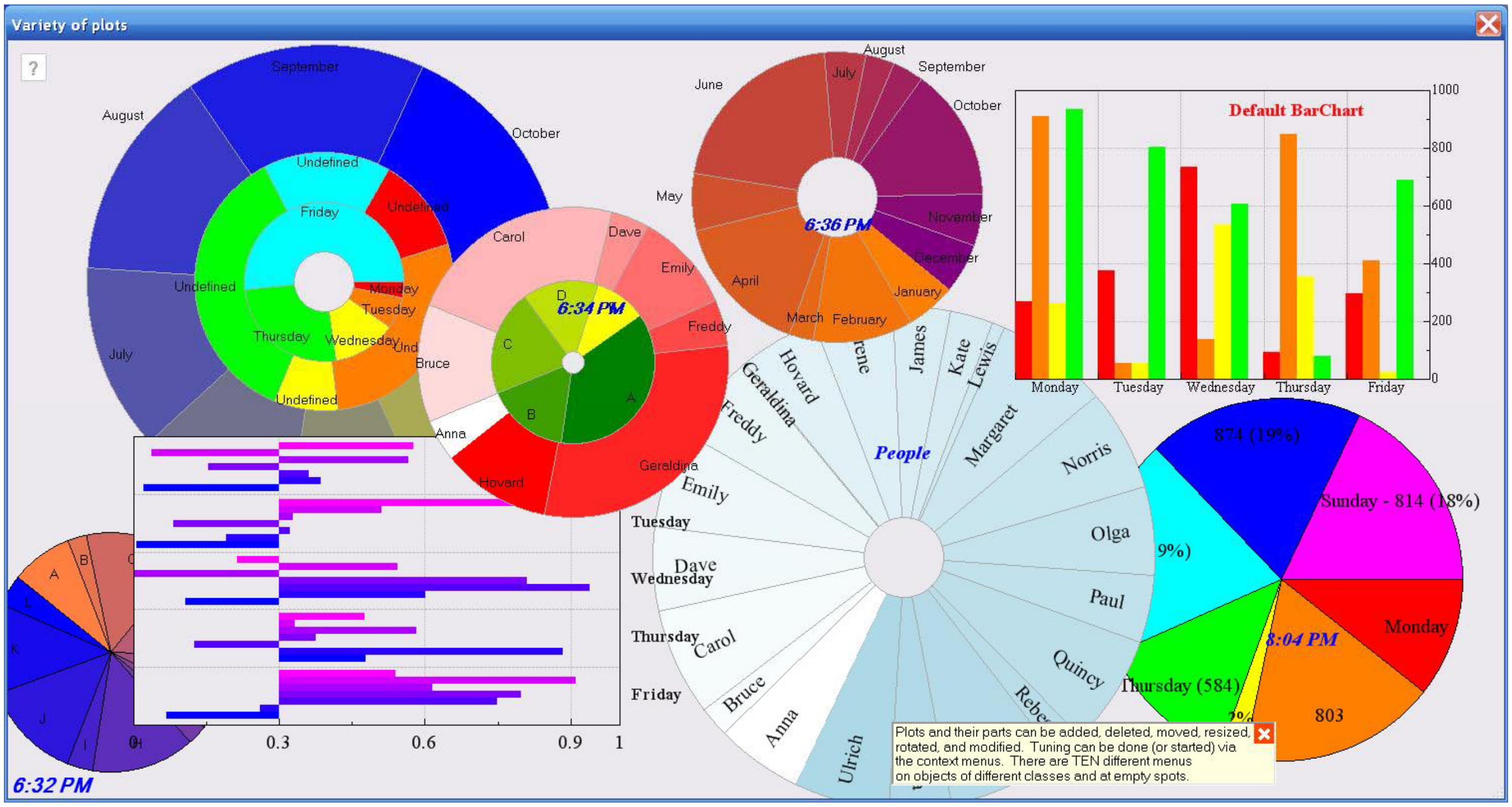

**Fig.17.38** Any number of different plots can appear in the **Form_PlotsVariety.cs**. All these elements can be moved, rotated, resized, tuned, hidden, and unveiled again.

Three classes of plots can be seen in the **Form_PlotsVariety.cs**: `BarChart`, `PieChart`, and `RingSet`. Instead of mentioning all three classes in many parts of the code, I combined them under the `SingleElement` class.

```
public class SingleElement
{
    MedleyElem elemtype;
    PieChart m_piechart = null;
    RingSet m_ringset = null;
    BarChart m_barchart = null;
```

Objects of the `SingleElement` class can be shown in one of those three views, but are organized in the single `List` of elements, used in the form.

```
public partial class Form_PlotsVariety : Form
{
    List<SingleElement> elems = new List<SingleElement> ();
```

Whenever any new plot is organized, it is included into this list. (Only part of the method is shown in the code below.)

```
private void DefaultView ()
{
    RingSet rs = new RingSet (this, new Point (ClientRectangle .Width / 4,
          ClientRectangle .Height / 4), 80, 30, new double [] { 1, 2, 4, 8, 16 });
    elems .Add (new SingleElement (rs));
    PieChart chart = new PieChart (this, new Point (ClientRectangle .Width / 5,
             ClientRectangle .Height * 2 / 3), Math .Max (ClientRectangle .Width,
                                  ClientRectangle .Height) / 4, fRandVal);
    elems .Insert (0, new SingleElement (chart));
    BarChart barchart = new BarChart (this, rc, fVals, Side.S, Auxi_Common.strDays,
                                  TextsDrawingDirection .LTtoRB, Side .E,
                                  1000.0, 0.0, GridOrigin .ByStep, 200.0);
    elems .Insert (0, new SingleElement (barchart));
}
```

Any number of plots can be organized inside the form; in addition there is an informational panel of the ClosableInfo class (info) and a small button (btnHelp) to return this information back into view, if it was previously hidden. All the plots and these two objects are included into the mover's queue. According to the rule, the button must be the first element in queue, as it is the only control in the form; I also decided to show the information atop all the plots.

```
private void RenewMover ()
{
    mover .Clear ();
    for (int i = elems .Count - 1; i >= 0; i--)
    {
        elems [i] .IntoMover (mover, 0);
    }
    if (info .Visible)
    {
        mover .Insert (0, info);
    }
    mover .Insert (0, btnHelp);
}
```

The OnMouseDown() method would include a single line with the call of the Mover.Catch() method, but I selected this form to be one of the few, which can be moved around by any inner point. There are only three such forms in this big Demo application; the **Form_PlotsVariety.cs** is one of them. Such move of the whole form was already explained in the chapter *Applications for science and engineering*; if you press the form anywhere inside but not on any of its movable elements, then you can move the whole form around the screen. The additional code to start such a movement constitutes the bigger part of the OnMouseDown() method.

```
private void OnMouseDown (object sender, MouseEventArgs e)
{
    ptMouse_Down = e .Location;
    if (!mover .Catch (e .Location, e .Button, bShowAngle))
    {
        if (e .Button == MouseButtons .Left)
        {
            bFormInMove = true;
            sizeMouseShift = new Size (PointToScreen (ptMouse_Down).X - Location.X,
                                        PointToScreen (ptMouse_Down) .Y - Location .Y);
        }
    }
    ContextMenuStrip = null;
}
```

Only one small remark to the OnMouseDown() method. There are a lot of elements in the form, which can be involved in rotations, but there is no trace of anything like StartRotation() method for any of the classes. Yet, without using such method there would be an obvious twitching of an object at the starting moment of rotation. There is no twitching of objects in the **Form_PlotsVariety.cs**, so somewhere the appropriate method for each involved class is used correctly, when an object is caught by the right button. The place, where the needed StartRotation() method is called, is the Mover class. Exactly for three involved classes – TextMR (all the comments are derived from this method), PieChart, and RingSet – the mover has all the needed information and starts the rotation. For these three classes everything is automated, but if you would decide to add some other objects with rotation, you would have to add several lines, as it was described in the chapter *Rotation*.

The OnMouseMove() method is also extremely simple. In a standard situation there would be a single call to one of the mover's methods; moving the form by any inner point added several lines to the code.

```
private void OnMouseMove (object sender, MouseEventArgs e)
{
    if (mover .Move (e .Location))
    {
        Invalidate ();
    }
    else
    {
```

```
                if (bFormInMove)
                {
                    Location = PointToScreen (e .Location) - sizeMouseShift;
                }
            }
        }
```

The `OnMouseUp()` method is a bit longer, but not more complicated. Some increase in the size of this method is due to the number of classes, used in the form, and also because further actions depend on whether an object was previously caught and now released by the left or by the right button. I want to emphasize that any standard moving or resizing of any object in the form is done automatically by the `Mover.Move()` method. Only the special situations must be described.

For the release of an object by the left button there are two special situations.

- If the small cross was clicked in the corner of the information panel, then this information must be closed.
- If a plot was moved for not more than three pixels, then the identification of the released object must be done and the plot must be moved on top.

```
private void OnMouseUp (object sender, MouseEventArgs e)
{
    ptMouse_Up = e .Location;
    double nDist = Auxi_Geometry .Distance (ptMouse_Down, ptMouse_Up);
    if (e .Button == MouseButtons .Left)
    {
        int iWasObject, iWasNode;
        if (mover .Release (out iWasObject, out iWasNode))
        {
            GraphicalObject grobj = mover [iWasObject] .Source;
            if (grobj is ClosableInfo)
            {
                if (iWasNode == 0)
                {
                    (grobj as ClosableInfo) .Visible = false;
                    btnHelp .Enabled = true;
                    RenewMover ();
                    Invalidate ();
                }
            }
            else if (grobj is PieChart || grobj is RingSet || grobj is BarChart)
            {
                if (nDist <= 3)
                {
                    Identification (iWasObject);
                    PopupElement (iElement);
                }
            }
        }
    }
```

When the right button is released, then everything is about calling one or another menu. This part of the `OnMouseUp()` method can be also divided into two branches.

- When the mouse was clicked anywhere at an empty spot, then the special menu for all empty places is called.
- When any object is released by the right button, then one of many menus must be called. The menu to be called depends on the class of the released object; in addition the view of the opened menu may depend on the parameters of the released object, so the full identification is needed before opening a menu. The case of the `ClosableInfo` object is looked at as a special case, because its menu contained exactly one line; instead of calling such a menu and selecting its solitary command, I put the direct call of the tuning form for this object.

```
private void OnMouseUp (object sender, MouseEventArgs e)
{
    … …
```

```
        else if (e .Button == MouseButtons .Right)
        {
            if (mover .Release ())
            {
                if (nDist <= 3)
                {
                    if (mover .WasCaughtSource is ClosableInfo)
                    {
                        info .ParametersDialog (this, RenewMover, ParamsChanged,
                                                null, PointToScreen (ptMouse_Up));
                    }
                    else
                    {
                        MenuSelection (mover .WasCaughtObject);
                    }
                }
            }
            else
            {
                if (nDist <= 3)
                {
                    ContextMenuStrip = menuOnEmpty;
                }
            }
        }
        bFormInMove = false;
    }
```

Six different forms for tuning the parameters of the complex objects can be called in the **Form_PlotsVariety.cs**; this is done from inside the `OnMousedoubleClick()` method. The case of the `ClosableInfo` object is again looked at as a special case; all other cases start with the identification of the clicked object.

```
    private void OnMouseDoubleClick (object sender, MouseEventArgs e)
    {
        if (mover .Catch (e .Location, MouseButtons .Left))
        {
            Point ptScreen = PointToScreen (e .Location);
            int iInMover = mover .CaughtObject;
            GraphicalObject grobj = mover [iInMover] .Source;
            if (grobj is ClosableInfo)
            {
                Info .ParametersDialog (this, RenewMover, ParamsChanged, null,
                                        ptScreen);
            }
            else
            {
                Identification (iInMover);
                if (iElement >= 0)
                {
                    BarChart chart;
                    if (grobj is BarChart || grobj is PieChart || grobj is RingSet)
                    {
                        ElementMainParametersDialog (iElement, ptScreen);
                    }
                    else if (grobj is Scale)
                    {
                        chart = elems [iElement] .BarChart;
                        chart .NumScale .ParametersDialog (this, RenewMover,
                                     ParamsChanged, null, chart .Title, ptScreen);
                    }
                    else if (grobj is TextScale)
```

```
                    {
                        chart = elems [iElement] .BarChart;
                        chart .TextScale .ParametersDialog (this, RenewMover,
                                     ParamsChanged, null, chart .Title, ptScreen);
                    }
                }
            }
        }
    }
```

The wide variety of commands is available via the 10 different context menus; some of the commands were already discussed in this chapter; others are obvious and do not need any explanation.

The saving of all objects and all the parameters of visualization are paired with the procedures of restoration, so users are not going to lose a bit of their efforts in changing the view of this form. Everything is organized according to the rules of user-driven applications.

# Calculators

There were a lot of complex graphical objects in the previous chapters. In this chapter I am going to use some graphical objects, but mostly it is about the controls. And also about the problem of turning good working programs, which were designed in an old style, into some type of user-driven applications.

Up to this place all the chapters of this book were devoted to the explanation of how the movable / resizable objects of different shapes and purposes can be organized and how to design the applications on the basis of such objects. The standard procedure is to understand the purpose and requirements of the new application, to design the classes needed to fulfil the task, and to construct an application. These are the standard steps in design of any NEW program.

But what to do with all those programs, which are already in use? Many of them are good enough to be used for some time more, but it would be nice to add the new features to them. Several questions jump out immediately, when you think about this problem.

1. Can the proposed technique be applied to the existing applications?
2. How difficult is it to include the new features (movability and resizability) into the existing programs?
3. If the new features are added to the existing application, how difficult for the users will be to switch to the new version?

The last question is easier to answer than the first two, as I have mentioned it partly in the chapter *Applications for science and engineering*. All the currently used applications are designer-driven; all the older applications were always designer driven since the beginning of the time. There had never been a Golden Age for the users of the programs; they could always do only whatever was allowed by the developers. But when the clients get the user-driven applications, they quickly appreciate the flexibility of the new programs and enjoy it. In a very short time they accept the new rules of applications and take them for granted.

The third question is mostly about the psychological aspects of switching to the user-driven applications; let us turn to the first two questions, which deal with the programming problems.

To turn any class of graphical objects into the movable / resizable, three additional methods must be written: `DefineCover()`, `Move()`, and `MoveNode()`, so any class of graphical objects need some additional coding. Controls do not need any of those methods, so the transformation of the applications based on controls must be easier. I was looking for some well known program to use it as an example; my attention stopped on a Calculator.

File: **Form_OldCalculator.cs**
Menu position: *Applications – Facelift of the old calculator*

The *Calculator* application, if there are no aberrations of my memory, appeared with the very first version of the Windows system, so regardless of when you personally became familiar with this system, the *Calculator* was already there. Similar applications exist for all other operating systems. The main thing is that <u>everyone</u> is familiar with this application and can make his own opinion on the usefulness of turning it into a user-driven application without waiting for some authorized opinion.

I am not a priest to bestow my blessing by a simple hand waving; I have to do a bit of the code writing for it. But I have no access to the original *Calculator*, so at **figure 18.1** is the view of the **Form_OldCalculator.cs**. The default view of this form copies the simpler view of the original calculator from Microsoft.

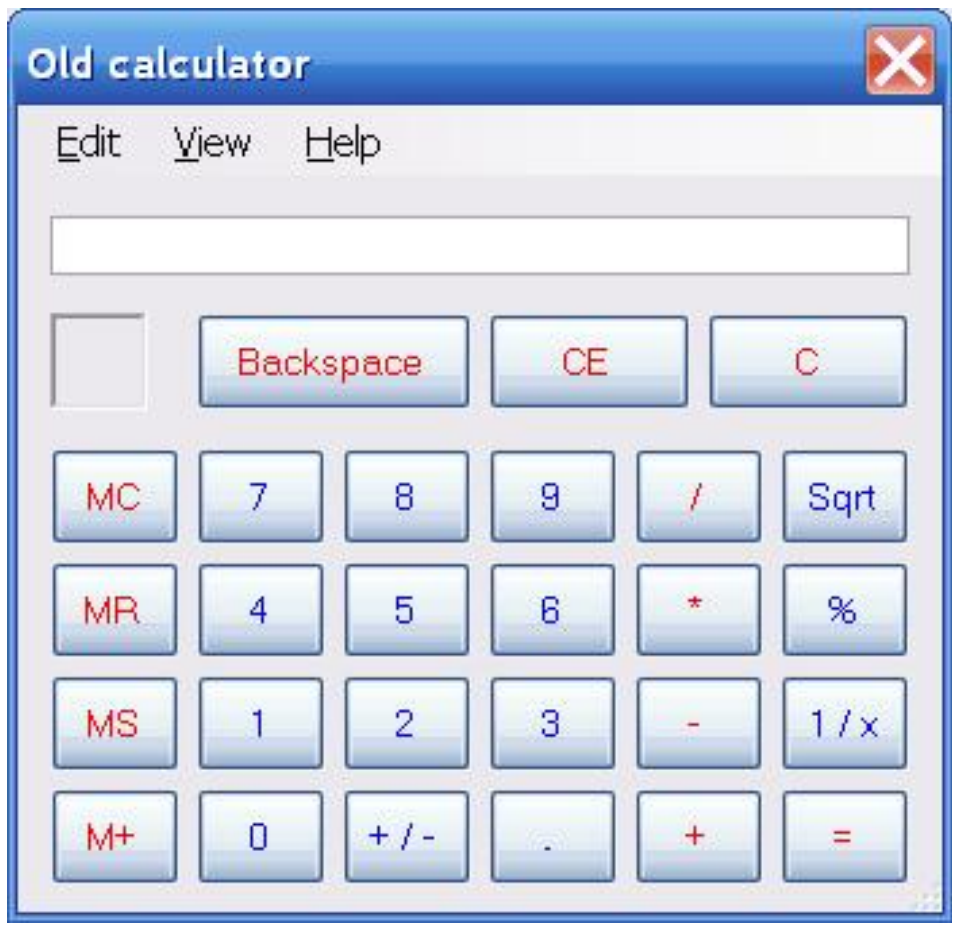

**Fig.18.1** Copy of the standard *Calculator* with very limited functionality, but with the movability of all the elements

I use the standard *Calculator* from Microsoft three or four times a year, when I need to divide two big numbers and too lazy at the moment to do it with a pen on a sheet of paper. On those rare occasions, when I have to use the standard *Calculator*, I have a problem with finding in it the needed numbers, because they are definitely not at the places, where I would put them as a designer. It is a big problem; if you use this *Calculator* as often, as I do. On each occasion you can spend several extra seconds for such a search, get the needed result, and then forget about the search problems for the next several months.

Then I decided to develop my own version of the Calculator, which allows to eliminate those search problems. The design is quickly done in Visual Studio; after it some coding must be done.

1. The mover must be declared and initialized. The original view of the form includes some controls, which I never need, for example, I never use the memory operations. The best way to get rid of the unneeded controls is to move them across the border, so the appropriate clipping must be used. The `Clipping.Safe` level allows to move elements across the right and bottom sides of the form.

```
Mover mover;
mover = new Mover (this);
mover .Clipping = Clipping .Safe;
```

2. The **Form_OldCalculator.cs** contains exclusively controls. Registering of all the elements in the mover's queue is simple.

```
foreach (Control control in Controls)
{
    mover .Add (control);
}
```

3. The whole process of moving the objects is organized with three mouse events. All three methods have the simplest possible form.

```
private void OnMouseDown (object sender, MouseEventArgs e)
{
    mover .Catch (e .Location, e .Button);
}
private void OnMouseUp (object sender, MouseEventArgs e)
{
    mover .Release ();
}
private void OnMouseMove (object sender, MouseEventArgs e)
{
    if (mover .Move (e .Location))
    {
        if (mover .CaughtSource is SolitaryControl)
        {
            Update ();
        }
        Invalidate ();
    }
}
```

With all these things implemented, I can move the unneeded controls over the border and change the places of other elements (**figure 18.2**). This application allows to perform only the basic arithmetical operations, but this is the only thing that I need from it. Certainly, I do not want to rearrange the view of this Calculator every time I am going to use it, so there are two simple methods for storing the whole view in the `Registry` and restoring it back from the `Registry` for the next use.

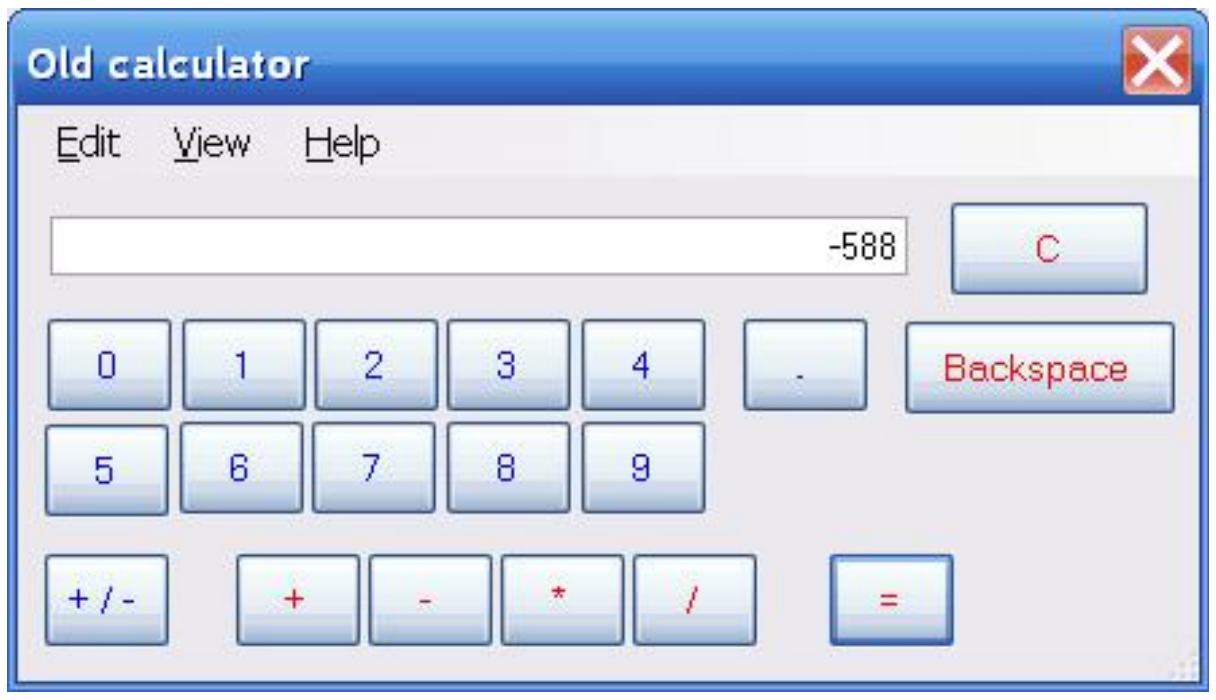

**Fig.18.2** The changed view of the same *Calculator*

```
private void SaveInfoToRegistry ()
{
    … …
    int nControls = Controls .Count;
    string [] strs = new string [nControls * 2 + 2];
    for (int i = 0; i < nControls; i++)
    {
        strs [i * 2] = Controls [i] .Left .ToString ();
        strs [i * 2 + 1] = Controls [i] .Top .ToString ();
    }
    strs [nControls * 2] = ClientSize .Width .ToString ();
    strs [nControls * 2 + 1] = ClientSize .Height .ToString ();
    regkey .SetValue (nameLocation, strs, RegistryValueKind .MultiString);
```

```
private void RestoreFromRegistry ()
{
    … …
    string [] strSizes = (string []) regkey .GetValue (nameLocation);
    if (strSizes != null && strSizes.Length == (Controls.Count * 2 + 2))
    {
        int nControls = Controls .Count;
        if (strSizes .Length == (Controls .Count * 2 + 2))
        {
            for (int i = 0; i < nControls; i++)
            {
                Controls[i].Location = Auxi_Convert .ToPoint (strSizes, i * 2);
            }
            ClientSize = Auxi_Convert.ToSize (strSizes, nControls * 2);
        }
    }
```

The above shown code restores the positions of the controls, but not their sizes, if they were changed. If you want the controls to be resizable, it can be easily done by setting the needed ranges through the `MinimumSize` and `MaximumSize` properties of those controls. But to restore the changed sizes of those controls, some changes of the code must be made. A `List` of the `SolitaryControl` objects must be organized and the whole work must be based on these objects. Some of the changes might look like the code samples below. I included these variants into the **Form_OldCalculator.cs** in the form of comments.

1. A `List` of all the movable / resizable elements is declared.

```
List<SolitaryControl> sc_S = new List<SolitaryControl> ();
```

2. Each control is turned into a `SolitaryControl` object.

```
foreach (Control control in Controls)
{
    sc_S .Add (new SolitaryControl (control));
}
```

3. All the `SolitaryControl` objects are registered with the mover.

```
for (int i = 0; i < sc_S .Count; i++)
{
    mover .Add (sc_S [i]);
}
```

4. All the `SolitaryControl` objects are saved.

```
private void SaveInfoToRegistry ()
{
    … …
    regkey.SetValue (nameLocation, new string [] {ClientSize.Width.ToString(),
              ClientSize.Height .ToString ()},RegistryValueKind .MultiString);
    for (int i = 0; i < sc_S .Count; i++)
    {
        sc_S [i] .IntoRegistry (regkey, "SC_" + i .ToString ());
    }
```

5. The size of the **Form_OldCalculator.cs** and all the `SolitaryControl` objects are restored.

```
private void RestoreFromRegistry ()
{
    … …
    string [] strSizes = (string []) regkey .GetValue (nameLocation);
    ClientSize = Auxi_Convert .ToSize (strSizes, 0);
    sc_S .Clear ();
    for (int i = 0; i < Controls .Count; i++)
    {
        SolitaryControl sc = SolitaryControl .FromRegistry (regkey,
```

```
                                         "SC_" + i .ToString (), Controls [i]);
            if (sc != null)
            {
                sc_S .Add (sc);
            }
        }
```

Both ways of working with the controls directly or organizing the List of SolitaryControl objects work fine. I have explained before that the controls are automatically wrapped in a SolitaryControl object even if you try to register them directly with the mover. The difference between two cases is only in saving and restoring the view. If you save only the position of control (the first case), then you can restore not more than this position, but the size of the control is going to be the one, which was set by the designer. Certainly, it can be changed by any user, but then the user would have to change those sizes every time the form is restored. I doubt that any user would appreciate such design. When the SolitaryControl objects are saved, then all their parameters are saved and can be restored. It is not only about the sizes; you can add more things to change the font and color; everything will be saved and restored. But this will lead us much farther from the original Calculator. To explore these things, let us take the next example.

File: **Form_Calculator.cs**
Menu position: *Applications – Calculator*

I had no intention of doing anything else with the *Calculator* than I had already shown, but one day nearly a year ago during a conversation with one of the colleagues I heard a complaint: having a poor vision, the colleague had big troubles with the standard *Calculator* program, because even the combined efforts of several specialists did not reveal any way to increase the font, used by this application. Another mystery from Microsoft. So I sat down and developed a *Calculator*, which that colleague could use. As it was not for me, then the program had to work as a normal *Calculator* with operations and functions. Certainly, changing of the font was not the only thing that I allowed to do in this program; it was designed according to all the rules of the user-driven applications. **Figure 18.3** demonstrates the default view of the **Form_Calculator.cs**.

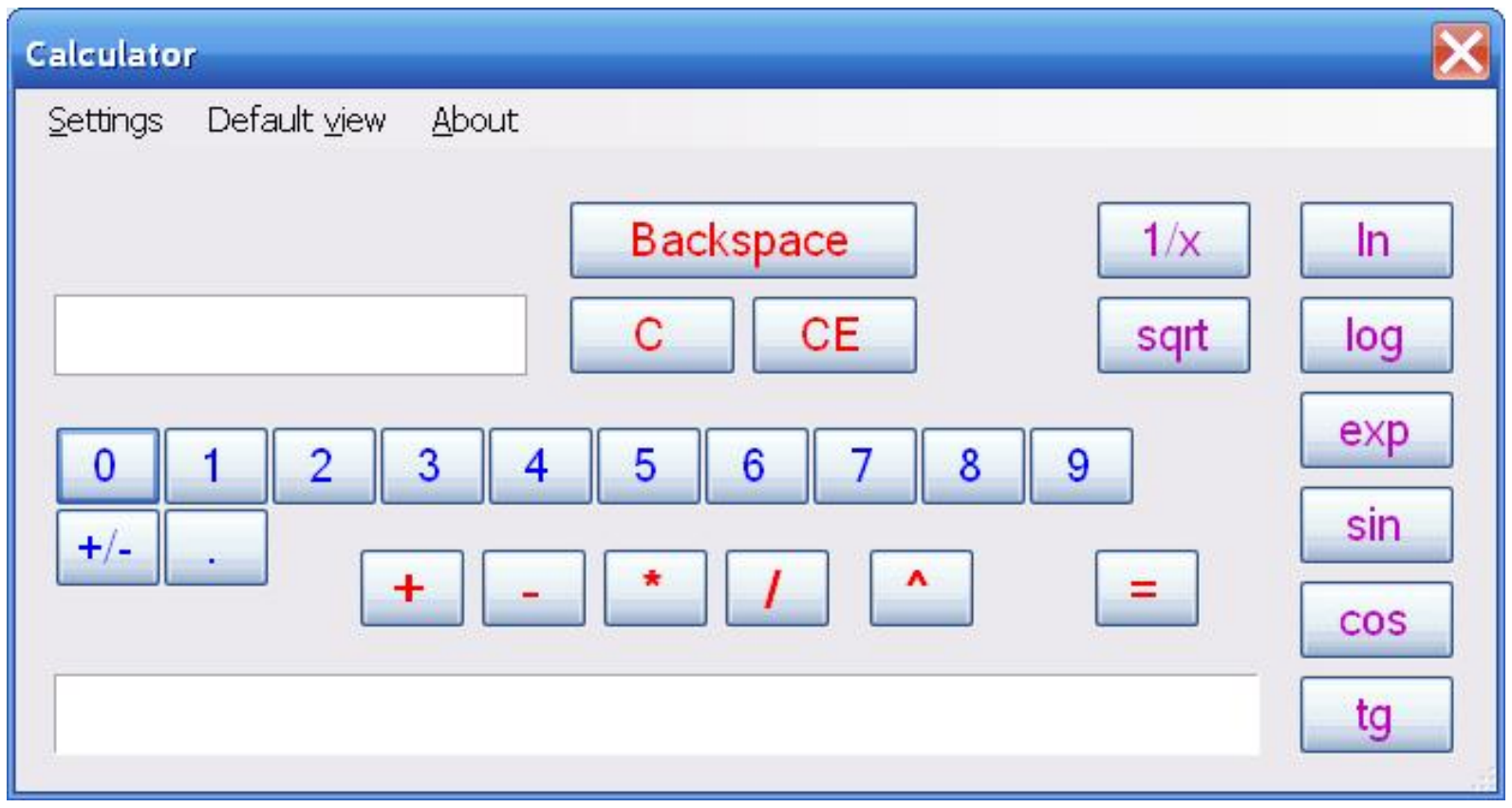

**Fig.18.3** The default view of the **Form_Calculator.cs**

There are still only controls in this form. There are no controls for memory operations, but there are some controls for often used functions. And there are two controls (not one!) to show the input values and results; I decided that it would be more informative for the work. All the controls of the form are divided into several groups; three groups play bigger role and their controls are marked out in default view by different colors.

- Numbers are shown at **figure 18.3** in *Blue* color. From the point of mathematics, dot and sign are definitely not numbers, but they are used as a part of notation, while writing numbers, so in this program the two corresponding buttons are included into this group.

```
Control [] controlDigit = new Control [] {  button_0, button_1, button_2,
                         button_3, button_4, button_5, button_6, button_7,
                         button_8, button_9, btnSign, btnDot };
```

- Operations are shown in *Red*, but only six small buttons, positioned in one line, belong to this group. Other three *Red* buttons in the upper part are not the members of this group.

```
Control [] controlOperation = new Control [] { btnPlus, btnMinus,
                         btnMultiply, btnDivide, btnDegree, btnEqual };
```

- Eight functions, which are used in this program, are shown in *Violet*. I could easily add more functions, but the person, for whom I designed this program, said that this set of functions would be enough. You can add more functions into this program, only look carefully through all the places, where the group is used.

```
Control [] controlFunction = new Control [] { btnLn, btnLog, btnExp,
                         btnSin, btnCos, btnTg, btnInverse, btnSqrt };
```

For the purpose of better showing the results, there are two different information controls; one of them is used to show the current number, another shows the whole expression.

*Calculator* is designed to fulfil its main purpose. But this is a user-driven application with all the possibilities of moving, resizing, tuning, saving, and restoring. Let us start with the moving of all the elements.

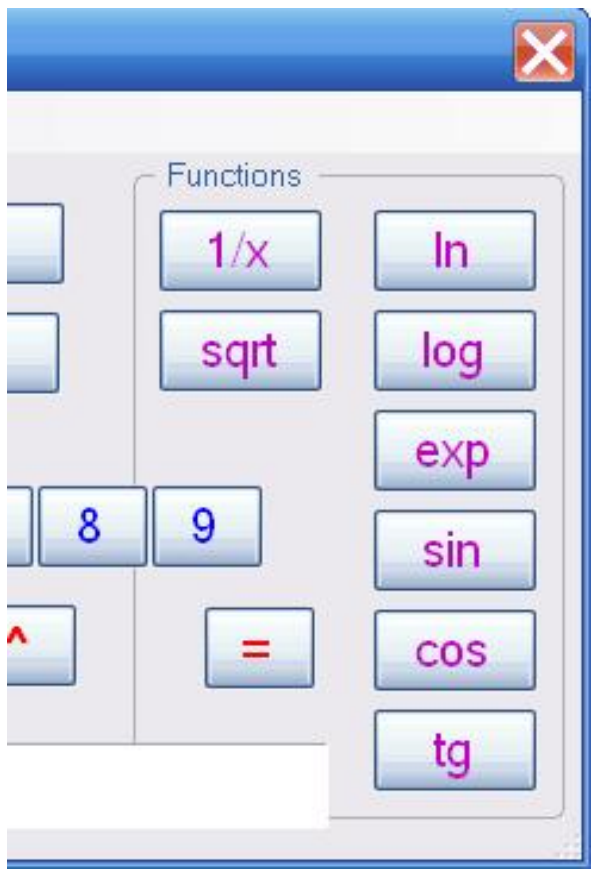

**Fig.18.4** Part of the form with the *Functions* group

Moving can be applied to individual controls, to the controls of those predetermined groups (numbers, operations, and functions), and to the temporary group of controls. The moving of a single control is organized in a standard way: press the left button somewhere close to the border of a control and move it to the new location. The area around the border of a control is used both for moving and resizing the control, but the change of a cursor tells you, which of these actions can be started at each particular spot.

A group around numbers, operations, or functions (**figure 18.4**) can be organized via the menu; I will write about the menus a bit later. Thus organized "standard" group includes only the controls of that predetermined group and nothing else, even if some other controls happen to be inside the frame, when the group is called. So the *Functions* group always includes exactly eight controls with the names of the functions; two other controls inside the frame at **figure 18.4** (one with the digit and another with the equal sign) do not belong to the group.

In the chapter *Groups of elements*, I have demonstrated the temporary groups of controls; the same idea of a temporary group is used in *Calculator*. Press the left button at an empty spot and move it without releasing the button. During such movement, you will see the painted frame based on the initial point and the current mouse point. If at the moment of the mouse release there is more than one control inside this rectangle then a temporary group is organized (**figure 18.5**). The temporary group can include all or some controls from the same predetermined group, it can include controls from different predetermined groups, or even the controls, which do not belong to any of them. Thus the temporary group from **figure 18.5** includes seven controls from the *Numbers* group, two controls from the *Operations* group, and one information control.

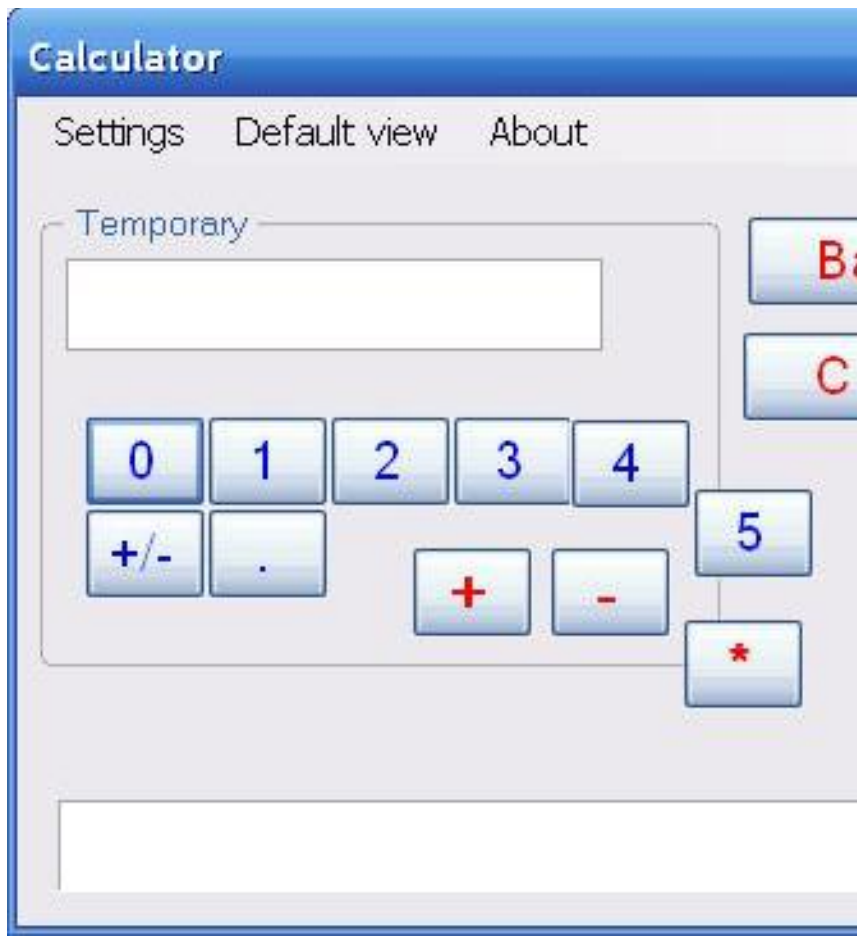

**Fig.18.5** Temporary group

Any group, whether it is a predetermined one or a temporary, belongs to the `ElasticGroup` class, so it can be moved by any inner point. If any inner element of the group is moved, the frame adjusts its size to the positions of all the inner elements. I would say that any group, organized in *Calculator*, serves like a temporary one and only for the purpose of moving the whole group of elements. Though it is always an `ElasticGroup` object, there is no command to modify this group as any `ElasticGroup` object in other forms of the Demo application. Modifying of the elements is organized in the different way, which is described a bit further on.

The whole set of controls in the **Form_Calculator.cs** (the total number is 31, as you can see from **figure 18.3**) is divided between two `Lists`. The names of these `Lists` make it obvious that one of them includes all the controls of the currently used group; all other controls go into another `List`.

```
List<Control> controlsSingle = new List<Control> ();
List<Control> controlsInFrame = new List<Control> ();
```

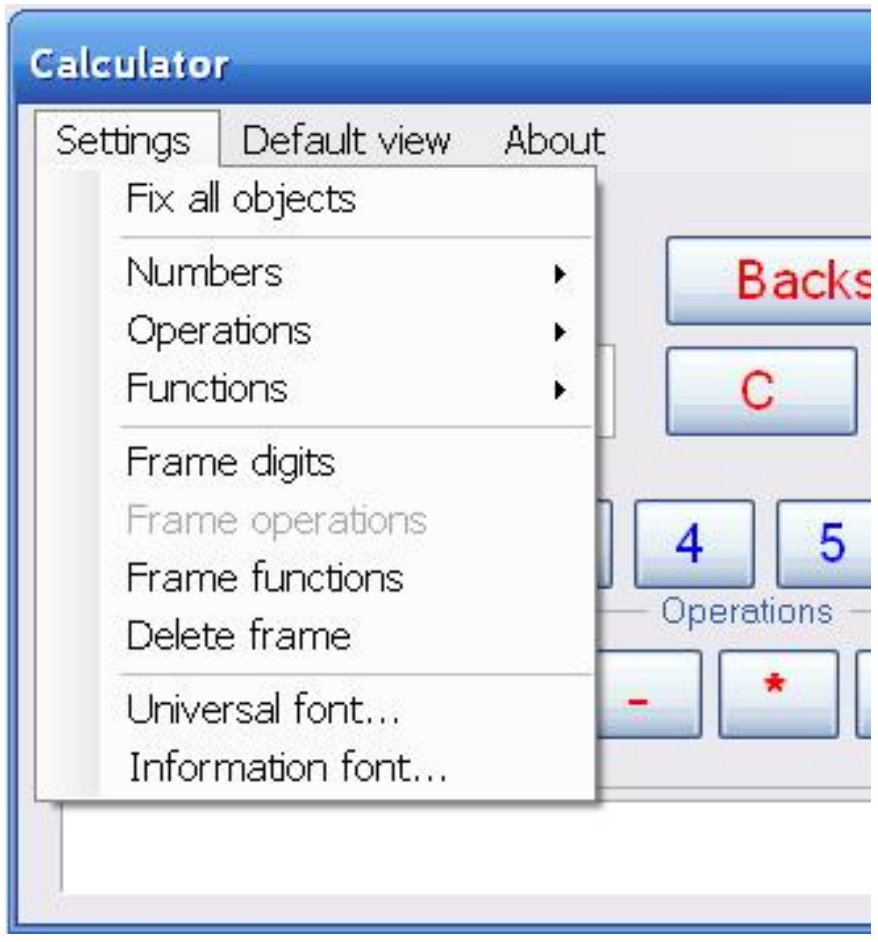

**Fig.18.6** The *Settings* menu

The distribution of the controls between the `Lists` is done either at the moment of calling an appropriate command from the *Settings* menu (**figure 18.6,** commands of the third group) or when a mouse is released with a drawn rectangle in view. In the first case the `SetGroup()` method is called with a parameter, which is the set of controls from the predetermined group. The controls from the predetermined group go into the `controlsInFrame`; all other controls go into the `controlsSingle.`

```
private void SetGroup (List<Control> ctrlsToFrame)
{
    controlsSingle .Clear ();
    controlsInFrame .Clear ();
    … …
    foreach (Control control in Controls)
```

```
    {
        if (ctrlsToFrame .Contains (control))
        {
            controlsInFrame .Add (control);
        }
        else
        {
            controlsSingle .Add (control);
        }
    }
```

In the second case, when a rectangle is painted by a mouse, another version of the `SetGroup()` method is called, which uses the selected rectangle as a parameter. All the controls are checked for a chance of being inside this rectangle.

```
private void SetGroup (Rectangle rc)
{
    controlsSingle .Clear ();
    controlsInFrame .Clear ();
    foreach (Control control in Controls)
    {
        if (rc .Contains (control .Bounds))
        {
            controlsInFrame .Add (control);
        }
        else
        {
            controlsSingle .Add (control);
        }
    }
    if (controlsInFrame .Count == 1)
    {
        controlsSingle .Add (controlsInFrame [0]);
        controlsInFrame .Clear ();
    }
    RenewMover ();
}
```

At the end of this method, after all the controls are checked and the distribution between two `Lists` is done, the number of controls inside the `controlsInFrame` is checked. This number is not allowed to be less than two; if there happened to be a single control inside, it is also moved to the `controlsSingle` and the frame is left empty. It shows another way of getting rid of any frame (group) in view: you can either use the *Delete frame* command from the *Settings* menu (**figure 18.6**) or draw the new rectangle which contains no controls or a single control. In each of these situations no group will be organized. It is not even needed to draw an empty rectangle; it is enough to click with the left button at any empty spot.

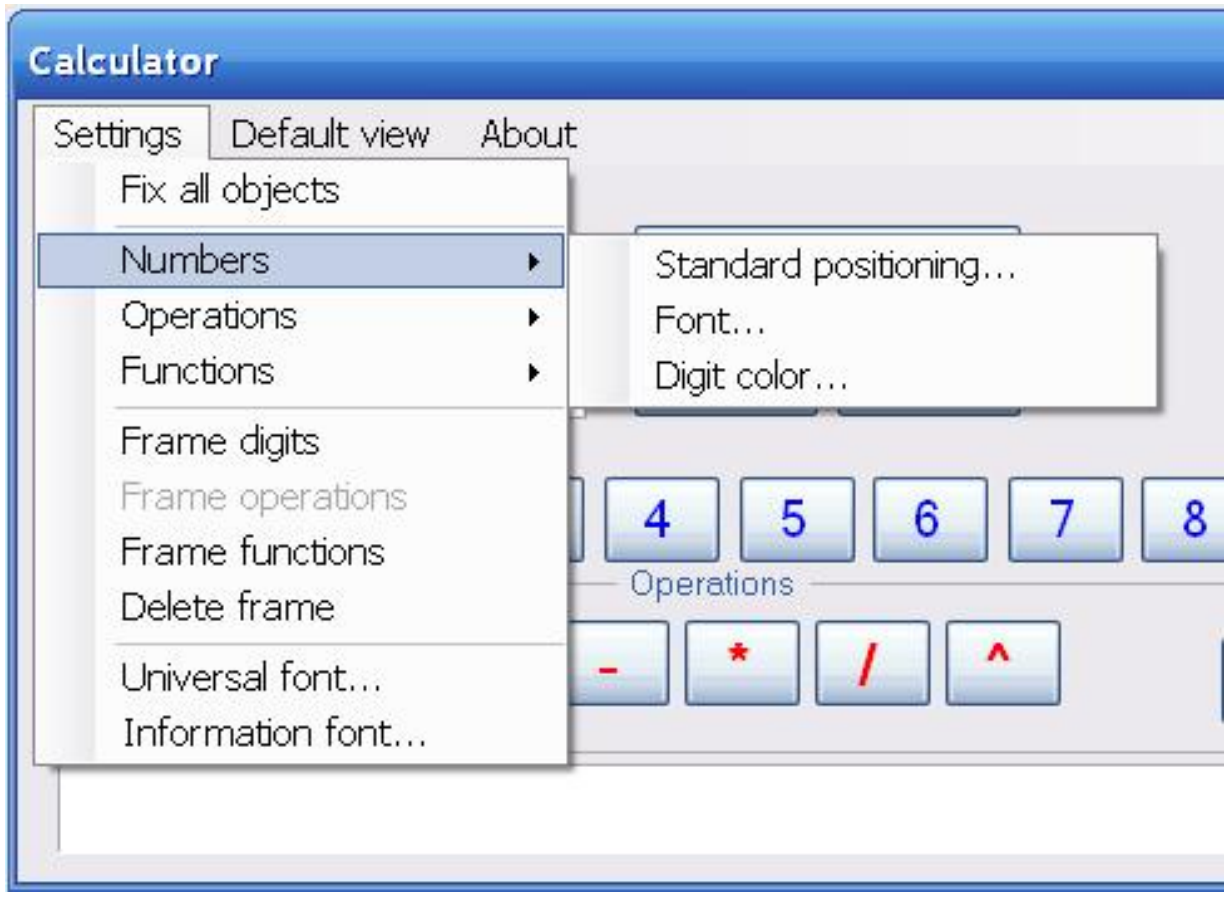

**Fig.18.7** The form, in which the standard positioning of the buttons inside the predetermined group can be selected, is achieved through the submenu

There is one more way to rearrange the positions of the elements, but this method can be applied only to the controls of three predetermined groups. **Figure 18.7** shows the submenu, which can be applied to the *Numbers* group; two other groups have similar submenus.

On pressing the *Settings – Numbers – Standard positioning* command, you open the **Form_BtnsPlacement_Numbers.cs**, in which you can select one of the views for the *Numbers* group (**figure 18.8**). Groups *Operations* and *Functions* have similar forms to select their views; only those groups include fewer controls, so they have fewer combinations to select from.

Each variant of the standard positioning is shown as a sketch with an additional check box to select this variant. Each sketch consists of a set of rectangles, representing the real controls. The rectangles are linked with the lines to show the order of controls in the group. Otherwise it would be impossible to distinguish the similar variants, in which the controls can be positioned either by rows or by columns. When the **Form_BtnsPlacement_Numbers.cs** is closed with the OK button, the buttons of the *Numbers* group in *Calculator* are lined according to the selected variant. It does not matter, whether the *Numbers* group has or has not the frame at this moment; the standard positioning works regardless of it. It is also possible to select the horizontal and vertical spaces between the buttons, which have to be lined in a standard way.

Let us continue with the *Calculator*, but after it I will return again to the discussion of sketches from **figure 18.8**.

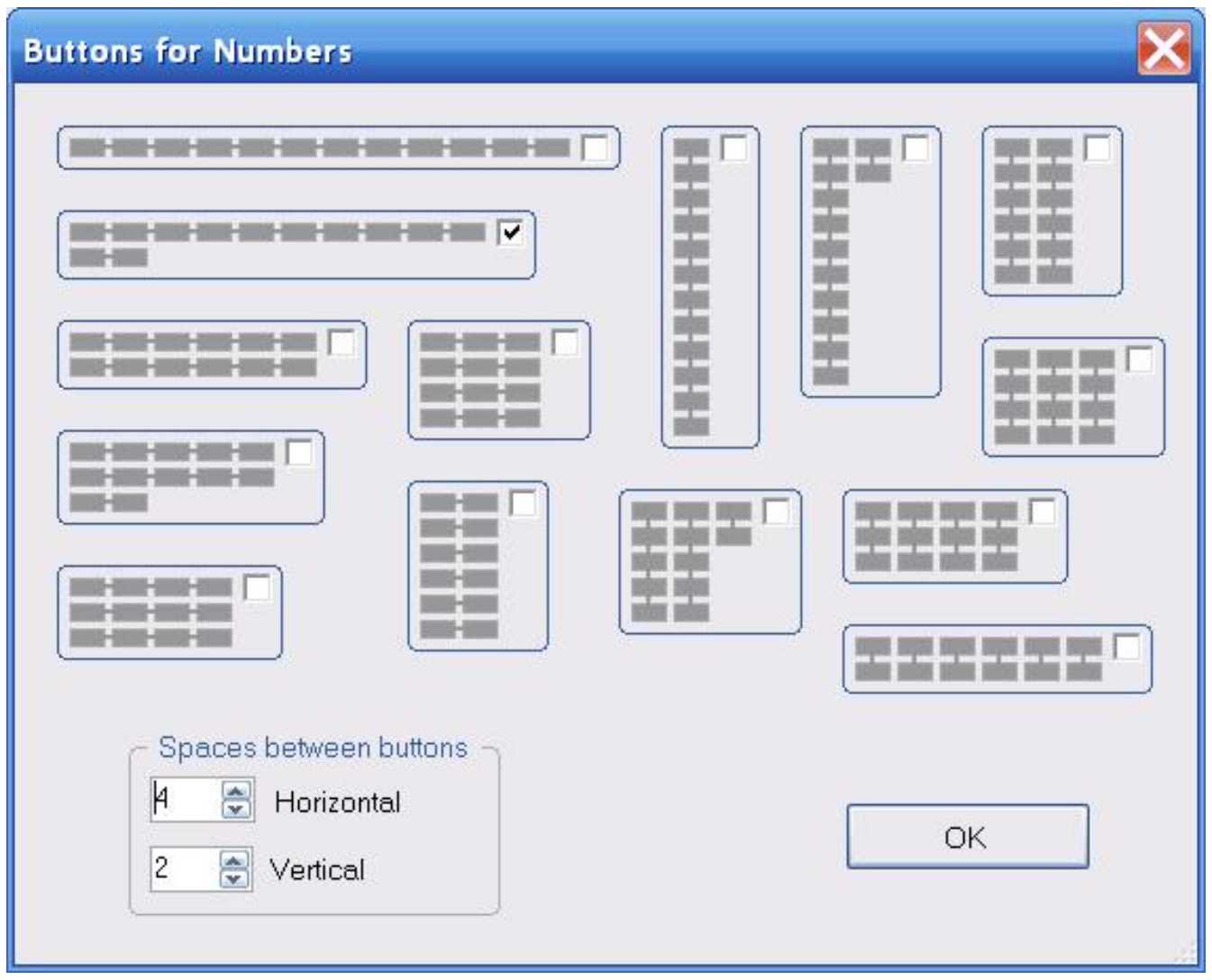

**Fig.18.8** The form **Form_BtnsPlacement_Numbers.cs** is used to select the standard positioning of controls inside the *Numbers* group

The resizing of all the elements in the **Form_Calculator.cs** can be done on an individual basis as each control can be resized by a mouse. But suppose that you resized one control with a number in such a way, which you would like it to have; what are you going to do next? To keep it different from other buttons with the numbers or to change the sizes of other 11 buttons in the *Numbers* group in exactly the same way? Are you going to do it manually? Some people may like to organize the growing size of the buttons according to the increase of the numbers on top or other extraordinary views (nothing is prohibited in the user-driven applications!), but somehow I have a feeling that the majority would prefer to see the controls of a group all having the same size. I also feel that manual standardizing of controls in a big group one after another wouldn't be a very popular operation.

Certainly, there is a better way, which was already used in the **Form_PlotsVaraiety.cs** and described in the previous chapter: an element can be used as a sample for all the siblings. The variations are only in choosing the parameters for spreading and also in decision about the group of controls, on which to spread the selected parameter.

**Figure 18.9** shows the menu that can be called on any of the controls from the *Numbers* group. Nearly all the elements in the *Calculator* are buttons. There are only three parameters which can be changed for such an element: size, font of the text, and color of the text.* Font and color can be changed via the commands of this menu; the new parameters are automatically spread on all the controls of the group. The menu also allows to spread the sizes of the changed control on all the siblings and even a bit wider: on the controls of two groups.

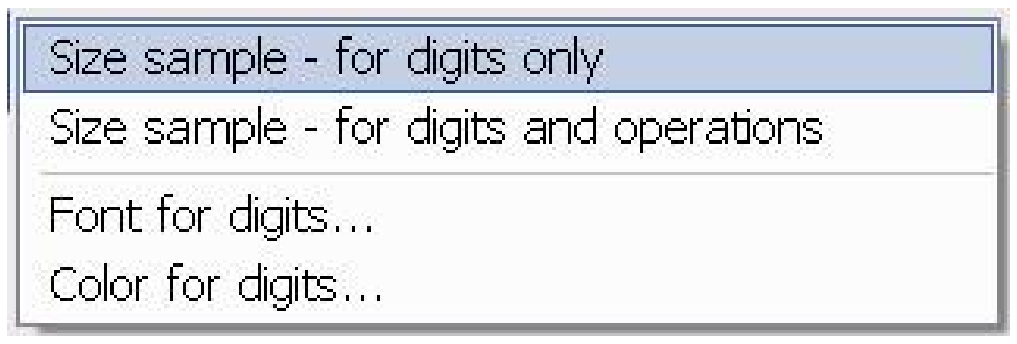

**Fig.18.9** Menu on numbers

Similar menus can be called for the buttons of the *Operations* and *Functions* groups.

While describing different types of plotting in the previous chapters, I showed that the tuning of many classes was started from different context menus. In the case of graphical objects, the selection of the appropriate menu is done by analysing the class of object, caught by mover. In the *Calculator* there are two different mechanisms of calling the context menus, but even the one, which is connected with the mover, is slightly different. In the **Form_Calculator.cs** a lot of objects, which need different menus, belong to the same class, so the menu selection starts with the number of the caught object in the mover's queue.

```
private void OnMouseUp (object sender, MouseEventArgs e)
{
    ptMouse_Up = e .Location;
    double dist = Auxi_Geometry .Distance (ptMouse_Down, ptMouse_Up);
    if (mover .Release ())
    {
        … …
```

* At the very end of writing and checking this book I understood that I forgot about one more possible tunable parameter: the control's background color. It is easy to do, but I do not want to make such changes at the last moment. If you want, you can add the change of the background color to this menu. Maybe I will do similar changes for the next version.

```
                else if (e .Button == MouseButtons .Right && dist <= 3)
                {
                    bMenuByMover = true;
                    ContextMenuSelection (mover .WasCaughtObject);
                }
```

Different menus are called on the controls of the predetermined groups, but all these controls were turned into `SolitaryControl` objects, so the decision about the menu to call cannot be based on the class of the pressed object. Instead, the number of the pressed object in the mover's queue helps to find the control itself, which has to be checked against the predetermined groups of controls. The belonging of the pressed control to one of these groups determines the needed menu to call.

```
        private void ContextMenuSelection (int iInMover)
        {
            GraphicalObject grobj = mover [iInMover] .Source;
            if (grobj is SolitaryControl)
            {
                Control ctrl = (grobj as SolitaryControl) .Control;
                if (listDigits .Contains (ctrl))
                {
                    controlTouched = ctrl;
                    ContextMenuStrip = menuOnDigitBtn;
                }
                else if (listOperations .Contains (ctrl))
                {
                    controlTouched = ctrl;
                    ContextMenuStrip = menuOnOperationBtn;
                }
                else if (listFunctions .Contains (ctrl))
                {
                    controlTouched = ctrl;
                    ContextMenuStrip = menuOnFunctionBtn;
                }
            }
        }
```

The second mechanism of calling the context menus for the elements in *Calculator* is specific. The menu can be mentioned in the `ContextMenuStrip` property of the control at the design stage; this would be enough for automatic call of this menu. I mentioned that this is a specific way of calling a context menu, because it can be used with the controls, but not with the graphical objects. As my programs deal mostly with different graphical object, then this way of calling menus can be hardly found anywhere in my applications. In the big Demo program, accompanying this book, the `ContextMenuStrip` property is used only in two examples: in the **Form_PersonalData.cs** and in the **Form_Calculator.cs**.

When a menu is called by pressing the button, there is no problem in identifying this button, so the method itself is very simple. But because of two different mechanisms of calling the menus, I had to add into the code of the form an additional field (`bMenuByMover`) to prevent calling the menu twice at the same moment. Here is the method for calling the menu on any control from the *Numbers* group; similar methods exist for the *Functions* and *Operations* groups.

```
        private void MouseDown_numbers (object sender, MouseEventArgs e)
        {
            if (e .Button == MouseButtons .Right)
            {
                controlTouched = sender as Control;
                ContextMenuStrip = menuOnDigitBtn;
            }
        }
```

By moving and resizing the controls, you have organized the perfect view of the *Calculator* and you do not want to damage this view even by a single accidental move of any element. A lot of previous examples include different commands to fix / unfix their elements. In some of the examples such type of command can be applied to one group of elements or another; in the *Calculator* the fixing / unfixing is applied to all the controls simultaneously via the menu at any empty spot. The command switches value of the `bMovable` field, which is used in the construction of all the `SolitaryControl` objects. The `SolitaryControl` class is designed in such a way that the cover for the resizable, but unmovable object is still visible by mover, so the context menu on such an object is available regardless of its movability.

```
private void RenewMover ()
{
    mover .Clear ();
    foreach (Control ctrl in controlsSingle)
    {
        mover .Add (new SolitaryControl (ctrl, bMovable));
    }
```

I mentioned a bit earlier that I would return to the discussion of the sketches from the **Form_BtnsPlacement_Numbers.cs** (**figure 18.8**); now is the time to do it. Sketches of the same class are used in two more forms (**Form_BtnsPlacement_Functions.cs** and **Form_BtnsPlacement_Operations.cs**); the further explanation is more about the `ButtonsSketch` class, but also about those forms. All three forms are designed according to the rules of user-driven applications. All the elements inside are movable, so you can rearrange the view and position those sketches in any way you want. There is also a context menu, which can be called anywhere outside the sketches.

I also mentioned before that you can add your own functions to the *Calculator*. If you are going to do this, do not forget that the needed additions must be done not only in the **Form_Calculator.cs**, but also in the **Form_BtnsPlacement_Functions.cs**. If there are more controls in the *Functions* group, then you will need to change all the sketches in the **Form_BtnsPlacement_Functions.cs** and maybe add the new sketches.

Whenever you are developing the user-driven applications, you have to design from time to time the movable objects with some interesting and non-standard features, because the requirements are non-standard. I think it is the most interesting in programming, when you have to think out something non-trivial.

The `ButtonsSketch` class has several requirements. When we have to organize the sketches for positioning of the controls for some group, then all these sketches are for the groups with exactly the same number of controls inside; let us call this number N. The class is described in the **ButtonsSketch.cs** file, so the code samples ahead are from this file. For better understanding I will show another figure with the sketches (**figure 18.10**) closer to the rules of their design.

- Each control is represented on the sketch by a small rectangle; at the moment it is 16*8 pixels (`w*h`)
- The distances between the neighbouring "buttons" on the sketch are always the same and equal to 4 pixels (`spaceInside`).
- The sketch has a rounded frame; the distance between the frame and the "buttons" is 6 pixels (`spaceOnBorders`).
- The sketch can be selected by checking it. The sketch is a graphical object without the use of any controls, so the check box, which you see on the sketch, is also part of the graphical object. This special check box occupies a rectangle with the side of 14 pixels (`sideSelectSquare`).

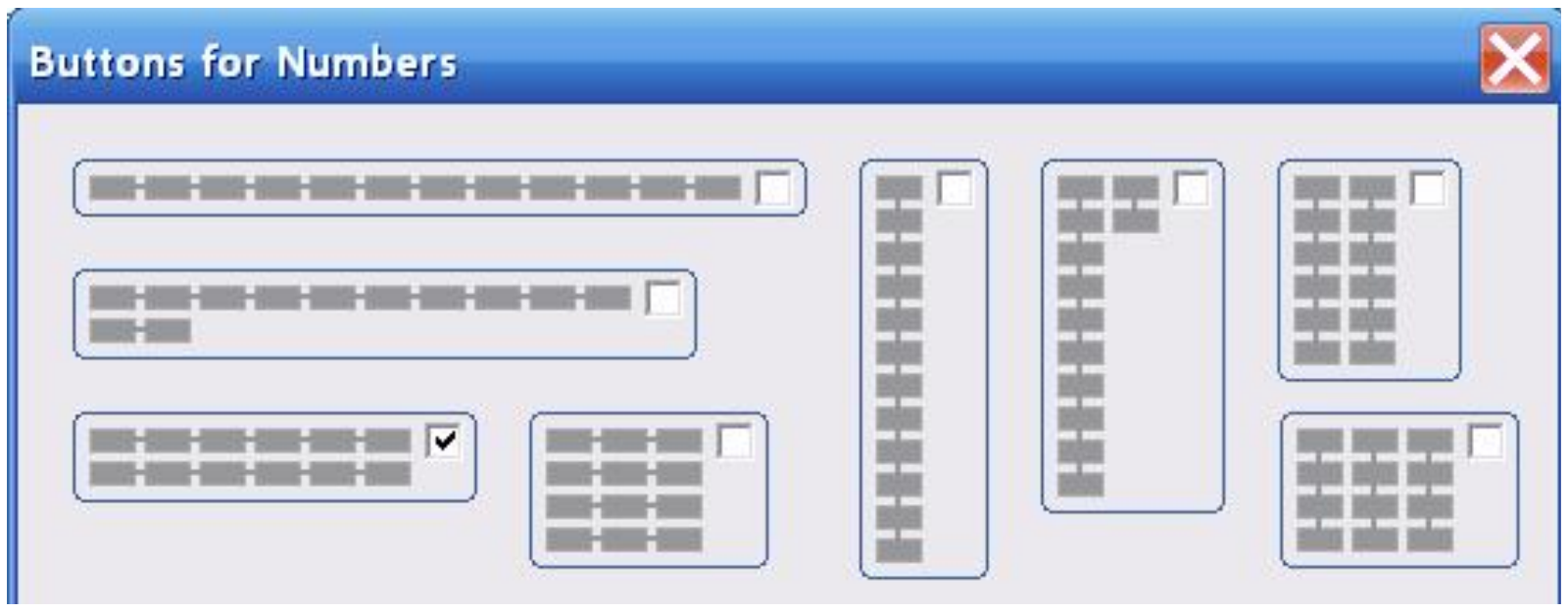

**Fig.18.10** Several sketches for the case of N = 12

- The sketch for N "buttons" can show them all in one row or in one column. To describe the position of the "buttons" on the sketch, I use a matrix of N*N. The numbering of the positions starts with 0 in the top left corner and goes to the right along the upper row; then it goes downstairs row by row; in each row the numbering starts on the left position and increases to the right. Thus, the left position in the second row has the number N; the position in the bottom right corner of the matrix has the biggest possible number of N*N-1. For example, all the sketches from the **figure 18.10** are designed for the case of N = 12. An array of positions is one of the parameters for initialization the sketch, so here you can see an array of integers, which describes the positions for the checked sketch from **figure 18.10**; it is the sketch in the bottom left corner.

```
ButtonsSketch sketch_2rows = new ButtonsSketch (nCode, false,
                        new Point (rect .Left, rect .Bottom + 20),
                        new int [] { 0, 1, 2, 3, 4, 5, 12, 13, 14, 15, 16, 17 },
                        brushBtns);
```

- Sketches may have the same number of rows and columns, but differ by the order of elements. For example, two sketches from **figure 18.10** have four rows and three columns each, but in the left sample the elements have to be positioned along the rows, while in another one – along the columns. To distinguish such cases visually, the consecutive elements on the sketch are connected by the lines, if these consecutive elements happen to be in the same row or the same column.

```
public void Draw (Graphics grfx)
{
    Pen penConnect = new Pen (brushBtns .Color, 2);
    Auxi_Drawing .CurvedFrame (grfx, rc, penFrame);
    for (int i = 0; i < ptBtn .Length; i++)
    {
        grfx .FillRectangle (brushBtns, ptBtn [i] .X, ptBtn [i] .Y, w, h);
        grfx .DrawRectangle (Pens .DarkGray, ptBtn [i] .X, ptBtn [i] .Y, w, h);
        if (i > 0)
        {
            if (ptBtn[i].X == ptBtn [i - 1].X || ptBtn[i].Y == ptBtn [i - 1].Y)
            {
                grfx .DrawLine (penConnect,
                         ptBtn [i] .X + w / 2, ptBtn [i] .Y + h / 2,
                         ptBtn [i - 1] .X + w / 2, ptBtn [i - 1] .Y + h / 2);
            }
        }
    }
    ControlPaint .DrawCheckBox (grfx, rcSelect,
                     bSelected ? ButtonState .Checked : ButtonState .Normal);
}
```

Several fields of the `ButtonsSketch` class are for drawing, moving, and other operations.

```
Rectangle rc;            // the whole area of an object
Rectangle rcSelect;      // the rectangle of a fictional check box
Point [] ptBtn;          // positions of all the "buttons" on the sketch
```

The sketch must be movable by any inner point, except for the area of the check box, so the cover consists of two rectangular nodes. As usual, the smaller node with the special purpose must precede the bigger one, which is responsible for moving the whole object.

```
public override void DefineCover ()
{
    CoverNode [] nodes = new CoverNode [] {
                                   new CoverNode (0, rcSelect, Cursors .Hand),
                                   new CoverNode (1, rc, Cursors .SizeAll) };
    cover = new Cover (nodes);
}
```

Moving of a sketch is done only if the bigger node (with number 1) is pressed.

```
public override bool MoveNode (int i, int dx, int dy, Point ptM, MouseButtons btn)
{
    bool bRet = false;
    if (btn == MouseButtons .Left)
    {
        if (i == 1)
        {
            Move (dx, dy);
            bRet = true;
        }
    }
    return (bRet);
}
```

All the sketches are non-resizable. All the fields, describing the positions of the inner parts (those three fields, mentioned above), must be moved synchronously in the `Move()` method.

```
        public override void Move (int dx, int dy)
        {
            rc .X += dx;
            rc .Y += dy;
            for (int i = 0; i < ptBtn .Length; i++)
            {
                ptBtn [i] .X += dx;
                ptBtn [i] .Y += dy;
            }
            rcSelect .X += dx;
            rcSelect .Y += dy;
        }
```

These are all the interesting things, which are going inside the ButtonsSketch class. One more important thing, related to the class, is organized in the form, where the sketches are used, for example, in the **Form_BtnsPlacement_Numbers.cs.** This is about the selection of one of the sketches. The selection is done by clicking the sketch inside the special area, which is shown in the form of a check box. It is not a real check box, but only its picture, which is covered by the small node. The correct identification of the pressed sketch is done in the standard way through the order of the pressed object in the mover's queue; the number of the pressed node is also checked, as the check box in every ButtonsSketch object is covered by the node zero.

```
    private void OnMouseUp (object sender, MouseEventArgs e)
    {
        ptMouse_Up = e .Location;
        double dist = Auxi_Geometry .Distance (ptMouse_Down, ptMouse_Up);
        int iObject, iNode;
        if (mover .Release (out iObject, out iNode))
        {
            GraphicalObject grobj = mover [iObject] .Source;
            if (e .Button == MouseButtons .Left)
            {
                if (grobj is ButtonsSketch && iNode == 0)
                {
                    ButtonsSketch sketch = grobj as ButtonsSketch;
                    for (int i = 0; i < sketches .Count; i++)
                    {
                        sketches [i] .Selected = false;
                    }
                    sketch .Selected = true;
                    viewSelected = (Calculator_DigitsGroupView) (sketch .Code);
                    places = sketch .Places;
                    Invalidate ();
                }
            }
```

# An exercise in painting

All the previous chapters were also filled with painting, but only in this one the main goal of the example is painting. It turned out that the rules of such an application are exactly the same, as for the complex scientific programs or any other complex applications.

Throughout the last several years I have prepared several programs to demonstrate the design and use of absolutely different movable objects. All the time I try to show that the algorithm is not aimed at the specific form of objects or at any area of design. Yes, the algorithm was born, because I understood the huge demand for movable / resizable objects in design of the scientific / engineering applications, but shortly after it I came to the understanding that the use of such objects is going to change the design of applications in many areas. In order to check how the use of movable / resizable elements can affect the programs in different areas, I tried to include the examples, which were far away from each other. One of such examples is simply an exercise in painting, which includes the drawing of different buildings. There were two examples of such type. In one of them the buildings looked like the apartment buildings, so that example was called *Town*. It is still available in the **Test_MoveGraphLibrary** application; see chapter *Programs and documents*. Another example includes several types of buildings; all of them are more suitable for the country style, so the example is called *Village*. This is the example, which I am going to discuss in this chapter.

File: **Form_Village.cs**
Menu position: *Applications – Village*

The village from **figure 19.1** is definitely not an attractive one; either there was no architect at all or a very bad one. But this picture shows nearly all the objects, which are important for discussion of the programming ideas, which are used in the **Form_Village.cs**. I try to include into this book the examples with some features, which were not used before, but which, from my point of view, can be interesting for design of other applications.

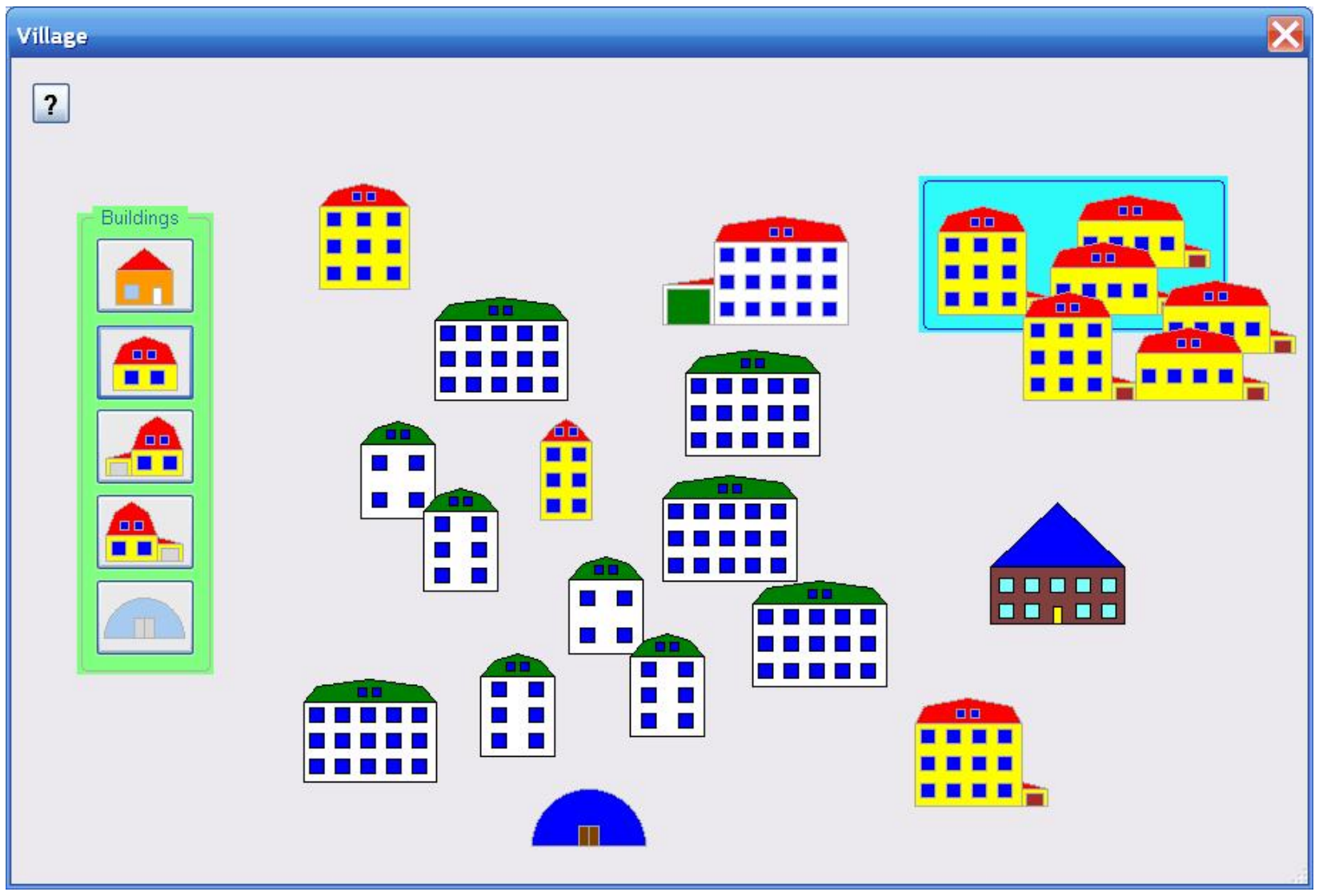

**Fig.19.1** This village was constructed without any plan and even without any sense. You can do much better.

Five types (classes) of different buildings are used in this example; all of them are derived from the base class `Building`. This base class is abstract, so you will never see an object of this class. Buildings of all five types can be seen on the buttons inside the *Buildings* group on the left side of **figure 19.1**.

A `PrimitiveHouse` can be from one to three storeys high and has between two and five windows on each floor; at the ground floor one window is substituted by the door (**figure 19.2**). The sizes of windows and door are fixed, but the house itself is resizable. The main (rectangular) part of this building can be resized by any side and any corner; also the height of the roof can be changed by moving the apex of the roof up or down.

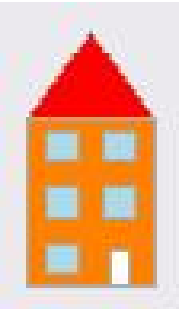

**Fig.19.2**

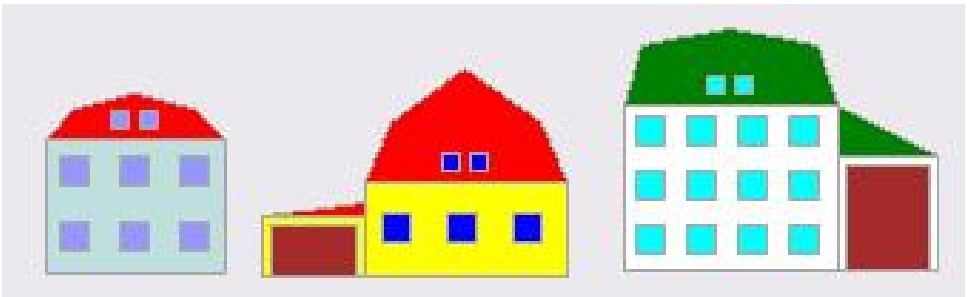

**Fig.19.3** Three types of similar houses

Houses of the `RuralHouse`, `RuralPlusLeft`, and `RuralPlusRight` classes are similar in design of the main part. The names of two last classes contain the information about the side, on which the garage is added (**figure 19.3**). The main (rectangular) part of each building can be resized by all sides and corners; the height of the roof can be changed by moving the upper point plus there are two additional points on the edge of the roof, which can be moved to change the shape of the roof. The move of each of these points is duplicated by another one, so the roof is always symmetrical. Size of a garage can be changed by moving the sides or corners; the angle of the roof above the garage can be changed by moving its upper point up or down.

The shape of the `Hangar` objects (**figure 19.4**) is absolutely different from the previous buildings. A hangar can be resized by any point of its roof, but the size of the door is fixed. It is not difficult to make it resizable, but I do not want to make the changes at the very last moment. Maybe later, though there is a chance that I will forget about it again.

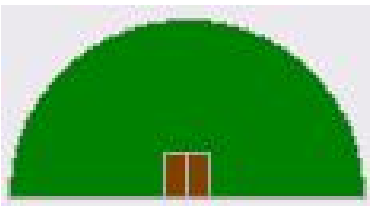

**Fig.19.4**

The design of covers for the first four classes of buildings is similar; only the number of nodes increases from class to class.

The cover of the `PrimitiveHouse` class consists of 10 nodes. As was described at the beginning of the book with the examples of rectangles, the smallest circular nodes in the corners must go first, then the stripes along the movable sides, and at the end the polygonal node that covers the whole area of an object.

```
public override void DefineCover ()
{
    PointF [] corner = Auxi_Geometry .CornersOfRectangle (rcHouse);  // LT,RT,RB,LB
    CoverNode [] nodes = new CoverNode [4 + 1 + 4 + 1];
    nodes [0] = new CoverNode (0, corner [0], Cursors .SizeNWSE);    // LT corner
    nodes [1] = new CoverNode (1, corner [1], Cursors .SizeNESW);    // RT corner
    nodes [2] = new CoverNode (2, corner [2], Cursors .SizeNWSE);    // RB corner
    nodes [3] = new CoverNode (3, corner [3], Cursors .SizeNESW);    // LB corner
    nodes [4] = new CoverNode (4, ptRidge);                          // roof top
    nodes [5] = new CoverNode (5, corner [0], corner [3], Cursors .SizeWE);
    nodes [6] = new CoverNode (6, corner [0], corner [1], Cursors .SizeNS);
    nodes [7] = new CoverNode (7, corner [1], corner [2], Cursors .SizeWE);
    nodes [8] = new CoverNode (8, corner [3], corner [2], Cursors .SizeNS);
    nodes [9] = new CoverNode (9, new PointF [] { corner [0], ptRidge, corner [1],
                                                  corner [2], corner [3] });
    cover = new Cover (nodes);
}
```

Of these 10 nodes only the last one is used for moving the whole object; all other nodes are used for resizing. With such a number of nodes, the `MoveNode()` method is too long to show it here, but the code for each node is simple; the proposed movements have to be checked against the minimum and maximum allowed sizes of the house. Here is the part of the `PrimitiveHouse.MoveNode()` method for moving the top left corner of the house (node number zero).

```
public override bool MoveNode (int i, int dx, int dy, Point ptM, MouseButtons btn)
{
    … …
    else if (i == 0)        // LT corner
    {
        hNew = rcHouse .Height - dy;
        if (minHouseHeight <= hNew && hNew <= maxHouseHeight)
        {
            MoveCeiling (dy);
            bRet = true;
        }
        wNew = rcHouse .Width - dx;
        if (minHouseWidth <= wNew && wNew <= maxHouseWidth)
        {
            MoveLeftSide (dx);
            bRet = true;
        }
    }
```

In the case of the `RuralHouse` class there are two more points on the slopes, where the circular nodes must be used, so the total number of nodes increases to 12.

The `RuralPlusLeft` and `RuralPlusRight` classes have to add more nodes for resizing of the garages; the covers for such objects consist of 19 nodes.

Objects of the `Hangar` class have absolutely different shape, so they have different covers and demonstrate two interesting aspects in their design. First, these objects have a curved border, so the N-node covers must be used. Second, to cover a half circle area, a circular node is used but preceded by a transparent rectangular node which ”cuts out” another half. Such technique was demonstrated earlier with the sectors of a circle in the chapter *Transparent nodes*.

```
public override void DefineCover ()
{
    CoverNode [] nodes = new CoverNode [nNodesOnPerimeter + 2];
    PointF [] pt = new PointF [nNodesOnPerimeter];
```

```
    for (int i = 0; i < nNodesOnPerimeter; i++)
    {
        pt [i] = Auxi_Geometry .PointToPoint (ptCenter,
                                  Math .PI * i / (nNodesOnPerimeter - 1), radius);
    }
    for (int i = 0; i < nNodesOnPerimeter; i++)
    {
        nodes [i] = new CoverNode (i, pt [i], nrSmall);
    }
    nodes [nNodesOnPerimeter] = new CoverNode (nNodesOnPerimeter,
            new Rectangle (ptCenter .X - radius, ptCenter .Y, 2 * radius, radius),
            Behaviour .Transparent);
    nodes [nNodesOnPerimeter + 1] = new CoverNode (nNodesOnPerimeter + 1,
                                           ptCenter, radius, Cursors .SizeAll);
    cover = new Cover (nodes);
}
```

When you start the *Village* application for the first time, you do not see all those buildings in the form. The form is empty with only one *Buildings* group in view. This group contains five buttons with the pictures of different buildings (**figure 19.5**). This is a group of the `ElasticGroup` class with all the standard features of this class. There is also a menu that can be called inside this group (**figure 19.6**). One of the commands from this menu allows to call the standard form for modifying an `ElasticGroup` object; other commands can be also useful. All these commands change the view and behaviour only of the group itself; they have no effect on the buildings painted in the form.

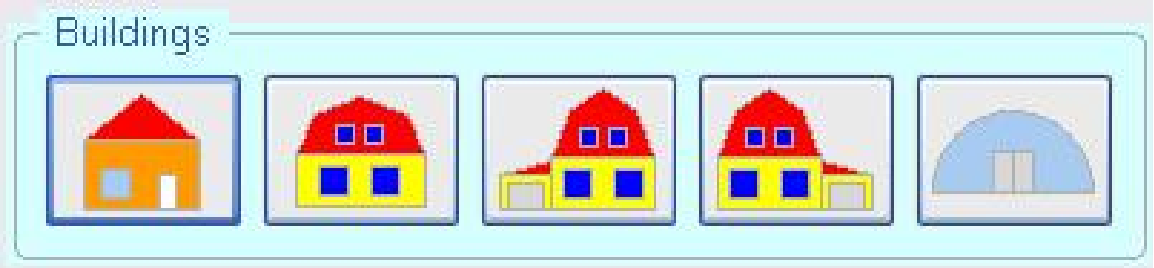

**Fig.19.5** *Buildings* group

By clicking any button in the *Buildings* group, you add a building of the appropriate type into the form. The new building initially appears in a small size in the middle of the form; from there it can be moved to any place, resized, and used in a lot of other operations.

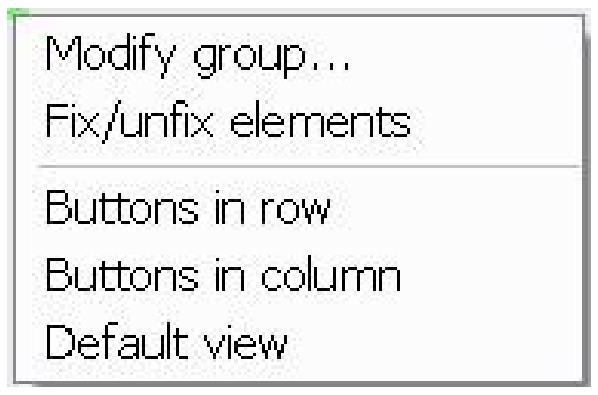

**Fig.19.6** Menu on the *Buildings* group

There are three ways of changing the whole picture of the form:

- Individual moving / resizing of the buildings.
- Using the commands from the menu that can be called on any building.
- Using a special group for any number of buildings.

The individual moving and resizing are organized in a standard way, as all the buildings are constructed in such a way that they can be moved by any inner point and resized by any border point. (The owners of the real houses can only dream about such a thing.)

While looking through the commands that are available in the menu on any building (**figure 19.7**), you can find a lot of similarities with the menu on plots from the **Form_Functions.cs** (**figure 16.5**). Though the objects in two applications are absolutely different, a lot of these commands are based on the general expectations in the case of a form with a lot of elements in view.

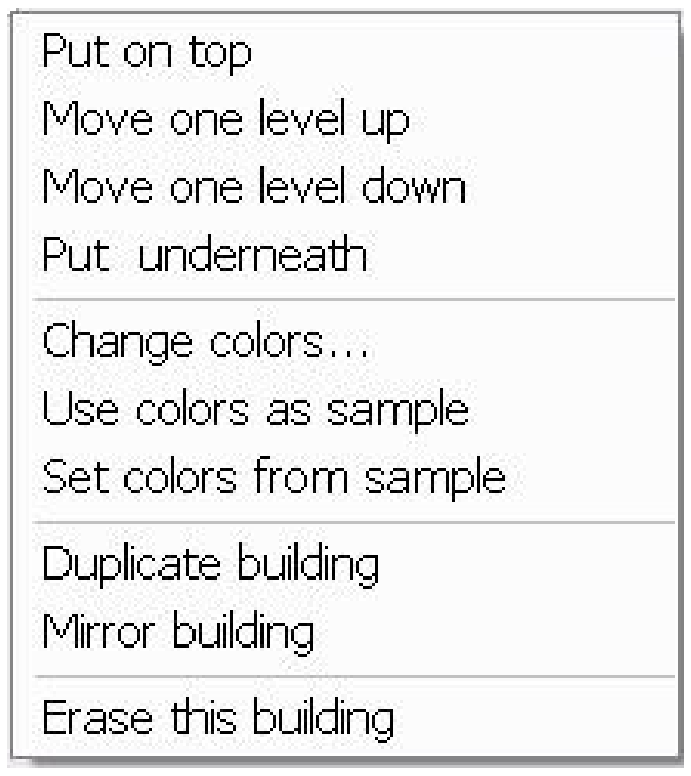

**Fig.19.7** Menu on buildings

When you have a form with many movable objects inside, these objects can be moved across each other and in many situations they will overlap. It is natural to have some commands for changing the order of these objects in view (the order of their drawing). Each object moves on its own level, so the standard set of commands allows to move an object one level up or down and to move it either to the highest or to the lowest level.

Nearly all the classes of buildings use the same set of colors to paint different parts: main building, roof, windows, door, and edges. (Because they are used in the majority of cases, the corresponding fields are placed in the base `Building` class.) A set of colors from any building can be used as a sample for painting other objects of the same class; two commands of the menu allow to declare and use such a sample of colors. Another command allows to change the colors of any building on an individual basis, but I will return to this process a bit later.

It would be a tiresome process if the only way to construct a village of 30 or 40 buildings would be to add them one by one and set their sizes and colors individually. There are two ways to simplify this process.

- A building can be duplicated by using the menu command (**figure 19.7**).
- Any set of buildings can be united into a group and several commands can be applied to this group.

The design of such a group of buildings is pretty simple and standard. Press the left button at an empty spot and move it without releasing the button. During such movement, you will see the painted rectangular frame, based on the initial point and the current mouse position. If at the moment of the mouse release there is more than one building inside this rectangle, then a temporary group is organized. This temporary group belongs to the `BuildingsGroup` class and it is always shown as a painted rectangle with a frame; at **figure 19.1** it is shown as a cyan rectangle with a blue frame. Not only the background color of the group is fixed, but its transparency is set to 0.2; the need for transparency will become obvious from the further discussion of the behaviour of this group. Whenever the group is shown, its background is painted with a `brushBack`.

```
brushBack = new SolidBrush (Color .FromArgb (204, Color .Cyan));
                              //    transparency = 0.2    A = 255 * (1 - coef)
```

Let us check the organizing of the group through the code.

When the mouse is pressed at an empty place, which means without catching any object by a mover, then the value of the `bTemporaryFrameInView` is set to **`true`** and signals the need for painting a temporary frame.

```
private void OnMouseDown (object sender, MouseEventArgs e)
{
    ptMouse_Down = e .Location;
    if (!mover .Catch (e .Location, e .Button))
    {
        if (e .Button == MouseButtons .Left)
        {
            bTemporaryFrameInView = true;
        }
    }
    ContextMenuStrip = null;
}
```

When the value of the `bTemporaryFrameInView` signals the need for painting a temporary frame, this rectangular frame is based on the initial point of the mouse press and the current mouse position.

```
private void OnPaint (object sender, PaintEventArgs e)
{
    Graphics grfx = e .Graphics;
    … …
    if (bTemporaryFrameInView)
    {
        grfx .DrawRectangle (Pens .Blue,
                             Math .Min (ptMouse_Down .X, ptMouse_Move .X),
                             Math .Min (ptMouse_Down .Y, ptMouse_Move .Y),
                             Math .Abs (ptMouse_Move .X - ptMouse_Down .X),
                             Math .Abs (ptMouse_Move .Y - ptMouse_Down .Y));
    }
```

When the left button is released without releasing any object by mover (the temporary frame is not a movable object, so this can be our case), then the selected rectangle must be checked for a possibility of organizing a `BuildingsGroup` object.

```
private void OnMouseUp (object sender, MouseEventArgs e)
{
    ptMouse_Up = e .Location;
    if (e .Button == MouseButtons .Left)
    {
        int iWasObject, iWasNode;
        if (mover .Release (out iWasObject, out iWasNode))
        {
            … …
        }
        else
```

```
        {
            if (bTemporaryFrameInView)
            {
                RectangleF rc = new RectangleF (Math .Min (ptMouse_Down .X, e.X),
                                                Math .Min (ptMouse_Down .Y, e.Y),
                                                Math .Abs (e.X - ptMouse_Down .X),
                                                Math .Abs (e.Y - ptMouse_Down .Y));
                SetBuildingsGroup (rc);
                RenewMover ();
                bTemporaryFrameInView = false;
                Invalidate ();
            }
        }
```

The group is organized (or can be organized) by the `SetBuildingsGroup()` method.

- Not more than one such group can exist in the **Form_Village.cs** at any moment, so if there is such a group at the moment of calling this method, its existence must be terminated.
- All the existing buildings must be checked against the area of rectangle; all the buildings, which are found inside, are included into the special `List` of buildings.
- If there are two or more buildings in this `List`, then the group is organized.

```
        private void SetBuildingsGroup (RectangleF rc)
        {
            if (groupHouses != null)
            {
                groupHouses = null;
                RenewMover ();
            }
            List<Building> elems = new List<Building> ();
            for (int i = 0; i < houses .Count; i++)
            {
                if (rc .Contains (houses [i] .RectAround))
                {
                    elems .Add (houses [i]);
                }
            }
            if (elems .Count > 1)
            {
                groupHouses = new BuildingsGroup (elems);
            }
        }
```

The drawing of the temporary frame is not the only reason for calling the `SetBuildingsGroup()` method. In the chapter *Calculators* I have demonstrated a similar group which has a set of elements fixed at the moment of the construction of the group. In the example of Calculator, even if the group was moved and released in such a place that some other controls happened to be inside the frame, it did not change the set of elements of the group. In the **Form_Vilalge.cs**, in case of the `BuildingsGroup` object, I purposely use different logic.

- Any building can be moved from outside and dropped inside the group thus including it into the group.
- When a group is moved and released at a new location, any building, which was not in a group before but happens to be inside the frame now, is added to the group. This is the main reason to make this group slightly transparent so it would be possible to look through and see what can be added to the group at one moment or another.

Both cases of adding new buildings to the group are based on the reconstruction of the group at the moment, when either a single building or a group is released.

```
    private void OnMouseUp (object sender, MouseEventArgs e)
    {
        … …
        if (e .Button == MouseButtons .Left)
```

```
        {
            int iWasObject, iWasNode;
            if (mover .Release (out iWasObject, out iWasNode))
            {
                GraphicalObject grobj = mover [iWasObject] .Source;
                … …
                if (groupHouses != null && (grobj is Building ||
                                            grobj is BuildingsGroup))
                {
                    SetBuildingsGroup (groupHouses .Frame);
                    RenewMover ();
                }
```

Moving a building from outside into a group and moving a group to the new location can change the set of buildings, belonging to the group, but there is one more possibility to change this set. This can happen as a result of using one of the commands from the menu for this group (**figure 19.8**). The second command of this menu allows to duplicate all the buildings of the group. To make the duplication obvious for the users, the copies of the existing buildings are placed not at exactly the same location but are moved aside both horizontally and vertically from the originals. The currently used shift is equal to 60 pixels along each direction, but you can change these values in the code.

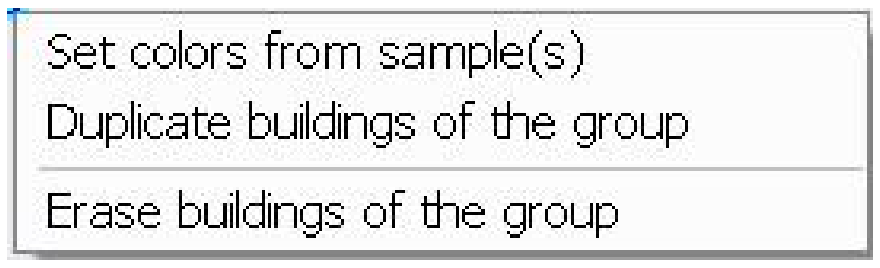

**Fig.19.8** Menu for a group of buildings

```
    private void Click_miDuplicateBuildings (object sender, EventArgs e)
    {
        Point ptCur, pt;
        for (int i = groupHouses .Elements .Count - 1; i >= 0; i--)
        {
            ptCur = groupHouses .Elements [i] .Location ();
            pt = new Point (ptCur .X + 60, ptCur .Y + 60);
            Building newBuilding = groupHouses .Elements [i] .Copy (pt);
            houses .Insert (0, newBuilding);
        }
        SetBuildingsGroup (groupHouses .Frame);
        RenewMover ();
        Invalidate ();
    }
```

The duplication produces the copy of each building from the group, but there are some other results, which depend on the size of the original group, its content, and the relative positions of the inner buildings (the "density" of the group). The distance between the inner elements and the frame depends on the used constructor of the group, but usually it is around 10 pixels. The copies of those inner buildings that happened to be close to the right or bottom parts of the frame before the command for duplication will be placed either on the frame or even beyond the frame and will be not included into the group on the final checking. But there can be other buildings in the group, whose copies are organized inside the frame; these copies are added to the group. Obviously, not all the copies will be added, but some of them can. If you use the command for copying twice in a row without any individual movements of buildings or any other commands in between, then it is possible that a set of buildings to be duplicated on these commands would differ.

**Figure 19.9** shows an organized group of three buildings plus another four buildings placed on the frame; of those four buildings the two are over the group and other two are under the group. This is a good example to write about the organization of the mover's queue, which is not the standard one in this **Form_Vilalge.cs**.

There is one main rule for organizing the mover's queue: the controls must precede the graphical objects. According to this rule, the ordering of nearly all the elements of the form is obvious. Only the information about this form (a `ClosableInfo` object) is not shown at **figure 19.1**, but this information, if it has to be shown, is painted atop all the buildings. Thus we receive such an order of objects.

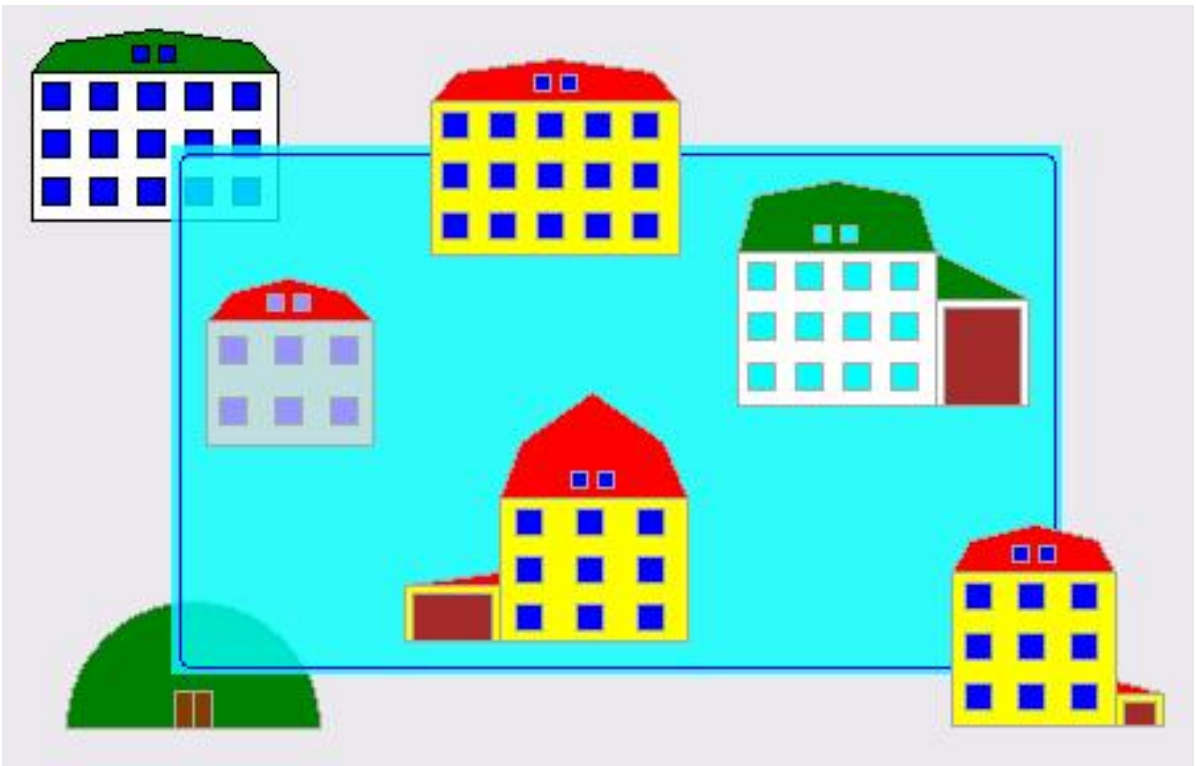

**Fig.19.9** A part of the form with some buildings included into the group, while others are not

1. The small button to call the information panel.
2. The *Buildings* group with all its buttons.
3. The information, if it has to be shown.
4. All the buildings. The buildings can change their relative order, but are never moved ahead of anything else.

So the part of the `RenemMover()` method to place the above mentioned elements into the mover's queue is simple.

```
private void RenewMover ()
{
    mover .Clear ();
    if (info .Visible)
    {
        mover .Add (info);
    }
    groupBtns .IntoMover (mover, 0);
    mover .Insert (0, btnHelp);
    for (int i = 0; i < houses .Count; i++)
    {
        mover .Add (houses [i]);
    }
```

And only one question is left: what is the correct place in the mover's queue for the group of buildings? Not for the buildings of the group; the order of buildings does not depend on whether some of them are included into a group or not. Every newly constructed building is placed at the head of the whole list of buildings, so that the new one is never closed from view even partly but appears atop all others. The question is about the order of that colored in cyan and slightly transparent rectangular area with a frame around.

The first suggestion is that the group must stay behind all its inner elements; in such a case any inner building can be easily grabbed and moved around. Does it mean that the group must be placed at the end of the queue behind all the buildings? It is a possible solution, but not a good one, when you have a densely populated area and somewhere in the middle you want to move several buildings without any other changes. There can be a lot of variants, but there are cases (I know, because I ran into them), when it would be very difficult to find a spot, by which to move such a group, if all the buildings are always shown atop the group.

After some consideration I came to another solution. I check the `List` of buildings starting from the end; when I find the first of the buildings belonging to the group, I put the group into the mover's queue behind this building. This guarantees that all the buildings of the group precede the group itself in the queue. At the same time the group is not always blocked by all the buildings, but some of them may be placed behind the group. Here is the final part of the `RenewMover()` method.

```
private void RenewMover ()
{
    … …
    if (groupHouses != null)
    {
        for (int i = mover .Count - 1; mover [i] .Source is Building; i--)
        {
            if (groupHouses .Elements .Contains (mover [i].Source as Building))
            {
                mover .Insert (i + 1, groupHouses);
                break;
            }
        }
    }
}
```

You can move the group around and release it at such a place, where some previously not included building can be seen dimly through the group. On releasing the group, you see the immediate change of the visibility of this building as it appears above the group. This does not mean the change in the order of buildings. It happened because the set of buildings in the group was changed, the new place for the group in the mover's queue was calculated, and this new place is closer to the end of the queue than it was before.

Shortly before, while showing the menu on buildings (**figure 19.7**) and discussing two commands from that menu, which allow to change the colors via samples, I mentioned the possibility of individual tuning of colors for any building. Let us look into this process. It will take us one level dipper inside the application, but there is at least one particular thing that is worthy mentioning.

When you click the *Change colors* command in the menu, which was called on any building, an additional small form is opened. This form contains an image of the building, which is similar to the one, on which the menu was called. The view can be slightly different, as it only represents the buildings of the same class; the number of storeys and windows can be different, but the colors will be definitely from the original. Each class of buildings from the **Form_Village.cs** has its own form to change the colors. Up till the last moment all these forms worked in exactly the same way; now there are two different ideas of their design.

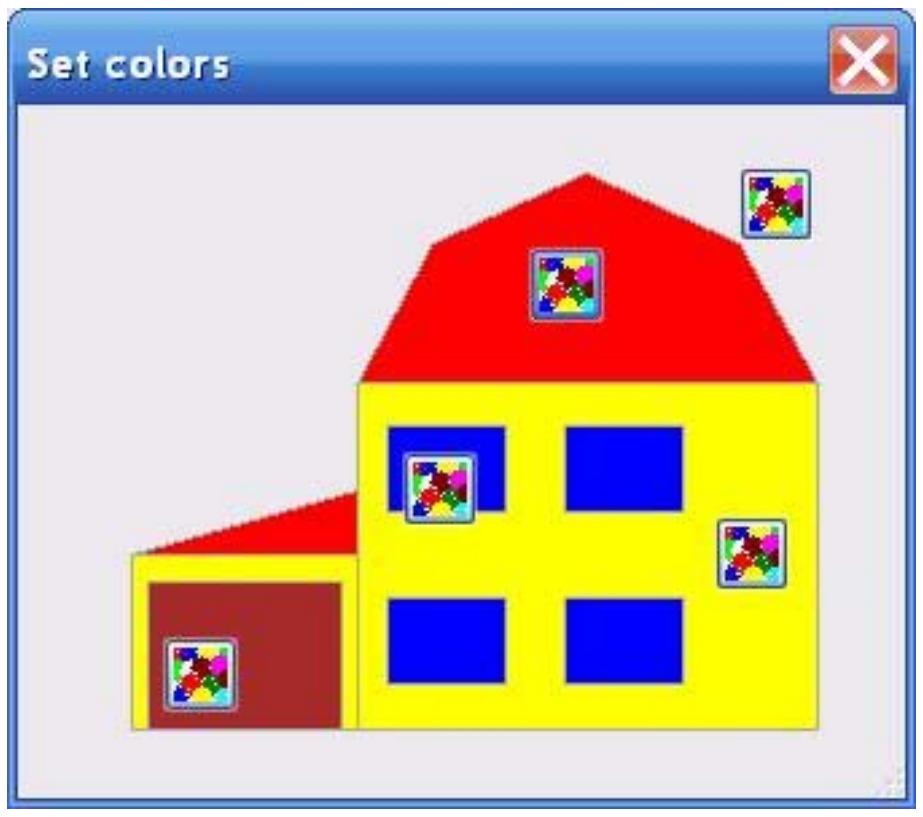

**Fig.19.10** The "old style" form to change the colors of a building

**Figure 19.10** shows the **Form_Colors_PlusLeft.cs** for changing the colors of the `RuralPlusLeft` objects. This is an "old style" form to fulfil such a task. The **Form_Colors_PlusRight.cs** for changing the colors of the `RuralPlusRight` objects and the **Form_Colors_Hangar.cs** for changing the colors of the `Hangar` objects are designed in the same way. Five colors can be changed for each building of the `RuralPlusLeft` class. You can see at the shown figure five buttons with the same image on them; throughout all my applications clicking of such buttons mean the opening of the standard `ColorDialog`. I hope that the positions of these five buttons make it obvious, which of them is responsible for changing the color of the roof, windows, door, and so on. The change of any color has the immediate effect not only on the sample, but also on the original building in the **Form_Village.cs**, for which this additional tuning form was called.

The house in this small **Form_Colors_PlusLeft.cs** is only movable, but not resizable, so the cover for such a `Sample_RuralPlusLeft` object is simple: the whole area must be covered by the nodes; two polygonal nodes would be enough. One node cannot solve the problem, as such a polygon would be non-convex, but two different polygonal nodes for the area of the main house and the garage would be the easiest solution.

```
public override void DefineCover ()
{
    CoverNode [] nodes = new CoverNode [2];
    nodes [0] = new CoverNode (0, new Point [] {rcHouse .Location,
        new Point (ptRidge.X – sizeSlopes .Width, ptRidge.Y + sizeSlopes .Height),
        ptRidge,
        new Point (ptRidge.X + sizeSlopes .Width, ptRidge.Y + sizeSlopes .Height),
        new Point (rcHouse .Right, rcHouse .Top),
        new Point (rcHouse .Right, rcHouse .Bottom),
        new Point (rcHouse .Left, rcHouse .Bottom) });
    nodes [1] = new CoverNode (1, new Point [] {
        new Point (rcGarage .Left, rcGarage .Top),
        ptGarageRoof,
        new Point (rcGarage .Right, rcHouse .Bottom),
        new Point (rcGarage .Left, rcHouse .Bottom)});
    cover = new Cover (nodes);
}
```

**Figure 19.11** shows the **Form_Colors_RuralHouse.cs** for changing the colors of the `RuralHouse` objects. This is a "new style" form for changing the colors; the **Form_Colors_PrimitiveHouse.cs** is designed in similar way for tuning the `PrimitiveHouse` objects. I especially picked for comparison the two forms with similar buildings, but the difference in the design of two forms is obvious immediately: in the second form there are no buttons to call the standard `ColorDialog`. But this dialog has to be called for change of colors. How is it done now? By the double click at any part of the sample, where you need to change the color. I think that such design is better for users. You may have different opinion, that is why I did not switch to the same style for all five tuning forms, but decided to demonstrate both styles and discuss them. The differences in view demanded the differences in covers and coding.

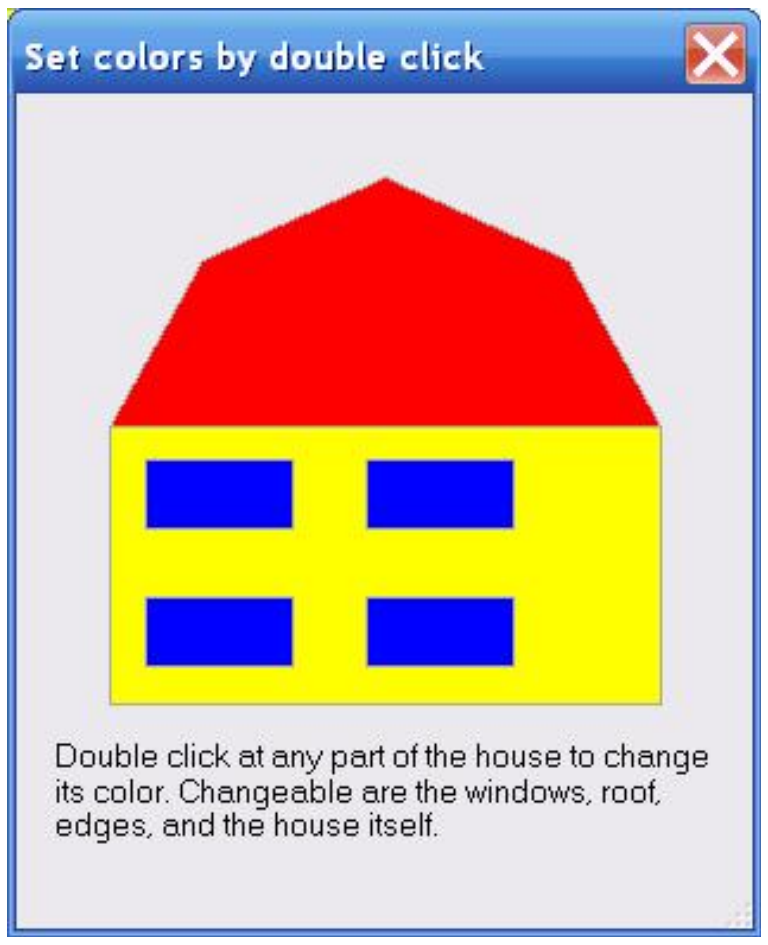

**Fig.19.11** The "new style" form to change the colors

First, this change required the change of the cover for such an object. In the "old style" form it would be enough to have one polygonal node covering the whole area of such a building. In the **Form_Colors_RuralHouse.cs** (**figure 19.11**) the double click on different parts must lead to change of different colors. Mover can identify different parts of an object only through the numbers of nodes, so the parts, associated with different colors, must be covered by different nodes. It does not matter that in a traditional way – for moving the sample - all those nodes would be used in an identical way (this sample is also only movable, but not resizable). The sample has to have nodes to cover the windows and has to have nodes on all the edges. Thus the number of nodes increases. The `Sample_RuralHouse.DefineCover()` method does not become more complicated; it is as simple, as it was under the "old style", but it is longer because of the number of nodes.

Usually the smaller nodes are placed ahead of any bigger nodes, but this is crucial in case of the overlapping nodes. When the nodes definitely stay away from each other, then other ideas can be taken into consideration. In the case of the `Sample_RuralHouse` class the nodes over the four windows are moved to the head; all other nodes are placed according to their increasing area. This is the full order of nodes in the cover of this class:

1. Four rectangular nodes over the windows.
2. Seven strip nodes over the segments of the perimeter of a building.
3. One more strip node over the ceiling, as this line is painted in the same way as other borders.
4. Polygonal node over the main rectangular part of the house.
5. Polygonal node to cover the roof.

The easiest way to write the `DefineCover()` method for this class is to have an array of all special points on perimeter and then use this array to construct the nodes.

```
public override void DefineCover ()
{
    Point [] pts = new Point [] {rcHouse .Location,        // 0
        new Point (ptRidge.X - sizeSlopes.Width, ptRidge.Y + sizeSlopes.Height),//1
        ptRidge,                                           // 2
        new Point (ptRidge.X + sizeSlopes.Width, ptRidge.Y + sizeSlopes.Height),//3
        new Point (rcHouse .Right, rcHouse .Top),          // 4
        new Point (rcHouse .Right, rcHouse .Bottom),       // 5
        new Point (rcHouse .Left, rcHouse .Bottom) };      // 6
    CoverNode [] nodes = new CoverNode [] {
        new CoverNode (0, Auxi_Geometry .CornersOfRectangle (rcWindows [0])),
        new CoverNode (1, Auxi_Geometry .CornersOfRectangle (rcWindows [1])),
        new CoverNode (2, Auxi_Geometry .CornersOfRectangle (rcWindows [2])),
        new CoverNode (3, Auxi_Geometry .CornersOfRectangle (rcWindows [3])),
        new CoverNode (4, pts [0], pts [1]),
        new CoverNode (5, pts [1], pts [2]),
        new CoverNode (6, pts [2], pts [3]),
        new CoverNode (7, pts [3], pts [4]),
        new CoverNode (8, pts [4], pts [5]),
        new CoverNode (9, pts [5], pts [0]),
        new CoverNode (10, pts [6], pts [1]),
        new CoverNode (11, pts [0], pts [4]),
        new CoverNode (12, Auxi_Geometry .CornersOfRectangle (rcHouse)),
        new CoverNode (13, new Point [] {pts[0], pts[1], pts[2], pts[3], pts[4]})
                                             };
    cover = new Cover (nodes);
}
```

The interesting and important thing that must be mentioned is the use of the `OnMouseDoubleClick()` method in the **Form_Colors_RuralHouse.cs**. The method is not complicated, but pay attention to the setting of the clipping area for cursor.

```
    private void OnMouseDoubleClick (object sender, MouseEventArgs e)
    {
        if (mover .Catch (e .Location) && e .Button == MouseButtons .Left)
        {
            if (mover .CaughtSource is Sample_RuralHouse)
```

```
            {
                Cursor .Clip = Rectangle .Empty;
                int iNode = mover .CaughtNode;
                ColorDialogWithTitle dlg = new ColorDialogWithTitle ();
                switch (iNode)
                {
                    case 12:
                        dlg .Color = house .HouseColor;
                        dlg .Title = "House";
                        break;
                    … …
```

I mentioned several times that the `Mover` class uses three different levels of clipping, depending on what you want to provide. In the chapter *Movements restrictions* I have already explained that the restrictions of the mouse movement are imposed at the moment when any object is caught by the mover and eliminated when an object is released. In the standard case of grabbing an object inside the `OnMouseDown()` method by using the `Mover.Catch()` and releasing it inside the `OnMouseUp()` method with the `Mover.Release()` you do not need to think about clipping at all as everything is automated inside those two mover's methods.

However, inside that `OnMouseDoubleClick()` method, the code of which is shown above, this standard procedure is broken. An object (the sample of a house) is caught by the `mover.Catch()` call; automatically the clipping is imposed up to the sides of the small enough **Form_Colors_RuralHouse.cs,** because the mover in this form was initialized with the `Clipping.Visual`. But as the reaction on double click the color dialog is opened.

```
                ColorDialogWithTitle dlg = new ColorDialogWithTitle ();
```

This is a dialog, derived from the standard `ColorDialog` and allowing to set the text of its title. This dialog appears somewhere on the screen; the exact location is decided by the system. Chances are high that without special changing of the clipping area you will have no chances even to move the cursor to this dialog, because the clipping was imposed by the mover to the boundaries of the **Form_Colors_RuralHouse.cs**. To solve this strange and arriving from nowhere problem, the clipping area must be changed before opening the dialog. The next statement makes it possible to use the opened `ColorDialogWithTitle` dialog regardless of its location.

```
                Cursor .Clip = Rectangle .Empty;
```

The last remark on the **Form_Village.cs**. This form demonstrates one more case of saving and restoring the needed data not only through the `Registry`, but also through the binary files. Menu, which can be called at any empty spot (**figure 19.12**), includes three commands, related to this process. Usually it is enough to have two commands – *Save* and *Read* - for the whole process; why are there three in this case?

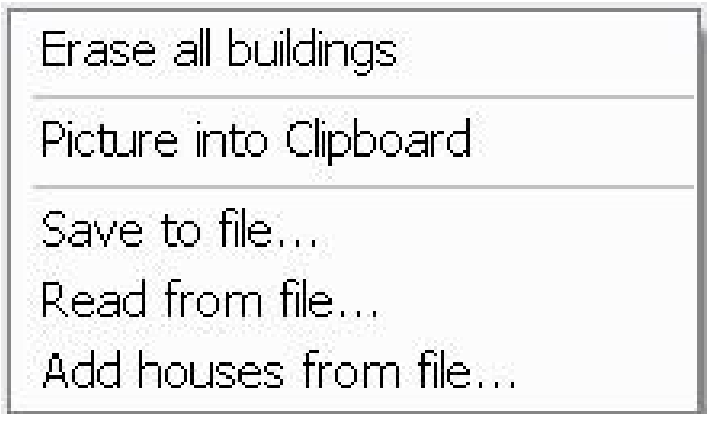

**Fig.19.12** Menu, which can be called at any empty place inside the **Form_Village.cs**

On saving, two types of information are written into a file.

- General information, which includes the size of the form, the position of the small button, the *Buildings* group, and the samples of colors for all five classes of buildings.
- Information about all the buildings from the form.

Two commands provide two different types of possible reading. The first one – *Read from file* – reads everything, so the form gets exactly the same view as it had at the moment of saving. The second one – *Add houses from the file* – gets only the information about the houses. In this case the general information from the file is skipped and does not change anything in the form; only the data about the buildings is read and the new buildings are added to already existing in the form.

# Summary

This summary consists of three parts: rules of cover design, rules for organizing moving / resizing of objects, and the rules of user driven applications.

## Covers

- A cover consists of an arbitrary number of nodes. The minimum number of nodes is 1, so a cover may consist of a single node. The number of nodes is unlimited. Covers of a special type may include hundreds of identical small nodes; such covers are called the *N-node* covers.
- Usually each node is used either for moving or resizing of an object; it is possible to use the same node for both operations, but then they are started by different buttons. If an object must be involved both in moving and resizing and both of them are started by the same button, then the minimum number of nodes is two.
- Nodes can be of different shapes: circular, a strip with semicircles at the ends, or a convex polygon. Circular nodes are defined by the central point and radius. Strip nodes are defined by two middle points at the ends of a strip and radius of semicircles; the width of a strip is equal to the diameter of semicircles. Convex polygon is defined by the vertices.
- Nodes of a cover may overlap, be placed side by side or stay apart. In some cases the order of nodes in the cover is absolutely unimportant; in other cases it can be very important as nodes are checked for moving according to this order. In the areas of overlapping, moving / resizing of an object is determined by the first selected node.
- Nodes do not duplicate the shape of an object and they are not required to be only inside the area of an object.
- Use of the transparent nodes can make the design of covers for the nontrivial areas much simpler. With the use of transparent nodes, the area of cover may significantly differ from the area of object.
- Nodes can be moved individually thus allowing to reconfigure an object.
- Nodes can be enlarged to cover as much of the area of object as possible. Such enlarged nodes are often used for moving the whole object.
- A cover may consist of the nodes that are not moved individually, but an attempt to move such a node results in the moving of the whole object (the `MoveNode`() method only calls the `Move`() method). This makes an object movable, but not resizable.
- Each of the nodes has its own parameters. It is easy to allow resizing along one direction but prohibit it along another; this means organizing a limited reconfiguration.
- One of the parameters of node is the shape of a cursor above it; the mouse cursor gives the information about the possible actions with an object underneath.
- It does not matter that some covers are associated with the graphical objects and others with controls or groups of controls. All covers are treated in the same way thus allowing the user to change easily the inner view of any application.
- Covers can be visualized, though the best design makes moving / resizing of objects obvious without such visualization. Visualization of cover includes possible filling of the inner area of nodes and drawing of the perimeter of nodes.

## Moving and resizing

- To organize the moving / resizing process, there must be an object of the `Mover` class.

```
Mover mover;
```

- To prevent the accidental moving of elements out of view, the mover must be initialized with an additional parameter. This parameter describes the bigger object (form, panel, etc.), on which the movable elements must be located.

```
mover = new Mover (this);
```

- Three levels of moving the objects across the borders of the form (panel, etc.) can be organized: moving outside is not allowed, moving allowed only across the right and lower borders, or moving allowed across all four borders.
- Any graphical object and any control can be turned into movable / resizable, but different procedures are used for turning into movable the elements of these two types.

- To make any graphical object movable / resizable, its class must be derived from the `GraphicalObject` class and three methods must be written for the new class: `DefineCover`(), `Move`(), and `MoveNode`().
- `DefineCover`() method defines the cover of an object as a set of nodes. Each node has an individual number used for identification. Numbers start from zero and go up with an increment of one. In the cover, consisting of N nodes, their identification numbers use all the values from the [0, N-1] range.
- `Move`() method describes the forward movement of an object as a whole. In reality it means the simple change of one or several primitives (points, rectangles), on which the drawing of an object is based.
- `MoveNode`() method describes the individual movement of the nodes. If the moving of a node results in the moving of a whole object, then the `Move()` method is called from inside the `MoveNode`() method. An individual movement of a node may cause the relocation of some or even all other nodes; in such cases the `DefineCover`() method is often called from inside the `MoveNode`() method. Some restrictions on calling the `DefineCover`() method from inside the `MoveNode`() method may exist for N-node covers.
- To make any control movable / resizable, it is enough to include it into the mover's queue. Those three methods are not needed for controls.
- To make any control resizable, the appropriate values must be determined in its `MinimumSize` and `MaximumSize` properties.
- Mover has a queue of movable objects and supervises the whole moving / resizing process only for the objects that are included into this queue. For registering the simple graphical objects and the controls, the standard methods are used.

```
mover .Add (…);
mover .Insert (…);
```

- Any combination of elements can organize a set of synchronously moving objects.
- For the complicated objects, consisting of the parts, which can be moved both synchronously and independently, and for the objects, for which the set of such parts can be changed, it is much better to develop an `IntoMover`() method, which is used instead of manual registering of all the movable parts and which guarantees the correct registering of an object and all its parts regardless of a set of constituents.
- Moving and resizing are done with a mouse; the whole process is organized via three standard mouse events: `MouseDown`, `MouseUp` and `MouseMove`.
- `MouseDown` starts moving / resizing of an object by grabbing one of the nodes in its cover. The only mandatory line of code in this method is

```
mover .Catch (…);
```

- `MouseMove` moves the whole object or a part of it. There is one mandatory line of code in this method, but in order to see the movement, the `Paint` method must be called

```
if (mover .Move (e .Location))
{
    Invalidate ();
}
```

- `MouseUp` ends moving / resizing by releasing any object that could be involved in the process. The only mandatory line of code in this method is

```
mover .Release ();
```

- The moving / resizing of an object starts, when the mouse is pressed and the mover catches an object. There can be several objects at the point of the mouse press; the mover decides about an object to catch by checking the objects in its queue, so the order of objects (their covers) in the mover's queue is very important.
- If several objects overlap at the place, where the mouse is pressed, any user would expect the upper one to be caught by mover, so the upper object on a screen must be the first in the mover's queue. Controls appear on the screen atop all the graphical objects, so all the controls must precede all the graphical objects in the mover's queue. Controls have to be registered in the mover's queue according to their Z-order; this will guarantee the catching of the upper control in case of their overlapping. Drawing of the graphical objects must be organized in the opposite order to their positions in the mover's queue; this will guarantee that the upper graphical object will be caught in case of their overlapping.

- Mover can provide a lot of information about the object, which is currently moved or just released, and even about an object that is simply underneath the mouse cursor. This data can be used to change the order of objects on the screen, to call the context menus, and so on. Mover can provide the information about the objects at any point of the form, in which it works.
- If covers must be shown, then one of the available drawing methods can be used, for example,

```
mover .DrawCovers (grfx);
```

- If needed, several `Mover` objects can be used to organize the whole moving / resizing process. Each mover deals only with the objects from its own queue.

## *User-driven applications*

- All the elements are movable.
- All the parameters of visibility must be easily controlled by the users.
- The users' commands on moving / resizing of objects or on changing the visibility parameters must be implemented exactly as they are; no additions or expanded interpretation by developer are allowed.
- All the parameters must be saved and restored.
- The above mentioned rules must be implemented at all the levels beginning from the main form and up to the farthest corners.

# Conclusion

This book is about the programs of the new type; about the applications that change the whole set of relations along the chain “developer – application – users”. *User-driven applications* are not some improvement of the currently used programs, but are their alternative from the point of users’ control over the applications. Visually the programs are the same and they continue to fulfil their main purposes. But they are designed according to another philosophy; the step from the currently used applications to the user-driven applications is bigger (for USERS!) than from DOS to Windows (or similar systems). The switch to multi-window operating systems, which happened 25 years ago, significantly increased the flexibility of our work with computers, but did not change the programmer – user relations: users continue to do only whatever was written for them in the designer’s scenario. The new programs give the main role to the users. The idea of **user-driven applications** is a new programming paradigm.

It is the standard practice of an explanation to describe the proposed changes with the examples from different areas. A comparison from another area is thought out by an author as a good example of similarity, though not all readers may agree with it. In any way, a comparison is an artificial model of the discussed process, but simplified to highlight only the most important features. Here are three comparisons of the transformation from currently used programs to the user-driven applications.

**Comparison 1.**

There are autocratic regimes and democracy. The level of freedom in the first case greatly depends on the personality of the ruler. It can be extremely bad, or it can be good enough, so that people would not complaint about it. But in nay case, if the people of such country want something new, they have to write the petition and ask for highest authorization of their request. If the ruler agrees with the proposal, he makes the needed adjustments. Citizens have to accept whatever is imposed.

> The currently used programs are designed by somebody and are ruled by the developers. If the users want some changes, they can send the proposals to the authors or speak at the users’ conference. The author of the program can agree to make some changes, but these changes may differ from what was asked. In any way, the users get the new version with the changes made by the author and have to accept it.

*Under the democracy the changes come via the voting process and election of those, who put into laws the expectations of the majority. The process can be not straightforward, but the general movement is in the direction of the citizens’ requirements. Citizens may not be the best specialists on one or another process, but usually they get whatever they want at the moment. If later they change their mind, there is another election to occur earlier than the next millennium will come.*

> *User-driven applications are ruled by users. Changes are made according to the user’s wish. It is possible that some changes are not the best, but then another change can be made at any moment. At different moments user may wish different things; he has a chance to make different changes in the program.*

**Comparison 2.**

The bus has a predetermined route and the fixed number of stops along its route. You can widen the picture to the whole public transportation system, which is a collection of those predetermined routes. With a good public transportation system you have a variety of choices, but only from the list that was thought out and organized by some department.

> *You can drive your car to any place; the route and the places to stop are decided by you. You are the driver, all the decisions are yours.*

**Comparison 3.**

For many years (decades, centuries...) and in many countries around the world the stores worked only during the fixed hours and not all days a week. There were variations from place to place and throughout the duration of time, but the schedule was fixed. Often the schedule was fixed not for a short period of time, but for decades along the time scale in both directions. There were some inconveniences, but they never raised a question of the fairness of the whole process. It was organized in the same way during the parent’s life and the grandparent’s life; there were no different examples, so there was nothing to compare with to raise any questions.

> *Now there are stores, which work 24 hours a day and 7 days a week; people do not complaint about such schedule. It is the best for everyone, because each individual can decide about the best procedure personally for him. Personal circumstances can change; the individual will adjust his schedule according to them.*

People got so used to this freedom of personal decisions that even a short return to the old style causes the huge inconvenience and immediately raises the questions of its foolishness.

You may agree or disagree with these examples from different areas. You can continue to insist that there is no sense in switching from the applications ruled by developers to the user-driven applications. From my point of view, it would be difficult to insist that there is no sense in user-driven applications if you have tried any of them. You have the right to have your own opinion and I would like people to form there own opinion by trying the user-driven applications.

Programming is one of many forms of peoples' activity. As in any other area, there are those, who produce something (programmers), and there are those, who use the results (users). Programming is one of those areas, in which the same person plays both roles. Programmer is always a user! I wrote this book for programmers to show the way of developing the new products – user-driven applications. While reading this book, you were estimating the usefulness of those ideas as a PROGRAMMER. But you have also to look at the results of implementation of those ideas as a USER. Would you like to see those rules implemented in the programs that you use every day from morning till night? I want to remind those basic rules.

**Rule 1**. All the elements are movable.

**Rule 2**. All the viewing parameters must be easily controlled by the users.

**Rule 3**. The users' commands on moving / resizing of objects or on changing the parameters of visibility must be implemented exactly as they are; no additions or expanded interpretation by developer are allowed.

**Rule 4**. All the parameters must be saved and restored.

These rules have to be used at all levels of an application from the main form to the smallest element in the rarely used part. The user-driven applications differ from the currently used programs not by some tiny or unimportant details, but by the main ideas:

- Designer has to provide an instrument to solve users' problems. This instrument consists of a correctly working engine (calculation part) and a set of tools to put data inside, to get the results, and to visualize them.
- Designer has to provide a very flexible system of visualization.
- Designer has no right to decide instead of a user, how this instrument has to be used.
- Users get an instrument for solving their tasks and all the possibilities of rearranging the view of this instrument in any way they want.

Both sides – designers and users – can be involved in transformation from the designer-driven to user-driven applications and both sides can (and will) react to such transformation.

I think that **rule 3** would be the most controversial one and would be the most criticized by programmers. This rule may cause an outcry because the opposite thing was hammered into the programmers' heads for years by the use of the dynamic layout, which makes a mess of the fool-proof programming and the design of interfaces. The fool-proof programming is one of the axioms of our work; it is really hard to doubt one of the axioms. In reality, I do not argue with this axiom, I only want it to be used exactly in the range, for which it must be an axiom. Originally it was an axiom for behaviour of applications, which meant only <u>calculations;</u> for this area it is absolutely correct. The spreading of the same axiom on the interface (that is one of the origins of dynamic layout) was definitely a mistake. So, continue to develop the fool-proof programs from the purpose of any of them, but do not implement fool-proof ideas, as YOU understand them, into the interface design. Let the programs work absolutely correctly, but be user-driven. An attempt to design user-driven applications with some kind of developers' control of the users' changes in interface would be a mistake. (Some out-of-date doctors continue to insist that it is impossible to be pregnant partly.)

For users the transformation from the old style programs to the new goes easily and quickly. Users do not need the long instructions of how to deal with the user-driven applications. They have to know only one main rule: EVERYTHING is movable and tunable. From there they can work exactly as before, but with all the new possibilities. Users accept the new rules very quickly and none of them ever talk about going back or about an overwhelming burden of adjusting the applications according to their preferences. Each user selects the level, at which he uses the full control of an application; he also knows that he can change the level of this control at any moment according to his own wish.

I am convinced that after the introduction of the user-driven applications to the significant amount of users there would be no way back to the fixed applications. It may look a bit strange and unusual at the beginning, but it will become normal very quickly. Exactly like it happened with the transition from DOS to multi windows systems.

The decision about making a step from the designer-driven to the user-driven applications must be made personally by each developer. If the majority of programmers decide to develop the user-driven applications, we will find ourselves in another programming world, which gives us – USERS – another level of possibilities.

Dr. Sergey Andreyev ( andreyev_sergey@yahoo.com )

May - October 2010

# Programs and documents

Several applications are developed for demonstration of design and use of movable / resizable objects. In most cases these applications were developed as accompanying programs for some publications. There are also some other documents with a helpful information. All files are available at www.sourceforge.net in the project **MoveableGraphics** (names of projects are case sensitive there!). I renew the files there from time to time, when the new (better) version of the library or a new demonstration program is ready. Usually a newer version of DLL may appear once a month; the preparation of the big book "***World of Movable Objects***" delayed the new version for six months. I try to organize these files in such a way that all the programs with all their source codes use the same latest version of DLL, which is currently on display. As there are people who strongly oppose using the DOC format, I try to include all the documents both in DOC and PDF formats. Some of the documents may appear on other sites, but www.sourceforge.net is the only place, where I place the DLL and all the demonstration projects with all the available source files.

| | |
|---|---|
| **WorldOfMoveableObjects.zip** | File contains the book "***World of Movable Objects***" in DOC format (this is the current document) and the program to accompany this book (the whole project with all the codes in C#). All the examples in the book are from this project. If you want only to run this application, then two files are needed: **WorldOfMoveableObjects.exe** and **MoveGraphLibrary.dll.** |
| **Book_WorldOfMoveableObjects.zip** | File contains the book both in DOC and PDF formats. 368 pages |
| **TheoryOfUserDrivenApplications.zip** | File contains an article "***User-Driven Applications***" (in DOC and PDF formats; 34 pages) and an accompanying application (the whole project with all the codes in C#). To run an application, only two files are needed: **TheoryOfUserDrivenApplications.exe** and **MoveGraphLibrary.dll.** |
| **OnTheTheoryOfMoveableObjects.zip** | File contains an article "***On the Theory of Moveable Objects***" (in DOC and PDF formats; 31 pages) and an accompanying application (the whole project with all the codes in C#). To run an application, only two files are needed: **TheoryOfMoveableObjects.exe** and **MoveGraphLibrary.dll.** |
| **MoveGraphLibrary_Classes.zip** | File contains the description of classes, included into the **MoveGraphLibrary.dll**. Two versions are available (DOC and PDF, 149 pages). |
| **MoveGraphLibrary.dll** | The library. |
| **LiveCalculator.zip** | Calculator, which is included as one of the examples into the **WorldOfMoveableObjects.exe**, but presented here as a separate program. To use this application, it must be accompanied by the **MoveGraphLibrary.dll**, which is also inside the zip file. |
| **Test_MoveGraphLibrary.zip** | Before the book "***World of Movable Objects***" was written, it was the biggest collection of the examples with their explanations. Now it became redundant, but I will keep this file for some time more because of some of the included examples. File contains a big article "***Moveable and Resizable Objects***" (in DOC and PDF formats; 97 pages) and an accompanying application (the whole project with all the codes in C#). To run an application, only two files are needed: **Test_MoveGraphLibrary.exe** and **MoveGraphLibrary.dll.** |
| **TuneableGraphics.exe** | The older application, which demonstrates the movable / resizable objects from absolutely different areas. Some of the used examples were not included into later Demo applications. |
| **MoveGraphLibrary_OlderClasses.doc** | Description of the older classes, still included into **MoveGraphLibrary.dll**. |

# Appendix A. Visualization of covers

*Visualization of covers* is both an interesting and a strange title, because I think that covers must not be visualized at all and the good design must avoid any type of visualization of covers. Looking at the examples of the accompanying application, you can find out that the visualization of covers can be found only in those of them, which are used for explanation of the cover design. Examples from chapters 2 – 9 explain the covers for the graphical objects and chapters 12 – 14 are about controls; outside those chapters and examples you are not going to see the 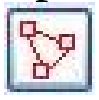 button, which switches the visualization of covers ON and OFF. After the design of covers is explained, you do not see the covers any more, because it is not needed any more. You do not see covers anywhere in the real applications from the second part of this book; you do not need to show covers in any real applications you would write yourself. However, if at one moment or another there would be a request for cover visualization, you have to know how to organize it. That is why I decided to explain it in this appendix, but while reading it do not forget that all the questions of visualization are important only when there is such a request. The normal situation is when the covers are not shown.

If you look throughout the codes of the examples, in which the covers can be shown, you will find that mostly it is done in two ways. The first one can be seen, for example, in the **Form_SolitaryControls.cs** (**figure 12.1**), where the covers for all the elements in the form are shown by a single mover's method `Mover.DrawCovers()`.

```
private void OnPaint (object sender, PaintEventArgs e)
{
    Graphics grfx = e .Graphics;
    … …
    if (bShowCovers)
    {
        mover .DrawCovers (grfx);
    }
    … …
}
```

Another solution can be even more popular. In such case, like in the **Form_RegPoly_Variants.cs** (**figure 6.1**), each cover is painted by a separate call to a similar method of the `MovableObject` class.

```
private void OnPaint (object sender, PaintEventArgs e)
{
    Graphics grfx = e .Graphics;
    for (int i = mover .Count - 1; i >= 0; i--)
    {
        … …
        if (bShowCovers)
        {
            mover [i] .DrawCover (grfx);
        }
    }
}
```

The first case is used in the situations, when you do not care too much about the order of drawing the covers and draw them on top of all the objects. The second case is much better for situations, when you want to draw each object with its own cover, so that you will see exactly the picture, which mover has to analyse, while picking up the objects for moving and resizing. What is common for both cases, that not the movable / resizable objects are responsible for painting their own covers, but the mover. It is really a strange thing, because the good programming practice (at least, how it is understood today) requires the `Draw`() method to be included into the list of methods of any visualized class, so that any object of such class can draw itself. Cover belongs to an object and all the demonstrated classes have their own `Draw`() methods, but there is not a single `DrawCover`() method in any of these classes. If the whole process of cover visualization is organized in such an unusual way, then it was definitely done so on purpose.

Cover belongs to an object, but it is needed and used only for the purpose of moving / resizing an object. It is an additional layer of parameters for object and it never interfere with any other parameters; cover is defined by other parameters (for example, by sizes) and strongly depends on them, but it is a one way relation: cover never rules other parameters. The main thing was already mentioned: cover is an additional, auxiliary parameter. Though a cover is designed at the moment when the object is initialized, you may look at the situation with its cover in such a way: it is activated, when an object is registered with a mover. You can look through all the examples and you will not find a single object with any kind of indication of either it is registered with any mover or not. No object knows if it is movable / resizable at the moment or not!

Any object, derived from the `GraphicalObject` class, can be used as unmovable or movable; this status can be easily changed, while the application is running, for example, by including an object into the mover's list or excluding an object from this list. Mover is the only one, who knows the list of movable objects at any moment; because of this, mover is responsible for drawing of all the covers.

Since the invention of the whole algorithm only the mover was responsible for drawing the covers. As the real applications never use the drawing of covers, I had no problems at all with this situation. However, while preparing more and more complex examples for one of the articles, I ran again and again into situations, when I needed to draw the covers not for all objects from the mover's queue, but only for several. It looks like the use of the `MovableObject.DrawCover`() method can easily solve this problem, but in some situations it is not so. There are many cases when the number and the order of movable objects in the mover's queue can easily change; it is difficult to calculate the order of the needed object in the mover's queue and then it is not obvious, which object of the `MoveableObject` class must be asked to draw its cover. As a result, I added the `GraphicalObject.DrawCover()` method. Every movable / resizable graphical object is derived from the `GraphicalObject` class, so now any object can draw its own cover. There are even two different `GraphicalObject.DrawCover()` methods: one of them uses the appropriate color from mover, another receives the color for drawing as a parameter..

Several examples in the Demo application demonstrate the use of the `GraphicalObject.DrawCover()` method; one of them is the **Form_CircleSector_OneMoveableSide.cs** (**figure 8.8**).

```
private void OnPaint (object sender, PaintEventArgs e)
{
    Graphics grfx = e .Graphics;
    for (int i = sectors .Count - 1; i >= 0; i--)
    {
        sectors [i] .Draw (grfx);
        if (bShowCovers)
        {
            sectors [i] .DrawCover (grfx, mover);
        }
    }
}
```

The most interesting questions around the visualization of covers are:

1. How to change the parameters of visualization?
2. How to visualize only part of the cover?
3. How to select only some of the covers for viewing?

Four different classes can be involved in the visualization of covers: `Mover`, `MovableObject`, `Cover`, and `CoverNode`.

Covers consist of the `CoverNode` objects, so the visualization of cover is in reality the visualization of all its nodes or only part of them. The shape of any node can be a circle, a strip, or a polygon; any of them has inner area and a border, so the choices for drawing nodes are limited:

- To fill or not the inner area of the node with some color.
- To draw or not the border of the node.

The default colors for this drawing are:

- White – to fill the area of node.
- Red – to draw the border of node. A border is painted by a solid line of one pixel width.

The color to fill the area of node can be set for each node individually; there is a `CoverNode.Color` property, but I never found a single chance to use it and if I have to fill the area of node, then it is always white.

Much more important and widely used is the setting of the flag which commands to clear (to fill) the area of node or not. The selection of the flag is usually done inside the `DefineCover`() method; the setting of the flag can be done with a `CoverNode.Clearance` property, but in many cases the default value is used. The default value of the clearance flag depends on the shape of a node: for circles and strips it is true, for polygons – false. These default values are based on the predominant use of the nodes of each shape. The circular nodes are mostly used in small size at the points, which are used for resizing or reconfiguring; the strip nodes are mostly used as narrow areas on the borders of objects for their resizing. Usually the nodes of these two shapes cover a small part of the object and do not significantly destroy the objects view, even when visualized. Because of the small sizes, it is better to clear the area of node; in such way the nodes are much better seen on visualization. On the contrary, polygonal nodes are usually big and cover the significant part of object or even its whole area. If cleared, they will simply wipe out the whole object, so, by default, they are not cleared.

If the node of any shape has an unusual size, then the possibility is high that the change of its Clearance property is needed; usually it is done in the DefineCover() method. There are different ways to change the clearance.

- It can be changed individually for any node with the CoverNode.Clearance property.
- It can be changed for all the nodes of the cover with the Cover.Clearance() method.
- It can be changed for the nodes of the particular shape with another variant of the Cover.Clearance() method.

In all the demonstrated examples of the big Demo application, there are only four cases of using the CoverNode.Clearance property. One of them is in the **Form_RegPoly_CircleInside.cs** (**figure 8.2**), which deals with the regular polygons with a big circular node inside. The cover for the RegPoly_CircleInside class uses the nodes of all three possible shapes. Nearly all of the nodes are used in their typical sizes: small circles, thin strips, and a big polygon. But there is one more node – a big circle, covering the hole in the middle. Because of its unusual size, the clearance of this circular node must be changed; otherwise the whole picture of an object with its cover will be destroyed. This circular node is the second from the end in the cover.

```
public override void DefineCover ()
{
    PointF [] pts = Vertices;
    … …
    nodes [k] = new CoverNode (k, center, Convert .ToInt32 (radiusCircle),
                               Behaviour .Transparent);
    nodes [k + 1] = new CoverNode (k + 1, pts);
    nodes [k] .Clearance = false;       // nodes .Length - 2
    cover = new Cover (nodes);
}
```

While the color for filling the area of node can be set individually for each node, the situation with the color for borders of nodes is opposite – usually it is set once for all nodes in all the covers. It is a very rare situation that this color is declared at all; usually the default color is used and in nearly all the examples, where the cover is visualized, all the borders are shown as red lines. In reality, each MovableObject may have the individual color for its cover, which can be set by the MovableObject.Color property. Though it can be done, I cannot imagine the situation, when the visualization of covers would require personal color for each of them. When any object is registered with a mover, color for its cover is set to whatever is currently set in the mover. The color for all the covers can be set by the Mover.Color property. Throughout all the examples of this Demo application there only ONE, in which I decided to demonstrate the use of this property, and even this was done only for the purpose of better explanation. In the **Form_StandardPanel.cs** (**figure 14.1**) two different movers are used; one of them for moving the panels, another – for moving several objects on one of the panels. To make it obvious that the moving of different objects is supervised by two different movers, I changed the color for covers for the objects on the panel.

```
moverInner .Color = Color .Blue;
```

If you need to draw the cover for one object, then there is the MovableObject.DrawCover() method. The nodes are drawn in reverse order, so the first one will be on top of all others – this is exactly the situation, which mover looks at, while making a decision about the moving / resizing.

If you need to draw all the covers, there is the method Mover.DrawCover(). Nodes of each cover are drawn in the reverse order of nodes; the covers are also drawn in reverse order (from the end of mover's queue to the beginning), so this is the exact situation with the covers and nodes as mover sees it and analyses.

This chapter is about the visualization of covers, but there are situations, when you do not want the cover for one or another class to be visualized even on request. It is not about hiding your ideas from anyone else; it is only for convenience. For example, I have explained in details the design of cover for the Plot class. Those objects can be moved by any inner point and resized by any border point; these are the only things users need to know to work with them, so there is absolutely no sense in visualizing their covers. If for any reason you want to exclude the visualization of cover for any class, add a short ShowCover property to this class.

```
public override sealed bool ShowCover
{
    get { return (false); }
}
```

In such a way the ShowCover property is used in the majority of classes, included into the library. However, it is not used for the TextMR class, but that was done only for the purpose of explanation; all the comments in the **MoveGraphLibrary.dll**, derived from the TextMR class, have such property.

# Appendix B. How many movers are needed?

File: **Form_ManyMovers.cs**
Menu position: *Graphical objects – Many movers*

Numerous examples, discussed throughout the book, demonstrate a wide variety of graphical objects and controls, involved in moving and resizing. In some cases there are only few simple objects; others include many complex objects, consisting of many different parts. In all these situations a single mover is enough to organize the moving / resizing process regardless of the number of the participants. However, there is no such rule as "one mover per form", and if you think that it would be easier to have several movers to deal with different objects, it is possible to organize the whole process in such a way. But in order to use one or several movers in the form, you have to understand the advantages and disadvantages of both cases.

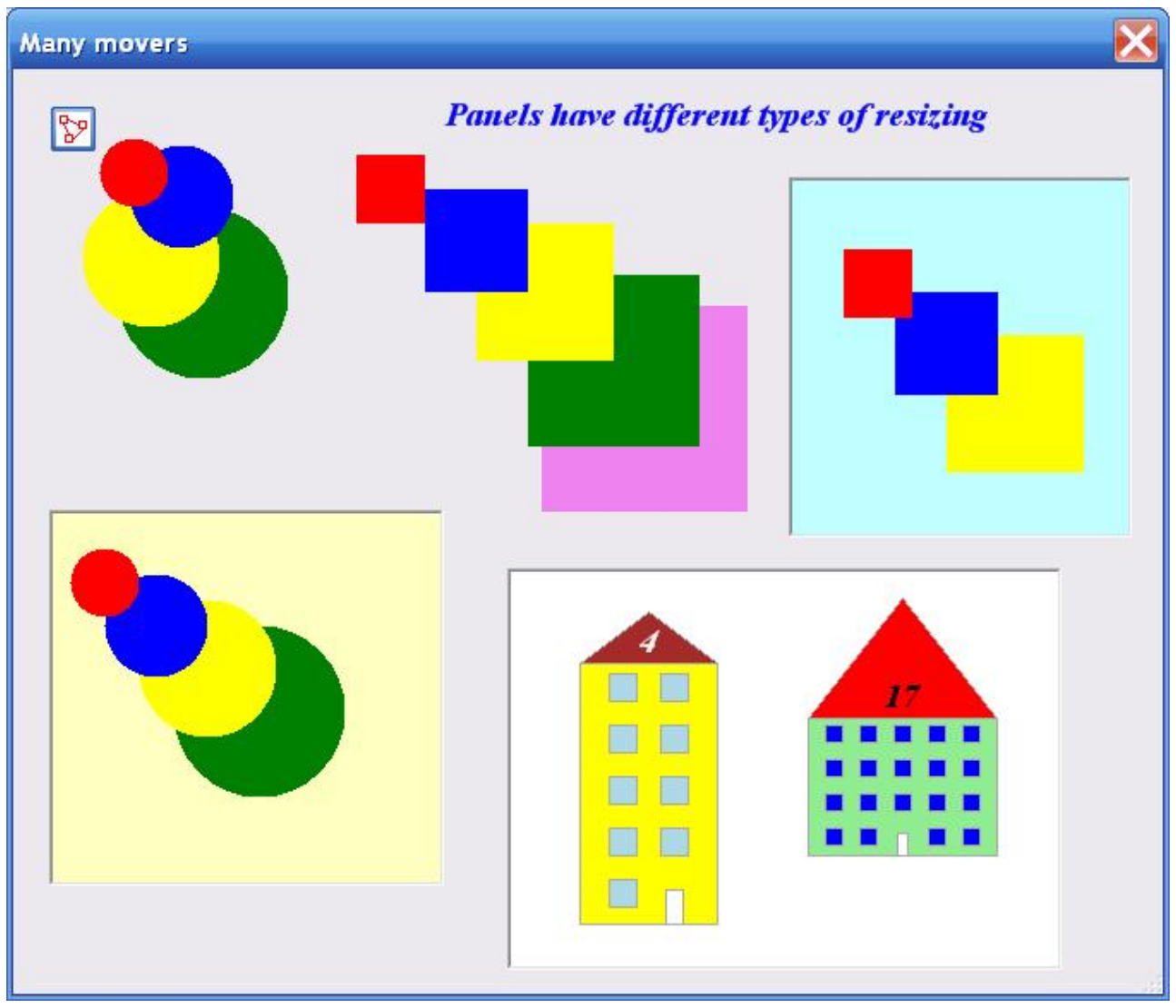

**Fig.B.1** Form with six movers

The **Form_ManyMovers.cs** contains three panels, a small button to switch the covers ON / OFF, a short information of the `TextMR` class, and graphical objects of three different classes (**figure B.1**). Each panel contains the graphical objects of only one type (circles, squares, or houses); panels are organized with different types of resizing. There are SIX movers in this form; each panel has its own mover; other three movers work in the client area of the form itself. The distribution of objects among the movers is organized in such a way.

- There is a mover for graphical objects on each panel; the names of the movers tell where they work.

```
Mover moverCirclesOnPanel, moverSquaresOnPanel, moverHouses;
```

- In the client area of the form there is one mover for each type of movable objects. Movers' name inform with which objects each of them work; the `TextMR` object is registered with the `moverControls`.

```
Mover moverControls, moverCircle, moverSquare;
```

Each mover is characterized by the level of clipping, which it provides for moving the objects, registered in its queue. The needed level can be changed at any moment by the `Mover.Clipping` property. Three different levels exist; two of these levels can be set initially, when one or another mover's constructor is used. In order to demonstrate the clipping variants, six movers in the **Form_ManyMovers.cs** are constructed in different ways.

```
moverControls = new Mover (this);
moverCircle = new Mover (this);
moverSquare = new Mover ();
moverCirclesOnPanel = new Mover ();
moverSquaresOnPanel = new Mover (panelSquares);
moverHouses = new Mover ();
```

For simplicity, all squares and circles are movable, but not resizable; houses are movable and resizable. Regardless of whether the objects exist on the panels or in the main client area, their moving and resizing are organized by using three standard events `MouseDown`, `MouseMove` and `MouseUp`. Objects on each panel are supervised by its own mover, so there is absolutely nothing new in organizing their moving and resizing.

Here are those three methods for the panel with the squares. The mover for this panel - `moverSquaresOnPanel` - is initialized in the standard way with the panel passed as a parameter. In such way the mover automatically gets the `Clipping` property set to `Visual`, so the squares on the panel can be moved only inside the visible area of the panel.

```
private void MouseDown_panelSquares (object sender, MouseEventArgs e)
{
    moverSquaresOnPanel .Catch (e .Location, e .Button);
}
private void MouseUp_panelSquares (object sender, MouseEventArgs e)
{
```

```
            moverSquaresOnPanel .Release ();
        }
        private void MouseMove_panelSquares (object sender, MouseEventArgs e)
        {
            if (moverSquaresOnPanel .Move (e .Location))
            {
                (sender as Panel) .Invalidate ();
            }
        }
```

When a square on the `panelSquares` is caught by mover, it cannot be entirely moved across any border, but such a square still can find itself out of view beyond the border. This may happen as a result of the squeezing of panel. The `MinimumSize` and `MaximumSize` properties of this panel get such values that the panel can change only its height. On squeezing the panel, you can easily leave the squares beyond the bottom of the panel; to return such squares back into view you would have to increase the height of the panel.

The mover for the panel with the circles - `moverCirclesOnPanel` – uses the constructor without any parameter. As a result, its `Clipping` property is set to `Clipping.Unsafe` and those circles can be easily moved across the borders of the panel. The panel with the circles is resizable in all directions, so if you move a circle across the right or lower border of the panel, you will have a chance to return it back into view by enlarging the panel; there is no way to see the circles again, if they are dropped anywhere beyond the left or upper border of the panel. This is correct only if the methods for mouse events on this panel would be written in the same way, as shown above for the panel with squares. If you turn into code several lines, which are now commented inside the `MouseDown_panelCircles()` method, then you can prevent the circles from being moved outside the panel. This shows that if you want, you can set your own clipping different from what is declared automatically by mover.

```
        private void MouseDown_panelCircles (object sender, MouseEventArgs e)
        {
            if (moverCirclesOnPanel .Catch (e .Location, e .Button))
            {
                Rectangle rcClip = (sender as Panel) .ClientRectangle;
                rcClip .Inflate (-4, -4);
                Cursor .Clip = (sender as Panel) .RectangleToScreen (rcClip);
            }
        }
```

For objects on each panel there is a single mover, so there is nothing special in organizing the movements on those panels. The situation in the client area of the form is much more interesting, as there are three different movers, operating in the same area, so different effects can be produced, depending on some code variations. Any mover does not know anything about the objects in the form; it is dealing only with the covers of the objects, registered in its queue. If you simply write the `OnMouseDown()` method in such a way, as shown below, then in those places, where the covers for objects of different types overlap, you have a chance to grab simultaneously two or three objects, because each mover catches an object, registered in its queue.

```
    private void OnMouseDown (object sender, MouseEventArgs e)
    {
        moverControls .Catch (e .Location, e .Button);
        moverSquare .Catch (e .Location, e .Button);
        moverCircle .Catch (e .Location, e .Button);
    }
```

As the idea of this form was not to organize such a strange movement of objects, but an ordinary one – one object at a time, then this code must be changed. Movers do not check each others status, but the `Mover.Catch`() method returns a value, indicating if anything was caught or not; this returned value can be very helpful. In the next version of the `OnMouseDown()` method, if any mover has caught an object, then other movers are not going even to try. One moving object at a time.

```
    private void OnMouseDown (object sender, MouseEventArgs e)
    {
        if (!moverControls .Catch (e .Location, e .Button)) {
            if (!moverSquare .Catch (e .Location, e .Button)) {
                moverCircle .Catch (e .Location, e .Button);
            }
        }
    }
```

While finishing any movement (**MouseUp** event), there is no need in checking, which mover was working and if any of them was working at all at the moment; simply release them all.

```
private void OnMouseUp (object sender, MouseEventArgs e)
{
    moverControls .Release ();
    moverSquare .Release ();
    moverCircle .Release ();
}
```

The same thing with the code of the **MouseMove** event; do not forget to mention all the movers, working in the client area of the form.

```
private void OnMouseMove (object sender, MouseEventArgs e)
{
    moverControls .Move (e .Location);
    moverSquare .Move (e .Location);
    moverCircle .Move (e .Location);
    if (moverControls .Caught || moverSquare .Caught || moverCircle .Caught)
    {
        Invalidate ();
    }
}
```

In such a way everything in the **Form_ManyMovers.cs** works fine, but in some situations you may notice something strange with the mouse cursor. The covers of the movable objects in this form use different cursor shapes to inform about their readiness for movement: circles and squares use the `Cursors.Hand`, the controls show different cursors, but for the ordinary forward movement they show `Cursors.SizeAll`. Anyone expects that if you press the mouse at such a place, where two objects overlap and each of them can be grabbed for moving, then the upper object would be caught and start moving. The shape of a cursor has to inform about the possible movement even before it really starts.

Move any panel so that its border is placed on any colored square. Panels are always atop all the graphical objects; thus positioned panel will partly close the square from view. Now move the cursor close to the border of this control, but over the square and press the mouse; the control will move and not the square underneath. The program works correctly, because the decision about the possible movement is made on the correct order of movers: first `moverControls`, then `moverSquare`, and then `moverCircle`. But look at the mouse cursor just before you pressed the button to start the movement; the cursor would have the `Cursors.Hand` shape. This is the wrong type of cursor, because it signals beforehand about the readiness of the square to be moved. The wrong information about the possible movement, though the movement itself would be correct. Why the information was wrong?

Look into the code of the `OnMouseMove`() method. The movers are checked for the possibility of movement in the correct order, but all of them are checked one after another, so the final shape of the cursor is decided by the last checked mover – `moverCircle`. As a result, the cursor in the form of a hand is shown. To avoid this discrepancy, the code of the `OnMouseMove`() method must be slightly changed in such a way.

```
private void OnMouseMove (object sender, MouseEventArgs e)
{
    if (!moverControls .Move (e .Location) && !moverControls .Sensed)
    {
        if (!moverSquare .Move (e .Location) && !moverSquare .Sensed)
        {
            moverCircle .Move (e .Location);
        }
    }
}
```

In this variant of the `OnMouseMove()` method the `Mover.Sensed` property is used. I have mentioned before, but demonstrated only in a single example of the track bars (chapter *Complex objects*) that mover can reveal the information about the movable objects under the cursor or even at any other point in the form. In our current situation with several movers I do not need any other information but the fact that mover has sensed something underneath. The `moverControls` is not involved in the moving of an object (panel) yet, but if it has sensed this panel, then all other movers would be not involved in the further checking. As a result, the shape of a cursor would be decided by the `moverControls.Move()` method and the cursor would inform that `moverControls` is ready to grab and move something.

The **Form_ManyMovers.cs** is a simple, but an artificial example of using several movers in one form. Occasionally I design forms with more than one mover. There were situations, when I designed the same form with one or two movers just to check, if one of the variants would be better for the programming or not. Users never saw any difference in one or another variant, because those versions with different numbers of movers worked absolutely identically. The difference was only in the code, so these situations can be decided by each programmer personally. There are two rules that help me to make the decision about the needed number of movers in the form.

- If there are movable / resizable objects only on "one level" in the form, then I try to use one mover.
- If there are movable / resizable objects on the "second level" (on a panel), then I use separate mover for these objects; one mover per objects on each panel.

These are not the strict rules, but only the suggestions. If I would develop the same **Form_ManyMovers.cs** not as an example, but as a part of the real application, I would use four different movers instead of six: one per panel and one for all graphical objects and controls in the main area; there are no advantages at all of having three movers there.

This short remark at the end of *Appendix B* is not about the preferable number of movers for the shown form, but about some strange feeling after using this form. Nearly all the elements in view can be moved around by any inner point: press and move. The only exceptions are those three panels; this highlights again the problem with the standard panels in applications. Three panels are painted in different colors and they look like other colored rectangles in view (**figure B.1**). Trying to move around all the existing objects, you do not know about difference in their origin and you do not expect such a difference in their behaviour. You try to press and move one after another; it is like a shock, when some of them do not want to react in a proper way to such a primitive command. This object has to move as any other; if it does not want to, then it is definitely a mistake in design. Who is the author? Hang him up! Or at least fire him.

I wrote about moving the standard panels in the chapter *Groups of elements*; I also mentioned that I stopped using them in my applications exactly because of this problem with their moving. All these panels look exactly like graphical rectangles, but rectangles have no problems with their moving by any inner point. If the task would be only to demonstrate the moving of circles, squares, and houses, then there would be no panels in this task. No chances! Everything would be graphical. That is one more example in my advocating for getting rid of the controls and using only graphical objects.